%% file: emdynar.tex
\documentclass[fleqn,usenatbib]{mnras}

\usepackage{newtxtext,newtxmath}

\usepackage[T1]{fontenc}
\usepackage{ae,aecompl}

\usepackage{graphicx}	
\usepackage{amsmath}	
\usepackage{amssymb}	
\usepackage[squaren,Gray]{SIunits}
\usepackage{numprint}
\usepackage{grffile}

\newcommand{\rLarmor}{r_{\rm B}}
\newcommand{\rlight}{r_{\rm L}}

\newcommand{\BQ}{B_{\rm qed}}

\newcommand{\me}{m_{\rm e}}

\title[Pulsar gamma-ray emission]{Pulsar gamma-ray emission in the radiation reaction regime}

\author[J. P\'etri]{
J. P\'etri,\thanks{E-mail: jerome.petri@astro.unistra.fr}
\\
Universit\'e de Strasbourg, CNRS, Observatoire astronomique de Strasbourg, UMR 7550, F-67000 Strasbourg, France.
}

\date{Accepted XXX. Received YYY; in original form ZZZ}

\pubyear{2019}

\begin{document}
\label{firstpage}
\pagerange{\pageref{firstpage}--\pageref{lastpage}}
\maketitle

\begin{abstract}
Since the era of the Fermi/LAT and atmospheric Cerenkov telescopes, pulsars are known to emit high and very high-energy photons, in the MeV-GeV range and sometimes up to TeV. To date, it is still unclear where and how these photons are produced. Nevertheless gamma-ray photons require particle acceleration to ultra-relativistic speeds. In this paper, we compute single particle trajectories for leptons in an arbitrary strong electromagnetic field in the so-called radiation reaction limit. In this picture, particle velocity only depends on the local electromagnetic field which we assume to follow the vacuum dipole rotator. From this velocity field, we compute the curvature radiation spectrum and light-curves. Sky maps and phase-resolved spectra are then deduced accounting for realistic pulsar periods and magnetic field strengths. Emission sites within the pulsar magnetosphere where most of radiation emanates are then localized. For standard parameters of millisecond and normal pulsars, we show that a break in the spectrum occurs at several GeV in agreement with the Fermi/LAT second pulsar catalogue. A sample of representative phase-resolved spectra and sky-maps are shown. A pair multiplicity of several tenths to several thousands is required to account for the total gamma-ray luminosity. Moreover depending on the geometry, single or double-peaked light-curves are found. Our model shows that minimalist assumptions are already able to reproduce salient features of pulsar emission. 
\end{abstract}




\section{Introduction}

Within the last decade, important progresses have been made towards a better and deeper understanding of pulsar magnetospheric physics, particle acceleration and radiation properties. We have witnessed breakthroughs in numerical simulations of the full non-linear problem of force-free \citep{contopoulos_axisymmetric_1999, spitkovsky_time-dependent_2006, parfrey_introducing_2012, petri_pulsar_2012} and magnetohydrodynamic (MHD) magnetospheres \citep{komissarov_simulations_2006, tchekhovskoy_time-dependent_2013}. Accurate and detailed multi-wavelength observations of pulsar light-curves especially in the gamma-ray band are now available \citep{abdo_second_2013}. Gamma-ray pulsars are believed to furnish a faithful indirect view of the pulsar engine because most of their spindown luminosity goes into pulsed gamma-ray emission seen by a distant observer.

On the theoretical side, numerical simulations are now able to include dissipative effects in an heuristic way \citep{li_resistive_2012, kalapotharakos_toward_2012} in order to incorporate self-consistently the acceleration zones and therefore localizing the emission sites within the magnetosphere. Nevertheless these simulations are inherently unable to follow single particle acceleration, preventing the building of power law distribution functions as required to fit observations. Recently, Particle In Cell (PIC) codes emerged to fully account for this single particle acceleration and its feedback onto the electromagnetic topology \citep{cerutti_particle_2015}. Unfortunately, in these simulations, the neutron star size is unrealistically large with $\rlight=3\,R$, with $\rlight$ the light-cylinder radius and $R$ the neutron star radius. These parameters thus correspond to a sub-millisecond pulsar rotation period. However, the plasma magnetization is more realistic with $\sigma \approx \numprint{e3}$. Such simulations showed that particles reach Lorentz factors up to \numprint{e3}-\numprint{e4}. However PIC simulations are still unable to catch neutron star electrodynamics for magnetic fields as high as those present in normal pulsars, that is about $10^8$~T. Even millisecond pulsars, believed to harbour fields of only $10^5$~T, are difficult to follow faithfully with current simulation techniques because of the large gap between the cyclotron frequency and the pulsar rotation frequency among others. Indeed, at the surface of the star, the ratio between gyro frequency~$\omega_B$ and stellar rotation~$\Omega$ is about
\begin{equation}
\frac{\omega_B}{\Omega} = \frac{e\,B}{\me\,\Omega} = \numprint{2.8e18} \, \left( \frac{P}{1~\SIunits{\second}} \right) \, \left( \frac{B}{\numprint{e8}~\SIunits{\tesla}} \right) .
\end{equation}
$e$ is the electron charge, $\me$ its mass, $P=2\,\upi/\Omega$ the pulsar period and $B$ its magnetic field strength. Moreover, the Larmor radius associated to these fields for a particle with Lorentz factor~$\gamma$ is
\begin{equation}
\rLarmor = \frac{\gamma\,m\,c}{e\,B} = \numprint{1.7e-5}~\SIunits{\meter} \, \left( \frac{\gamma}{\numprint{e6}} \right)  \, \left( \frac{B}{\numprint{e8}~\SIunits{\tesla}} \right)^{-1}
\end{equation}
where $c$ is the speed of light. This lengthscale remains much smaller than the typical size of a neutron star estimated to be about $R = 12$~\SIunits{\kilo\meter}. Thus the ratio between Larmor radius and neutron star radius is about $\epsilon = \rLarmor / R = \numprint{e-10}$, allowing to separate both scale. This clearly show that PIC codes will have tough time to solve the full span of dynamical ranges. Farther away from the star, the situation could get better. Actually, the magnetic field strength decreases quickly with radius like $B\propto r^{-3}$ because of its dipolar nature but for a millisecond pulsar, the ratio at the light cylinder is still
\begin{equation}
\frac{\omega_B}{\Omega} = \numprint{2.8e9} .
\end{equation}
The same ratio applies for a normal pulsar. Thus even at the light-cylinder, the microscopic and macroscopic timescales are too disparate to be caught by standard numerical techniques.

In order to circumvent these severe limitations, the magnetic field intensity is usually artificially decreased by several orders of magnitude to alleviate the stringent requirement about the time step for integrating the equation of motion. Unfortunately, such artefacts drastically minder the electric field strength too, electric field induced by the rotating dipole, thus disabling particle acceleration to ultra-relativistic speeds with $\gamma \gg 10^4$. The highest Lorentz factors obtained so far are usually around $\numprint{e3}-\numprint{e4}$. Consequently, photons are produced with artificially low energies. \cite{kalapotharakos_three-dimensional_2018} were able to approach realistic values, using a pulsar period $P=0.1$~s but a still too low magnetic field of $B=\numprint{e2}$~T. In order to approach realistic $B$~fields of $B=\numprint{e8}$~T, they followed the high-energy tail of their particle distribution functions, assuming that their orbits are geometrically correct independently of~$B$, integrating the energy conservation equation where electric acceleration is counterbalanced by radiation reaction. This should help to track the Lorentz factor evolution in time for real particles with $\gamma \approx \numprint{e8}$. Nevertheless some correcting factors are introduced to renormalize all energy scales including photon energies, a technique that is at least questionable. This represents the major flaw of direct numerical simulations of neutron star magnetospheres intended to compute realistic spectra from first principle particle in cell simulations.

From a geometrical point of view, several popular emission sites like the outer gaps, the slot gaps and the polar caps were often hypothesized to be privileged regions for producing photons. It is then possible to compare the merit of each zone and test their ability to reproduce the observed light-curves \citep{dyks_two-pole_2003, dyks_relativistic_2004}. The presence of a plasma partially or completely screening the electric field shows up in distortions of the light-curves from a vacuum rotator (see for instance sky maps in \cite{bai_uncertainties_2010} but who surprisingly also assumed some force-free prescription for vacuum fields!) compared to a force-free model \citep{bai_modeling_2010}. Often in the vacuum field investigations, widely used in the literature, the associated accelerating electric fields are not taken into account self-consistently. Let us however mention the work of \cite{kalapotharakos_gamma-ray_2012} and \cite{kalapotharakos_gamma-ray_2014}, who indeed used resistive plasma models with low conductivity to mimic almost vacuum electromagnetic fields. They also produced sky-maps and light-curves taking into account the accelerating electric field.

\cite{watters_atlas_2009} compiled an atlas of geometric light curves for young pulsars showing the essential characteristics of gamma-ray profiles depending on viewing angle and obliquity. \cite{romani_constraining_2010} then designed a tool to constrain the magnetospheric structure from these gamma-ray light curves. \cite{venter_probing_2009} investigated the special population of millisecond gamma-ray pulsars showing that two-pole caustics and outer gap models are favored. See also 
\citet{pierbattista_light-curve_2015, pierbattista_young_2016} for a large sample of pulsars fitted with several emission models and \citet{johnson_constraints_2014} for a similar study about millisecond pulsars. Obviously, more constraints can be obtained from simultaneous radio and gamma-ray fitting. \citet{harding_gamma-ray_2011}  produced atlases of two-pole caustics and outer gap emission models in force-free and vacuum retarded dipole field geometry to compare light curve features in symmetric and asymmetric slot gap cavities.

PIC codes are now able to follow particle trajectories including radiation reaction correction and therefore producing sky maps and light-curves assuming synchrotron emission. Unfortunately, the magnetic field strength as already pointed out is artificially decreased to too low values.  \cite{cerutti_modelling_2016} got acceleration only up to $\gamma \approx \numprint{e2}$. This is clearly not enough to explain MeV or GeV photons produced by synchrotron radiation. The current sheet and the Y-point become the preferred site to produce high-energy gamma-ray photons \citep{philippov_ab-initio_2018}, showing light-curve features in agreement with Fermi/LAT observations \citep{abdo_second_2013}. However the maximum polar cap potential drop they used was set to get Lorentz factor at most of $\gamma=500$. This threshold is many orders of magnitude below any realistic pulsar acceleration efficiency. The neutron star period is also slighty to high with $R=4\,\rlight$. PIC simulations in their current development stage are unable to deal with real pulsar parameters. The derived spectra and light-curves are therefore also unrealistic as long as the down scaling operates to extrapole dangerously to $\numprint{e5}-\numprint{e9}$~\SIunits{\tesla}. In all PIC simulations, the hierarchy of time scales is obviously respected but unfortunately not their ratio. We believe that such strong extrapolations must at least be verified on simple problems before dealing with the full complexity of a pulsar magnetosphere.

The second Fermi gamma-ray pulsar catalogue \citep{abdo_second_2013} contains plenty of information about gamma-ray pulsars spectra and light-curves. The gamma-ray peak separation clusters around $\Delta \approx 0.5$ and the radio peak usually leads the first gamma-ray peak but with some outliers. Force-free or ideal MHD computations are unable to self-consistently accelerate particles and localize the emission site. Some kind of dissipation of the electromagnetic field is required in order to produce a signal detectable on earth. So dissipation within the magnetosphere and/or wind must occur, but the precise mechanism and its efficiency are difficult to predict from first principles. Nevertheless, some dissipative magnetospheres, called FIDO and introduced by \cite{kalapotharakos_toward_2012}, were used by \citet{brambilla_testing_2015} for computing the phase-averaged and phase-resolved $\gamma$-ray spectra of eight of the brightest Fermi pulsars. They used billions of test particles trajectories to compute curvature radiation spectra in realistic fields of $\numprint{e7}-\numprint{e9}$~T. Based on this work \cite{kalapotharakos_fermi_2017} constrained the dissipation mechanism by looking at curvature radiation in the equatorial current sheet outside the light-cylinder, using Fermi/LAT spectral data. This could put limits on the strength of the accelerating electric field. They used test particle integration in the radiation reaction limit regime in the global force-free dissipative magnetosphere which is basically a fluid description avoiding the stringent strong field constrain faced by PIC codes. In such a way, they were able to deduce realistic spectra for realistic pulsar field strength and period. Nevertheless, starting from PIC simulations, \cite{kalapotharakos_three-dimensional_2018} recently found a relation between the particle injection rate and the spindown luminosity. This work shows the fruitful feedback between simulations and observations to extract useful information about the nature of particle acceleration and dissipation of the relativistic magnetized flow. \cite{harding_gamma-ray_2016} and \cite{venter_high-energy_2018} gave recent reviews of the successful interplay between magnetospheric modeling and gamma-ray observations.

Other attempts to fit particular pulsars were carried out by other groups. For instance \cite{takata_polarization_2007} and \cite{hirotani_outer-gap_2008} used the vacuum retarded dipole to model the outer gap of the Crab pulsar. \citet{du_gamma-ray_2011} performed computation in the annular gap context for the Vela pulsar whereas \citet{du_radio--tev_2012} did it for the Crab pulsar. Several millisecond pulsars were also fitted by \citet{du_radio_2013} using a static dipole.

Global magnetospheric simulations converge to a stationary picture of a corotating electromagnetic field and particle distribution function. Nevertheless the paradigm of pulsars being stable and constant broadband emitter in time has been invalidated in radio since their discovery fifty years ago. However, gamma-ray pulsars were though to still remain steady emitter. But this picture has recently also been challenged by some gamma-ray variability reported for instance in PSR~J2021+4026 by \cite{allafort_psr_2013}.

All the above investigations started from more or less sophisticated numerical simulations of neutron star magnetospheres according to force-free, MHD, resistive/dissipative or PIC approximations. Observational signatures are then post-processed or self-consistently included for comparison with existing data in radio and gamma-rays. This is always the starting point to support any model of pulsar magnetosphere. Force-free magnetosphere simulations give quick and accurate answers to the global electromagnetic field produced by ideal presssureless and massless plasmas. It corresponds to the ultra-strong field limit where particles move at the speed of light. Unfortunately, these simulations cannot resolve for individual particle acceleration. PIC codes are therefore intended to catch all the physics, from macroscopic scales to microscopic scales, self-consistently. This formidable and laudable task is however hampered by the span in time and length scales. This forces simulations to run with unrealistically low values of the electromagnetic fields, which is the major drawback of full particle approaches. There is so far no way out to satisfactorily treat single particle acceleration with radiation reaction self-consistently in ultra-strong electromagnetic fields. This represents a major task towards a deeper and closer investigation of realistic pulsar electrodynamics but so far no numerical technique is able to deal with such regimes.

In the present paper, we decided to start the study of pulsar high-energy emission from a different perspective, trying as much as possible to shortcut any large and time consuming plasma simulations in realistic field strengths. Instead, we require a minimal amount of assumptions putting special emphasize to computing light-curves and spectra with realistic values of the electromagnetic field of about \numprint{e8}~\SIunits{\tesla} for normal radio pulsars and \numprint{e5}~\SIunits{\tesla} for millisecond pulsars. In the radiation reaction limit, particles follow a velocity field solely prescribed by the local electromagnetic field itself. It is sometimes called Aristotelian dynamics or zero mass dynamics \citep{gruzinov_aristotelian_2013} but it is a simple consequence of particle motion with friction in the ultra-relativistic limit \citep{mestel_stellar_2012}. In section \ref{sec:RadiationLimit}, we remind the velocity prescription derived from the radiation reaction limit and the method to compute light-curves and spectra. The limit of applicability of the classical curvature radiation formula is briefly discussed. In section \ref{sec:Results} we show detailed results about spectra and sky maps for millisecond pulsars and normal pulsars. Possible future detection in the sub-TeV range from CTA is also briefly investigated. The limit of our approach is discussed in details in Sec.~\ref{sec:Discussion}. Conclusions are then drawn in Sec.~\ref{sec:Conclusion}.

\section{Magnetospheric emission model}
\label{sec:RadiationLimit}

We start with a description of the minimalist model used to compute light-curves, sky maps and spectra. Our primary target is to refrain from adding excessive a priori unconstrained parameters into the model in order to catch the essential physics required to fit gamma-ray pulsar data compiled in the second Fermi gamma-ray pulsar catalogue. The master physical quantities are the neutron star period~$P$, its period derivative~$\dot P$  (from which we deduce a fiducial magnetic field strength~$B$ at the equator) and the inclination angle between its rotation and magnetic dipole axis depicted by the obliquity~$\chi$. Apart from this obliquity~$\chi$ which is not constrained by observations, $P$ and $\dot P$ are well quantified by pulsar timing campaigns. However, some other inputs are required like the particle distribution function and the extent of the emitting volume. We recall these inputs in the following paragraphs.

\subsection{Radiation reaction}

Pulsar magnetospheres are filled with ultra-relativistic electron/positron pairs copiously radiating while accelerated by the electric field. It is safe to assume in a first stage that they reach an equilibrium state between acceleration and braking, called radiation reaction limit regime. The photon back reaction onto the particle motion is therefore important. The radiative friction brakes the particle in a direction opposite to its motion such that in the ultra-relativistic limit its velocity depends only on the local value of the electromagnetic field $\mathbf{B}$ and $\mathbf{E}$ (also sometimes called Aristotilean electrodynamics). Electrons and positrons will not react the same way to the electric field $\mathbf{E}$ thus two expressions for the velocity are required. It can be shown, assuming that both particle species speeds are equal to $c$, that the velocity is given by \citep{gruzinov_aristotelian_2013, petri_general-relativistic_2018} 
\begin{equation}
\label{eq:VRR}
 \bmath v_\pm = \frac{\bmath E \wedge \bmath B \pm ( E_0 \, \bmath E / c + c \, B_0 \, \bmath B)}{E_0^2/c^2+B^2}
\end{equation}
where the plus sign corresponds to positrons and the minus sign to electrons. Moreover, we introduced the two electromagnetic invariants $E_0$ and $B_0$ such that 
\begin{subequations}
\begin{align}
 \bmath E^2 - c^2 \, \bmath B^2 & = E_0^2 - c^2 \, B_0^2 \\
 \bmath E \cdot \bmath B & = E_0 \, B_0
\end{align}
\end{subequations}
with the subsidiary condition $E_0 \geqslant 0$ ensuring that the radiation reaction force is always directed oppositely to the velocity direction. As explained in \cite{petri_general-relativistic_2018} these invariants are related to the electromagnetic field strength in a frame where $\mathbf{E}$ and $\mathbf{B}$ are parallel. The lepton motion can be decomposed into an electric drift part $\bmath E \wedge \bmath B$, a motion along magnetic field lines $\mathbf{B}$ and a motion along electric field lines  $\mathbf{E}$. This last part of the motion is responsible for dissipation because the power of the Lorentz force is $q\,(\bmath E + \bmath v_\pm \wedge \bmath B) \cdot \bmath v_\pm = q \, \bmath v_\pm \cdot \bmath E \geq 0$ where $q=\pm e$ depending on the charge, positron or electron.

In the near field zone, i.e. close to the neutron star surface, where $E\ll c\,B$, the particle velocity simplifies into a motion solely along $\mathbf{B}$ such that
\begin{equation}
\label{eq:VitesseB}
\bmath v_\pm =  \pm c \, \frac{( \bmath E \cdot \bmath B ) \, \bmath B}{E_0 \, (E_0^2/c^2+B^2)} .
\end{equation}
This expression can be reduced to
\begin{equation}
 \bmath v_\pm = \pm c \, \textrm{sign}(B_0) \, \frac{\bmath B }{B}
\end{equation}
by noting that in this weak electric field limit the magnitude of $\bmath B$ is almost equal to the invariant $B_0$, namely $B^2 \approx B_0^2$. Particles are accelerated mostly by the electric component parallel to the magnetic field. The surface $\bmath E \cdot \bmath B=0$ are of particular interest because the velocity change sign when the particle cross this region. It is called a force-free surface and represents trapping regions for those particles \citep{finkbeiner_effects_1989}. We return to this important point in Sec.~\ref{sec:Discussion}.

\subsection{Curvature radiation}

Particle trajectories can be computed from the velocity field prescription given in eq.~(\ref{eq:VRR}). These trajectories are obviously bent, leading to curved paths and therefore curvature radiation. The curvature radius~$\rho_c$ is computed according to the acceleration following the expression
\begin{equation}
\label{eq:Acceleration}
 \bmath a_\pm = \frac{d\bmath v_\pm}{dt} = \frac{c^2}{\rho_c}.
\end{equation}
This acceleration is evaluated by a simple second order finite difference scheme. The associated curvature radiation spectrum for a particle with Lorentz factor~$\gamma$ is given in \cite{jackson_electrodynamique_2001} by
\begin{equation}
 \frac{dI}{d\omega} = \frac{\sqrt{3} \, e^2}{4\,\upi\,\varepsilon_0\,c} \, \gamma \, \frac{\omega}{\omega_c} \, \int_{\omega/\omega_c}^{+\infty} K_{5/3}(x) \, dx
\end{equation}
where $K_{5/3}$ is the modified Bessel function of order $5/3$, $\varepsilon_0$ the vacuum permittivity, $I$ the intensity and $\omega$ the angular frequency. The fundamental frequency is $\omega_0 = c/\rho_c$ and the characteristic curvature photon frequency therefore reads
\begin{equation}
\label{eq:OmegaCut}
 \omega_c = \frac{3}{2} \, \gamma^3 \, \frac{c}{\rho_c}
\end{equation}
from which we deduce the curvature power as
\begin{equation}
 P_c = \frac{e^2}{6\,\upi\,\varepsilon_0} \, \gamma^4 \, \frac{c}{\rho_c^2} .
\end{equation}
The curvature emissivity depends on the observation frequency~$\omega$ as well as on the location in the magnetosphere~$\mathbf{r}$. This emissivity is given by
\begin{equation}
 j_{\rm cur}(\mathbf{r}, \omega) = \frac{\sqrt{3}}{2\,\upi} \, \alpha_{\rm sf} \, \frac{\hbar \, c}{\rho_c(\mathbf{r})} \, \gamma \, F\left(\frac{\omega}{\omega_c(\mathbf{r})} \right)
\end{equation}
showing explicitly the spatial dependence of this emissivity. $\alpha_{\rm sf}$ is the fine structure constant defined by
\begin{equation}
\alpha_{\rm sf} = \frac{e^2}{4\,\upi\,\varepsilon_0\,\hbar\,c}
\end{equation}
with $\hbar$ the reduced Planck constant. Curvature radiation is very similar to synchrotron radiation for which the function F is usually defined by
\begin{equation}
 F(x) = x \,  \int_{x}^{+\infty} K_{5/3}(t) \, dt .
\end{equation}
The spectra and cut off frequency in both cases are described in \cite{jackson_electrodynamique_2001}. 
In the radiation reaction limit, the power exerted by the electric field is simply $\pm e \, \mathbf{v}_\pm \cdot \mathbf{E} = e \, c \, E_0 \geqslant 0$. The work done is always positive as it should be for a dissipative force. According to curvature radiation losses, the maximum Lorentz factor an electron or a positron can reach is
\begin{equation}
\label{eq:GammaRR}
 \gamma^4 = \frac{6\,\upi\,\varepsilon_0}{e} \, E_0 \, \rho_c^2 .
\end{equation}
This equilibrium Lorentz factor~$\gamma$ weakly depends on the electric field and curvature radius. Aristotelian electrodynamics implies a particle speed exactly equal to the speed of light, by definition and construction of the velocity given in expression~(\ref{eq:VRR}). Thus, technically, the Lorentz factor~$\gamma$ is computed from the knowledge of the curvature radius deduced from the acceleration of ultra-relativistic particles, eq.~(\ref{eq:Acceleration}). We will show that the actual Lorentz factors are $\gamma \gtrsim 10^8$ thus widely justifying the approximation of taking $v=c$. In Aristotelian electrodynamics, particles do not have memory about their past trajectory because the velocity is computed according to only the local current electromagnetic field at their position. This locality remains true as long as the particles are able to accelerate due to the electric field or decelerate due to radiation reaction on a length scale~$\ell$ much smaller than electromagnetic field gradient and curvature radius. The distance required to gain energy up to $\gamma\,\me\,c^2$ is
\begin{equation}
\ell = \frac{\gamma\,\me\,c^2}{e\,E_0} .
\end{equation}
With the radiation reaction limiting Lorentz factor, we find
\begin{subequations}
\begin{align}
 \frac{\ell}{\rho_c} & = (6\,\upi\,\varepsilon_0)^{1/4} \, \frac{\me\,c^2}{e^{5/4}\,E_0^{3/4}\,\rho_c^{1/2}} \\
 & = \numprint{4.7e-6} \, \left( \frac{E_0}{\numprint{e12}~\SIunits{\volt\per\meter}} \right)^{-3/4} \, \left( \frac{\rho_c}{12~\SIunits{\kilo\meter}} \right)^{-1/2}.
\end{align}\end{subequations}
This ratio is always much less than one for realistic pulsar parameters. In other words, particle emission at some location is not affected by the electric field the particle encountered at another position. It loses its memory within a short distance much smaller than any macroscopic length scale.

The total luminosity radiated by the magnetosphere is therefore
\begin{equation}
 \frac{dI_{\rm tot}}{d\omega\,dt} = \iiint_{V} j_{\rm cur}(\mathbf{r}, \omega) \, n(\mathbf{r}) \, d^3\mathbf{r}
\end{equation}
where $n$ is the particle density number and integration goes along the emitting volume~$V$. The inner and outer boundary of the integration is not specified. A natural choice for the minimum radius is the neutron star size and a possible maximum radius is the light-cylinder although other prescription are conceivable. For instance, photon production outside the light-cyliner is another interesting possibility.

The density of leptons is another important unknown parameters. As we want to stay minimalist in our model, we assume a spherically symmetric profile with a decrease in radius according to
\begin{equation}
 n(\mathbf{r}) = n_0 \, \left(\frac{R}{r}\right)^q
\end{equation}
where $n_0$ is a normalisation factor and $q$ the exponent of the power law decrease in radius. In the same spirit of simplicity, we do not introduce any power law particle distribution function but straightforwardly choose the local Lorentz factor according to the radiation reaction limit regime prescribed in eq.~(\ref{eq:GammaRR}).


\subsection{Quantum corrections}

Radiation processes are usually derived in a non QED framework where quantum corrections to emission are neglected. Such expressions remain valid as long as the magnetic field strength stays well below the quantum critical field of $\BQ = \numprint{4.4e9}$~T. Quantum corrections arises because of the particle recoil and when the photon wavelength becomes comparable to the particle Compton wavelength 
\begin{equation}
\lambdabar_c = \frac{\hbar}{\me\,c} .
\end{equation}
Let us quantify when QED sets in to modify the photon spectra. Curvature radiation is very similar to synchrotron radiation. Both processes originate from the radiation of a charged particle subject to acceleration. The associated photon spectra are therefore similar if the cyclotron gyro-frequency is replaced by the instantaneous rotation frequency of the particle along its curved path. Due to the general law of conservation of energy, no charge can radiate more than its kinetic energy. It is well known that quantum synchrotron sets in whenever the following parameter reaches s close to unity \citep{erber_high-energy_1966,aharonian_astrophysics_2013}
\begin{equation}
  \chi_{\rm sync} = \frac{3}{2} \, \gamma \, \frac{B}{\BQ} \approx 1 .
\end{equation}
We stress that because of the emitting particle Lorentz factor intervening in the above expression, quantum effects manifest already at field strengths much less that $\BQ$. This is of primary importance in pulsar magnetospheres because as will be shown later, $\gamma$ can go up to \numprint{e8}-\numprint{e9}. Consequently, quantum synchrotron radiation is at work up to very large distances compared to the neutron star radius. In the same vain, looking for the curvature radiation, quantum effects become perceptible whenever the parameter
\begin{equation}
  \chi_{\rm curv} = \frac{3}{2} \, \gamma^2 \, \frac{\lambdabar_c}{\rho_c}
\end{equation}
approaches unity. Now the $\chi$ parameter is even more sensitive to the Lorentz factor. We will check a posteriori that $\chi_{\rm curv}$ remains weak or at least $\chi_{\rm curv} \lesssim 1$ in all our computations.

\subsection{Normalisation}

In order to simulate realistic value of electromagnetic field strengths, electron/positron energies and photon energies, we normalise the fundamental quantities of the problem. The magnetospheric distances are normalised to the light-cylinder radius~$\rlight = c/\Omega$. Velocities are normalised to the speed of light~$c$. The magnetic field normalisation is performed according to the critical field $\tilde B = B / \BQ \equiv b$. The electric field typical value is given by Schwinger value of
\begin{equation}
 E_{\rm Schw} = \frac{\me^2 \, c^3}{e\,\hbar} = 10^{18} ~ \SIunits{\volt\per\meter}
\end{equation}
such that the normalised value of the electric field becomes $\tilde{E} = E_0 / E_{\rm Schw}$.
The characteristic curvature photon energy in normalised units is conveniently written in units of the electron rest mass energy such that
\begin{equation}
 k_c = \frac{\hbar \, \omega_c}{\me\,c^2} = \frac{3}{2} \, \gamma^3 \, \frac{\lambdabar_c}{\rho_c} = \gamma \, \chi_{\rm curv}.
\end{equation}
The Lorentz factor balancing exactly acceleration against radiation therefore becomes
\begin{equation}
 \gamma^4 = \frac{3}{2} \, \frac{\tilde{E}}{\alpha_{\rm sf}} \, \frac{\rho_c^2}{\lambdabar_c^2} .
\end{equation}
The curvature power emitted by a single particle accelerated in the electric field is
\begin{equation}
 P_c = \frac{\me^2 \, c^4}{\hbar} \, \tilde{E} \approx \numprint{6.36e7} ~ \SIunits{\watt}  \, \tilde{E}.
\end{equation}
In orders of magnitude, the normalised electric field strength is
\begin{equation}
\label{eq:Enormee}
 \tilde{E} = \tilde{B} \, \frac{R}{\rlight} .
\end{equation}
As a characteristic particle number density, we use the expression deduced from the force-free condition, the Goldreich-Julian density, given by
\begin{equation}
 n_0 = \frac{2\,\varepsilon_0\,\Omega\,B}{e} = \frac{1}{2\,\upi\,\alpha_{\rm sf} \, \lambdabar_c^2} \, \frac{\tilde{B}}{\rlight} \approx \numprint{1.46e21} ~ \SIunits{\meter^{-3}} \, \left( \frac{\rlight}{10^5 ~ \SIunits{\meter}} \right)^{-1} \, \tilde{B} .
\end{equation}
For the radiative properties, normalizing energies also to the electron rest mass energy, the curvature emissivity is given by
\begin{equation}
 \frac{d\tilde{I}}{d\tilde{\omega} \, d\tilde{t}} = \frac{\sqrt{3}}{2\,\upi} \, \alpha_{\rm sf} \, \frac{\lambdabar_c}{\rho_c} \, \gamma \, F\left(\frac{\omega}{\omega_c} \right) .
\end{equation}
We introduced normalised frequency and time such that $\tilde{\omega}=\hbar\,\omega/\me\,c^2$ and $c\,t=\lambdabar_c\,\tilde{t}$.
In normalised units, the luminosity becomes by introducing the multiplicity factor~$\kappa$
\begin{equation}
 \frac{d\tilde{I}_{\rm tot}}{d\tilde{\omega} \, d\tilde{t}} = \kappa \, \iiint_V n(\mathbf{r}) \, \frac{d\tilde{I}}{d\tilde{\omega} \, d\tilde{t}} \, d^3\mathbf{r} 
\end{equation}
or explicitly with the spatial dependence of curvature radius and Lorentz factor
\begin{equation}
 \frac{d\tilde{I}_{\rm tot}}{d\tilde{\omega} \, d\tilde{t}} = \frac{\sqrt{3}}{4\,\upi^2} \, \kappa \,  \frac{\rlight^2}{\lambdabar_c} \, \tilde{B} \, \iiint_V n(\mathbf{r}) \, \frac{\gamma(\mathbf{r})}{\rho_c(\mathbf{r})} \, F\left(\frac{\omega}{\omega_c(\mathbf{r})} \right) \, d^3\mathbf{r} .
\end{equation}
With the normalisation of the density $n_0$ we finally get
\begin{equation}
\label{eq:Spectre}
 \frac{d\tilde{I}_{\rm tot}}{d\tilde{\omega} \, d\tilde{t}} = \frac{\sqrt{3}}{4\,\upi^2} \, \kappa \ \frac{\rlight}{\lambdabar_c} \, \tilde{B} \, \iiint_V \tilde{n}(\mathbf{\tilde{r}}) \, \gamma(\mathbf{\tilde{r}}) \, \frac{ \rlight}{\rho_c(\mathbf{\tilde{r}})} \, F\left(\frac{\omega}{\omega_c(\mathbf{\tilde{r}})} \right) \, d^3\mathbf{\tilde{r}} .
\end{equation}
This last expression is used to compute the whole information about emission in the magnetosphere. In particular, spectra and light-curves shown in the following section are derived from eq.~(\ref{eq:Spectre}).
The flux restored with SI units therefore becomes
\begin{multline}
\label{eq:SpectreSIunits}
 \hbar\,\omega \, \frac{dI_{\rm tot}}{d(\hbar\,\omega)\, dt} = \frac{\me^2\,c^4}{\hbar} \frac{\sqrt{3}}{4\,\upi^2} \, \kappa \ \frac{\rlight}{\lambdabar_c} \, \tilde{B} \, \tilde{\omega} \, \times \\
  \iiint_V \tilde{n}(\mathbf{\tilde{r}}) \, \gamma(\mathbf{\tilde{r}}) \, \frac{ \rlight}{\rho_c(\mathbf{\tilde{r}})} \, F\left(\frac{\omega}{\omega_c(\mathbf{\tilde{r}})} \right) \, d^3\mathbf{\tilde{r}} .
\end{multline}
For the neutron star radius, we take a fiducial value of 12~\SIunits{\kilo\meter} \citep{ozel_masses_2016}. The pair multiplicity is fixed to $\kappa=1$ if not otherwise specified. The radial boundary radii are normalized to the light-cylinder, $r_{\rm in} = R_{\rm in}/\rlight$ and $r_{\rm out} = R_{\rm out}/\rlight$. Because in the following section there is no confusion possible between electric field~$E$ and photon spectra~$I_{\rm tot}$, we restore the usual notation, replacing $I_{\rm tot}$ by $E=\hbar\,\omega$ when discussing spectra. We therefore use the conventional notation again like~$E^2\,dN/dE\,dt$ in eq.~(\ref{eq:SpectreSIunits}). Next we show a detailed analysis of the pulsed emission characteristics.

\section{Simulations}
\label{sec:Results}

High energy emission emanates from regions close to the neutron star surface because the electromagnetic field is largest there and therefore the invariant field quantity~$E_0$ as well as the Lorentz factor required in the radiation reaction limit regime too. On one hand, TeV photons are produced in the innermost part of the magnetosphere. As a general comment, for normal pulsars, the field is strong enough to disintegrate these photons into electron/positron pairs, rendering the medium opaque to this light. Therefore the effective TeV photon flux, if any, is much weaker than in the case of a magnetically optically thin magnetosphere. On the other hand, sub-GeV and MeV photons are produced close to the light-cylinder and freely escape the magnetosphere with a low probability interaction with the magnetic field.

In this section, we show some typical mean and phase-resolved spectra, sky maps and light curves for realistic magnetic field strengths, rotation periods and geometries when particles radiate in the radiation reaction limit regime. We also discuss the cut-off energy and the gamma-ray luminosity dependence on these fundamental parameters. Results are shown for two archetypal classes of pulsars: millisecond pulsars with typical period of~$P=5$~ms and normal pulsars with typical period of~$P=100$~ms. The magnetic field strength is given in units of the critical field~$\BQ$ such that we used the normalized field given by the parameter $b=B/\BQ$. The Lorentz factor used for beaming in the direction of motion of particle as imposed by eq.~(\ref{eq:VRR}) is set to~$\Gamma=10$. It should actually be beamed into a cone of opening angle $\propto 1/\gamma\ll1$ but this would require a fantastic angular resolution in the volume integration of eq.~(\ref{eq:Spectre}). In any case, for $\gamma\gg1$ light-curves and spectra become insensitive to the precise value of $\gamma$. They are shaped by the electromagnetic field topology that is a macroscopic scale.

\subsection{High-energy spectra}

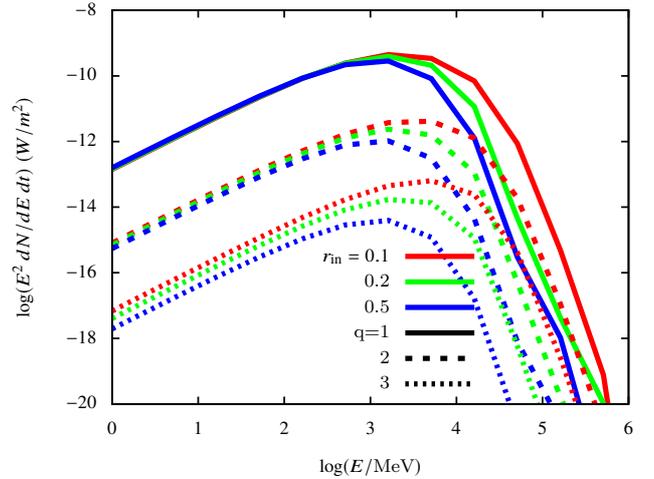
\begin{figure}
	\centering
	\resizebox{0.5\textwidth}{!}{\input{spectre_moyen_r00025_b-3_g10_dx}}
	\caption{Mean spectra for a 100~ms pulsar, density profile $n(r) \propto r^{-q}$ with $q\in\{1,2,3\}$ with respectively solid, dashed and dotted lines. The magnetic field strength is $b=10^{-3}$ and $\chi=60$\degr. The inner boundary of the emission volume is given by $r_{\rm in} = \{0.1,0.2,0.5\}$ and the outer boundary by $r_{\rm out}=1$. Fluxes are evaluated at a distance of 1~kpc.}
	\label{fig:spectre_moyen_r00025_b-3_g10_dx}
\end{figure}

High-energy spectra are easily compiled by computing the energy flux~$E^2 \, d^2N/dt\,dE$ for different energy bands. A typical example of spectra for a normal pulsar is shown in Fig.~\ref{fig:spectre_moyen_r00025_b-3_g10_dx}.
The pulsar obliquity is set to~$\chi=60\degr$ and the magnetic field strength to~$b=10^{-3}$. The particle density profile~$n(r)$ is spherically symmetric and decreases with radius according to $n(r) \propto r^{-q}$ with $q$ an arbitrary constant taken for concreteness within the set $q\in\{1,2,3\}$. The spherical symmetry is clearly a crude approximation of the spatial distribution of particles. A better description would require a deeper understanding of pair creation within the magnetosphere. In our minimalistic approach, we bypass such refinements. Moreover, we assume a Goldreich-Julian corotation density normalisation at the surface such that $e \, n(R) = 2\varepsilon_0\,\Omega \, B(R)$. The energy flux is measured at a distance of 1~kpc. Emissivity occurs within a spherical shell of inner radius~$R_{\rm in}$ and outer radius~$R_{\rm out}$ not necessarily equal to the light-cylinder radius~$\rlight$. Emissivity is integrated within the volume located between $R_{\rm in}$ and $R_{\rm out}$. Because of magnetic photo-absorption efficiency close to the surface, very high energy photons preferentially come from outer regions such that $r \gtrsim 0.5\,\rlight$ rather than from regions $r \lesssim 0.1\,\rlight$. Emission from high altitude magnetospheric sites is also preferred following current wisdom. Although the particle distribution is not monoenergetic (the monoenergetic distribution we enforce is spatially variable due to the variability of the electromagnetic field and the corresponding radiation reaction rate), the volume integrated spectra resemble a monoenergetic distribution, typically a power law with an exponential cut-off. Moreover, the cut-off energy lies around several~GeV as seen in the Fermi/LAT pulsar catalog \citep{abdo_second_2013} for our special choice of $b=10^{-3}$. The slope of the power law below the cut-off agrees with the $1/3$ exponent of a single particle curvature radiation spectrum. The maximum energy flux at its peak depends on the size of the emission volume as expected, proportional to the total number of particles. Indeed, by inspection of Fig.~\ref{fig:spectre_moyen_r00025_b-3_g10_dx}, we deduce that increasing $q$ reduced the total energy flux because the particle density decreases also faster for $q=3$ compared to $q=1$. The location of the inner radius $R_{\rm in}$ impacts only on the shape of the exponential cut-off. This is because the most energetic photons are produced in the strongest accelerating field that is close to the surface. Cutting the emission volume at higher altitude removes these photons from the spectra as expected (compare the red, green and blue lines).

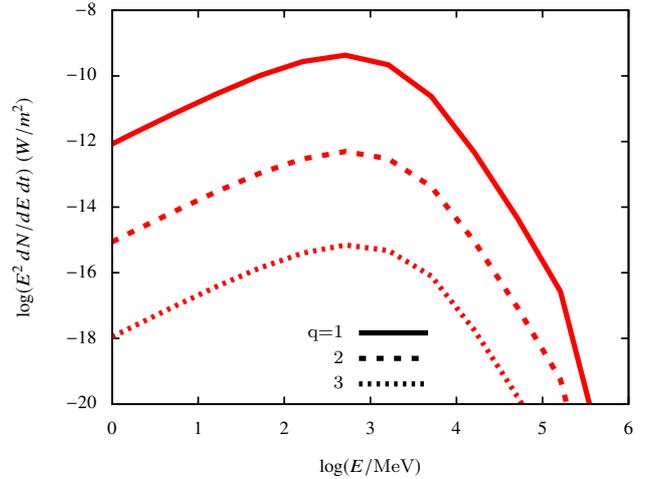
\begin{figure}
	\centering
	\resizebox{0.5\textwidth}{!}{\input{spectre_moyen_r00025_b-3_g10_dx_onde}}
	\caption{Mean spectra for a 100~ms pulsar, density profile $n(r) \propto r^{-q}$ with $q\in\{1,2,3\}$ with respectively solid, dashed and dotted line. The magnetic field strength is $b=10^{-3}$ and $\chi=60$\degr. The inner boundary of the emission volume is given by $r_{\rm in} = 1$ and the outer boundary by $r_{\rm out}=5$. Fluxes are evaluated at a distance of 1~kpc.}
	\label{fig:spectre_moyen_r00025_b-3_g10_dx_onde}
\end{figure}

In the wave zone, outside the light-cylinder, when radiation emanates from distances $r>\rlight$, the average spectra remain very similar to those produced inside the light-cylinder,  Fig.~\ref{fig:spectre_moyen_r00025_b-3_g10_dx_onde}. The cut-off energy is slightly less outside but still very close to several GeV. Consequently, spectral features are insensitive to the precise extent of the emission regions. However, as will be shown later, light-curves shapes and pulse profiles are sensitive to the location of the photon production sites.

\begin{figure}
 \centering
 \resizebox{0.5\textwidth}{!}{\input{spectre_moyen_r005_b-6_g10_dx}}
 \caption{Mean spectra for a 5~ms pulsar, density profile density $n(r) \propto r^{-q}$ with $q\in\{1,2,3\}$ with respectively solid, dashed and dotted lines. The magnetic field strength is $b=10^{-6}$ and $\chi=60$\degr. The inner boundary of the emission volume is given by $r_{\rm in} = \{0.1,0.2,0.5\}$ and the outer boundary by $r_{\rm out}=1$. Fluxes are evaluated at a distance of 1~kpc.}
 \label{fig:spectre_moyen_r005_b-6_g10_dx}
\end{figure}
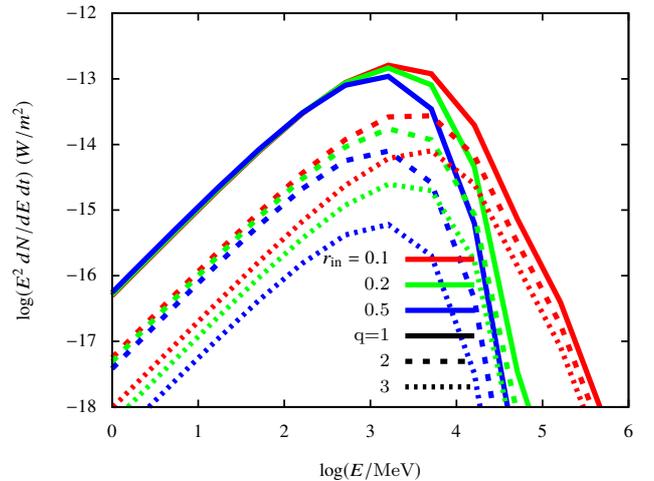

A second example of mean spectra is shown in Fig.~\ref{fig:spectre_moyen_r005_b-6_g10_dx} for a 5~ms period pulsar with a lower magnetic field of~$b=10^{-6}$. Again a power-law with exponential cut-off is observed but with a sharp extinction above the cut-off energy of several~GeV. The density profile mainly impacts on the maximum intensity level whereas the extension of the emitting region controlled by $R_{\rm in}$ slightly shapes the cut-off behaviour as for normal pulsars.

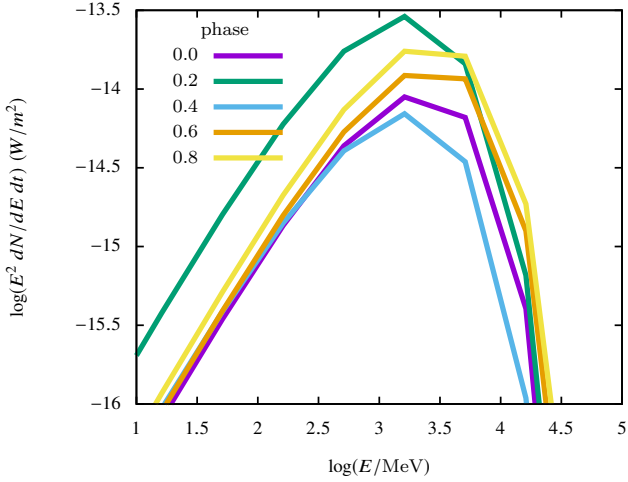
\begin{figure}
	\centering
	\resizebox{0.5\textwidth}{!}{\input{spectre_resolu_r00025_b-3_ri0.2_g10_d3}}
	\caption{Phase-resolved spectra for a 100~ms pulsar, $r_{\rm in} = 0.2$, $r_{\rm out} = 1$ and $n(r) \propto r^{-3}$. The magnetic field strength is $b=10^{-3}$, $\chi=60\degr$ and $\zeta=60\degr$.}
	\label{fig:spectre_resolu_r00025_b-3_ri02_g10_d3}
\end{figure}

\begin{figure}
 \centering
 \resizebox{0.5\textwidth}{!}{\input{spectre_resolu_r005_b-6_ri0.2_g10_d3}}
 \caption{Phase-resolved spectra for a 5~ms pulsar, $r_{\rm in} = 0.2$, $r_{\rm out} = 1$ and $n(r) \propto r^{-3}$ and $n\propto r^{-3}$. The magnetic field strength is $b=10^{-6}$, $\chi=60\degr$ and $\zeta=60\degr$.}
 \label{fig:spectre_resolu_r005_b-6_ri02_g10_d3}
\end{figure}
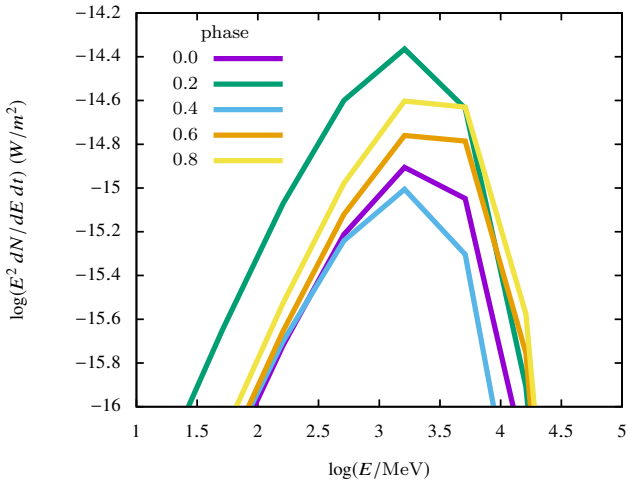

Mean spectra highlight the general trend of magnetospheric emission. For a peculiar pulsar, phase-resolved spectra offer valuable insight into the emission region, its shape and geometry within the magnetosphere. Therefore, Fig.~\ref{fig:spectre_resolu_r00025_b-3_ri02_g10_d3} shows a phase-resolved spectrum for a 100~ms pulsar with $\chi=60$\degr, $b=10^{-3}$ and $r_{\rm in}=0.2$.
The total flux variation between the off-pulse and on-pulse peak intensity is about one order of magnitude. The spectral shape remains substantially the same for all phases during its rotation. Very similar results are found for a 5~ms pulsar for which the phase-resolved spectra are given in Fig.~\ref{fig:spectre_resolu_r005_b-6_ri02_g10_d3}. We essentially observe the same trend as for the 100~ms pulsar that is similar spectra for all phases but with a shift in magnitude of at most one decade.
After this brief survey on the pulsed spectral features, we dig into the geometrical properties of the light-curves as depicted in sky map diagrams.

\subsection{Sky maps}

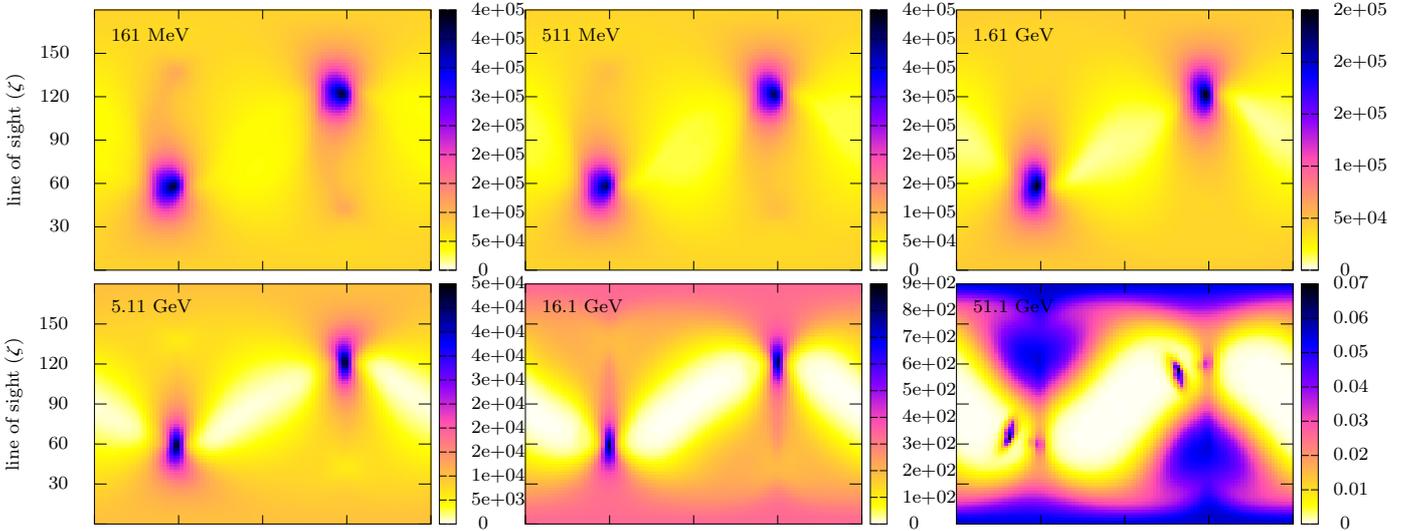
\begin{figure*}
\centering
\resizebox{\textwidth}{!}{\input{skymaps_cl_r00025_b-3_c60_ri0.2_ro1_g10_d1}}
\caption{Sky maps for a 100~ms pulsar, $r_{\rm in}=0.2$, $r_{\rm out} = 1$ and $n\propto r^{-1}$. The magnetic field strength is $b=10^{-3}$, $\chi=60\degr$.}
\label{fig:skymaps_cl_r00025_b-3_c60_ri0.2_ro1_g10_d1}
\end{figure*}

\begin{figure*}
 \centering
 \resizebox{\textwidth}{!}{\input{skymaps_cl_r005_b-6_c60_ri0.2_ro1_g10_d3}}
 \caption{Sky maps for a 5~ms pulsar, $r_{\rm in}=0.2$, $r_{\rm out} = 1$ and $n\propto r^{-3}$. The magnetic field strength is $b=10^{-6}$, $\chi=60\degr$.}
 \label{fig:skymaps_cl_r005_b-6_c60_ri0.2_ro1_g10_d3}
\end{figure*}
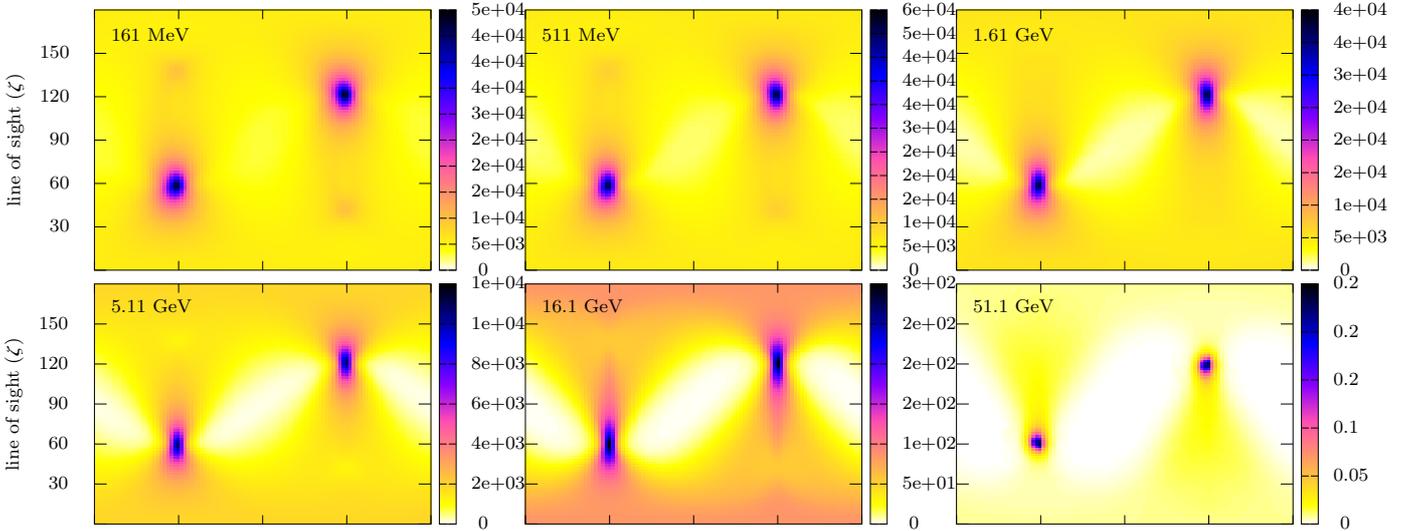

Sky maps are useful graphical representations of the light-curve profiles depending on the obliquity~$\chi$ and inclination of the line of sight~$\zeta$. Several pertinent subset of sky maps are shown in the following figures. First, Fig.~\ref{fig:skymaps_cl_r00025_b-3_c60_ri0.2_ro1_g10_d3} shows a sample of sky maps for a 100~ms pulsar with appropriate magnetic field strength as given in the previous paragraph. The density profile sharply decreases with $q=3$. In each plot of the panel, the $x$ axis depicts the phase of the pulsar (one period normalized to unity) and the $y$ axis depicts the line of sight inclination angle~$\zeta$ going from $0\degr$ to $180\degr$.
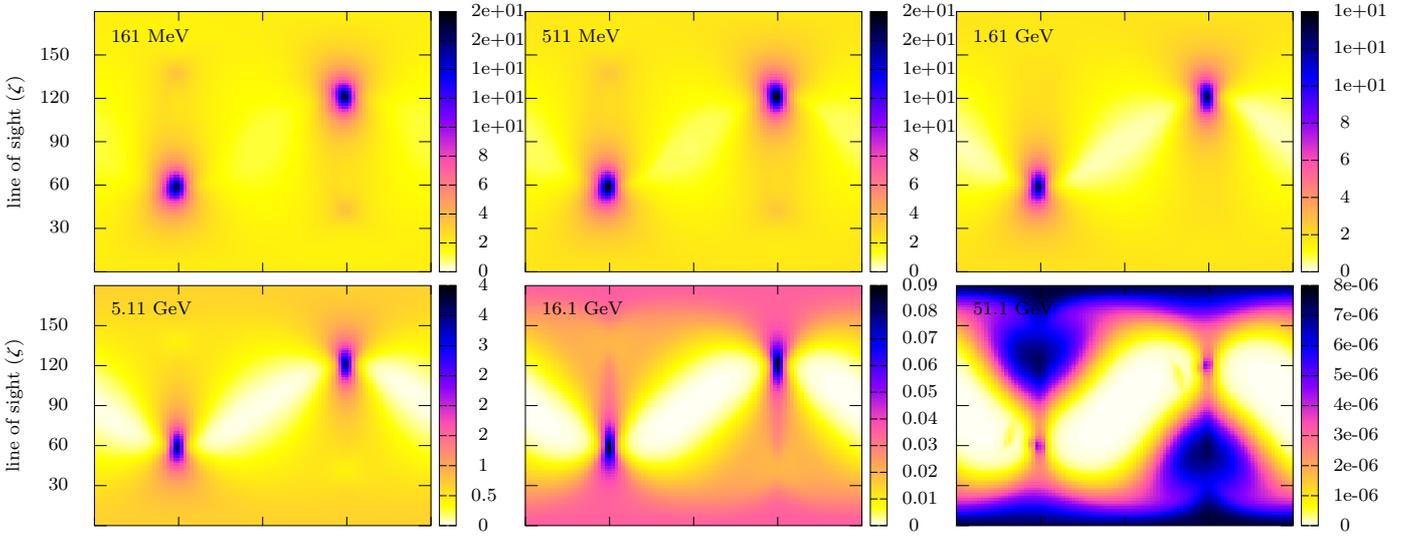
\begin{figure*}
	\centering
	\resizebox{\textwidth}{!}{\input{skymaps_cl_r00025_b-3_c60_ri0.2_ro1_g10_d3}}
	\caption{Sky maps for a 100~ms pulsar, $r_{\rm in}=0.2$, $r_{\rm out} = 1$ and $n(r)\propto r^{-3}$. The magnetic field strength is $b=10^{-3}$ and $\chi=60\degr$.}
	\label{fig:skymaps_cl_r00025_b-3_c60_ri0.2_ro1_g10_d3}
\end{figure*}
Each plot is given for a specified energy, increasing from top left to bottom right. Precise values are shown in the top left labels of each map. The intensity level of each map is shown in the colorbox legend on the right. In the power law regime of curvature radiation scaling as $\omega^{1/3}$, sky maps are more or less the same at all energies below the cut-off energy of several~GeV, but around and above this cut-off, the light-curves appreciably change their profile due to the exponential tail until a very low imperceptible flux and eventually an extinction at very high energies. A notable difference with respect to all other magnetospheric or wind emission models is that the position in phase of the two peaks is insensitive to the inclination angle~$\zeta$. Moreover, the peak separation in phase remains invariably equal to 0.5. This is not typical for Fermi/LAT pulsars. However, such patterns reflects the symmetry of the electromagnetic field dictated by the rotating star. Shifting the location of the magnetic dipole with respect to the geometrical centre of the star would introduce an asymmetry in the field topology around the north and south poles, alleviating the phase separation exactly equal to half a period observed in the present work. Such extension of our model is left for future work but these asymmetries have already been reported in our previous works, showing asymmetric polar cap shapes \citep{kundu_pulsed_2017}, asymmetric wind structures \citep{petri_radiation_2016} as well as asymmetries in the polarization pattern \citep{petri_polarized_2017}. In our particular set-up, we fixed the origin of phase~$\phi_0$ in order to locate the peaks around phase~$\phi=0.25$ and $\phi=0.75$. This value of the phase origin~$\phi_0$ is chosen such that both peaks stay visible well within the phase interval $\phi \in [0,1]$. This artificial lag prevents an unintended cut at phase zero or one, that is, right in the middle of a pulse. It is performed just for graphical purposes, avoiding to plot light-curves on two periods as sometimes done for better readability. Obviously, this phase~$\phi_0$ is arbitrary but with absolutely no impact on light-curves and spectra. Two strong spots are visible around $\zeta \approx \chi$ and $\zeta = \upi - \chi$ when observing below the cut-off. They reflect the location of both polar caps. The situation reverses at the highest energies above the cut-off. The two spots become invisible letting emerge a more diffuse emission away from the polar caps. The emission sites for high energy spread around the outer part of the light cylinder. Note also the drastic decrease in flux of several decades with respect to the low energy part. A second example of sky maps is shown in Fig.~\ref{fig:skymaps_cl_r00025_b-3_c60_ri0.2_ro1_g10_d1} for a slowly decreasing density profile with $q=1$. In such a scenario, the high energy flux remains significant and the emission appears less diffuse than for the case $q=3$. But for $q=2$ as shown in Fig.~\ref{fig:skymaps_cl_r00025_b-3_c60_ri0.2_ro1_g10_d2} the change in sky maps above several GeV is already apparent and resembles the $q=3$ case. The impact of $R_{\rm in}$ on these same sky maps is also investigated by inspection of Fig.~\ref{fig:skymaps_cl_r00025_b-3_c60_ri0.1_ro1_g10_d3} for which $r_{\rm in}=0.1$ and Fig.~\ref{fig:skymaps_cl_r00025_b-3_c60_ri0.5_ro1_g10_d3} for which $r_{\rm in}=0.5$. When emission is shifted to the outer parts of the light-cylinder, like in the case $r_{\rm in}=0.5$, the peaks broaden and show a shift with respect to the polar cap location, especially at highest energies above several GeV.
\begin{figure*}
\centering
\resizebox{\textwidth}{!}{\input{skymaps_cl_r00025_b-3_c60_ri0.2_ro1_g10_d2}}
\caption{Sky maps for a 100~ms pulsar, $r_{\rm in}=0.2$, $r_{\rm out} = 1$ and $n\propto r^{-2}$. The magnetic field strength is $b=10^{-3}$, $\chi=60\degr$.}
\label{fig:skymaps_cl_r00025_b-3_c60_ri0.2_ro1_g10_d2}
\end{figure*}
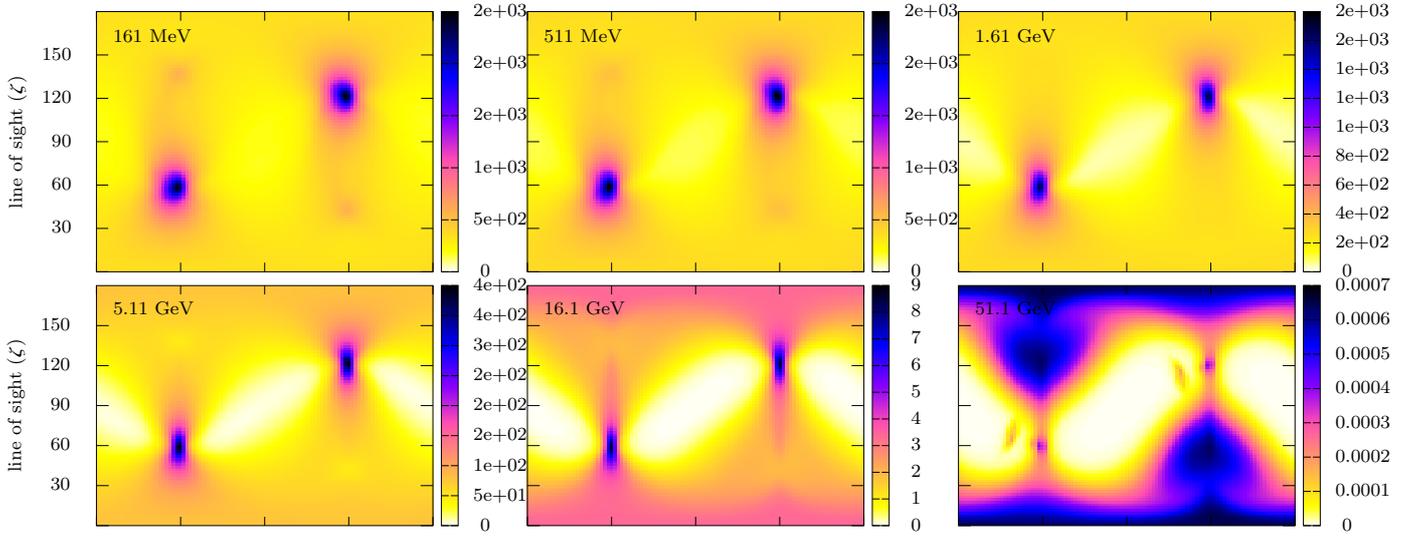
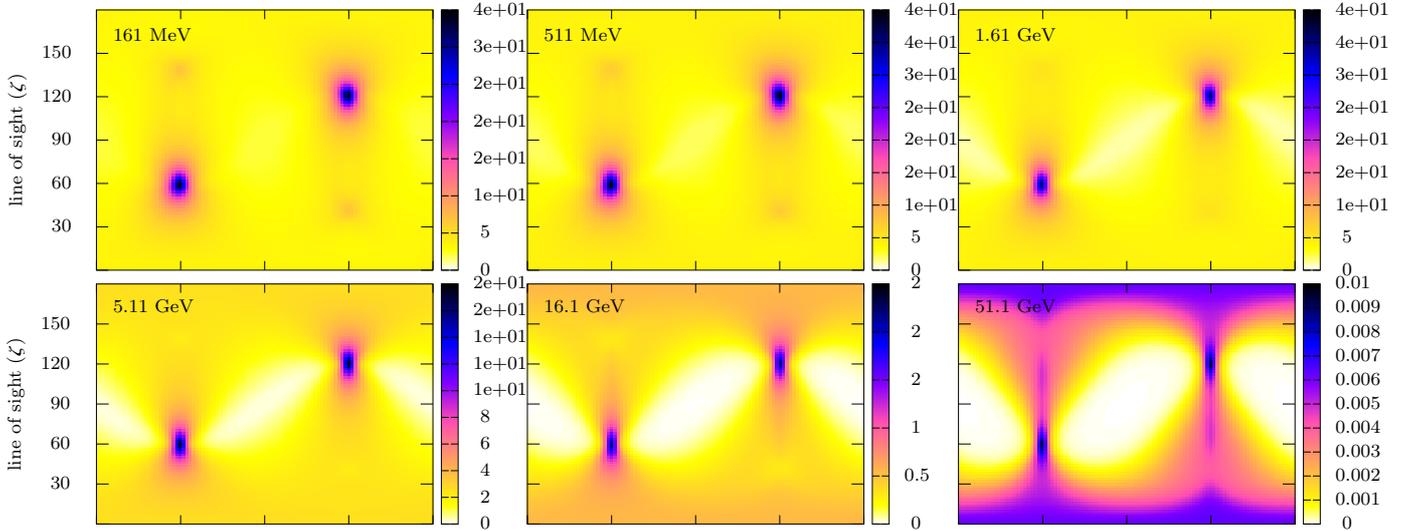
\begin{figure*}
\centering
\resizebox{\textwidth}{!}{\input{skymaps_cl_r00025_b-3_c60_ri0.1_ro1_g10_d3}}
\caption{Sky maps for a 100~ms pulsar, $r_{\rm in}=0.1$, $r_{\rm out} = 1$ and $n\propto r^{-3}$. The magnetic field strength is $b=10^{-3}$, $\chi=60\degr$.}
\label{fig:skymaps_cl_r00025_b-3_c60_ri0.1_ro1_g10_d3}
\end{figure*}
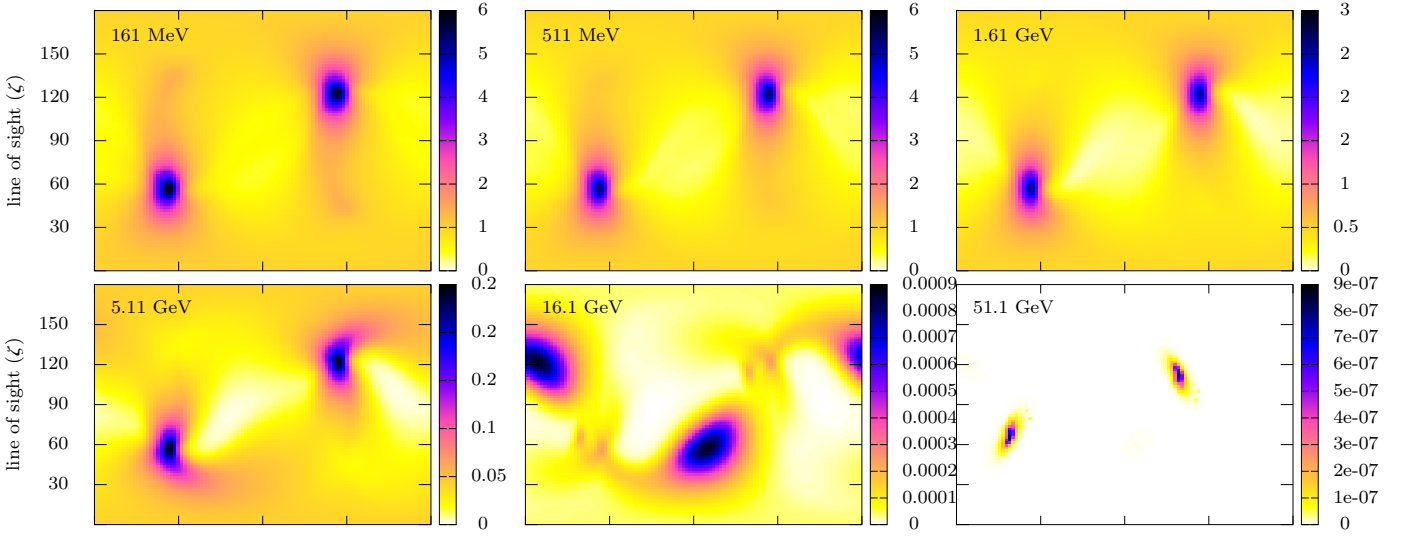
\begin{figure*}
\centering
\resizebox{\textwidth}{!}{\input{skymaps_cl_r00025_b-3_c60_ri0.5_ro1_g10_d3}}
\caption{Sky maps for a 100~ms pulsar, $r_{\rm in}=0.5$, $r_{\rm out} = 1$ and $n\propto r^{-3}$. The magnetic field strength is $b=10^{-3}$, $\chi=60\degr$.}
\label{fig:skymaps_cl_r00025_b-3_c60_ri0.5_ro1_g10_d3}
\end{figure*}
A last example of sky maps is shown in Fig.~\ref{fig:skymaps_cl_r005_b-6_c60_ri0.2_ro1_g10_d3} for a millisecond pulsar. However there is no noticeable discrepancy to discuss between millisecond and normal pulsars.

We finish our discussion about sky maps by exploring their dependence on the inner and outer boundaries $R_{\rm in}$ and  $R_{\rm out}$. Indeed, photons of given energies are produced in different radial shells. The sky-maps shown above are the sum of the radiation from all these shells. To better understand the physiognomy of this radiation, we separate the contribution from each spherical shell, assuming a thickness of $\Delta r/\rlight = 0.1$ for each shell. A sample is shown in Fig.~\ref{fig:skymaps_cl_r00025_b-3_c60_g10_d3_om6} for a 100~ms pulsar and a range of $r_{\rm in}$ from 0.1 to 0.9 and $R_{\rm out} = R_{\rm in} + \Delta r$ with $E=511$~MeV. The same plot around the cut-off energy $E=5.1$~GeV is shown in Fig.~\ref{fig:skymaps_cl_r00025_b-3_c60_g10_d3_om8} and well , at $E=51$~GeV in Fig.~\ref{fig:skymaps_cl_r00025_b-3_c60_g10_d3_om10}.
\begin{figure*}
	\centering
	\resizebox{\textwidth}{!}{\input{skymaps_cl_r00025_b-3_c60_g10_d3_om6}}
	\caption{Sky maps for a 100~ms pulsar, the range $r_{\rm in}-r_{\rm out}$ as shown in the labels and $n\propto r^{-3}$. The magnetic field strength is $b=10^{-3}$, $\chi=60\degr$ and $E=511$~MeV.}
	\label{fig:skymaps_cl_r00025_b-3_c60_g10_d3_om6}
\end{figure*}
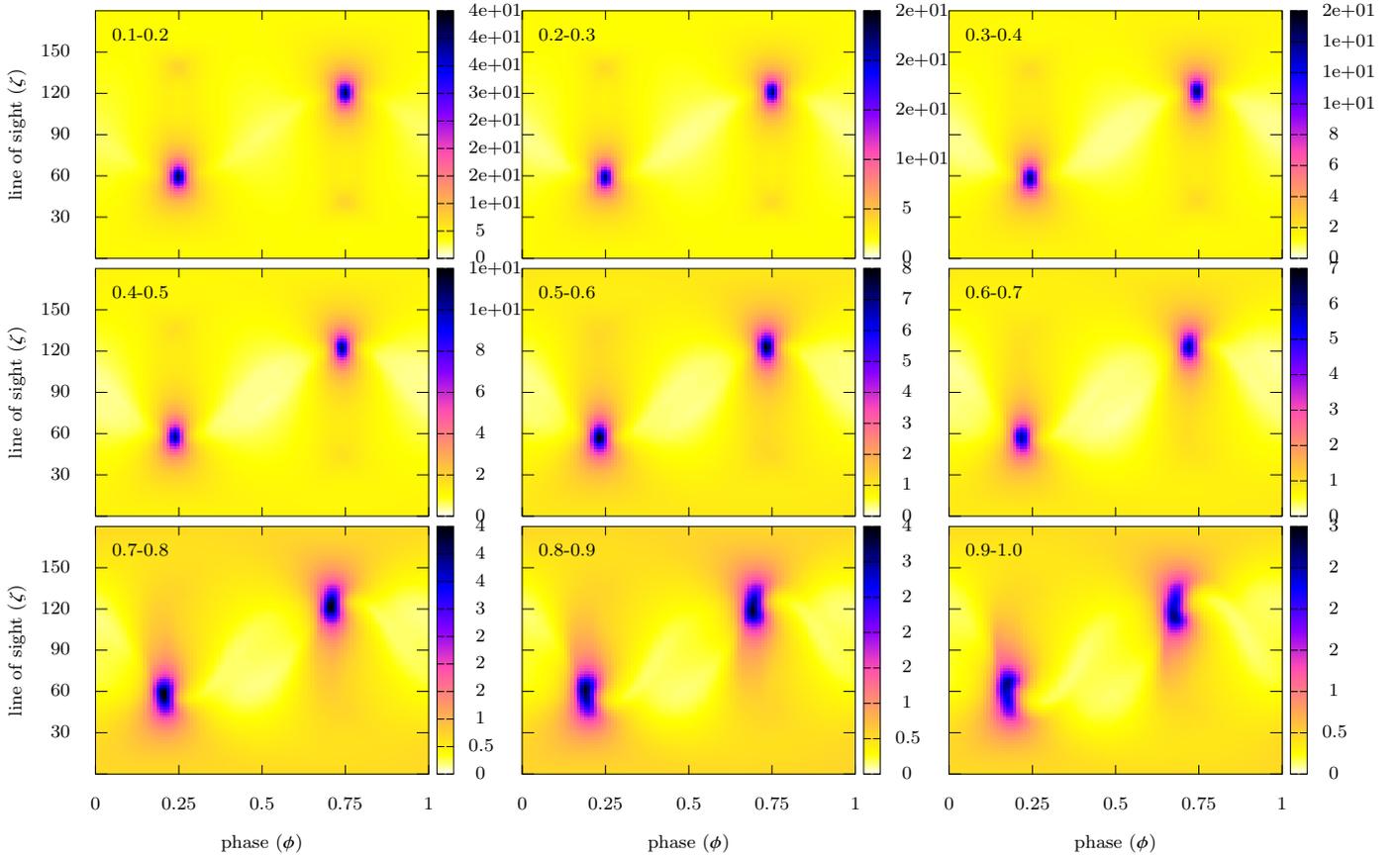
\begin{figure*}
	\centering
	\resizebox{\textwidth}{!}{\input{skymaps_cl_r00025_b-3_c60_g10_d3_om8}}
	\caption{Sky maps for a 100~ms pulsar, the range $r_{\rm in}-r_{\rm out}$ as shown in the labels and $n\propto r^{-3}$. The magnetic field strength is $b=10^{-3}$, $\chi=60\degr$ and $E=5$~GeV.}
	\label{fig:skymaps_cl_r00025_b-3_c60_g10_d3_om8}
\end{figure*}
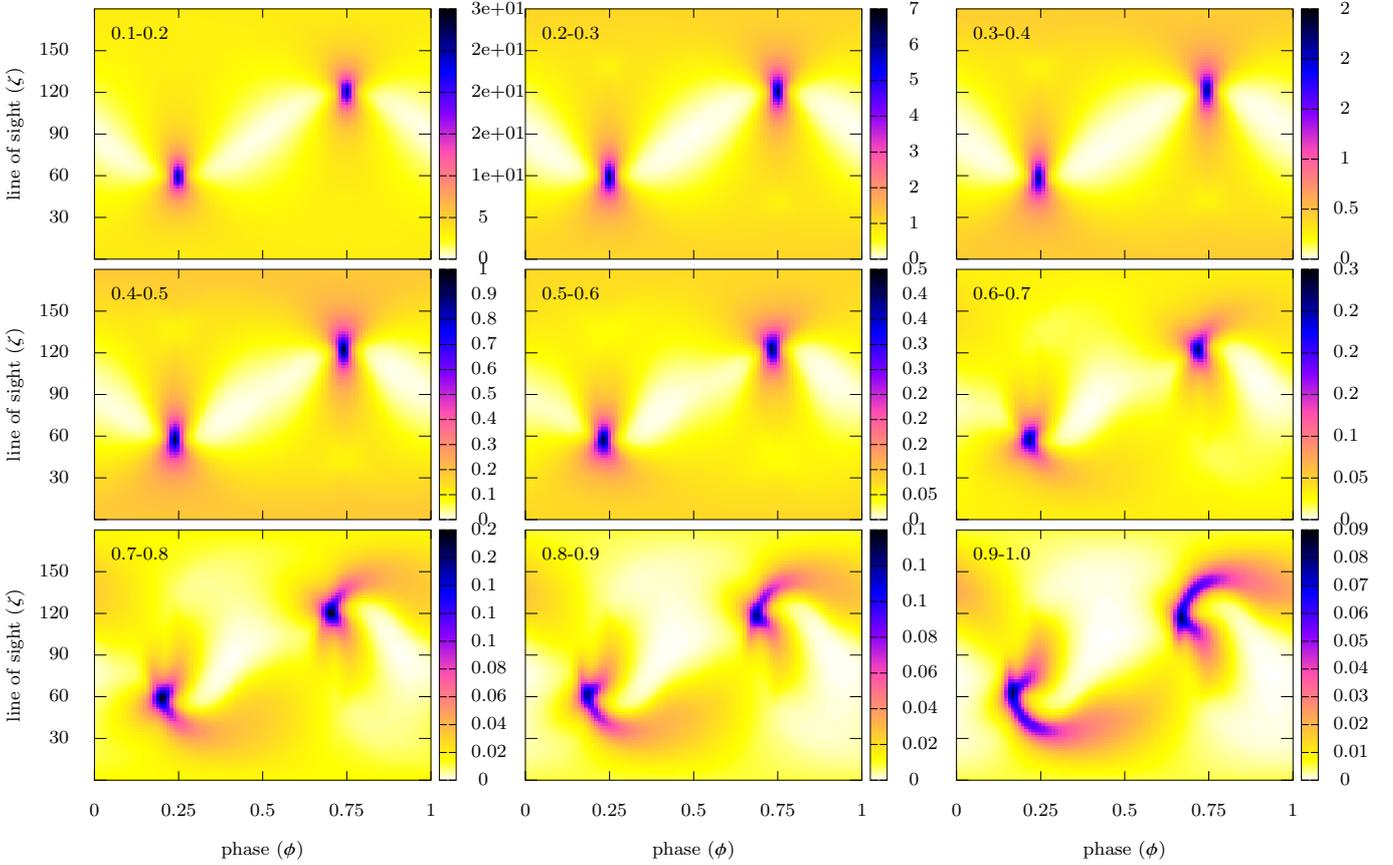
\begin{figure*}
	\centering
	\resizebox{\textwidth}{!}{\input{skymaps_cl_r00025_b-3_c60_g10_d3_om10}}
	\caption{Sky maps for a 100~ms pulsar, the range $r_{\rm in}-r_{\rm out}$ as shown in the labels and $n\propto r^{-3}$. The magnetic field strength is $b=10^{-3}$, $\chi=60\degr$ and $E=51$~GeV.}
	\label{fig:skymaps_cl_r00025_b-3_c60_g10_d3_om10}
\end{figure*}
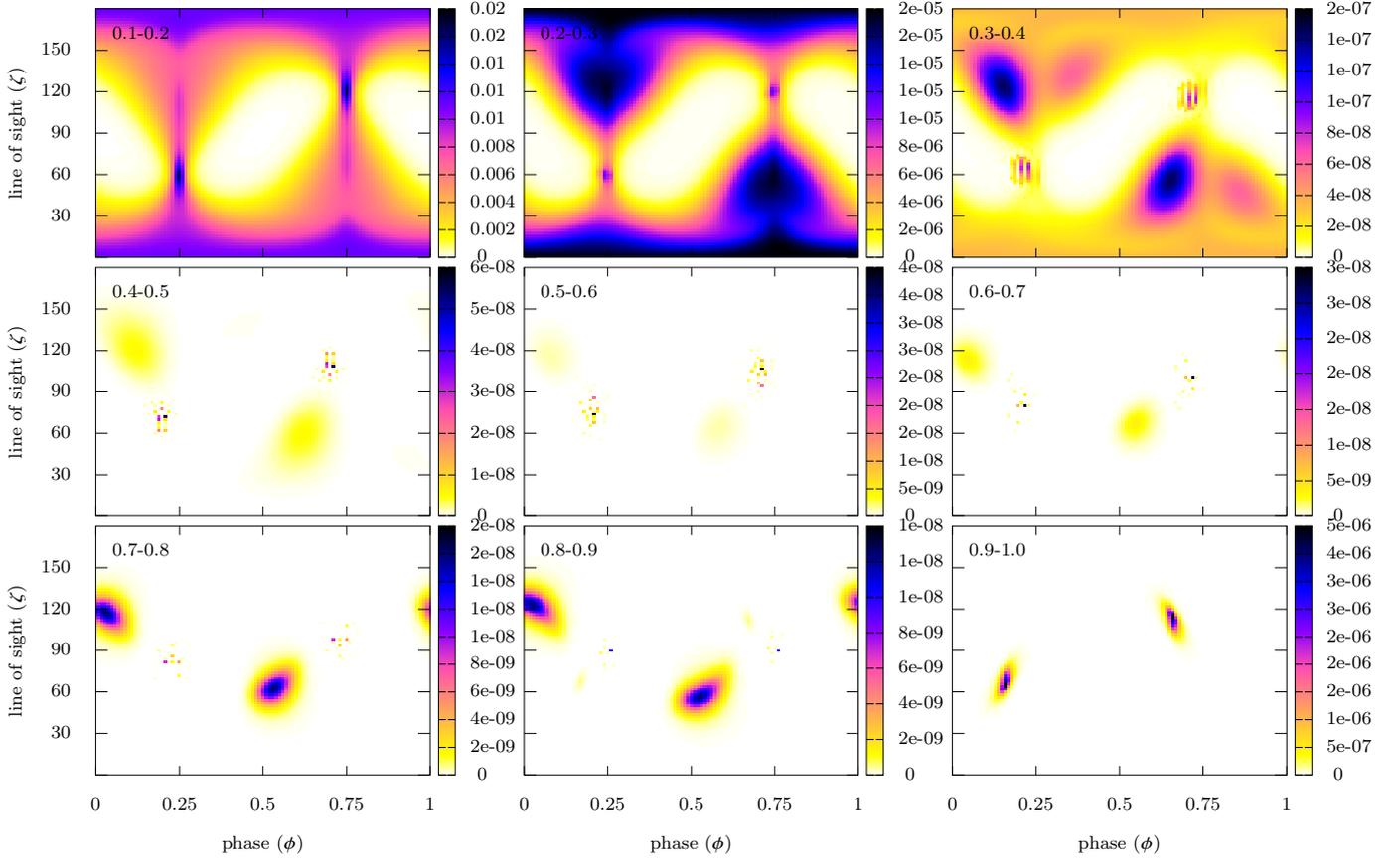
Below the cut-off energy, sky maps are very similar, whatever the location of the emitting shell. The overall flux decreases but the peak intensity remains close to the position of the polar caps. However a small spread in the peak profiles is observed when approaching the light-cylinder. In the vicinity of this light-cylinder, electrons and positrons do not contribute symmetrically to the light-curve because the electric field becomes comparable in intensity to the magnetic field. Therefore their velocity field differ significantly, leading to different individual light-curves, see bottom right panel of Fig.~\ref{fig:skymaps_cl_r00025_b-3_c60_g10_d3_om6}. The situation is even more prominent at $E=5.1$~GeV, Fig.~\ref{fig:skymaps_cl_r00025_b-3_c60_g10_d3_om8}. Above a height of $r/\rlight>0.7$, significant emission is produced outside the polar cap window, leading to S-shape intensity maps, see the case $r_{\rm in}=0.8$. This leads to a possible phase lag between radio and gamma-ray peaks. At the highest energies, $E=51$~GeV, Fig.~\ref{fig:skymaps_cl_r00025_b-3_c60_g10_d3_om10}, a new pulsed component different from the polar cap region appears, especially close to the light cylinder, for $r/\rlight>0.7$.

Because light-curves are almost energy insensitive across the spectrum, except around the cut-off frequency, it is worth to compute light-curves at a given typical energy of the power law band. Moreover, the contribution from electrons and positrons to the total intensity are usually not symmetrical. Therefore we also show their respective light-curves for an energy in the $\omega^{1/3}$ regime with $r_{\rm in}=0.2$ and obliquities $\chi=\{30\degr,60\degr,90\degr\}$. A normal pulsar is shown in Fig.~\ref{fig:skymaps_cl_r00025_b-3_ri0.2_ro1_g10_d3_om8} and a millisecond pulsar is shown in Fig.~\ref{fig:skymaps_cl_r005_b-6_ri0.2_ro1_g10_d3_om8} with respectively $\chi=30\degr$ in the first row, $\chi=60\degr$ in the second row and $\chi=90\degr$ in the third row. The total intensity is shown in red, the electron contribution in green and the positron contribution in blue. Electrons and positrons contribute similarly to the total flux. Three main light-curve profiles are observed. A first class of profiles showing an almost constant intensity where both electrons and positrons contribute in a symmetric manner. Pulsation is therefore very difficult to detect. A second class of profiles showing a prominent single pulse is observed mainly for obliquity much less than $\chi=90\degr$. A third class of double peaked structure is always seen when $\zeta \approx 90\degr$.
\begin{figure*}
 \centering
 \resizebox{\textwidth}{!}{\input{skymaps_cl_r00025_b-3_ri0.2_ro1_g10_d3_om8}}
 \caption{Sky maps for a 100~ms pulsar and $n\propto r^{-3}$, $r_{\rm in}=0.2$, $r_{\rm out}=1$, $E=5.1$~GeV. The total intensity is shown in red, the electron contribution in green and the positron contribution in blue.}
 \label{fig:skymaps_cl_r00025_b-3_ri0.2_ro1_g10_d3_om8}
\end{figure*}
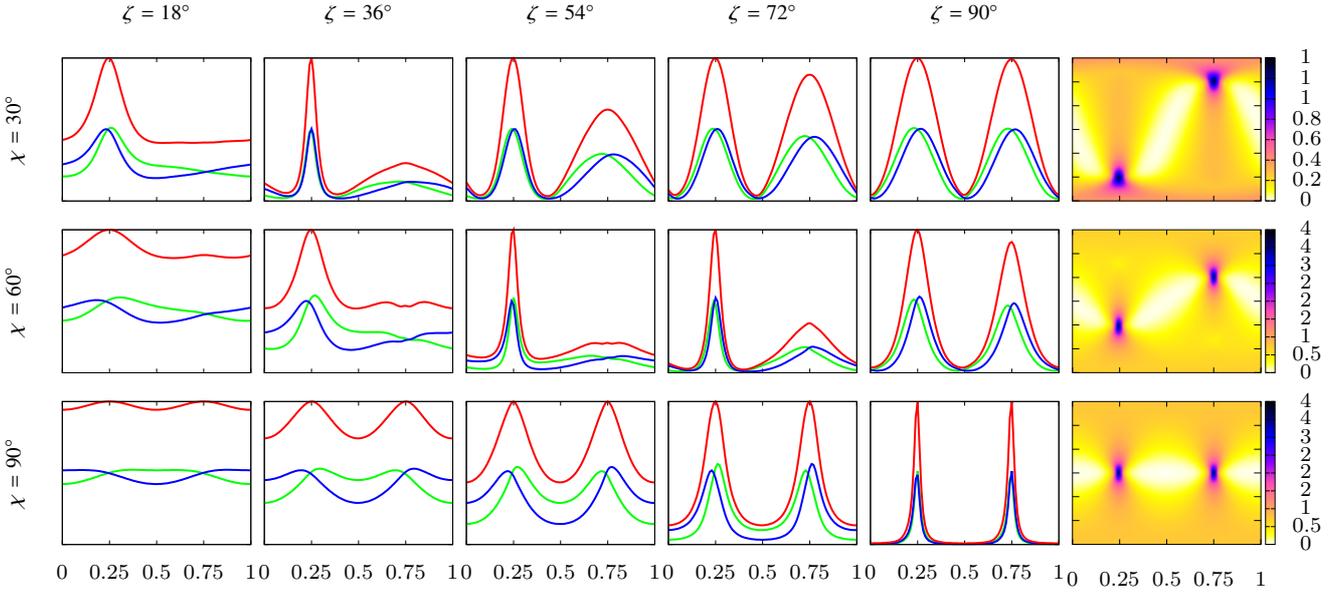
\begin{figure*}
 \centering
 \resizebox{\textwidth}{!}{\input{skymaps_cl_r005_b-6_ri0.2_ro1_g10_d3_om8}}
 \caption{Sky maps for a 5~ms pulsar and $n\propto r^{-3}$, $r_{\rm in}=0.2$, $r_{\rm out}=1$, $E=5.1$~GeV. The total intensity is shown in red, the electron contribution in green and the positron contribution in blue.}
 \label{fig:skymaps_cl_r005_b-6_ri0.2_ro1_g10_d3_om8}
\end{figure*}
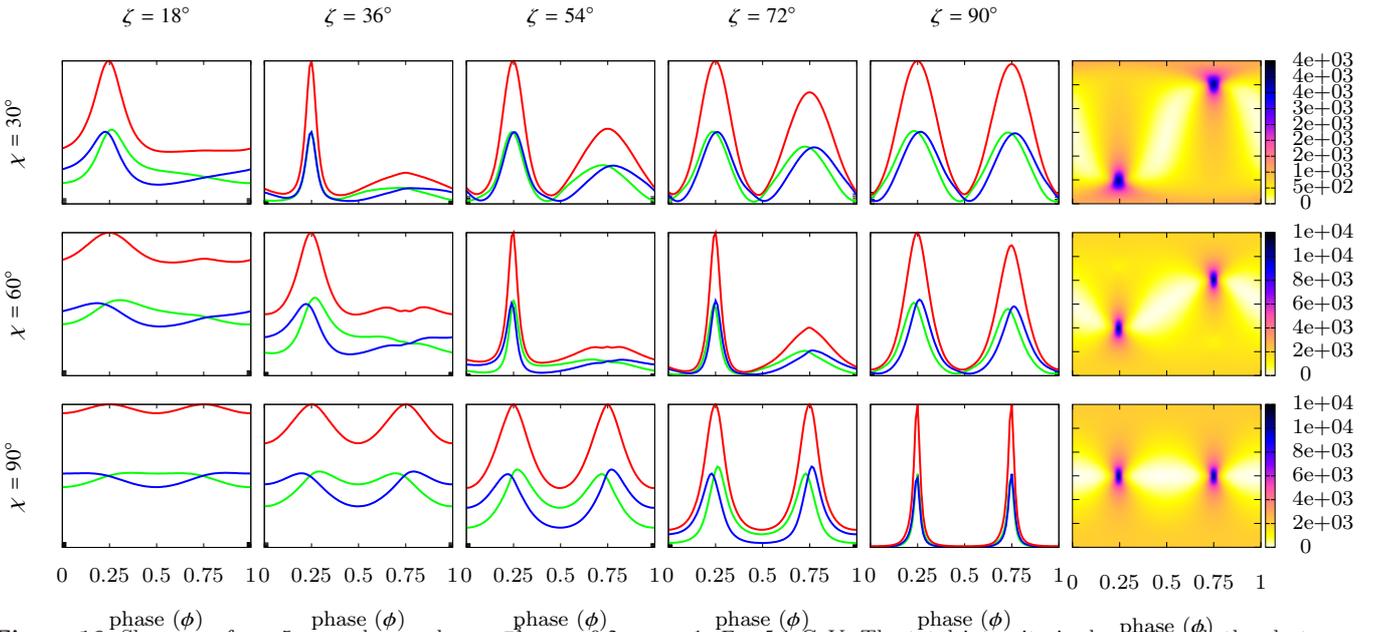
Millisecond and normal pulsars show similar trends. Moreover, electrons and positrons almost give the same contribution to the light curves, although some small discrepancies are seen in these particular cases.

If the inner boundary is shifted around the light-cylinder, then the pulses are no more aligned with the location of the magnetic poles. For instance, in Fig.~\ref{fig:skymaps_cl_r00025_b-3_ri1_ro5_g10_d3_om8}, the dominant gamma-ray peak can lead or trail the radio peaks located at phase 0.25 and 0.75 for $E=5.1$~GeV. Electron and positron contributions also differ drastically, leading to highly asymmetric pulse profiles. In almost all cases, emission deviates from zero only around the peaks. Below the cut-off, at $511$~MeV, the light-curves look even more complex, Fig.~\ref{fig:skymaps_cl_r00025_b-3_ri1_ro5_g10_d3_om6}, reflecting the complicated velocity field of the leptons.
\begin{figure*}
	\centering
	\resizebox{\textwidth}{!}{\input{skymaps_cl_r00025_b-3_ri1_ro5_g10_d3_om8}}
	\caption{Sky maps for a 100~ms pulsar and $n\propto r^{-3}$, $r_{\rm in}=1$, $r_{\rm out}=5$, $E=5.1$~GeV. The total intensity is shown in red, the electron contribution in green and the positron contribution in blue.}
	\label{fig:skymaps_cl_r00025_b-3_ri1_ro5_g10_d3_om8}
\end{figure*}
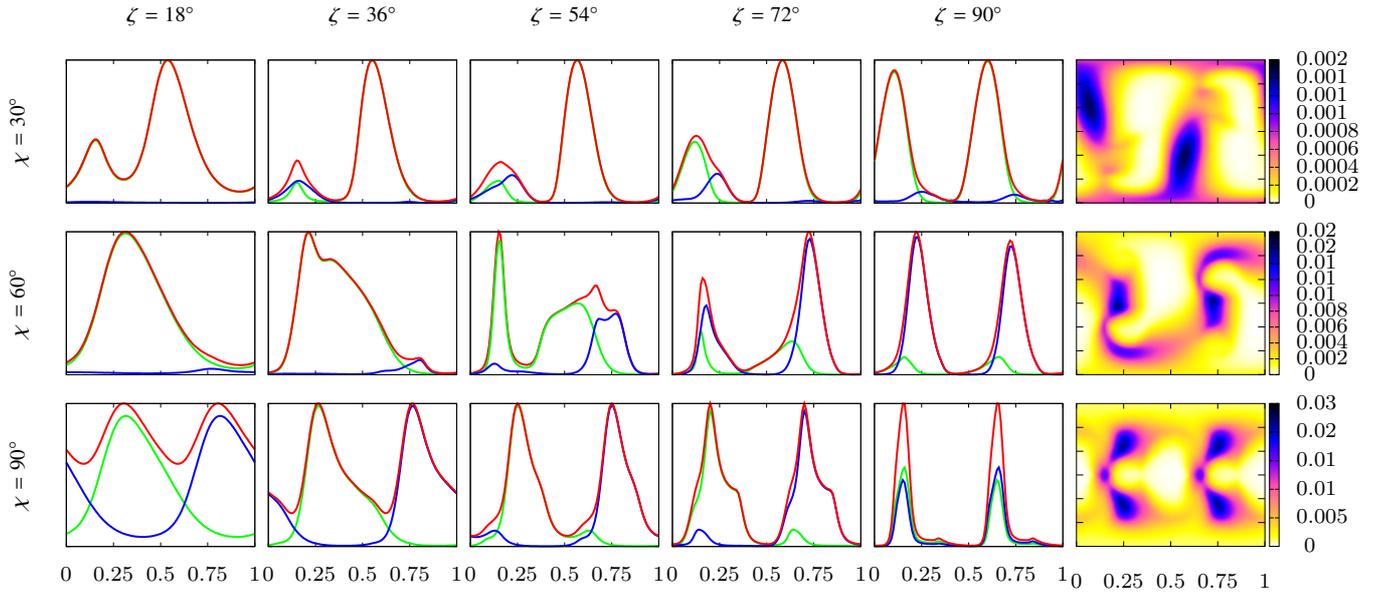
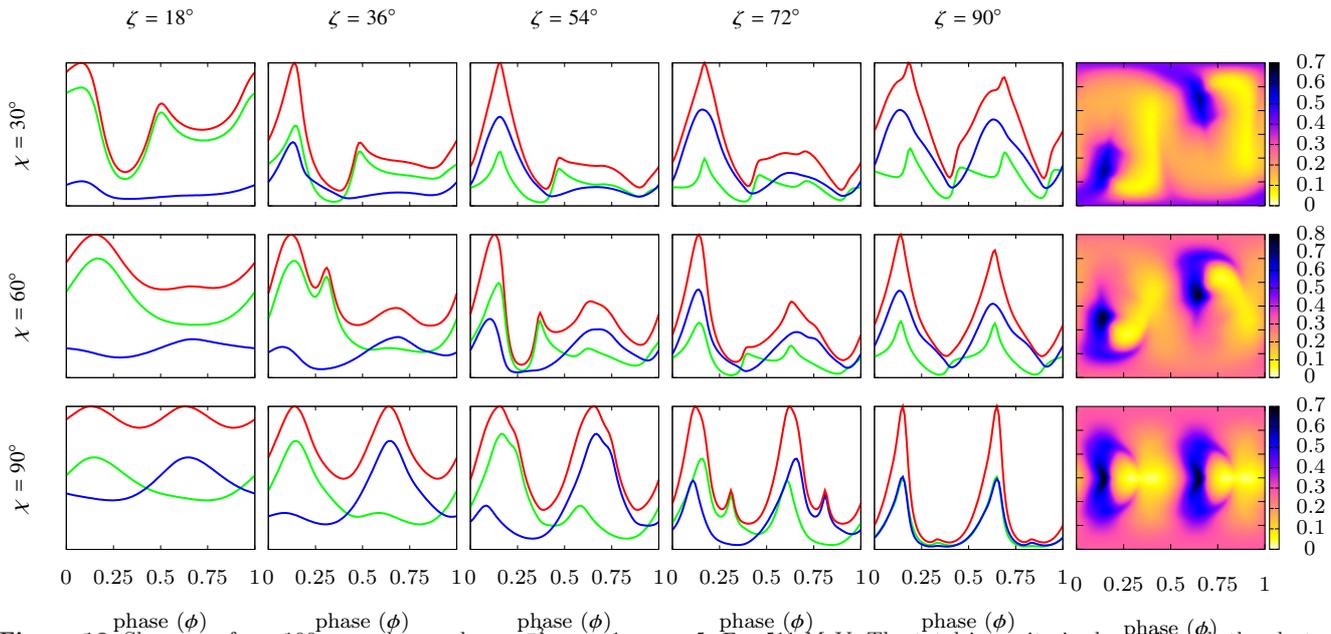
\begin{figure*}
	\centering
	\resizebox{\textwidth}{!}{\input{skymaps_cl_r00025_b-3_ri1_ro5_g10_d3_om6}}
	\caption{Sky maps for a 100~ms pulsar and $n\propto r^{-3}$, $r_{\rm in}=1$, $r_{\rm out}=5$, $E=511$~MeV. The total intensity is shown in red, the electron contribution in green and the positron contribution in blue.}
	\label{fig:skymaps_cl_r00025_b-3_ri1_ro5_g10_d3_om6}
\end{figure*}

From the point of view of possible light-curve profiles, our model can reproduce the same shapes as the one obtained by the competing models mentioned previously. Nevertheless, our new model naturally computes the evolution of a light curve with respect to energy and with realistic magnetic field strengths and rotation periods. Next we explore multi-wavelength light-curves in the following section.

\subsection{Multi-wavelength light-curves}

Fermi/LAT has shown that double peaked gamma-ray pulsar light-curves evolve with increasing energy towards a dominance of one pulse over the other and a possible shrinking of the pulse width. In the context of our model, we investigate the light-curve evolution with energies from MeV up to sub-TeV, picking out a subset of simulation parameters similar to the previous ones such that $\chi=60\degr$, $r_{\rm in}=0.2$, $r_{\rm out}=1$, $q=\{1,3\}$ and $\zeta=40\degr$. For a normal pulsar, the evolution with photon energy~$E$ is shown in Fig.~\ref{fig:courbe_lumiere_r00025_b-3_ri0.2_ro1_g10_dx_z40}. In this particular case, the second peak is dominant at low energy $E \lesssim 10$~GeV, both peaks become equal in intensity at $E\approx10$~GeV and the first peak dominates above $E=50$~GeV. Note also that the first pulse, almost undetectable and wide at lowest energy becomes intense and sharper at the highest energies. These conclusions do not depend on the density profile, the $q=1$ case on the right column shows similar trend as the $q=3$ case on the left column.
\begin{figure*}
 \centering
 \resizebox{0.9\textwidth}{!}{\input{courbe_lumiere_r00025_b-3_c60_ri0.2_ro1_g10_dx_z60}}
 \caption{Light curves depending on photon energy~$E$ for a 100~ms pulsar with $\chi=60\degr$, $r_{\rm in}=0.2$, $r_{\rm out}=1$, $q=\{1,3\}$ on the \{rihgt, left\} column and $\zeta=60\degr$.}
 \label{fig:courbe_lumiere_r00025_b-3_ri0.2_ro1_g10_dx_z40}
\end{figure*}

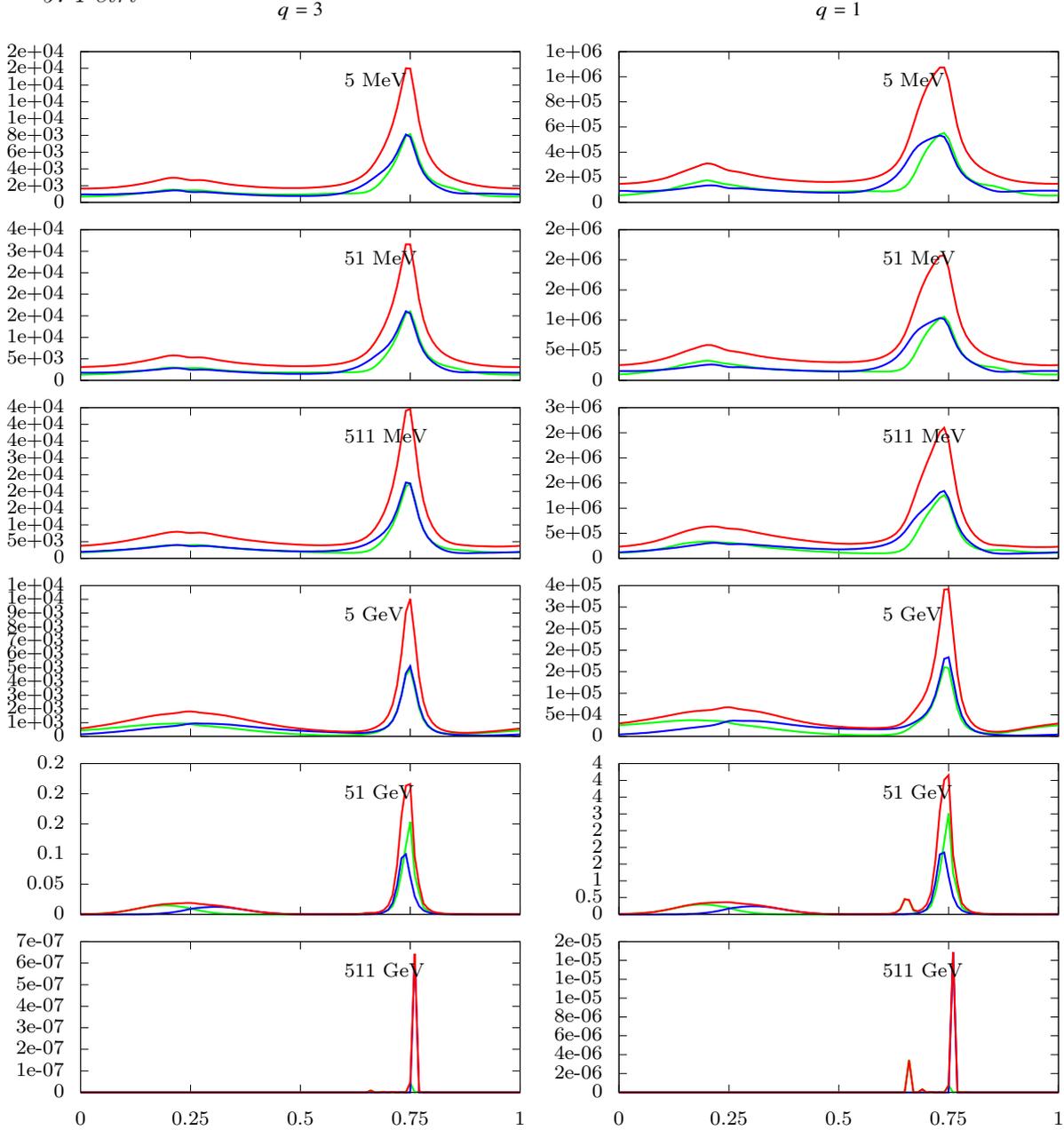
\begin{figure*}
 \centering
 \resizebox{0.9\textwidth}{!}{\input{courbe_lumiere_r005_b-6_c60_ri0.2_ro1_g10_dx_z60}}
 \caption{Light curves depending on photon energy~$E$ for a 5~ms pulsar with $\chi=60\degr$, $r_{\rm in}=0.2$, $r_{\rm out}=1$, $q=\{1,3\}$ on the \{right, left\} column and $\zeta=60\degr$.}
 \label{fig:courbe_lumiere_r005_b-6_ri0.2_ro1_g10_dx_z60}
\end{figure*}

For a millisecond pulsar, the first peak intensity increases with photon energy too, as for the normal pulsar, but at the highest energies, the second pulse remains dominant, the first one almost disappearing, Fig.~\ref{fig:courbe_lumiere_r005_b-6_ri0.2_ro1_g10_dx_z60}. Here also, not much differences are reported between the $q=1$ and the $q=3$ cases. Consequently, several scenarios are possible for main pulse/interpulse dominance depending on period and magnetic field strength.
The variation in the peak intensity ratio is a consequence of the particular behaviour of the phase-resolved spectra presented in section~3.1. In Fig.~\ref{fig:spectre_resolu_r00025_b-3_ri02_g10_d3} for the normal pulsar, the second peak overtakes the first peak in the last visible point in $\log(E/MeV)$ between [4,4.5]. Such overtaking is not seen in the millisecond pulsar case plotted in Fig.~\ref{fig:spectre_resolu_r005_b-6_ri02_g10_d3}.

The values of magnetic field strength and rotation period are adjusted to fit the second Fermi/LAT pulsar gamma-ray catalogue, especially the cut-off energy. But how are these cut-off energies related to the fundamental parameters of our model? This is the question we investigate in the next section.

\subsection{Cut-off energy}

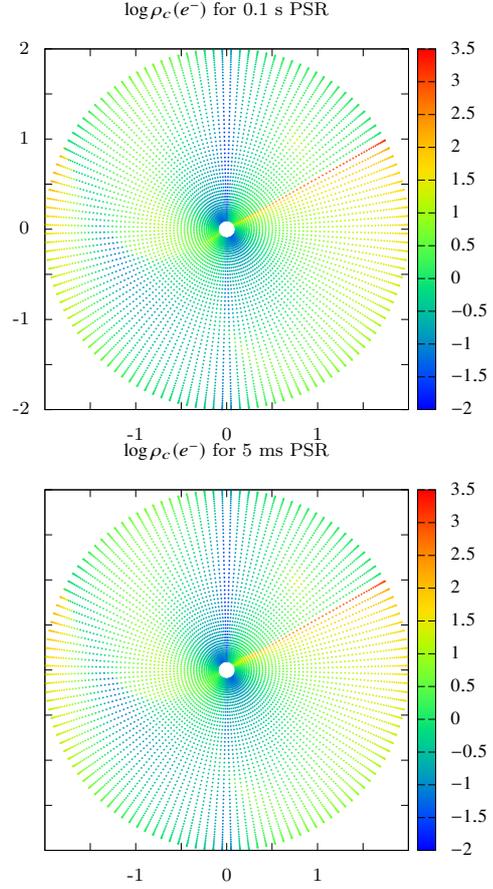
\begin{figure}
\centering
\resizebox{0.5\textwidth}{!}{\input{courbure_s_ms_c60}}
\caption{Electron curvature radius~$\rho_{\rm c}$ for a 100~millisecond pulsar, upper panel and a 5~millisecond pulsar, lower panel, with both $\chi=60\degr$.}
\label{fig:courbure_s_ms_c60}
\end{figure}

The cut-off energy~$E_{\rm cut}$ is usually obtained by fitting the Fermi/LAT spectra by a power-law with exponential or sometimes sub-exponential cut-off. As such fits seem not very robust against the sub-exponential coefficient and are subject to errors related to the power law index used below the cut-off, we prefer to use a more robust value simply by looking for the maximum in the spectral energy distribution. This definition is independent of any assumption about the fit. This maximum in the spectral energy distribution is similar to the apex energy defined by \cite{renault-tinacci_phase_2015}. Results for the energy at the maximum flux~$E_{\rm cut}=k_{\rm cut}\,\me\,c^2$ are shown in Fig.~\ref{fig:Ecut_r00025_c60_g10_nr64} for a normal pulsar and in Fig.~\ref{fig:Ecut_r005_c60_g10_nr64} for a millisecond pulsar. 
\begin{figure}
 \centering
 \resizebox{0.5\textwidth}{!}{\input{Ecut_r0.0025_c60_q0_g10_nr64}}
 \caption{Cut-off energy~$E_{\rm cut}$ for a 100~ms pulsar with $q=3$ and $\chi=60\degr$. The 3/4 power law for $\tilde{B}$ is also shown.}
 \label{fig:Ecut_r00025_c60_g10_nr64}
\end{figure}
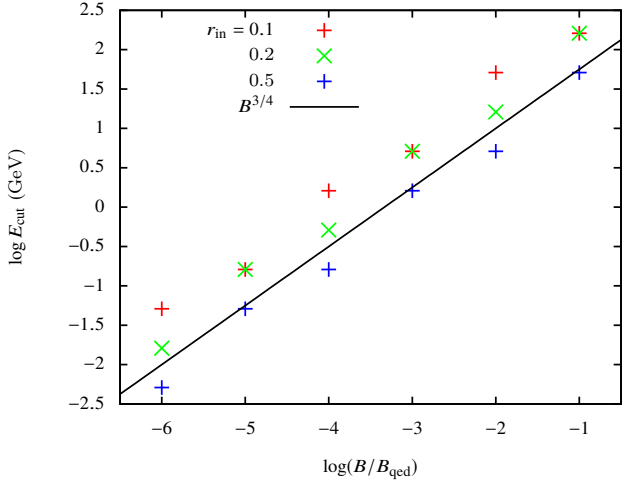
\begin{figure}
 \centering
 \resizebox{0.5\textwidth}{!}{\input{Ecut_r0.05_c60_q0_g10_nr64}}
 \caption{Cut-off energy~$E_{\rm cut}$ for a 5~ms pulsar with $q=3$ and $\chi=60\degr$. The 3/4 power law for $\tilde{B}$ is also shown.}
 \label{fig:Ecut_r005_c60_g10_nr64}
\end{figure}
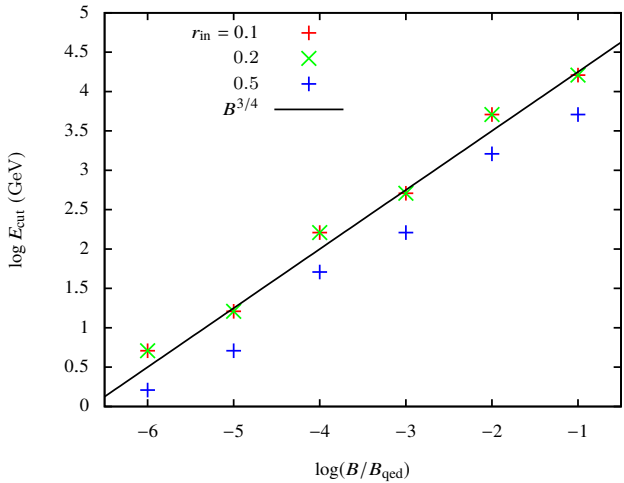
Both cut-off energies follow the law
\begin{equation}
\label{eq:CutOff}
 k_{\rm cut} = \left(\frac{3}{2}\right)^{7/4} \, \left(\frac{R}{\alpha_{\rm sf} \, \rlight}\right)^{3/4} \, \left(\frac{\rho_c}{\lambdabar_c}\right)^{1/2} \, \tilde{B}^{3/4}
\end{equation}
derived from eq.~(\ref{eq:OmegaCut}), (\ref{eq:GammaRR}) and (\ref{eq:Enormee}). The zig-zag curve is an artefact due to the finite number of frequency points~$\epsilon$ used in regular intervals of $1/2$ in a logarithmic scale of $\log(\epsilon/\me\,c^2)$.
Both cut-off energies follow the $\tilde{B}^{3/4}$ law but with different numerical constant in front of it related to the $R/\rlight$ ratio and to the curvature~$\rho_c$. Note also that the curvature $\rho_c$ is insensitive to the pulsar period or in other words insensitive to the ratio $R/\rlight$ because it is related to the electromagnetic field topology that is almost the same whatever the pulsar period constrained by the ratio $R/\rlight$. At large distances, it reduces to a plane electromagnetic wave as shown in \cite{petri_multipolar_2015} with a relative amplitude between electric part and magnetic part independent of $\Omega$. Indeed, inspecting Fig.~\ref{fig:courbure_s_ms_c60} where two maps of curvature radius~$\rho_{\rm c}$ are shown, one for a 100~ms pulsar on the upper panel and one for a 5~ms pulsar on the lower panel, we do not observe any significant difference, whether close to the surface nor around the light-cylinder or beyond it. Consequently, the cut-off scale as $k_{\rm cut} \propto (B/\rlight)^{3/4} \propto (\Omega\,B)^{3/4}$ for any pulsar and because this product $\Omega\,B$ is very similar for all pulsars, we do not expect a large spread in cut off energies. Noting that the product $\Omega\,B$ is proportional to $\sqrt{\dot P/P}$ (assuming that $B \propto \sqrt{P\dot P}$), the cut off scales with the characteristics~$\tau_{\rm c}$ according to $k_{\rm cut} \propto \tau_{\rm c}^{-3/8}$.

To finish our discussion, we need to adjust the particle density number to arrive at the correct energy flux and at the total gamma-ray luminosity. This is done in the next section, following the data from the second gamma-ray pulsar catalogue of Fermi/LAT (2PC).

\subsection{Luminosity}

Pulsar gamma-ray luminosities in the 2PC are estimated by integrating the flux in the energy range 100~MeV-100~GeV. Because of the sharp spectral cut-off around 1-5~GeV, the upper bound of this range, when well above this cut-off energy, does not impact on the total luminosity. A crude guess of the gamma-ray luminosity is simply given by the total flux radiated at this cut-off energy. In order to compare with Fermi/LAT data, we compute the total gamma-ray luminosity between 161~MeV and 161~GeV. The upper bound difference has no impact on the real gamma-ray luminosity as the cut-off is well below 100~GeV. The lower bound taken to be 161~MeV instead of 100~MeV has also little impact on the luminosity estimate as the spectrum peaks around 1-5~GeV. The value of 161 comes from our energy discretization grid which is uniform on a $\log$ scale thus getting sky-maps computed at energies in the form~$10^{a/2} \times \me\,c^2$ where $a$ is an integer. Thus energies are decades in $\sqrt{10} \, \me\,c^2 = 1.61$~MeV or decades in $\me\,c^2 = 511$~keV.

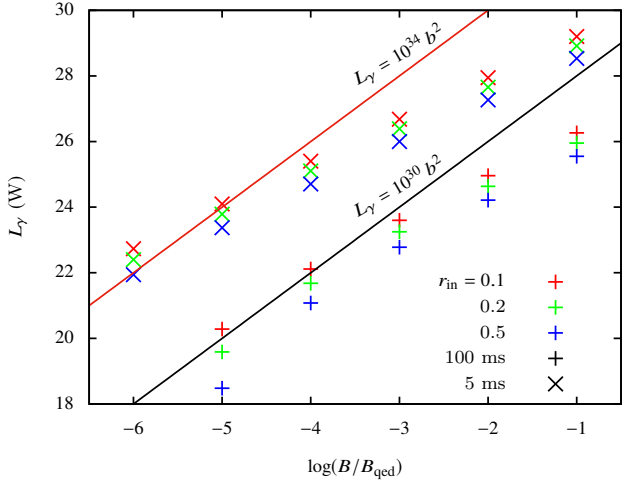
\begin{figure}
 \centering
 \resizebox{0.5\textwidth}{!}{\input{luminosite_rx_c60_g10_nr128}}
 \caption{Gamma-ray luminosity for normal and millisecond pulsars depending on the magnetic field strength~$b=B/\BQ$ and with a density profile~$q=3$. Some dependence on $b^2$ are also shown for reference.}
 \label{fig:luminosite_rx_c60_g10_nr128}
\end{figure}

In order to compute the total kinetic rotational energy losses, we remember that the spindown luminosity depends on the period~$P$, its derivative~$\dot P$ and the stellar moment of inertia~$I$ according to
\begin{equation}
 \dot E = 4 \, \upi^2 \, I \, \dot P \, P^{-3} .
\end{equation}
From the magnetodipole losses, the magnetic field strength is related to pulsar timing by
\begin{equation}
 B \approx \numprint{e8} \, \SIunits{\tesla} \, \sqrt{ \left(\frac{P}{1~\SIunits{\second}}\right) \, \left(\frac{\dot P}{\numprint{e-15}}\right) } 
\end{equation}
thus the spindown is evaluated to
\begin{equation}
 \dot E = 4 \, \upi^2 \, \frac{I}{\numprint{e31}} \, B^2 \, P^{-4} \approx \numprint{7.7e27} ~ \SIunits{\watt} \left( \frac{B}{\BQ} \right)^2 \, \left( \frac{P}{1~\SIunits{\second}}\right)^{-4}.
\end{equation}
Fig.~\ref{fig:luminosite_rx_c60_g10_nr128} shows the gamma-ray luminosity for normal and millisecond pulsars depending on the magnetic field strength. In all points, we used an obliquity~$\chi=60\degr$ and a density profile with~$q=3$. The plus sign $+$ depicts normal pulsars and the cross sign $\times$ depicts millisecond pulsars. The luminosity depends on $B^2$ for both kind of pulsars. For a fixed magnetic field strength, millisecond pulsars are more luminous than normal pulsars. This is because the cut-off energy in millisecond pulsar is unrealistically high compared to normal pulsars for a same magnetic field strength~$B$. Moreover, according to the Fermi/LAT gamma-ray pulsar catalogue \citep{abdo_second_2013}, millisecond pulsars luminosities are mostly in the range $10^{25}-10^{27}$~W, requiring magnetic field strengths of $B=10^{-5}-10^{-6}\,\BQ$, in accordance with observations. On the other side, normal pulsars possess luminosities mostly in the range $10^{26}-10^{30}$~W, requiring magnetic field strengths $B=10^{-2}-10^{-3}\,\BQ$, also in accordance with current wisdom. The inflection of the curve at low and high magnetic field strength~$b$ is an artefact due to the cut-off energy being respectively well below or well above the range used to computed the gamma-ray luminosity which is [161~MeV,161~GeV]. Correcting for this effect, the gamma-ray luminosity is proportional to the square of the magnetic field strength and shown by the line $L_\gamma = 10^\eta \, b^2~ \SIunits{\watt}$ in the plot, with $\eta \gtrsim 30$.

\begin{figure}
 \centering
 \resizebox{0.5\textwidth}{!}{\input{luminosite_Edot_c60_g10_d3_nr128}}
 \caption{Gamma-ray luminosity~$L_\gamma$ for normal and millisecond pulsars depending on the spindown luminosity~$\dot E$~with a density profile~$q=3$}.
 \label{fig:luminosite_Edot_c60_g10_d3_nr128}
\end{figure}
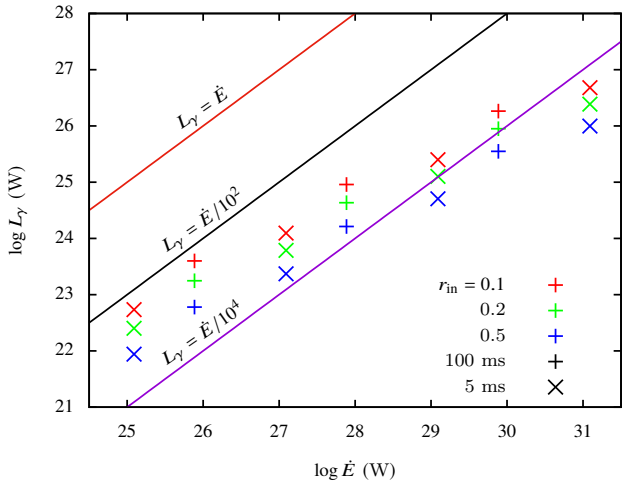

Fig.~\ref{fig:luminosite_Edot_c60_g10_d3_nr128} shows the gamma-ray luminosity for normal and millisecond pulsars depending on the spindown luminosity $\dot E$. Lines $L_\gamma = \eta \, \dot E$ are also shown for reference. For emission starting at an altitude $h>0.1\,\rlight$ the gamma-ray luminosity never exceeds the spindown power as required from the basic principle of energy conservation. 
It is actually 2 to 3 orders of magnitude less than $\dot E$. To increase $L_\gamma$, we can invoke the pair multiplicity factor~$\kappa$ constraining it to $\kappa = \numprint{e2}-\numprint{e4}$ to reconcile this plot with Fermi/LAT second catalogue, depending on the particular pulsar fitted. Consequently, our simple model reproduce the main spectral and timing properties of gamma-ray pulsars without any violation of basic physical principles. Here again, the inflection at high spindown luminosities~$\dot E$ is an artefact. Correcting for this effect, the gamma-ray luminosity is proportional to the spindown power for both millisecond and normal pulsars, shown by the line $L_\gamma = \eta \, \dot E$ in the plot, with $\eta \leq 1$.

\subsection{Detection of the VHE component}

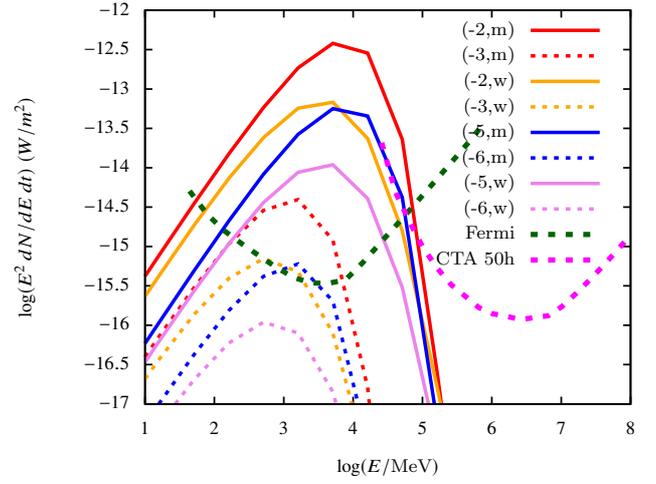
\begin{figure}
 \centering
  \resizebox{0.5\textwidth}{!}{\input{fermi_sensitivity}}
  \caption{Gamma-ray flux estimates for normal and millisecond pulsars from magnetospheric emission (m, $r_{\rm in}=0.5$ and $r_{\rm out}=1$) and wind emission (w, $r_{\rm in}=1$ and $r_{\rm out}=5$) with $\kappa=1$. $\log(b)=\{-2,-3\}$ for normal pulsars and  $\log(b)=\{-5,-6\}$ for millisecond pulsars as shown in the legend. Fermi and CTA sensitivities are plotted for the most optimistic configurations.}
  \label{fig:FermiCTAsensitivity}
\end{figure}

Fermi/LAT already reported more than 200 gamma-ray pulsars. It offers a good sample to extrapolate emission at sub-TeV energies in the window of atmospheric Cerenkov telescopes such as HESS-II who detected Vela above 20~GeV \citep{collaboration_first_2018} and the upcoming Cerenkov Telescope Array (CTA) observatory \citep{vercellone_next_2014}. It is therefore opportune to show expectation of gamma-ray fluxes in the sub-TeV range for emission in the radiation reaction limit. A representative sample of spectra for normal and millisecond pulsars is shown in Fig.~\ref{fig:FermiCTAsensitivity} with comparison to Fermi and CTA sensitivities. The pair multiplicity is $\kappa=1$ by default, the obliquity is $\chi=60\degr$. Emission within the magnetosphere is reported as~(m) meaning $r_{\rm in}=0.5$ and $r_{\rm out}=1$ whereas wind emission is reported as~(w) meaning $r_{\rm in}=1$ and $r_{\rm out}=5$. Normal and millisecond pulsars with respectively $\log(b)=-2$ and $\log(b)=-5$ are marginally detectable above 100~GeV with CTA South in 50h whatever the emission location, within the magnetosphere or within the wind. Lower magnetic fields drastically reduce the flux as well as the cut-off energy. A more realistic multiplicity $\kappa\gg1$ would increase the flux but not the cut-off. We do not expect any emission above several hundreds of GeV, irrespective of the leptons distribution function in space and momentum. Nevertheless our crude analysis still requires a precise and careful analysis for individual pulsars to make clear and sensible predictions in the sub-TeV range.

\section{Discussion}
\label{sec:Discussion}
 
The light-curves and spectra exposed in our simplistic view of pulsar magnetospheric acceleration and radiation mechanisms gives already interesting results worthwhile to extend although we used very little inputs and fitting parameters except for those well constrained by observations such as the period~$P$, its derivative~$\dot P$, and the neutron star radius~$R$. There is no doubt that our approach could benefit from some improvement to better fit particular pulsars. Nevertheless, in the next section we focus on some problems and non elucidated electrodynamics about pulsar magnetospheres related to the complex geometry of trajectories and particle behaviour within this magnetosphere, pointing out the limitation of our work.

Our model deals with realistic pulsar periods and magnetic fields, producing spectral high-energy features in agreement with gamma-ray observations performed by the Fermi/LAT \citep{abdo_second_2013}. We are also able to produce double-peaked light-curves with variable peak intensity ratio, variable shapes and widths. Nevertheless, these light-curve profiles rely heavily on the underlying spatial lepton distribution within the magnetosphere. We certainly took a too simplistic view of spherically symmetric repartition decreasing in radius with a simple power law in radius depicted by $n \propto r^{-q}$. Assuming emission emanating from any point is admittedly too a crude approximation. However, the energy balance between electric acceleration and radiation reaction is not impacted by the spatial distribution of particles except for some corrections due to the back reaction of the plasma onto the field. However, the electric current generated by the plasma flow close to the surface remains too weak to appreciably perturb the electromagnetic field. We conclude that the results obtained about spectral shape and cut-off is robust and insensitive to the geometry except for small changes imprinted by the curvature radius.

\subsection{Particle flow and trapping}

A proper account of particle flow within this magnetosphere according to the radiation reaction limit prescription requires to solve for the particle number conservation law supplemented with an appropriate source term of electron/positron pair formation. The arbitrariness of our spatial distribution would then be transposed to the arbitrariness of pair creation efficiency. We could get various spatial distribution by changing this source function. We could impose pair cascading only around the polar caps, or spread out over the whole neutron star surface or even within some special regions in the magnetosphere up to the light-cylinder. Clearly, such conclusions would not be as robust as on the energy budget. Solving for the pair formation is a difficult task about the microphysics which inevitably translates into a geometrical problem of localising the source of leptons. It immediately reflects into the light-curve profiles as an observable. We will not go into such refinement but stress that the damped motion implied by radiation reaction produces three kind of particle flows:
\begin{enumerate}
	\item outflowing particles escaping the neutron star and its magnetosphere, forming the base of the pulsar (striped) wind.
	\item trapped particles, staying in a defined region close to the surface for a long time with respect to the pulsar period. These regions are identified as $\mathbf{E} \cdot \mathbf{B} = 0$ surfaces or volumes and called force-free surface.
	\item returning particles that hit the neutron star surface.
\end{enumerate}
These classes of trajectories are found by a direct numerical integration of the velocity eq.~(\ref{eq:VRR}) assuming a background Deutsch field.

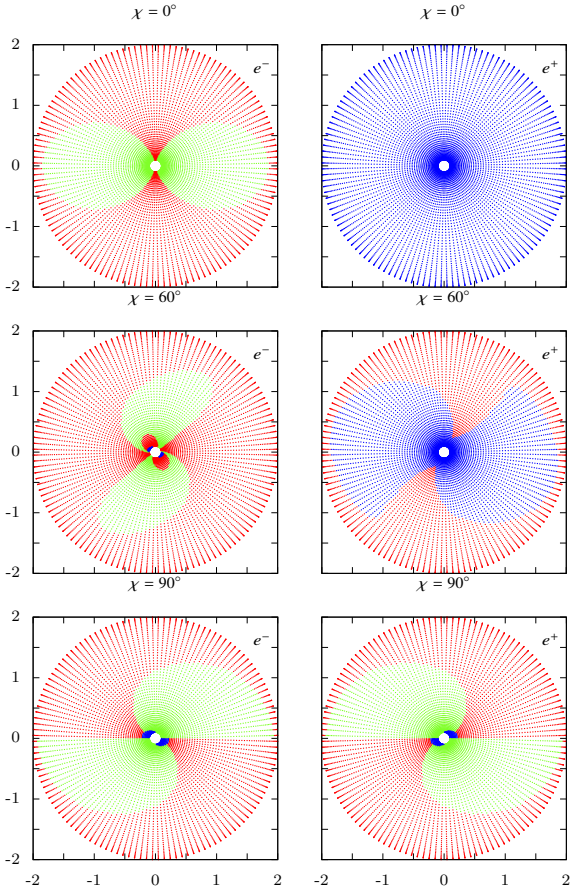
\begin{figure}
\begin{center}
\resizebox{0.9\columnwidth}{!}{\input{cavite_cx}}
\end{center}
\caption{Meridional cross section in the $xOz$ plane of outflowing (red points), trapping (green points) and returning (blue points) regions for different obliquities $\chi=\{0\degr,60\degr,90\degr\}$ and $R=0.1\,\rlight$. The left column denoted by $e^-$ is for electrons and the right column denoted by $e^+$ is for positrons.}
\label{fig:Cavite}
\end{figure}

The outflowing, trapped or returning motion depends on the initial position where the particle has been launched. The overall maps obtained by integration show large inner volumes where particles return to the star and sometimes are trapped. As a general trend however, for the outer most regions far from the stellar surface or well beyond the light cylinder, particles always escape to infinity because they essentially feel a plane vacuum electromagnetic wave with $\mathbf{E} \cdot \mathbf{B} \approx 0$. A typical cross section of these regions in the meridional plane $xOz$ is shown in Fig.~\ref{fig:Cavite} for several inclination angles with $\chi=\{0\degr,60\degr,90\degr\}$. The initial particle positions for escaping are shown with red dots, for trapping particles with green dots and for returning particles with blue dots.  Obviously, electrons (left column) and positrons (right column) do not share the same returning and outflowing regions. Trapped regions can be large too and being trapped is the privilege of only one species when the geometry is close to an aligned or to a counter-aligned rotator. We emphasize that these regions do not correspond to places where particles are actually trapped but to the starting point of the trajectory for which particles move to trapping regions. The same interpretation holds for escaping and returning trajectories.

Trapped regions are spread around the so-called force-free surface defined by $\mathbf{E} \cdot \mathbf{B} = 0$. The three different regions are not the same for electrons and positrons. We note however that the situation is symmetric with respect to the obliquity~$\chi$ and $\pi-\chi$, meaning that the cavities for electrons with obliquity~$\chi$ are the same as the cavities for positrons with obliquity~$\pi-\chi$. If the content in electrons and positrons within the magnetosphere is not exactly the same, it is in principle possible to differentiate the obliquity~$\chi$ from $\pi-\chi$ in the sky-maps and light-curves, departing therefore from the traditional symmetry with respect to the equatorial plane.

\subsection{Particle trajectories}

A detailed study of particle motion in the Deutsch electromagnetic field is out of the scope of this paper. However several interesting works about trajectories in this field or in the near quasi-static zone have been discussed in the literature. Some details can be found for instance in \cite{laue_acceleration_1986} for a perpendicular rotator with $\chi=90\degr$. The importance of trapping regions for arbitrary obliquity~$\chi$ is emphasized by \cite{finkbeiner_effects_1989}. It is possible that a fraction of particles stay in the electrospheric configuration given by a dome+torus geometry \cite{jackson_new_1976, krause-polstorff_electrosphere_1985, petri_global_2002}. \cite{zachariades_particle_1989} even found bounded trajectories outside the light-cylinder showing the complexity of defining particle motion and density in the surrounding of neutron stars when damping is included.

\begin{figure}
\begin{center}
	\resizebox{0.9\columnwidth}{!}{\input{carte_fuite_cx}}
\end{center}
\caption{Initial position on the stellar surface for outflowing (red points), trapping (green points) and returning (blue points) particles for $R=0.1\,\rlight$. The left column with $\chi=60\degr$ shows electrons~$e^-$ whereas the right column with $\chi=120\degr$ shows positrons~$e^+$.}
\label{fig:carte_fuite_cx}
\end{figure}
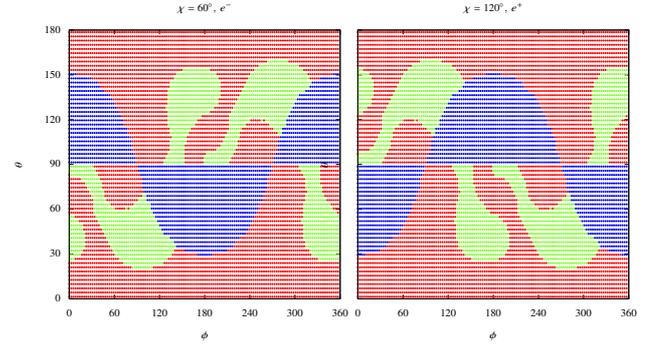

To better grasp the complexity of trajectories available for particles already in our simple vacuum model, we computed the long term motion of particles launched from the surface of the star. This should mimic the effect of pair creation in the vicinity of the star. Already at this surface, three kind of trajectories emerged as explained above: escaping, trapping and returning particles. This is shown in Fig.~\ref{fig:carte_fuite_cx} where the initial position of particles (on the stellar surface) is represented in spherical coordinates $(\phi,\theta)$. Escaping particles are depicted by red dots, trapping particles by green dots and returning particles by blue dots. Two symmetrical cases are shown, one for electrons~$e^-$ with $\chi=60\degr$, left column, and the other for positrons~$e^+$ with $\chi=120\degr$. The species of the opposite charge immediately returns to the star. We therefore obtained a dynamical structure that is not symmetric with respect to the angle $\chi$ and $\pi-\chi$. If the content in electrons and positrons differs to each other, we expect to get different magnetospheric configurations and therefore different emission properties (light-curves and spectra) with respect to the two geometries $\chi$ and $\pi-\chi$. The initial position of escaping, trapping and returning particles builds complicated shapes far from the standard polar cap geometry. It is actually related to the full electromagnetic field and not only to the magnetic part.

\begin{figure}
\begin{center}
	\resizebox{0.9\columnwidth}{!}{\input{trajectoire_fuite_cx}}
\end{center}
\caption{Equatorial $xOy$, left plot, and meridional $yOz$, right plot, projection of outflowing electrons for $\chi=60\degr$ and $R=0.1\,\rlight$.}
\label{fig:trajectoire_fuite_cx}
\end{figure}
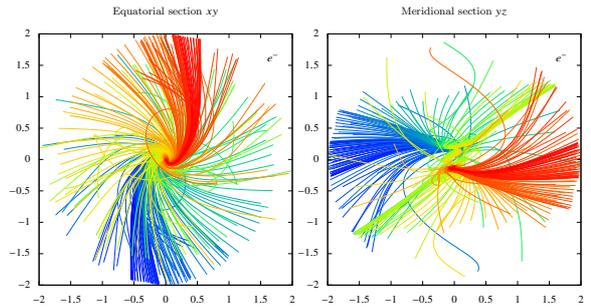

\begin{figure}
\begin{center}
	\resizebox{0.9\columnwidth}{!}{\input{carte_fuite_cx_onde}}
\end{center}
\caption{Angular distribution of escaping leptons at a distance $r=2\,\rlight$ for $R=0.1\,\rlight$. Left column for electrons~$e^-$ with $\chi=60\degr$ and right column for positrons~$e^+$ with $\chi=120\degr$. Low and high density regions for electrons and positrons are clearly visible.}
\label{fig:carte_fuite_cx_onde}
\end{figure}
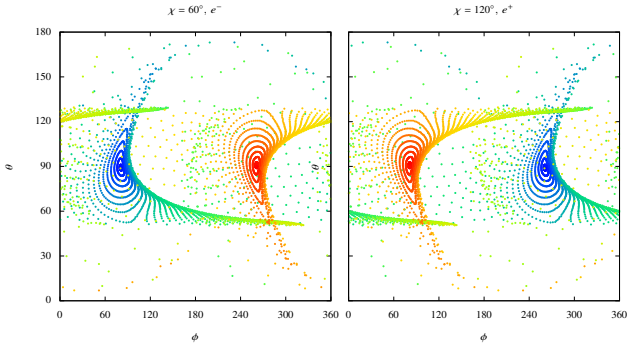

A sample of trajectories for escaping electrons with $\chi=60\degr$ is shown in Fig.\ref{fig:trajectoire_fuite_cx} where the anisotropic character of the filling is clearly visible. The left plot shows the projection onto the equatorial plane $xOy$ whereas the right plot shows the projection onto the meridional plane $yOz$. Close to the star, their trajectories can be complicated but outside the light-cylinder, they become almost radial as the electromagnetic wave tends to a plane wave propagating in the direction $\bmath e_{\rm r} \approx \bmath n \propto \bmath E \wedge \bmath B $.
At large distances, well outside the light cylinder, the distribution of leptons is concentrated in specific sky directions as shown in Fig.~\ref{fig:carte_fuite_cx_onde}. Some regions are devoid of electrons whereas other regions are devoid of positrons. Many electron trajectories tend to preferred directions in the sky as shown by these maps.

Studying single particle trajectories in a background electromagnetic field of a rotating magnetized neutron star is a full topic by itself. Such refinement must be included in a comprehensive description of pulsar electrodynamics but the scope of this paper was to focus mainly on high-energy emission from a simplistic model without resorting to large scale particle simulations.

\subsection{Invariants and particle dynamics}

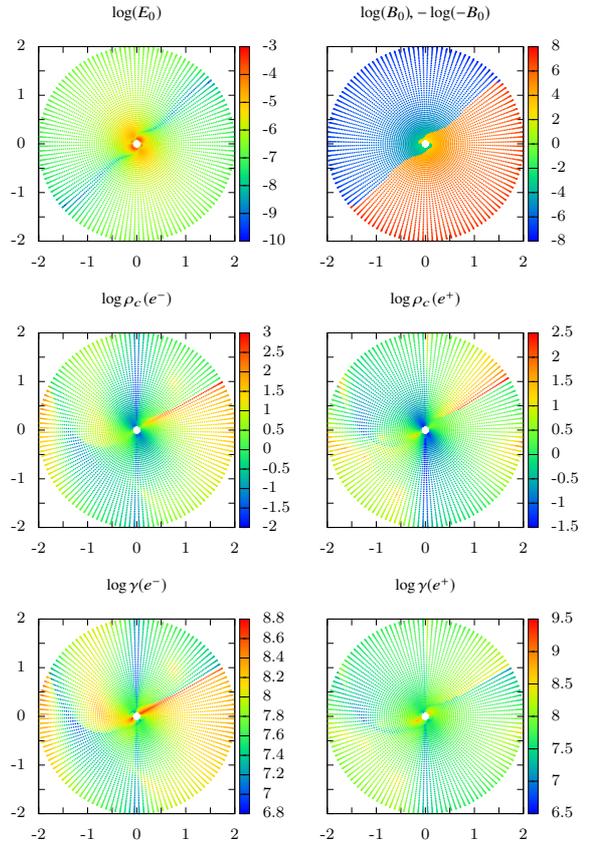
\begin{figure}
\begin{center}
	\resizebox{0.9\columnwidth}{!}{\input{invariant_c60}}
\end{center}
\caption{Important characteristics of a pulsar with $R=0.1\,\rlight$, $b=10^{-3}$ and $\chi=60\degr$. The electromagnetic invariants $E_0$ and $B_0$ are shown on a log scale in the upper panel in the left and right column respectively. The curvature radius, normalized to the light-cylinder radius, for electrons and positrons are shown in the middle panel, left and right column respectively. The Lorentz factor reached by these electrons and positrons are shown in a log scale in the lower panel, left and right column respectively.}
\label{fig:invariant_c60}
\end{figure}

In order to better understand the radiative properties of the magnetosphere, we plot several important geometrical and dynamical properties in the meridional plane $xOz$. Normalized units are used as explained in the previous section. The electromagnetic invariants, $E_0$ and $B_0$ intervening in the velocity field are shown in the upper panel of Fig.~\ref{fig:invariant_c60} on a log scale, for a pulsar with $R=0.1\,\rlight$, $b=10^{-3}$ and $\chi=60\degr$. Because $B_0$ can be of either sign, we plot $\log(B_0)$ for $B_0>0$ and $-\log(-B_0)$ for $B_0<0$. This helps to identify the location where $B_0$ abruptly changes sign. In the plot, a negative invariant~$B_0$ corresponds to $\log(-B_0)>0$. The highest values of $E_0$ and $B_0$ are observed close to the neutron star. $B_0$ changes sign in the region around $x=z$. This implies a discontinuity in the velocity field eq.~(\ref{eq:VRR}) along the direction of $\mathbf{B}$. The middle panel shows the curvature radius~$\rho_c$ normalized to the light-cylinder radius~$\rlight$, for electrons on the left and for positrons on the right. The curvature radius goes from $\rho_c/\rlight=10^{-2}$ to $\rho_c/\rlight=10^{3}$ for both species. It is shortest close to the star and along the rotation axis. The lower panel shows the Lorentz factor for electrons on the left and for positrons on the right. They attain similar speeds from $\gamma=10^{6.5}$ to $\gamma=10^{9.5}$, the highest values being obtained very close to the star along an axis inclined with respect to the rotation axis. From the knowledge of the curvature radius and the Lorentz factor, we compute $\chi_{\rm curv}$ and check that quantum corrections to curvature radiation remain negligible.

\subsection{Influence of the electric charge}

The total charge of the neutron star is an important parameter to determine the electric field at long distance. It produces a monopolar component decreasing very slowly with radius, therefore producing sensitive effects even around and outside the light-cylinder. Here we do not report on the full impact of this charge on the neutron star electrodynamics and radiation. We stress that assuming a dipolar field inside the star, an electric charge given by
\begin{equation}
Q_{\rm ns} = Q_{\rm c} \, \cos\chi.
\end{equation}
is located at the centre. The characteristic electric charge scale is
\begin{equation}
Q_{\rm c} = \frac{8\,\pi}{3} \, \varepsilon_0 \, \Omega\,B\,R^3.
\end{equation}
A net charge shifts the mean spectra to higher photon energies making the cut-off less sharp. A typical example is shown in Fig.~\ref{fig:spectre_moyen_r00025_b-3_q1_g10_dx} where such shift is clearly seen by comparison with Fig.~\ref{fig:spectre_moyen_r00025_b-3_g10_dx}. The net charge also influences light-curves and spectra. To remain brief, we show a small sample in Fig.~\ref{fig:skymaps_cl_r00025_b-3_c60_ri0.5_ro1_q1_g10_d3} to be compared with Fig.~\ref{fig:skymaps_cl_r00025_b-3_c60_ri0.5_ro1_g10_d3}. If high energy emission emanates from regions around or beyond the light-cylinder, following current wisdom, this charge must be included for a self-consistent picture of spectra and light-curves. Our findings urge us to better take care of this electric charge breathing when modelling pulsar radiation. It opens up another road towards a better understanding of neutron star electrodynamics and on its intrinsic temporal variability as already demonstrated by numerical simulations of time-dependent pair creation as found by \cite{timokhin_time-dependent_2010} and by \cite{timokhin_current_2013}.
\begin{figure}
	\centering
	\resizebox{0.5\textwidth}{!}{\input{spectre_moyen_r00025_b-3_q1_g10_dx}}
	\caption{Mean spectra for a 100~ms pulsar, density profile $n(r) \propto r^{-q}$ with $q\in\{1,2,3\}$ with respectively solid, dashed and dotted line. The magnetic field strength is $b=10^{-3}$ and $\chi=60$\degr. The inner boundary of the emission volume is given by $r_{\rm in} = \{0.1,0.2,0.5\}$ and the outer boundary by $r_{\rm out}=5$. The electric charge is $Q=Q_{\rm ns}$. Fluxes are evaluated at a distance of 1~kpc.}
	\label{fig:spectre_moyen_r00025_b-3_q1_g10_dx}
\end{figure}
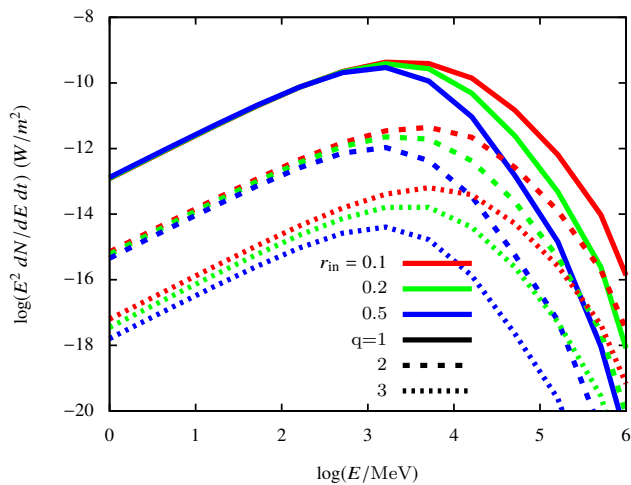
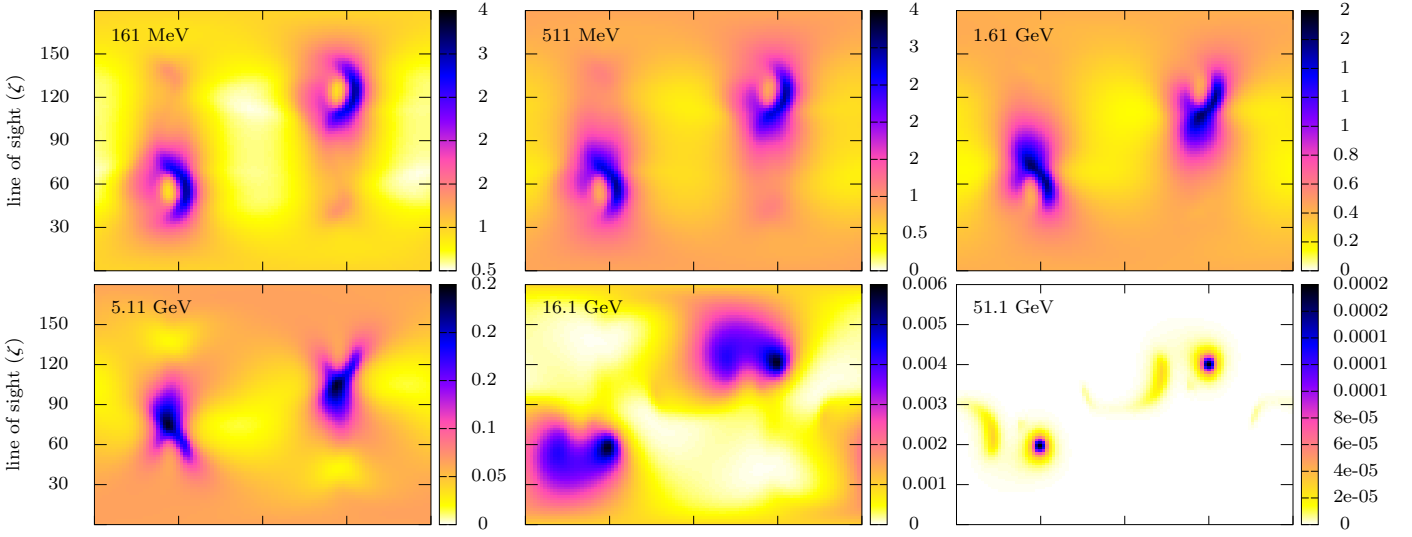
\begin{figure*}
	\centering
	\resizebox{\textwidth}{!}{\input{skymaps_cl_r00025_b-3_c60_ri0.5_ro1_q1_g10_d3}}
	\caption{Sky maps for a 100~ms pulsar, $r_{\rm in}=0.5$, $r_{\rm out} = 1$ and $n(r)\propto r^{-3}$. The magnetic field strength is $b=10^{-3}$ and $\chi=60\degr$. The electric charge is $Q=Q_{\rm ns}$.}
	\label{fig:skymaps_cl_r00025_b-3_c60_ri0.5_ro1_q1_g10_d3}
\end{figure*}

\subsection{Magnetic field strength estimates}

\begin{figure}
	\centering
	\resizebox{0.5\textwidth}{!}{\input{spectre_moyen_r00025_b-3_q0_g10_d3_ro2.0}}
	\caption{Mean spectra for a 100~ms pulsar with density profile $n(r) \propto r^{-3}$ and the outer boundary by $r_{\rm out}=2$. The magnetic field strength is $b=10^{-3}$ andthe obliquity  $\chi=60$\degr. The inner boundary of the emission volume is given by $r_{\rm in} = \{0.1,0.2,0.5, 1.0\}$. The electric charge is nul. Fluxes are evaluated at a distance of 1~kpc.}
	\label{fig:spectre_moyen_r00025_b-3_q0_g10_d3_ro2.0}
\end{figure}
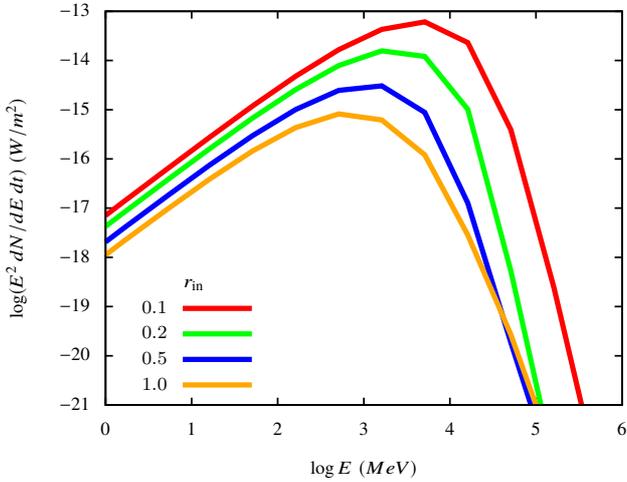

In order to obtain realistic spectra with cut-off energies around several GeV for gamma-ray pulsars, we need to fix the magnetic field strength at the surface. We showed that the cut-off scales as $B^{3/4}$ thus it is possible to retrieve any cut-off value by simply adjusting the magnetic field strength~$B$. In this work we used field strengths that seem slightly underestimated compared to what is usually assumed from magnetodipole losses. Actually, the cut-off also depends on the location of the inner boundary $r_{\rm in}$ where gamma-ray photons start to escape the magnetosphere without being magnetically absorbed. This effect is shown in Fig.~\ref{fig:spectre_moyen_r00025_b-3_q0_g10_d3_ro2.0} for a 100~ms pulsar with a density profile~$n(r) \propto r^{-3}$, a variable inner boundary~$r_{\rm in}$ and an outer boundary $r_{\rm out}=2$. The magnetic field strength is $b=10^{-3}$ and the obliquity~$\chi=60$\degr. It is seen that an increase in $r_{\rm in}$ implies a decrease in cut-off energy. Switching from $r_{\rm in}=0.1$ to $r_{\rm in}=1.0$ the cut-off decreases by one order of magnitude. Therefore if emission starts only around the light-cylinder, the magnetic field strength must be augmented by at least a factor ten. Thus our estimates become closer to traditional field estimates from vacuum magneto-dipole losses (although that such estimates are not necessarily realistic when plasma, wind and multipolar components are taken into account). Precise values of $B$ would require fitting spectra on a case-by-case basis for each pulsar. This is however left for future work.

\section{Conclusions}
\label{sec:Conclusion}

We have shown that a simple magnetospheric emission model can account for the gross features of gamma-ray pulsars light-curves and spectra. The spectral shape as reported in the second Fermi/LAT catalogue, cut-off energies and fluxes, are retrieved with realistic neutron star parameters such as its period and magnetic field strength consistent with millisecond and normal pulsars. Moreover light-curves with single or double peaked profiles are obtained depending on the viewing angle and obliquity. Although the gamma-ray luminosity falls well below the $L_\gamma = \dot E$ line in some cases, introducing a pair multiplicity factor~$\kappa$ much larger than unity increases the total power radiated by the magnetosphere. We found estimates of the order~$\kappa = \numprint{e2}-\numprint{e4}$. 

Single particle trajectories have been computed, showing the complexity of escaping, trapping and returning motion allowed within the magnetosphere. Such trajectories induce complicated geometries for possible vacuum gaps and filled regions that require further and deeper investigation to fully understand their impact on real pulsar electrodynamics. Because in the radiation reaction limit the velocity field depends also on the electric field, we expect variation in light-curves and spectra due to fluctuating electric charge within the magnetosphere. The particle outflow need not be stationary neither exactly compensating one charge escape by the other charge escape. Thus a kind of magnetospheric breathing is induced, impacting also on the pair formation rate. The total charge of the neutron star indeed affects the spectra and light-curve as shown in depth in \cite{petri_general-relativistic_2018}.

Moreover, this study suffers from several flaws that need to be fixed in forthcoming works. Firstly, a mono-energetic particle distribution function is only able to reproduce precisely a small samples of the Fermi/LAT gamma-ray spectra below the cut-off energy. This is because the measured power-law spectra require a power-law distribution of emitting particles not restricting the spectra to a simple $\omega^{-1/3}$ law. But this introduces one more free parameter to our model, thus opposite to the philosophy we followed here for our minimalist model. Nevertheless such studies are planed in the future to fit several samples of millisecond and normal pulsars that do not belong to fluxes depicted by mono-energetic lepton distribution functions. Secondly, back reaction of the plasma onto the intially vacuum rotating electromagnetic field must be added, especially close and behind the light-cylinder where this retroaction is preponderant. Kinetic simulations are therefore unavoidable but, unfortunately, such codes are still unable to catch the full span of length and time scales, going from the Larmor frequency to the neutron star rotation frequency. However, this is absolutely compulsory to reach Lorentz factors as high as $\gamma=\numprint{e9}$ and thus realistic photon energies.

Another interesting class of gamma-ray emitting pulsars are the soft gamma-ray pulsar population discussed in \citep{kuiper_soft_2015}. These pulsars must also be fitted by the same model, looking for spectra and light-curves. Moreover performing some predictions about phase-resolved polarization in high-energy, notably in X-rays in view of the coming IXPE mission \citep{weisskopf_imaging_2016} will better constrain the location an topology of the photon production sites.

\section*{Acknowledgements}

I am grateful to the anonymous referee for its useful insight and comments that improved the paper quality. This work has been published under the framework of the IdEx Unistra and benefits from a funding from the state managed by the French National Research Agency as part of the investments for the future program. It also benefited from grant No. ANR-13-JS05-0003-01 (project EMPERE) and from the computational facilities available at Equip@Meso (Universit\'e de Strasbourg). It also benefited from a CEFIPRA grant IFC/F5904-B/2018.




%


%
%
%
%
%


\bsp	
\label{lastpage}
\end{document}

%% file: spectre_moyen_r00025_b-3_g10_dx.tex
\begingroup
  \makeatletter
  \providecommand\color[2][]{%
    \GenericError{(gnuplot) \space\space\space\@spaces}{%
      Package color not loaded in conjunction with
      terminal option `colourtext'%
    }{See the gnuplot documentation for explanation.%
    }{Either use 'blacktext' in gnuplot or load the package
      color.sty in LaTeX.}%
    \renewcommand\color[2][]{}%
  }%
  \providecommand\includegraphics[2][]{%
    \GenericError{(gnuplot) \space\space\space\@spaces}{%
      Package graphicx or graphics not loaded%
    }{See the gnuplot documentation for explanation.%
    }{The gnuplot epslatex terminal needs graphicx.sty or graphics.sty.}%
    \renewcommand\includegraphics[2][]{}%
  }%
  \providecommand\rotatebox[2]{#2}%
  \@ifundefined{ifGPcolor}{%
    \newif\ifGPcolor
    \GPcolortrue
  }{}%
  \@ifundefined{ifGPblacktext}{%
    \newif\ifGPblacktext
    \GPblacktextfalse
  }{}%
  \let\gplgaddtomacro\g@addto@macro
  \gdef\gplbacktext{}%
  \gdef\gplfronttext{}%
  \makeatother
  \ifGPblacktext
    \def\colorrgb#1{}%
    \def\colorgray#1{}%
  \else
    \ifGPcolor
      \def\colorrgb#1{\color[rgb]{#1}}%
      \def\colorgray#1{\color[gray]{#1}}%
      \expandafter\def\csname LTw\endcsname{\color{white}}%
      \expandafter\def\csname LTb\endcsname{\color{black}}%
      \expandafter\def\csname LTa\endcsname{\color{black}}%
      \expandafter\def\csname LT0\endcsname{\color[rgb]{1,0,0}}%
      \expandafter\def\csname LT1\endcsname{\color[rgb]{0,1,0}}%
      \expandafter\def\csname LT2\endcsname{\color[rgb]{0,0,1}}%
      \expandafter\def\csname LT3\endcsname{\color[rgb]{1,0,1}}%
      \expandafter\def\csname LT4\endcsname{\color[rgb]{0,1,1}}%
      \expandafter\def\csname LT5\endcsname{\color[rgb]{1,1,0}}%
      \expandafter\def\csname LT6\endcsname{\color[rgb]{0,0,0}}%
      \expandafter\def\csname LT7\endcsname{\color[rgb]{1,0.3,0}}%
      \expandafter\def\csname LT8\endcsname{\color[rgb]{0.5,0.5,0.5}}%
    \else
      \def\colorrgb#1{\color{black}}%
      \def\colorgray#1{\color[gray]{#1}}%
      \expandafter\def\csname LTw\endcsname{\color{white}}%
      \expandafter\def\csname LTb\endcsname{\color{black}}%
      \expandafter\def\csname LTa\endcsname{\color{black}}%
      \expandafter\def\csname LT0\endcsname{\color{black}}%
      \expandafter\def\csname LT1\endcsname{\color{black}}%
      \expandafter\def\csname LT2\endcsname{\color{black}}%
      \expandafter\def\csname LT3\endcsname{\color{black}}%
      \expandafter\def\csname LT4\endcsname{\color{black}}%
      \expandafter\def\csname LT5\endcsname{\color{black}}%
      \expandafter\def\csname LT6\endcsname{\color{black}}%
      \expandafter\def\csname LT7\endcsname{\color{black}}%
      \expandafter\def\csname LT8\endcsname{\color{black}}%
    \fi
  \fi
    \setlength{\unitlength}{0.0500bp}%
    \ifx\gptboxheight\undefined%
      \newlength{\gptboxheight}%
      \newlength{\gptboxwidth}%
      \newsavebox{\gptboxtext}%
    \fi%
    \setlength{\fboxrule}{0.5pt}%
    \setlength{\fboxsep}{1pt}%
\begin{picture}(5760.00,4320.00)%
    \gplgaddtomacro\gplbacktext{%
      \csname LTb\endcsname
      \put(814,704){\makebox(0,0)[r]{\strut{}$-20$}}%
      \put(814,1270){\makebox(0,0)[r]{\strut{}$-18$}}%
      \put(814,1836){\makebox(0,0)[r]{\strut{}$-16$}}%
      \put(814,2402){\makebox(0,0)[r]{\strut{}$-14$}}%
      \put(814,2967){\makebox(0,0)[r]{\strut{}$-12$}}%
      \put(814,3533){\makebox(0,0)[r]{\strut{}$-10$}}%
      \put(814,4099){\makebox(0,0)[r]{\strut{}$-8$}}%
      \put(946,484){\makebox(0,0){\strut{}$0$}}%
      \put(1682,484){\makebox(0,0){\strut{}$1$}}%
      \put(2418,484){\makebox(0,0){\strut{}$2$}}%
      \put(3155,484){\makebox(0,0){\strut{}$3$}}%
      \put(3891,484){\makebox(0,0){\strut{}$4$}}%
      \put(4627,484){\makebox(0,0){\strut{}$5$}}%
      \put(5363,484){\makebox(0,0){\strut{}$6$}}%
    }%
    \gplgaddtomacro\gplfronttext{%
      \csname LTb\endcsname
      \put(198,2401){\rotatebox{-270}{\makebox(0,0){\strut{}$\log(E^2 \, dN/dE\,dt) ~ (W/m^2)$}}}%
      \put(3154,154){\makebox(0,0){\strut{}$\log(E/\textrm{MeV})$}}%
      \csname LTb\endcsname
      \put(3321,1977){\makebox(0,0)[r]{\strut{}$r_{\rm in}=0.1$}}%
      \csname LTb\endcsname
      \put(3321,1757){\makebox(0,0)[r]{\strut{}0.2}}%
      \csname LTb\endcsname
      \put(3321,1537){\makebox(0,0)[r]{\strut{}0.5}}%
      \csname LTb\endcsname
      \put(3321,1317){\makebox(0,0)[r]{\strut{}q=1}}%
      \csname LTb\endcsname
      \put(3321,1097){\makebox(0,0)[r]{\strut{}2}}%
      \csname LTb\endcsname
      \put(3321,877){\makebox(0,0)[r]{\strut{}3}}%
    }%
    \gplbacktext
    \put(0,0){\includegraphics{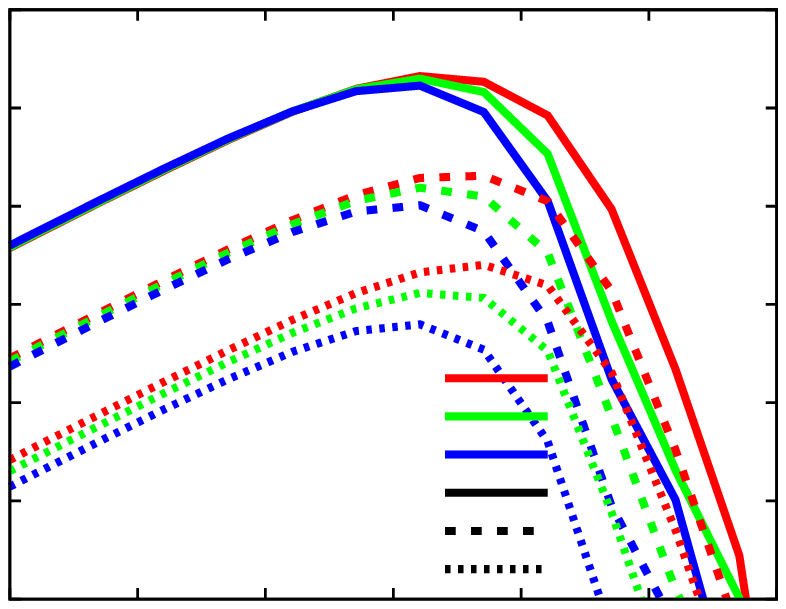}}%
    \gplfronttext
  \end{picture}%
\endgroup

%% file: spectre_moyen_r00025_b-3_g10_dx_onde.tex
\begingroup
  \makeatletter
  \providecommand\color[2][]{%
    \GenericError{(gnuplot) \space\space\space\@spaces}{%
      Package color not loaded in conjunction with
      terminal option `colourtext'%
    }{See the gnuplot documentation for explanation.%
    }{Either use 'blacktext' in gnuplot or load the package
      color.sty in LaTeX.}%
    \renewcommand\color[2][]{}%
  }%
  \providecommand\includegraphics[2][]{%
    \GenericError{(gnuplot) \space\space\space\@spaces}{%
      Package graphicx or graphics not loaded%
    }{See the gnuplot documentation for explanation.%
    }{The gnuplot epslatex terminal needs graphicx.sty or graphics.sty.}%
    \renewcommand\includegraphics[2][]{}%
  }%
  \providecommand\rotatebox[2]{#2}%
  \@ifundefined{ifGPcolor}{%
    \newif\ifGPcolor
    \GPcolortrue
  }{}%
  \@ifundefined{ifGPblacktext}{%
    \newif\ifGPblacktext
    \GPblacktextfalse
  }{}%
  \let\gplgaddtomacro\g@addto@macro
  \gdef\gplbacktext{}%
  \gdef\gplfronttext{}%
  \makeatother
  \ifGPblacktext
    \def\colorrgb#1{}%
    \def\colorgray#1{}%
  \else
    \ifGPcolor
      \def\colorrgb#1{\color[rgb]{#1}}%
      \def\colorgray#1{\color[gray]{#1}}%
      \expandafter\def\csname LTw\endcsname{\color{white}}%
      \expandafter\def\csname LTb\endcsname{\color{black}}%
      \expandafter\def\csname LTa\endcsname{\color{black}}%
      \expandafter\def\csname LT0\endcsname{\color[rgb]{1,0,0}}%
      \expandafter\def\csname LT1\endcsname{\color[rgb]{0,1,0}}%
      \expandafter\def\csname LT2\endcsname{\color[rgb]{0,0,1}}%
      \expandafter\def\csname LT3\endcsname{\color[rgb]{1,0,1}}%
      \expandafter\def\csname LT4\endcsname{\color[rgb]{0,1,1}}%
      \expandafter\def\csname LT5\endcsname{\color[rgb]{1,1,0}}%
      \expandafter\def\csname LT6\endcsname{\color[rgb]{0,0,0}}%
      \expandafter\def\csname LT7\endcsname{\color[rgb]{1,0.3,0}}%
      \expandafter\def\csname LT8\endcsname{\color[rgb]{0.5,0.5,0.5}}%
    \else
      \def\colorrgb#1{\color{black}}%
      \def\colorgray#1{\color[gray]{#1}}%
      \expandafter\def\csname LTw\endcsname{\color{white}}%
      \expandafter\def\csname LTb\endcsname{\color{black}}%
      \expandafter\def\csname LTa\endcsname{\color{black}}%
      \expandafter\def\csname LT0\endcsname{\color{black}}%
      \expandafter\def\csname LT1\endcsname{\color{black}}%
      \expandafter\def\csname LT2\endcsname{\color{black}}%
      \expandafter\def\csname LT3\endcsname{\color{black}}%
      \expandafter\def\csname LT4\endcsname{\color{black}}%
      \expandafter\def\csname LT5\endcsname{\color{black}}%
      \expandafter\def\csname LT6\endcsname{\color{black}}%
      \expandafter\def\csname LT7\endcsname{\color{black}}%
      \expandafter\def\csname LT8\endcsname{\color{black}}%
    \fi
  \fi
    \setlength{\unitlength}{0.0500bp}%
    \ifx\gptboxheight\undefined%
      \newlength{\gptboxheight}%
      \newlength{\gptboxwidth}%
      \newsavebox{\gptboxtext}%
    \fi%
    \setlength{\fboxrule}{0.5pt}%
    \setlength{\fboxsep}{1pt}%
\begin{picture}(5760.00,4320.00)%
    \gplgaddtomacro\gplbacktext{%
      \csname LTb\endcsname
      \put(814,704){\makebox(0,0)[r]{\strut{}$-20$}}%
      \put(814,1270){\makebox(0,0)[r]{\strut{}$-18$}}%
      \put(814,1836){\makebox(0,0)[r]{\strut{}$-16$}}%
      \put(814,2402){\makebox(0,0)[r]{\strut{}$-14$}}%
      \put(814,2967){\makebox(0,0)[r]{\strut{}$-12$}}%
      \put(814,3533){\makebox(0,0)[r]{\strut{}$-10$}}%
      \put(814,4099){\makebox(0,0)[r]{\strut{}$-8$}}%
      \put(946,484){\makebox(0,0){\strut{}$0$}}%
      \put(1682,484){\makebox(0,0){\strut{}$1$}}%
      \put(2418,484){\makebox(0,0){\strut{}$2$}}%
      \put(3155,484){\makebox(0,0){\strut{}$3$}}%
      \put(3891,484){\makebox(0,0){\strut{}$4$}}%
      \put(4627,484){\makebox(0,0){\strut{}$5$}}%
      \put(5363,484){\makebox(0,0){\strut{}$6$}}%
    }%
    \gplgaddtomacro\gplfronttext{%
      \csname LTb\endcsname
      \put(198,2401){\rotatebox{-270}{\makebox(0,0){\strut{}$\log(E^2 \, dN/dE\,dt) ~ (W/m^2)$}}}%
      \put(3154,154){\makebox(0,0){\strut{}$\log(E/\textrm{MeV})$}}%
      \csname LTb\endcsname
      \put(2925,1317){\makebox(0,0)[r]{\strut{}q=1}}%
      \csname LTb\endcsname
      \put(2925,1097){\makebox(0,0)[r]{\strut{}2}}%
      \csname LTb\endcsname
      \put(2925,877){\makebox(0,0)[r]{\strut{}3}}%
    }%
    \gplbacktext
    \put(0,0){\includegraphics{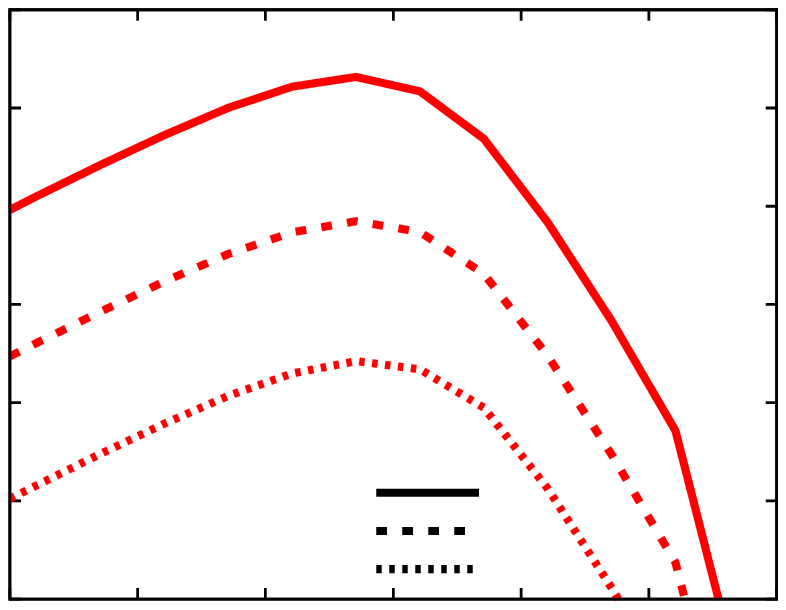}}%
    \gplfronttext
  \end{picture}%
\endgroup

%% file: spectre_moyen_r005_b-6_g10_dx.tex
\begingroup
  \makeatletter
  \providecommand\color[2][]{%
    \GenericError{(gnuplot) \space\space\space\@spaces}{%
      Package color not loaded in conjunction with
      terminal option `colourtext'%
    }{See the gnuplot documentation for explanation.%
    }{Either use 'blacktext' in gnuplot or load the package
      color.sty in LaTeX.}%
    \renewcommand\color[2][]{}%
  }%
  \providecommand\includegraphics[2][]{%
    \GenericError{(gnuplot) \space\space\space\@spaces}{%
      Package graphicx or graphics not loaded%
    }{See the gnuplot documentation for explanation.%
    }{The gnuplot epslatex terminal needs graphicx.sty or graphics.sty.}%
    \renewcommand\includegraphics[2][]{}%
  }%
  \providecommand\rotatebox[2]{#2}%
  \@ifundefined{ifGPcolor}{%
    \newif\ifGPcolor
    \GPcolortrue
  }{}%
  \@ifundefined{ifGPblacktext}{%
    \newif\ifGPblacktext
    \GPblacktextfalse
  }{}%
  \let\gplgaddtomacro\g@addto@macro
  \gdef\gplbacktext{}%
  \gdef\gplfronttext{}%
  \makeatother
  \ifGPblacktext
    \def\colorrgb#1{}%
    \def\colorgray#1{}%
  \else
    \ifGPcolor
      \def\colorrgb#1{\color[rgb]{#1}}%
      \def\colorgray#1{\color[gray]{#1}}%
      \expandafter\def\csname LTw\endcsname{\color{white}}%
      \expandafter\def\csname LTb\endcsname{\color{black}}%
      \expandafter\def\csname LTa\endcsname{\color{black}}%
      \expandafter\def\csname LT0\endcsname{\color[rgb]{1,0,0}}%
      \expandafter\def\csname LT1\endcsname{\color[rgb]{0,1,0}}%
      \expandafter\def\csname LT2\endcsname{\color[rgb]{0,0,1}}%
      \expandafter\def\csname LT3\endcsname{\color[rgb]{1,0,1}}%
      \expandafter\def\csname LT4\endcsname{\color[rgb]{0,1,1}}%
      \expandafter\def\csname LT5\endcsname{\color[rgb]{1,1,0}}%
      \expandafter\def\csname LT6\endcsname{\color[rgb]{0,0,0}}%
      \expandafter\def\csname LT7\endcsname{\color[rgb]{1,0.3,0}}%
      \expandafter\def\csname LT8\endcsname{\color[rgb]{0.5,0.5,0.5}}%
    \else
      \def\colorrgb#1{\color{black}}%
      \def\colorgray#1{\color[gray]{#1}}%
      \expandafter\def\csname LTw\endcsname{\color{white}}%
      \expandafter\def\csname LTb\endcsname{\color{black}}%
      \expandafter\def\csname LTa\endcsname{\color{black}}%
      \expandafter\def\csname LT0\endcsname{\color{black}}%
      \expandafter\def\csname LT1\endcsname{\color{black}}%
      \expandafter\def\csname LT2\endcsname{\color{black}}%
      \expandafter\def\csname LT3\endcsname{\color{black}}%
      \expandafter\def\csname LT4\endcsname{\color{black}}%
      \expandafter\def\csname LT5\endcsname{\color{black}}%
      \expandafter\def\csname LT6\endcsname{\color{black}}%
      \expandafter\def\csname LT7\endcsname{\color{black}}%
      \expandafter\def\csname LT8\endcsname{\color{black}}%
    \fi
  \fi
    \setlength{\unitlength}{0.0500bp}%
    \ifx\gptboxheight\undefined%
      \newlength{\gptboxheight}%
      \newlength{\gptboxwidth}%
      \newsavebox{\gptboxtext}%
    \fi%
    \setlength{\fboxrule}{0.5pt}%
    \setlength{\fboxsep}{1pt}%
\begin{picture}(5760.00,4320.00)%
    \gplgaddtomacro\gplbacktext{%
      \csname LTb\endcsname
      \put(814,704){\makebox(0,0)[r]{\strut{}$-18$}}%
      \put(814,1270){\makebox(0,0)[r]{\strut{}$-17$}}%
      \put(814,1836){\makebox(0,0)[r]{\strut{}$-16$}}%
      \put(814,2402){\makebox(0,0)[r]{\strut{}$-15$}}%
      \put(814,2967){\makebox(0,0)[r]{\strut{}$-14$}}%
      \put(814,3533){\makebox(0,0)[r]{\strut{}$-13$}}%
      \put(814,4099){\makebox(0,0)[r]{\strut{}$-12$}}%
      \put(946,484){\makebox(0,0){\strut{}$0$}}%
      \put(1682,484){\makebox(0,0){\strut{}$1$}}%
      \put(2418,484){\makebox(0,0){\strut{}$2$}}%
      \put(3155,484){\makebox(0,0){\strut{}$3$}}%
      \put(3891,484){\makebox(0,0){\strut{}$4$}}%
      \put(4627,484){\makebox(0,0){\strut{}$5$}}%
      \put(5363,484){\makebox(0,0){\strut{}$6$}}%
    }%
    \gplgaddtomacro\gplfronttext{%
      \csname LTb\endcsname
      \put(198,2401){\rotatebox{-270}{\makebox(0,0){\strut{}$\log(E^2 \, dN/dE\,dt) ~ (W/m^2)$}}}%
      \put(3154,154){\makebox(0,0){\strut{}$\log(E/\textrm{MeV})$}}%
      \csname LTb\endcsname
      \put(3321,1977){\makebox(0,0)[r]{\strut{}$r_{\rm in}=0.1$}}%
      \csname LTb\endcsname
      \put(3321,1757){\makebox(0,0)[r]{\strut{}0.2}}%
      \csname LTb\endcsname
      \put(3321,1537){\makebox(0,0)[r]{\strut{}0.5}}%
      \csname LTb\endcsname
      \put(3321,1317){\makebox(0,0)[r]{\strut{}q=1}}%
      \csname LTb\endcsname
      \put(3321,1097){\makebox(0,0)[r]{\strut{}2}}%
      \csname LTb\endcsname
      \put(3321,877){\makebox(0,0)[r]{\strut{}3}}%
    }%
    \gplbacktext
    \put(0,0){\includegraphics{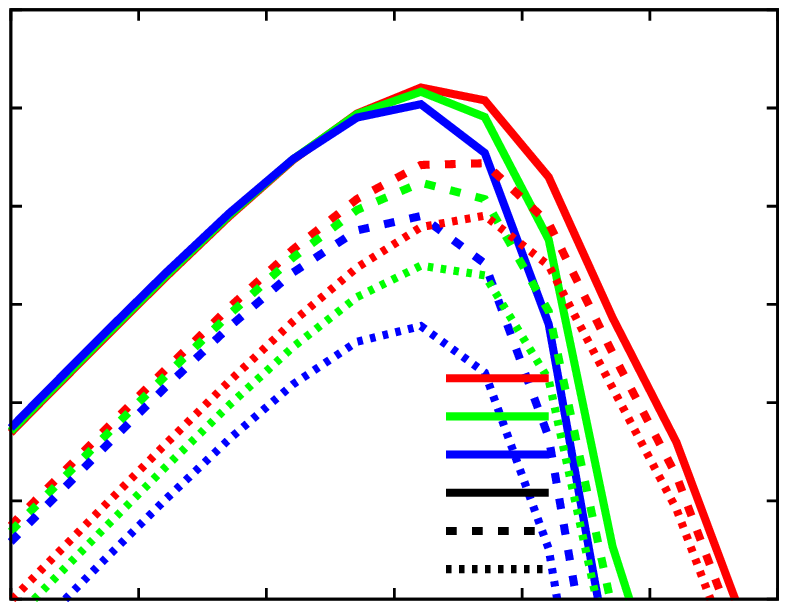}}%
    \gplfronttext
  \end{picture}%
\endgroup

%% file: spectre_resolu_r00025_b-3_ri0.2_g10_d3.tex
\begingroup
  \makeatletter
  \providecommand\color[2][]{%
    \GenericError{(gnuplot) \space\space\space\@spaces}{%
      Package color not loaded in conjunction with
      terminal option `colourtext'%
    }{See the gnuplot documentation for explanation.%
    }{Either use 'blacktext' in gnuplot or load the package
      color.sty in LaTeX.}%
    \renewcommand\color[2][]{}%
  }%
  \providecommand\includegraphics[2][]{%
    \GenericError{(gnuplot) \space\space\space\@spaces}{%
      Package graphicx or graphics not loaded%
    }{See the gnuplot documentation for explanation.%
    }{The gnuplot epslatex terminal needs graphicx.sty or graphics.sty.}%
    \renewcommand\includegraphics[2][]{}%
  }%
  \providecommand\rotatebox[2]{#2}%
  \@ifundefined{ifGPcolor}{%
    \newif\ifGPcolor
    \GPcolortrue
  }{}%
  \@ifundefined{ifGPblacktext}{%
    \newif\ifGPblacktext
    \GPblacktextfalse
  }{}%
  \let\gplgaddtomacro\g@addto@macro
  \gdef\gplbacktext{}%
  \gdef\gplfronttext{}%
  \makeatother
  \ifGPblacktext
    \def\colorrgb#1{}%
    \def\colorgray#1{}%
  \else
    \ifGPcolor
      \def\colorrgb#1{\color[rgb]{#1}}%
      \def\colorgray#1{\color[gray]{#1}}%
      \expandafter\def\csname LTw\endcsname{\color{white}}%
      \expandafter\def\csname LTb\endcsname{\color{black}}%
      \expandafter\def\csname LTa\endcsname{\color{black}}%
      \expandafter\def\csname LT0\endcsname{\color[rgb]{1,0,0}}%
      \expandafter\def\csname LT1\endcsname{\color[rgb]{0,1,0}}%
      \expandafter\def\csname LT2\endcsname{\color[rgb]{0,0,1}}%
      \expandafter\def\csname LT3\endcsname{\color[rgb]{1,0,1}}%
      \expandafter\def\csname LT4\endcsname{\color[rgb]{0,1,1}}%
      \expandafter\def\csname LT5\endcsname{\color[rgb]{1,1,0}}%
      \expandafter\def\csname LT6\endcsname{\color[rgb]{0,0,0}}%
      \expandafter\def\csname LT7\endcsname{\color[rgb]{1,0.3,0}}%
      \expandafter\def\csname LT8\endcsname{\color[rgb]{0.5,0.5,0.5}}%
    \else
      \def\colorrgb#1{\color{black}}%
      \def\colorgray#1{\color[gray]{#1}}%
      \expandafter\def\csname LTw\endcsname{\color{white}}%
      \expandafter\def\csname LTb\endcsname{\color{black}}%
      \expandafter\def\csname LTa\endcsname{\color{black}}%
      \expandafter\def\csname LT0\endcsname{\color{black}}%
      \expandafter\def\csname LT1\endcsname{\color{black}}%
      \expandafter\def\csname LT2\endcsname{\color{black}}%
      \expandafter\def\csname LT3\endcsname{\color{black}}%
      \expandafter\def\csname LT4\endcsname{\color{black}}%
      \expandafter\def\csname LT5\endcsname{\color{black}}%
      \expandafter\def\csname LT6\endcsname{\color{black}}%
      \expandafter\def\csname LT7\endcsname{\color{black}}%
      \expandafter\def\csname LT8\endcsname{\color{black}}%
    \fi
  \fi
    \setlength{\unitlength}{0.0500bp}%
    \ifx\gptboxheight\undefined%
      \newlength{\gptboxheight}%
      \newlength{\gptboxwidth}%
      \newsavebox{\gptboxtext}%
    \fi%
    \setlength{\fboxrule}{0.5pt}%
    \setlength{\fboxsep}{1pt}%
\begin{picture}(5760.00,4320.00)%
    \gplgaddtomacro\gplbacktext{%
      \csname LTb\endcsname
      \put(1078,704){\makebox(0,0)[r]{\strut{}$-16$}}%
      \put(1078,1383){\makebox(0,0)[r]{\strut{}$-15.5$}}%
      \put(1078,2062){\makebox(0,0)[r]{\strut{}$-15$}}%
      \put(1078,2741){\makebox(0,0)[r]{\strut{}$-14.5$}}%
      \put(1078,3420){\makebox(0,0)[r]{\strut{}$-14$}}%
      \put(1078,4099){\makebox(0,0)[r]{\strut{}$-13.5$}}%
      \put(1210,484){\makebox(0,0){\strut{}$1$}}%
      \put(1729,484){\makebox(0,0){\strut{}$1.5$}}%
      \put(2248,484){\makebox(0,0){\strut{}$2$}}%
      \put(2767,484){\makebox(0,0){\strut{}$2.5$}}%
      \put(3287,484){\makebox(0,0){\strut{}$3$}}%
      \put(3806,484){\makebox(0,0){\strut{}$3.5$}}%
      \put(4325,484){\makebox(0,0){\strut{}$4$}}%
      \put(4844,484){\makebox(0,0){\strut{}$4.5$}}%
      \put(5363,484){\makebox(0,0){\strut{}$5$}}%
    }%
    \gplgaddtomacro\gplfronttext{%
      \csname LTb\endcsname
      \put(198,2401){\rotatebox{-270}{\makebox(0,0){\strut{}$\log(E^2 \, dN/dE\,dt) ~ (W/m^2)$}}}%
      \put(3286,154){\makebox(0,0){\strut{}$\log(E/\textrm{MeV})$}}%
      \put(1967,3926){\makebox(0,0){\strut{}phase}}%
      \csname LTb\endcsname
      \put(1738,3706){\makebox(0,0)[r]{\strut{}0.0}}%
      \csname LTb\endcsname
      \put(1738,3486){\makebox(0,0)[r]{\strut{}0.2}}%
      \csname LTb\endcsname
      \put(1738,3266){\makebox(0,0)[r]{\strut{}0.4}}%
      \csname LTb\endcsname
      \put(1738,3046){\makebox(0,0)[r]{\strut{}0.6}}%
      \csname LTb\endcsname
      \put(1738,2826){\makebox(0,0)[r]{\strut{}0.8}}%
    }%
    \gplbacktext
    \put(0,0){\includegraphics{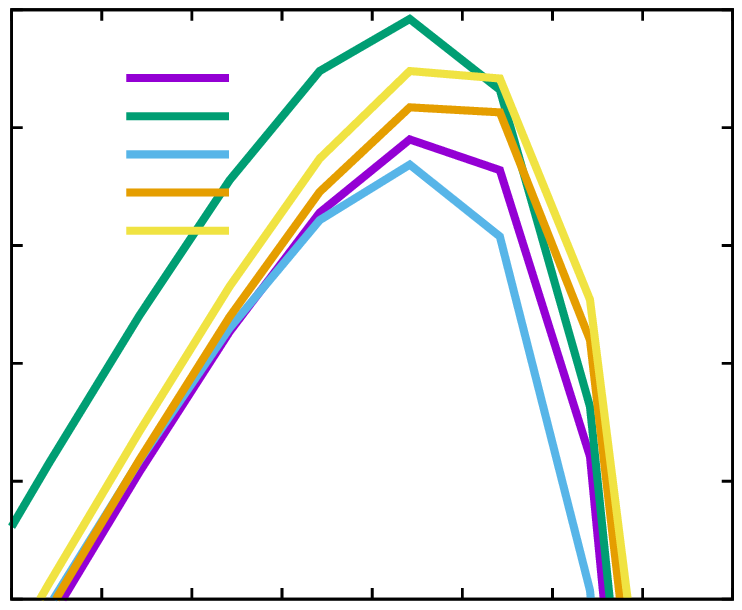}}%
    \gplfronttext
  \end{picture}%
\endgroup

%% file: spectre_resolu_r005_b-6_ri0.2_g10_d3.tex
\begingroup
  \makeatletter
  \providecommand\color[2][]{%
    \GenericError{(gnuplot) \space\space\space\@spaces}{%
      Package color not loaded in conjunction with
      terminal option `colourtext'%
    }{See the gnuplot documentation for explanation.%
    }{Either use 'blacktext' in gnuplot or load the package
      color.sty in LaTeX.}%
    \renewcommand\color[2][]{}%
  }%
  \providecommand\includegraphics[2][]{%
    \GenericError{(gnuplot) \space\space\space\@spaces}{%
      Package graphicx or graphics not loaded%
    }{See the gnuplot documentation for explanation.%
    }{The gnuplot epslatex terminal needs graphicx.sty or graphics.sty.}%
    \renewcommand\includegraphics[2][]{}%
  }%
  \providecommand\rotatebox[2]{#2}%
  \@ifundefined{ifGPcolor}{%
    \newif\ifGPcolor
    \GPcolortrue
  }{}%
  \@ifundefined{ifGPblacktext}{%
    \newif\ifGPblacktext
    \GPblacktextfalse
  }{}%
  \let\gplgaddtomacro\g@addto@macro
  \gdef\gplbacktext{}%
  \gdef\gplfronttext{}%
  \makeatother
  \ifGPblacktext
    \def\colorrgb#1{}%
    \def\colorgray#1{}%
  \else
    \ifGPcolor
      \def\colorrgb#1{\color[rgb]{#1}}%
      \def\colorgray#1{\color[gray]{#1}}%
      \expandafter\def\csname LTw\endcsname{\color{white}}%
      \expandafter\def\csname LTb\endcsname{\color{black}}%
      \expandafter\def\csname LTa\endcsname{\color{black}}%
      \expandafter\def\csname LT0\endcsname{\color[rgb]{1,0,0}}%
      \expandafter\def\csname LT1\endcsname{\color[rgb]{0,1,0}}%
      \expandafter\def\csname LT2\endcsname{\color[rgb]{0,0,1}}%
      \expandafter\def\csname LT3\endcsname{\color[rgb]{1,0,1}}%
      \expandafter\def\csname LT4\endcsname{\color[rgb]{0,1,1}}%
      \expandafter\def\csname LT5\endcsname{\color[rgb]{1,1,0}}%
      \expandafter\def\csname LT6\endcsname{\color[rgb]{0,0,0}}%
      \expandafter\def\csname LT7\endcsname{\color[rgb]{1,0.3,0}}%
      \expandafter\def\csname LT8\endcsname{\color[rgb]{0.5,0.5,0.5}}%
    \else
      \def\colorrgb#1{\color{black}}%
      \def\colorgray#1{\color[gray]{#1}}%
      \expandafter\def\csname LTw\endcsname{\color{white}}%
      \expandafter\def\csname LTb\endcsname{\color{black}}%
      \expandafter\def\csname LTa\endcsname{\color{black}}%
      \expandafter\def\csname LT0\endcsname{\color{black}}%
      \expandafter\def\csname LT1\endcsname{\color{black}}%
      \expandafter\def\csname LT2\endcsname{\color{black}}%
      \expandafter\def\csname LT3\endcsname{\color{black}}%
      \expandafter\def\csname LT4\endcsname{\color{black}}%
      \expandafter\def\csname LT5\endcsname{\color{black}}%
      \expandafter\def\csname LT6\endcsname{\color{black}}%
      \expandafter\def\csname LT7\endcsname{\color{black}}%
      \expandafter\def\csname LT8\endcsname{\color{black}}%
    \fi
  \fi
    \setlength{\unitlength}{0.0500bp}%
    \ifx\gptboxheight\undefined%
      \newlength{\gptboxheight}%
      \newlength{\gptboxwidth}%
      \newsavebox{\gptboxtext}%
    \fi%
    \setlength{\fboxrule}{0.5pt}%
    \setlength{\fboxsep}{1pt}%
\begin{picture}(5760.00,4320.00)%
    \gplgaddtomacro\gplbacktext{%
      \csname LTb\endcsname
      \put(1078,704){\makebox(0,0)[r]{\strut{}$-16$}}%
      \put(1078,1081){\makebox(0,0)[r]{\strut{}$-15.8$}}%
      \put(1078,1458){\makebox(0,0)[r]{\strut{}$-15.6$}}%
      \put(1078,1836){\makebox(0,0)[r]{\strut{}$-15.4$}}%
      \put(1078,2213){\makebox(0,0)[r]{\strut{}$-15.2$}}%
      \put(1078,2590){\makebox(0,0)[r]{\strut{}$-15$}}%
      \put(1078,2967){\makebox(0,0)[r]{\strut{}$-14.8$}}%
      \put(1078,3345){\makebox(0,0)[r]{\strut{}$-14.6$}}%
      \put(1078,3722){\makebox(0,0)[r]{\strut{}$-14.4$}}%
      \put(1078,4099){\makebox(0,0)[r]{\strut{}$-14.2$}}%
      \put(1210,484){\makebox(0,0){\strut{}$1$}}%
      \put(1729,484){\makebox(0,0){\strut{}$1.5$}}%
      \put(2248,484){\makebox(0,0){\strut{}$2$}}%
      \put(2767,484){\makebox(0,0){\strut{}$2.5$}}%
      \put(3287,484){\makebox(0,0){\strut{}$3$}}%
      \put(3806,484){\makebox(0,0){\strut{}$3.5$}}%
      \put(4325,484){\makebox(0,0){\strut{}$4$}}%
      \put(4844,484){\makebox(0,0){\strut{}$4.5$}}%
      \put(5363,484){\makebox(0,0){\strut{}$5$}}%
    }%
    \gplgaddtomacro\gplfronttext{%
      \csname LTb\endcsname
      \put(198,2401){\rotatebox{-270}{\makebox(0,0){\strut{}$\log(E^2 \, dN/dE\,dt) ~ (W/m^2)$}}}%
      \put(3286,154){\makebox(0,0){\strut{}$\log(E/\textrm{MeV})$}}%
      \put(1967,3926){\makebox(0,0){\strut{}phase}}%
      \csname LTb\endcsname
      \put(1738,3706){\makebox(0,0)[r]{\strut{}0.0}}%
      \csname LTb\endcsname
      \put(1738,3486){\makebox(0,0)[r]{\strut{}0.2}}%
      \csname LTb\endcsname
      \put(1738,3266){\makebox(0,0)[r]{\strut{}0.4}}%
      \csname LTb\endcsname
      \put(1738,3046){\makebox(0,0)[r]{\strut{}0.6}}%
      \csname LTb\endcsname
      \put(1738,2826){\makebox(0,0)[r]{\strut{}0.8}}%
    }%
    \gplbacktext
    \put(0,0){\includegraphics{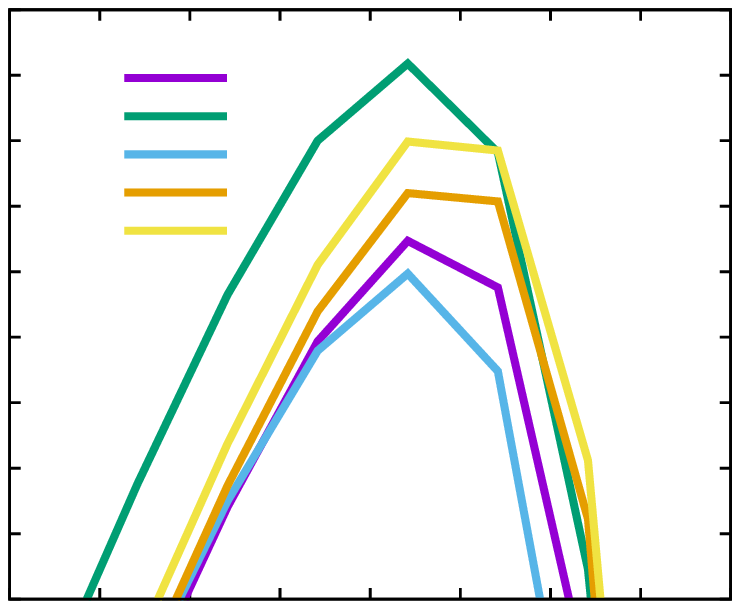}}%
    \gplfronttext
  \end{picture}%
\endgroup

%% file: skymaps_cl_r00025_b-3_c60_ri0.2_ro1_g10_d3.tex
\begingroup
  \makeatletter
  \providecommand\color[2][]{%
    \GenericError{(gnuplot) \space\space\space\@spaces}{%
      Package color not loaded in conjunction with
      terminal option `colourtext'%
    }{See the gnuplot documentation for explanation.%
    }{Either use 'blacktext' in gnuplot or load the package
      color.sty in LaTeX.}%
    \renewcommand\color[2][]{}%
  }%
  \providecommand\includegraphics[2][]{%
    \GenericError{(gnuplot) \space\space\space\@spaces}{%
      Package graphicx or graphics not loaded%
    }{See the gnuplot documentation for explanation.%
    }{The gnuplot epslatex terminal needs graphicx.sty or graphics.sty.}%
    \renewcommand\includegraphics[2][]{}%
  }%
  \providecommand\rotatebox[2]{#2}%
  \@ifundefined{ifGPcolor}{%
    \newif\ifGPcolor
    \GPcolortrue
  }{}%
  \@ifundefined{ifGPblacktext}{%
    \newif\ifGPblacktext
    \GPblacktextfalse
  }{}%
  \let\gplgaddtomacro\g@addto@macro
  \gdef\gplbacktext{}%
  \gdef\gplfronttext{}%
  \makeatother
  \ifGPblacktext
    \def\colorrgb#1{}%
    \def\colorgray#1{}%
  \else
    \ifGPcolor
      \def\colorrgb#1{\color[rgb]{#1}}%
      \def\colorgray#1{\color[gray]{#1}}%
      \expandafter\def\csname LTw\endcsname{\color{white}}%
      \expandafter\def\csname LTb\endcsname{\color{black}}%
      \expandafter\def\csname LTa\endcsname{\color{black}}%
      \expandafter\def\csname LT0\endcsname{\color[rgb]{1,0,0}}%
      \expandafter\def\csname LT1\endcsname{\color[rgb]{0,1,0}}%
      \expandafter\def\csname LT2\endcsname{\color[rgb]{0,0,1}}%
      \expandafter\def\csname LT3\endcsname{\color[rgb]{1,0,1}}%
      \expandafter\def\csname LT4\endcsname{\color[rgb]{0,1,1}}%
      \expandafter\def\csname LT5\endcsname{\color[rgb]{1,1,0}}%
      \expandafter\def\csname LT6\endcsname{\color[rgb]{0,0,0}}%
      \expandafter\def\csname LT7\endcsname{\color[rgb]{1,0.3,0}}%
      \expandafter\def\csname LT8\endcsname{\color[rgb]{0.5,0.5,0.5}}%
    \else
      \def\colorrgb#1{\color{black}}%
      \def\colorgray#1{\color[gray]{#1}}%
      \expandafter\def\csname LTw\endcsname{\color{white}}%
      \expandafter\def\csname LTb\endcsname{\color{black}}%
      \expandafter\def\csname LTa\endcsname{\color{black}}%
      \expandafter\def\csname LT0\endcsname{\color{black}}%
      \expandafter\def\csname LT1\endcsname{\color{black}}%
      \expandafter\def\csname LT2\endcsname{\color{black}}%
      \expandafter\def\csname LT3\endcsname{\color{black}}%
      \expandafter\def\csname LT4\endcsname{\color{black}}%
      \expandafter\def\csname LT5\endcsname{\color{black}}%
      \expandafter\def\csname LT6\endcsname{\color{black}}%
      \expandafter\def\csname LT7\endcsname{\color{black}}%
      \expandafter\def\csname LT8\endcsname{\color{black}}%
    \fi
  \fi
    \setlength{\unitlength}{0.0500bp}%
    \ifx\gptboxheight\undefined%
      \newlength{\gptboxheight}%
      \newlength{\gptboxwidth}%
      \newsavebox{\gptboxtext}%
    \fi%
    \setlength{\fboxrule}{0.5pt}%
    \setlength{\fboxsep}{1pt}%
\begin{picture}(11520.00,5760.00)%
    \gplgaddtomacro\gplbacktext{%
    }%
    \gplgaddtomacro\gplfronttext{%
      \csname LTb\endcsname
      \put(577,2662){\makebox(0,0){\strut{}}}%
      \put(1297,2662){\makebox(0,0){\strut{}}}%
      \put(2016,2662){\makebox(0,0){\strut{}}}%
      \put(2735,2662){\makebox(0,0){\strut{}}}%
      \put(3455,2662){\makebox(0,0){\strut{}}}%
      \put(364,2939){\makebox(0,0)[r]{\strut{}}}%
      \put(364,3313){\makebox(0,0)[r]{\strut{}30}}%
      \put(364,3687){\makebox(0,0)[r]{\strut{}60}}%
      \put(364,4061){\makebox(0,0)[r]{\strut{}90}}%
      \put(364,4435){\makebox(0,0)[r]{\strut{}120}}%
      \put(364,4809){\makebox(0,0)[r]{\strut{}150}}%
      \put(-98,4061){\rotatebox{-270}{\makebox(0,0){\strut{}line of sight ($\zeta$)}}}%
      \put(3803,2939){\makebox(0,0)[l]{\strut{} 0}}%
      \put(3803,3188){\makebox(0,0)[l]{\strut{} 2}}%
      \put(3803,3437){\makebox(0,0)[l]{\strut{} 4}}%
      \put(3803,3687){\makebox(0,0)[l]{\strut{} 6}}%
      \put(3803,3936){\makebox(0,0)[l]{\strut{} 8}}%
      \put(3803,4185){\makebox(0,0)[l]{\strut{}1e+01}}%
      \put(3803,4435){\makebox(0,0)[l]{\strut{}1e+01}}%
      \put(3803,4684){\makebox(0,0)[l]{\strut{}1e+01}}%
      \put(3803,4933){\makebox(0,0)[l]{\strut{}2e+01}}%
      \put(3803,5183){\makebox(0,0)[l]{\strut{}2e+01}}%
      \put(721,4959){\makebox(0,0)[l]{\strut{}161~MeV}}%
    }%
    \gplgaddtomacro\gplbacktext{%
    }%
    \gplgaddtomacro\gplfronttext{%
      \csname LTb\endcsname
      \put(4263,2662){\makebox(0,0){\strut{}}}%
      \put(4983,2662){\makebox(0,0){\strut{}}}%
      \put(5702,2662){\makebox(0,0){\strut{}}}%
      \put(6421,2662){\makebox(0,0){\strut{}}}%
      \put(7141,2662){\makebox(0,0){\strut{}}}%
      \put(4050,2939){\makebox(0,0)[r]{\strut{}}}%
      \put(4050,3313){\makebox(0,0)[r]{\strut{}}}%
      \put(4050,3687){\makebox(0,0)[r]{\strut{}}}%
      \put(4050,4061){\makebox(0,0)[r]{\strut{}}}%
      \put(4050,4435){\makebox(0,0)[r]{\strut{}}}%
      \put(4050,4809){\makebox(0,0)[r]{\strut{}}}%
      \put(7489,2939){\makebox(0,0)[l]{\strut{} 0}}%
      \put(7489,3188){\makebox(0,0)[l]{\strut{} 2}}%
      \put(7489,3437){\makebox(0,0)[l]{\strut{} 4}}%
      \put(7489,3687){\makebox(0,0)[l]{\strut{} 6}}%
      \put(7489,3936){\makebox(0,0)[l]{\strut{} 8}}%
      \put(7489,4185){\makebox(0,0)[l]{\strut{}1e+01}}%
      \put(7489,4435){\makebox(0,0)[l]{\strut{}1e+01}}%
      \put(7489,4684){\makebox(0,0)[l]{\strut{}1e+01}}%
      \put(7489,4933){\makebox(0,0)[l]{\strut{}2e+01}}%
      \put(7489,5183){\makebox(0,0)[l]{\strut{}2e+01}}%
      \put(4407,4959){\makebox(0,0)[l]{\strut{}511~MeV}}%
    }%
    \gplgaddtomacro\gplbacktext{%
    }%
    \gplgaddtomacro\gplfronttext{%
      \csname LTb\endcsname
      \put(7950,2662){\makebox(0,0){\strut{}}}%
      \put(8670,2662){\makebox(0,0){\strut{}}}%
      \put(9389,2662){\makebox(0,0){\strut{}}}%
      \put(10108,2662){\makebox(0,0){\strut{}}}%
      \put(10828,2662){\makebox(0,0){\strut{}}}%
      \put(7737,2939){\makebox(0,0)[r]{\strut{}}}%
      \put(7737,3313){\makebox(0,0)[r]{\strut{}}}%
      \put(7737,3687){\makebox(0,0)[r]{\strut{}}}%
      \put(7737,4061){\makebox(0,0)[r]{\strut{}}}%
      \put(7737,4435){\makebox(0,0)[r]{\strut{}}}%
      \put(7737,4809){\makebox(0,0)[r]{\strut{}}}%
      \put(11176,2939){\makebox(0,0)[l]{\strut{} 0}}%
      \put(11176,3259){\makebox(0,0)[l]{\strut{} 2}}%
      \put(11176,3580){\makebox(0,0)[l]{\strut{} 4}}%
      \put(11176,3900){\makebox(0,0)[l]{\strut{} 6}}%
      \put(11176,4221){\makebox(0,0)[l]{\strut{} 8}}%
      \put(11176,4541){\makebox(0,0)[l]{\strut{}1e+01}}%
      \put(11176,4862){\makebox(0,0)[l]{\strut{}1e+01}}%
      \put(11176,5183){\makebox(0,0)[l]{\strut{}1e+01}}%
      \put(8094,4959){\makebox(0,0)[l]{\strut{}1.61~GeV}}%
    }%
    \gplgaddtomacro\gplbacktext{%
    }%
    \gplgaddtomacro\gplfronttext{%
      \csname LTb\endcsname
      \put(577,484){\makebox(0,0){\strut{}}}%
      \put(1297,484){\makebox(0,0){\strut{}}}%
      \put(2016,484){\makebox(0,0){\strut{}}}%
      \put(2735,484){\makebox(0,0){\strut{}}}%
      \put(3455,484){\makebox(0,0){\strut{}}}%
      \put(355,750){\makebox(0,0)[r]{\strut{}}}%
      \put(355,1095){\makebox(0,0)[r]{\strut{}30}}%
      \put(355,1440){\makebox(0,0)[r]{\strut{}60}}%
      \put(355,1785){\makebox(0,0)[r]{\strut{}90}}%
      \put(355,2130){\makebox(0,0)[r]{\strut{}120}}%
      \put(355,2475){\makebox(0,0)[r]{\strut{}150}}%
      \put(-107,1785){\rotatebox{-270}{\makebox(0,0){\strut{}line of sight ($\zeta$)}}}%
      \put(3803,750){\makebox(0,0)[l]{\strut{} 0}}%
      \put(3803,1008){\makebox(0,0)[l]{\strut{}0.5}}%
      \put(3803,1267){\makebox(0,0)[l]{\strut{} 1}}%
      \put(3803,1526){\makebox(0,0)[l]{\strut{} 2}}%
      \put(3803,1785){\makebox(0,0)[l]{\strut{} 2}}%
      \put(3803,2043){\makebox(0,0)[l]{\strut{} 2}}%
      \put(3803,2302){\makebox(0,0)[l]{\strut{} 3}}%
      \put(3803,2561){\makebox(0,0)[l]{\strut{} 4}}%
      \put(3803,2820){\makebox(0,0)[l]{\strut{} 4}}%
      \put(721,2613){\makebox(0,0)[l]{\strut{}5.11~GeV}}%
    }%
    \gplgaddtomacro\gplbacktext{%
    }%
    \gplgaddtomacro\gplfronttext{%
      \csname LTb\endcsname
      \put(4263,484){\makebox(0,0){\strut{}}}%
      \put(4983,484){\makebox(0,0){\strut{}}}%
      \put(5702,484){\makebox(0,0){\strut{}}}%
      \put(6421,484){\makebox(0,0){\strut{}}}%
      \put(7141,484){\makebox(0,0){\strut{}}}%
      \put(4041,750){\makebox(0,0)[r]{\strut{}}}%
      \put(4041,1095){\makebox(0,0)[r]{\strut{}}}%
      \put(4041,1440){\makebox(0,0)[r]{\strut{}}}%
      \put(4041,1785){\makebox(0,0)[r]{\strut{}}}%
      \put(4041,2130){\makebox(0,0)[r]{\strut{}}}%
      \put(4041,2475){\makebox(0,0)[r]{\strut{}}}%
      \put(7489,750){\makebox(0,0)[l]{\strut{} 0}}%
      \put(7489,980){\makebox(0,0)[l]{\strut{}0.01}}%
      \put(7489,1210){\makebox(0,0)[l]{\strut{}0.02}}%
      \put(7489,1440){\makebox(0,0)[l]{\strut{}0.03}}%
      \put(7489,1670){\makebox(0,0)[l]{\strut{}0.04}}%
      \put(7489,1900){\makebox(0,0)[l]{\strut{}0.05}}%
      \put(7489,2130){\makebox(0,0)[l]{\strut{}0.06}}%
      \put(7489,2360){\makebox(0,0)[l]{\strut{}0.07}}%
      \put(7489,2590){\makebox(0,0)[l]{\strut{}0.08}}%
      \put(7489,2820){\makebox(0,0)[l]{\strut{}0.09}}%
      \put(4407,2613){\makebox(0,0)[l]{\strut{}16.1~GeV}}%
    }%
    \gplgaddtomacro\gplbacktext{%
    }%
    \gplgaddtomacro\gplfronttext{%
      \csname LTb\endcsname
      \put(7950,484){\makebox(0,0){\strut{}}}%
      \put(8670,484){\makebox(0,0){\strut{}}}%
      \put(9389,484){\makebox(0,0){\strut{}}}%
      \put(10108,484){\makebox(0,0){\strut{}}}%
      \put(10828,484){\makebox(0,0){\strut{}}}%
      \put(7728,750){\makebox(0,0)[r]{\strut{}}}%
      \put(7728,1095){\makebox(0,0)[r]{\strut{}}}%
      \put(7728,1440){\makebox(0,0)[r]{\strut{}}}%
      \put(7728,1785){\makebox(0,0)[r]{\strut{}}}%
      \put(7728,2130){\makebox(0,0)[r]{\strut{}}}%
      \put(7728,2475){\makebox(0,0)[r]{\strut{}}}%
      \put(11176,750){\makebox(0,0)[l]{\strut{} 0}}%
      \put(11176,1008){\makebox(0,0)[l]{\strut{}1e-06}}%
      \put(11176,1267){\makebox(0,0)[l]{\strut{}2e-06}}%
      \put(11176,1526){\makebox(0,0)[l]{\strut{}3e-06}}%
      \put(11176,1785){\makebox(0,0)[l]{\strut{}4e-06}}%
      \put(11176,2043){\makebox(0,0)[l]{\strut{}5e-06}}%
      \put(11176,2302){\makebox(0,0)[l]{\strut{}6e-06}}%
      \put(11176,2561){\makebox(0,0)[l]{\strut{}7e-06}}%
      \put(11176,2820){\makebox(0,0)[l]{\strut{}8e-06}}%
      \put(8094,2613){\makebox(0,0)[l]{\strut{}51.1~GeV}}%
    }%
    \gplbacktext
    \put(0,0){\includegraphics{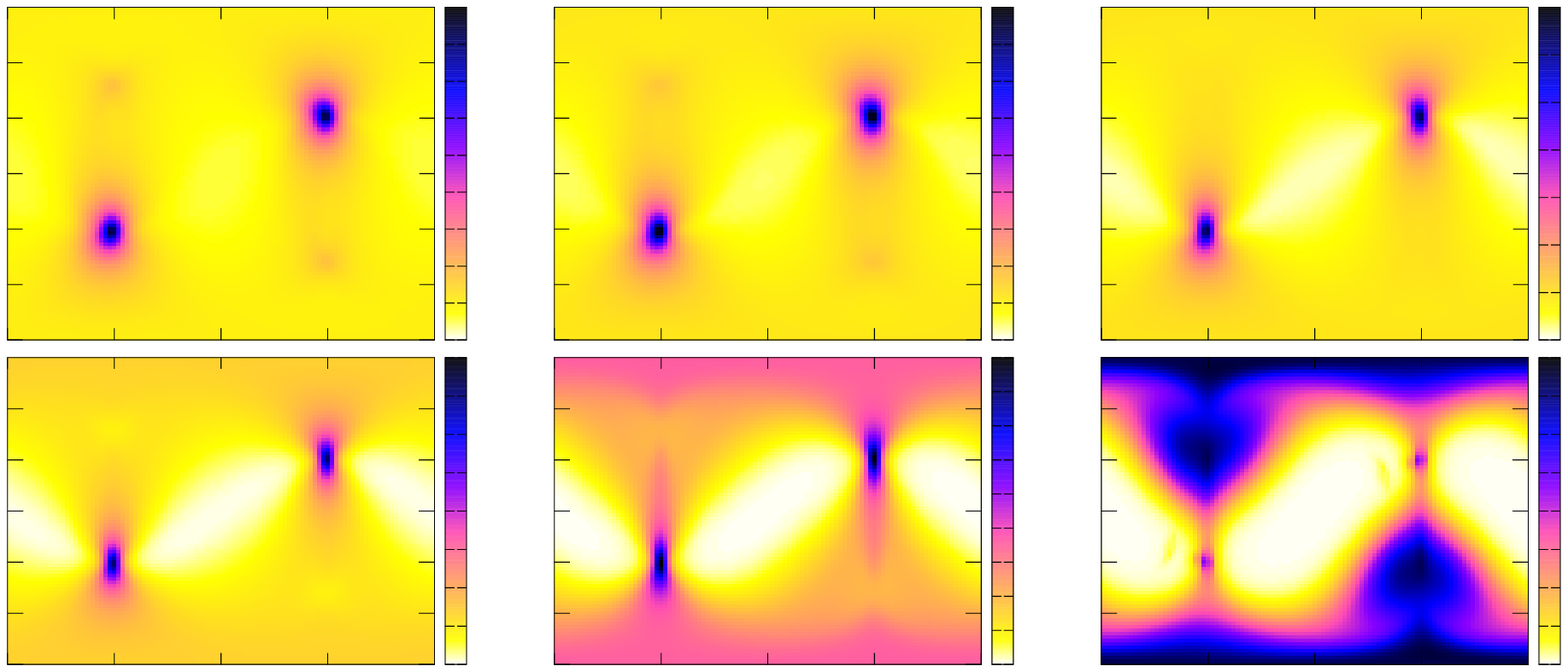}}%
    \gplfronttext
  \end{picture}%
\endgroup

%% file: skymaps_cl_r00025_b-3_c60_ri0.2_ro1_g10_d1.tex
\begingroup
  \makeatletter
  \providecommand\color[2][]{%
    \GenericError{(gnuplot) \space\space\space\@spaces}{%
      Package color not loaded in conjunction with
      terminal option `colourtext'%
    }{See the gnuplot documentation for explanation.%
    }{Either use 'blacktext' in gnuplot or load the package
      color.sty in LaTeX.}%
    \renewcommand\color[2][]{}%
  }%
  \providecommand\includegraphics[2][]{%
    \GenericError{(gnuplot) \space\space\space\@spaces}{%
      Package graphicx or graphics not loaded%
    }{See the gnuplot documentation for explanation.%
    }{The gnuplot epslatex terminal needs graphicx.sty or graphics.sty.}%
    \renewcommand\includegraphics[2][]{}%
  }%
  \providecommand\rotatebox[2]{#2}%
  \@ifundefined{ifGPcolor}{%
    \newif\ifGPcolor
    \GPcolortrue
  }{}%
  \@ifundefined{ifGPblacktext}{%
    \newif\ifGPblacktext
    \GPblacktextfalse
  }{}%
  \let\gplgaddtomacro\g@addto@macro
  \gdef\gplbacktext{}%
  \gdef\gplfronttext{}%
  \makeatother
  \ifGPblacktext
    \def\colorrgb#1{}%
    \def\colorgray#1{}%
  \else
    \ifGPcolor
      \def\colorrgb#1{\color[rgb]{#1}}%
      \def\colorgray#1{\color[gray]{#1}}%
      \expandafter\def\csname LTw\endcsname{\color{white}}%
      \expandafter\def\csname LTb\endcsname{\color{black}}%
      \expandafter\def\csname LTa\endcsname{\color{black}}%
      \expandafter\def\csname LT0\endcsname{\color[rgb]{1,0,0}}%
      \expandafter\def\csname LT1\endcsname{\color[rgb]{0,1,0}}%
      \expandafter\def\csname LT2\endcsname{\color[rgb]{0,0,1}}%
      \expandafter\def\csname LT3\endcsname{\color[rgb]{1,0,1}}%
      \expandafter\def\csname LT4\endcsname{\color[rgb]{0,1,1}}%
      \expandafter\def\csname LT5\endcsname{\color[rgb]{1,1,0}}%
      \expandafter\def\csname LT6\endcsname{\color[rgb]{0,0,0}}%
      \expandafter\def\csname LT7\endcsname{\color[rgb]{1,0.3,0}}%
      \expandafter\def\csname LT8\endcsname{\color[rgb]{0.5,0.5,0.5}}%
    \else
      \def\colorrgb#1{\color{black}}%
      \def\colorgray#1{\color[gray]{#1}}%
      \expandafter\def\csname LTw\endcsname{\color{white}}%
      \expandafter\def\csname LTb\endcsname{\color{black}}%
      \expandafter\def\csname LTa\endcsname{\color{black}}%
      \expandafter\def\csname LT0\endcsname{\color{black}}%
      \expandafter\def\csname LT1\endcsname{\color{black}}%
      \expandafter\def\csname LT2\endcsname{\color{black}}%
      \expandafter\def\csname LT3\endcsname{\color{black}}%
      \expandafter\def\csname LT4\endcsname{\color{black}}%
      \expandafter\def\csname LT5\endcsname{\color{black}}%
      \expandafter\def\csname LT6\endcsname{\color{black}}%
      \expandafter\def\csname LT7\endcsname{\color{black}}%
      \expandafter\def\csname LT8\endcsname{\color{black}}%
    \fi
  \fi
    \setlength{\unitlength}{0.0500bp}%
    \ifx\gptboxheight\undefined%
      \newlength{\gptboxheight}%
      \newlength{\gptboxwidth}%
      \newsavebox{\gptboxtext}%
    \fi%
    \setlength{\fboxrule}{0.5pt}%
    \setlength{\fboxsep}{1pt}%
\begin{picture}(11520.00,5760.00)%
    \gplgaddtomacro\gplbacktext{%
    }%
    \gplgaddtomacro\gplfronttext{%
      \csname LTb\endcsname
      \put(577,2662){\makebox(0,0){\strut{}}}%
      \put(1297,2662){\makebox(0,0){\strut{}}}%
      \put(2016,2662){\makebox(0,0){\strut{}}}%
      \put(2735,2662){\makebox(0,0){\strut{}}}%
      \put(3455,2662){\makebox(0,0){\strut{}}}%
      \put(364,2939){\makebox(0,0)[r]{\strut{}}}%
      \put(364,3313){\makebox(0,0)[r]{\strut{}30}}%
      \put(364,3687){\makebox(0,0)[r]{\strut{}60}}%
      \put(364,4061){\makebox(0,0)[r]{\strut{}90}}%
      \put(364,4435){\makebox(0,0)[r]{\strut{}120}}%
      \put(364,4809){\makebox(0,0)[r]{\strut{}150}}%
      \put(-98,4061){\rotatebox{-270}{\makebox(0,0){\strut{}line of sight ($\zeta$)}}}%
      \put(3803,2939){\makebox(0,0)[l]{\strut{} 0}}%
      \put(3803,3188){\makebox(0,0)[l]{\strut{}5e+04}}%
      \put(3803,3437){\makebox(0,0)[l]{\strut{}1e+05}}%
      \put(3803,3687){\makebox(0,0)[l]{\strut{}2e+05}}%
      \put(3803,3936){\makebox(0,0)[l]{\strut{}2e+05}}%
      \put(3803,4185){\makebox(0,0)[l]{\strut{}2e+05}}%
      \put(3803,4435){\makebox(0,0)[l]{\strut{}3e+05}}%
      \put(3803,4684){\makebox(0,0)[l]{\strut{}4e+05}}%
      \put(3803,4933){\makebox(0,0)[l]{\strut{}4e+05}}%
      \put(3803,5183){\makebox(0,0)[l]{\strut{}4e+05}}%
      \put(721,4959){\makebox(0,0)[l]{\strut{}161~MeV}}%
    }%
    \gplgaddtomacro\gplbacktext{%
    }%
    \gplgaddtomacro\gplfronttext{%
      \csname LTb\endcsname
      \put(4263,2662){\makebox(0,0){\strut{}}}%
      \put(4983,2662){\makebox(0,0){\strut{}}}%
      \put(5702,2662){\makebox(0,0){\strut{}}}%
      \put(6421,2662){\makebox(0,0){\strut{}}}%
      \put(7141,2662){\makebox(0,0){\strut{}}}%
      \put(4050,2939){\makebox(0,0)[r]{\strut{}}}%
      \put(4050,3313){\makebox(0,0)[r]{\strut{}}}%
      \put(4050,3687){\makebox(0,0)[r]{\strut{}}}%
      \put(4050,4061){\makebox(0,0)[r]{\strut{}}}%
      \put(4050,4435){\makebox(0,0)[r]{\strut{}}}%
      \put(4050,4809){\makebox(0,0)[r]{\strut{}}}%
      \put(7489,2939){\makebox(0,0)[l]{\strut{} 0}}%
      \put(7489,3188){\makebox(0,0)[l]{\strut{}5e+04}}%
      \put(7489,3437){\makebox(0,0)[l]{\strut{}1e+05}}%
      \put(7489,3687){\makebox(0,0)[l]{\strut{}2e+05}}%
      \put(7489,3936){\makebox(0,0)[l]{\strut{}2e+05}}%
      \put(7489,4185){\makebox(0,0)[l]{\strut{}2e+05}}%
      \put(7489,4435){\makebox(0,0)[l]{\strut{}3e+05}}%
      \put(7489,4684){\makebox(0,0)[l]{\strut{}4e+05}}%
      \put(7489,4933){\makebox(0,0)[l]{\strut{}4e+05}}%
      \put(7489,5183){\makebox(0,0)[l]{\strut{}4e+05}}%
      \put(4407,4959){\makebox(0,0)[l]{\strut{}511~MeV}}%
    }%
    \gplgaddtomacro\gplbacktext{%
    }%
    \gplgaddtomacro\gplfronttext{%
      \csname LTb\endcsname
      \put(7950,2662){\makebox(0,0){\strut{}}}%
      \put(8670,2662){\makebox(0,0){\strut{}}}%
      \put(9389,2662){\makebox(0,0){\strut{}}}%
      \put(10108,2662){\makebox(0,0){\strut{}}}%
      \put(10828,2662){\makebox(0,0){\strut{}}}%
      \put(7737,2939){\makebox(0,0)[r]{\strut{}}}%
      \put(7737,3313){\makebox(0,0)[r]{\strut{}}}%
      \put(7737,3687){\makebox(0,0)[r]{\strut{}}}%
      \put(7737,4061){\makebox(0,0)[r]{\strut{}}}%
      \put(7737,4435){\makebox(0,0)[r]{\strut{}}}%
      \put(7737,4809){\makebox(0,0)[r]{\strut{}}}%
      \put(11176,2939){\makebox(0,0)[l]{\strut{} 0}}%
      \put(11176,3387){\makebox(0,0)[l]{\strut{}5e+04}}%
      \put(11176,3836){\makebox(0,0)[l]{\strut{}1e+05}}%
      \put(11176,4285){\makebox(0,0)[l]{\strut{}2e+05}}%
      \put(11176,4734){\makebox(0,0)[l]{\strut{}2e+05}}%
      \put(11176,5183){\makebox(0,0)[l]{\strut{}2e+05}}%
      \put(8094,4959){\makebox(0,0)[l]{\strut{}1.61~GeV}}%
    }%
    \gplgaddtomacro\gplbacktext{%
    }%
    \gplgaddtomacro\gplfronttext{%
      \csname LTb\endcsname
      \put(577,484){\makebox(0,0){\strut{}}}%
      \put(1297,484){\makebox(0,0){\strut{}}}%
      \put(2016,484){\makebox(0,0){\strut{}}}%
      \put(2735,484){\makebox(0,0){\strut{}}}%
      \put(3455,484){\makebox(0,0){\strut{}}}%
      \put(355,750){\makebox(0,0)[r]{\strut{}}}%
      \put(355,1095){\makebox(0,0)[r]{\strut{}30}}%
      \put(355,1440){\makebox(0,0)[r]{\strut{}60}}%
      \put(355,1785){\makebox(0,0)[r]{\strut{}90}}%
      \put(355,2130){\makebox(0,0)[r]{\strut{}120}}%
      \put(355,2475){\makebox(0,0)[r]{\strut{}150}}%
      \put(-107,1785){\rotatebox{-270}{\makebox(0,0){\strut{}line of sight ($\zeta$)}}}%
      \put(3803,750){\makebox(0,0)[l]{\strut{} 0}}%
      \put(3803,957){\makebox(0,0)[l]{\strut{}5e+03}}%
      \put(3803,1164){\makebox(0,0)[l]{\strut{}1e+04}}%
      \put(3803,1371){\makebox(0,0)[l]{\strut{}2e+04}}%
      \put(3803,1578){\makebox(0,0)[l]{\strut{}2e+04}}%
      \put(3803,1785){\makebox(0,0)[l]{\strut{}2e+04}}%
      \put(3803,1992){\makebox(0,0)[l]{\strut{}3e+04}}%
      \put(3803,2199){\makebox(0,0)[l]{\strut{}4e+04}}%
      \put(3803,2406){\makebox(0,0)[l]{\strut{}4e+04}}%
      \put(3803,2613){\makebox(0,0)[l]{\strut{}4e+04}}%
      \put(3803,2820){\makebox(0,0)[l]{\strut{}5e+04}}%
      \put(721,2613){\makebox(0,0)[l]{\strut{}5.11~GeV}}%
    }%
    \gplgaddtomacro\gplbacktext{%
    }%
    \gplgaddtomacro\gplfronttext{%
      \csname LTb\endcsname
      \put(4263,484){\makebox(0,0){\strut{}}}%
      \put(4983,484){\makebox(0,0){\strut{}}}%
      \put(5702,484){\makebox(0,0){\strut{}}}%
      \put(6421,484){\makebox(0,0){\strut{}}}%
      \put(7141,484){\makebox(0,0){\strut{}}}%
      \put(4041,750){\makebox(0,0)[r]{\strut{}}}%
      \put(4041,1095){\makebox(0,0)[r]{\strut{}}}%
      \put(4041,1440){\makebox(0,0)[r]{\strut{}}}%
      \put(4041,1785){\makebox(0,0)[r]{\strut{}}}%
      \put(4041,2130){\makebox(0,0)[r]{\strut{}}}%
      \put(4041,2475){\makebox(0,0)[r]{\strut{}}}%
      \put(7489,750){\makebox(0,0)[l]{\strut{} 0}}%
      \put(7489,980){\makebox(0,0)[l]{\strut{}1e+02}}%
      \put(7489,1210){\makebox(0,0)[l]{\strut{}2e+02}}%
      \put(7489,1440){\makebox(0,0)[l]{\strut{}3e+02}}%
      \put(7489,1670){\makebox(0,0)[l]{\strut{}4e+02}}%
      \put(7489,1900){\makebox(0,0)[l]{\strut{}5e+02}}%
      \put(7489,2130){\makebox(0,0)[l]{\strut{}6e+02}}%
      \put(7489,2360){\makebox(0,0)[l]{\strut{}7e+02}}%
      \put(7489,2590){\makebox(0,0)[l]{\strut{}8e+02}}%
      \put(7489,2820){\makebox(0,0)[l]{\strut{}9e+02}}%
      \put(4407,2613){\makebox(0,0)[l]{\strut{}16.1~GeV}}%
    }%
    \gplgaddtomacro\gplbacktext{%
    }%
    \gplgaddtomacro\gplfronttext{%
      \csname LTb\endcsname
      \put(7950,484){\makebox(0,0){\strut{}}}%
      \put(8670,484){\makebox(0,0){\strut{}}}%
      \put(9389,484){\makebox(0,0){\strut{}}}%
      \put(10108,484){\makebox(0,0){\strut{}}}%
      \put(10828,484){\makebox(0,0){\strut{}}}%
      \put(7728,750){\makebox(0,0)[r]{\strut{}}}%
      \put(7728,1095){\makebox(0,0)[r]{\strut{}}}%
      \put(7728,1440){\makebox(0,0)[r]{\strut{}}}%
      \put(7728,1785){\makebox(0,0)[r]{\strut{}}}%
      \put(7728,2130){\makebox(0,0)[r]{\strut{}}}%
      \put(7728,2475){\makebox(0,0)[r]{\strut{}}}%
      \put(11176,750){\makebox(0,0)[l]{\strut{} 0}}%
      \put(11176,1045){\makebox(0,0)[l]{\strut{}0.01}}%
      \put(11176,1341){\makebox(0,0)[l]{\strut{}0.02}}%
      \put(11176,1637){\makebox(0,0)[l]{\strut{}0.03}}%
      \put(11176,1932){\makebox(0,0)[l]{\strut{}0.04}}%
      \put(11176,2228){\makebox(0,0)[l]{\strut{}0.05}}%
      \put(11176,2524){\makebox(0,0)[l]{\strut{}0.06}}%
      \put(11176,2820){\makebox(0,0)[l]{\strut{}0.07}}%
      \put(8094,2613){\makebox(0,0)[l]{\strut{}51.1~GeV}}%
    }%
    \gplbacktext
    \put(0,0){\includegraphics{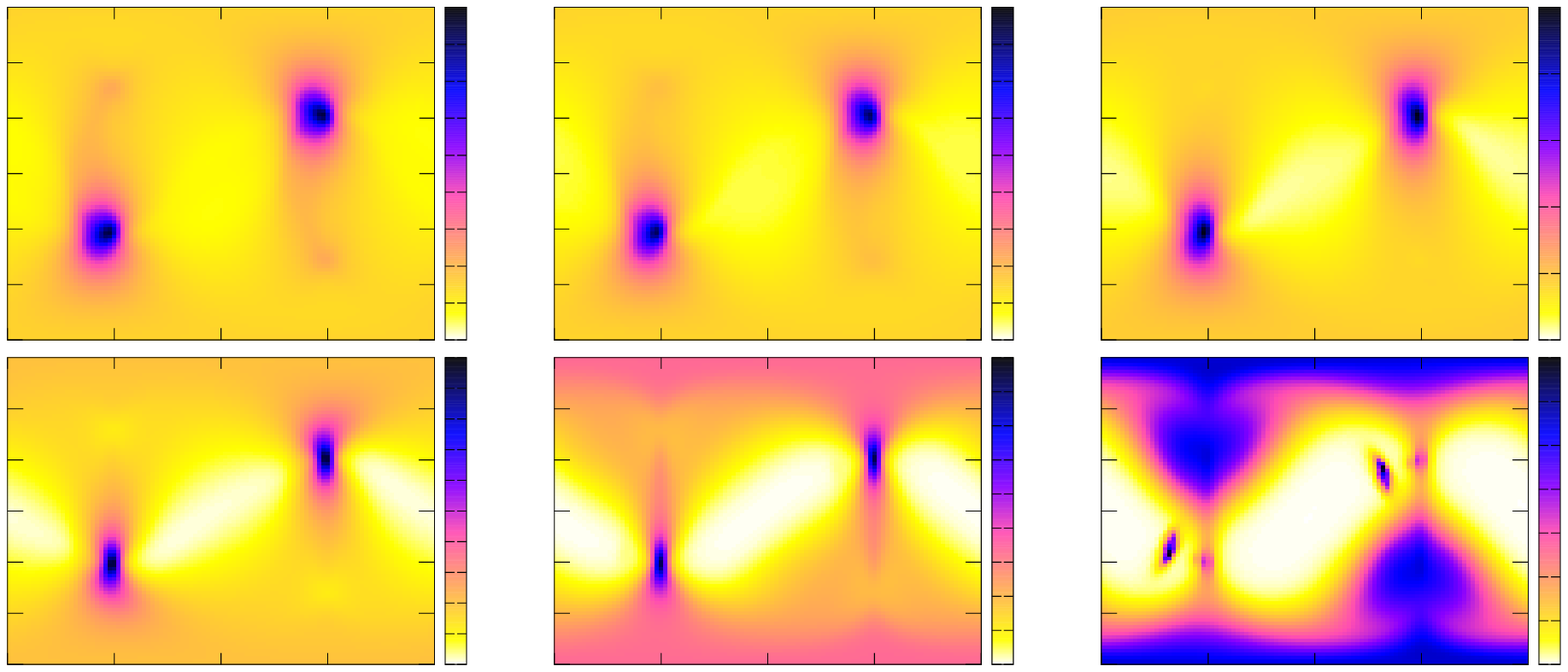}}%
    \gplfronttext
  \end{picture}%
\endgroup

%% file: skymaps_cl_r00025_b-3_c60_ri0.2_ro1_g10_d2.tex
\begingroup
  \makeatletter
  \providecommand\color[2][]{%
    \GenericError{(gnuplot) \space\space\space\@spaces}{%
      Package color not loaded in conjunction with
      terminal option `colourtext'%
    }{See the gnuplot documentation for explanation.%
    }{Either use 'blacktext' in gnuplot or load the package
      color.sty in LaTeX.}%
    \renewcommand\color[2][]{}%
  }%
  \providecommand\includegraphics[2][]{%
    \GenericError{(gnuplot) \space\space\space\@spaces}{%
      Package graphicx or graphics not loaded%
    }{See the gnuplot documentation for explanation.%
    }{The gnuplot epslatex terminal needs graphicx.sty or graphics.sty.}%
    \renewcommand\includegraphics[2][]{}%
  }%
  \providecommand\rotatebox[2]{#2}%
  \@ifundefined{ifGPcolor}{%
    \newif\ifGPcolor
    \GPcolortrue
  }{}%
  \@ifundefined{ifGPblacktext}{%
    \newif\ifGPblacktext
    \GPblacktextfalse
  }{}%
  \let\gplgaddtomacro\g@addto@macro
  \gdef\gplbacktext{}%
  \gdef\gplfronttext{}%
  \makeatother
  \ifGPblacktext
    \def\colorrgb#1{}%
    \def\colorgray#1{}%
  \else
    \ifGPcolor
      \def\colorrgb#1{\color[rgb]{#1}}%
      \def\colorgray#1{\color[gray]{#1}}%
      \expandafter\def\csname LTw\endcsname{\color{white}}%
      \expandafter\def\csname LTb\endcsname{\color{black}}%
      \expandafter\def\csname LTa\endcsname{\color{black}}%
      \expandafter\def\csname LT0\endcsname{\color[rgb]{1,0,0}}%
      \expandafter\def\csname LT1\endcsname{\color[rgb]{0,1,0}}%
      \expandafter\def\csname LT2\endcsname{\color[rgb]{0,0,1}}%
      \expandafter\def\csname LT3\endcsname{\color[rgb]{1,0,1}}%
      \expandafter\def\csname LT4\endcsname{\color[rgb]{0,1,1}}%
      \expandafter\def\csname LT5\endcsname{\color[rgb]{1,1,0}}%
      \expandafter\def\csname LT6\endcsname{\color[rgb]{0,0,0}}%
      \expandafter\def\csname LT7\endcsname{\color[rgb]{1,0.3,0}}%
      \expandafter\def\csname LT8\endcsname{\color[rgb]{0.5,0.5,0.5}}%
    \else
      \def\colorrgb#1{\color{black}}%
      \def\colorgray#1{\color[gray]{#1}}%
      \expandafter\def\csname LTw\endcsname{\color{white}}%
      \expandafter\def\csname LTb\endcsname{\color{black}}%
      \expandafter\def\csname LTa\endcsname{\color{black}}%
      \expandafter\def\csname LT0\endcsname{\color{black}}%
      \expandafter\def\csname LT1\endcsname{\color{black}}%
      \expandafter\def\csname LT2\endcsname{\color{black}}%
      \expandafter\def\csname LT3\endcsname{\color{black}}%
      \expandafter\def\csname LT4\endcsname{\color{black}}%
      \expandafter\def\csname LT5\endcsname{\color{black}}%
      \expandafter\def\csname LT6\endcsname{\color{black}}%
      \expandafter\def\csname LT7\endcsname{\color{black}}%
      \expandafter\def\csname LT8\endcsname{\color{black}}%
    \fi
  \fi
    \setlength{\unitlength}{0.0500bp}%
    \ifx\gptboxheight\undefined%
      \newlength{\gptboxheight}%
      \newlength{\gptboxwidth}%
      \newsavebox{\gptboxtext}%
    \fi%
    \setlength{\fboxrule}{0.5pt}%
    \setlength{\fboxsep}{1pt}%
\begin{picture}(11520.00,5760.00)%
    \gplgaddtomacro\gplbacktext{%
    }%
    \gplgaddtomacro\gplfronttext{%
      \csname LTb\endcsname
      \put(577,2662){\makebox(0,0){\strut{}}}%
      \put(1297,2662){\makebox(0,0){\strut{}}}%
      \put(2016,2662){\makebox(0,0){\strut{}}}%
      \put(2735,2662){\makebox(0,0){\strut{}}}%
      \put(3455,2662){\makebox(0,0){\strut{}}}%
      \put(364,2939){\makebox(0,0)[r]{\strut{}}}%
      \put(364,3313){\makebox(0,0)[r]{\strut{}30}}%
      \put(364,3687){\makebox(0,0)[r]{\strut{}60}}%
      \put(364,4061){\makebox(0,0)[r]{\strut{}90}}%
      \put(364,4435){\makebox(0,0)[r]{\strut{}120}}%
      \put(364,4809){\makebox(0,0)[r]{\strut{}150}}%
      \put(-98,4061){\rotatebox{-270}{\makebox(0,0){\strut{}line of sight ($\zeta$)}}}%
      \put(3803,2939){\makebox(0,0)[l]{\strut{} 0}}%
      \put(3803,3387){\makebox(0,0)[l]{\strut{}5e+02}}%
      \put(3803,3836){\makebox(0,0)[l]{\strut{}1e+03}}%
      \put(3803,4285){\makebox(0,0)[l]{\strut{}2e+03}}%
      \put(3803,4734){\makebox(0,0)[l]{\strut{}2e+03}}%
      \put(3803,5183){\makebox(0,0)[l]{\strut{}2e+03}}%
      \put(721,4959){\makebox(0,0)[l]{\strut{}161~MeV}}%
    }%
    \gplgaddtomacro\gplbacktext{%
    }%
    \gplgaddtomacro\gplfronttext{%
      \csname LTb\endcsname
      \put(4263,2662){\makebox(0,0){\strut{}}}%
      \put(4983,2662){\makebox(0,0){\strut{}}}%
      \put(5702,2662){\makebox(0,0){\strut{}}}%
      \put(6421,2662){\makebox(0,0){\strut{}}}%
      \put(7141,2662){\makebox(0,0){\strut{}}}%
      \put(4050,2939){\makebox(0,0)[r]{\strut{}}}%
      \put(4050,3313){\makebox(0,0)[r]{\strut{}}}%
      \put(4050,3687){\makebox(0,0)[r]{\strut{}}}%
      \put(4050,4061){\makebox(0,0)[r]{\strut{}}}%
      \put(4050,4435){\makebox(0,0)[r]{\strut{}}}%
      \put(4050,4809){\makebox(0,0)[r]{\strut{}}}%
      \put(7489,2939){\makebox(0,0)[l]{\strut{} 0}}%
      \put(7489,3387){\makebox(0,0)[l]{\strut{}5e+02}}%
      \put(7489,3836){\makebox(0,0)[l]{\strut{}1e+03}}%
      \put(7489,4285){\makebox(0,0)[l]{\strut{}2e+03}}%
      \put(7489,4734){\makebox(0,0)[l]{\strut{}2e+03}}%
      \put(7489,5183){\makebox(0,0)[l]{\strut{}2e+03}}%
      \put(4407,4959){\makebox(0,0)[l]{\strut{}511~MeV}}%
    }%
    \gplgaddtomacro\gplbacktext{%
    }%
    \gplgaddtomacro\gplfronttext{%
      \csname LTb\endcsname
      \put(7950,2662){\makebox(0,0){\strut{}}}%
      \put(8670,2662){\makebox(0,0){\strut{}}}%
      \put(9389,2662){\makebox(0,0){\strut{}}}%
      \put(10108,2662){\makebox(0,0){\strut{}}}%
      \put(10828,2662){\makebox(0,0){\strut{}}}%
      \put(7737,2939){\makebox(0,0)[r]{\strut{}}}%
      \put(7737,3313){\makebox(0,0)[r]{\strut{}}}%
      \put(7737,3687){\makebox(0,0)[r]{\strut{}}}%
      \put(7737,4061){\makebox(0,0)[r]{\strut{}}}%
      \put(7737,4435){\makebox(0,0)[r]{\strut{}}}%
      \put(7737,4809){\makebox(0,0)[r]{\strut{}}}%
      \put(11176,2939){\makebox(0,0)[l]{\strut{} 0}}%
      \put(11176,3188){\makebox(0,0)[l]{\strut{}2e+02}}%
      \put(11176,3437){\makebox(0,0)[l]{\strut{}4e+02}}%
      \put(11176,3687){\makebox(0,0)[l]{\strut{}6e+02}}%
      \put(11176,3936){\makebox(0,0)[l]{\strut{}8e+02}}%
      \put(11176,4185){\makebox(0,0)[l]{\strut{}1e+03}}%
      \put(11176,4435){\makebox(0,0)[l]{\strut{}1e+03}}%
      \put(11176,4684){\makebox(0,0)[l]{\strut{}1e+03}}%
      \put(11176,4933){\makebox(0,0)[l]{\strut{}2e+03}}%
      \put(11176,5183){\makebox(0,0)[l]{\strut{}2e+03}}%
      \put(8094,4959){\makebox(0,0)[l]{\strut{}1.61~GeV}}%
    }%
    \gplgaddtomacro\gplbacktext{%
    }%
    \gplgaddtomacro\gplfronttext{%
      \csname LTb\endcsname
      \put(577,484){\makebox(0,0){\strut{}}}%
      \put(1297,484){\makebox(0,0){\strut{}}}%
      \put(2016,484){\makebox(0,0){\strut{}}}%
      \put(2735,484){\makebox(0,0){\strut{}}}%
      \put(3455,484){\makebox(0,0){\strut{}}}%
      \put(355,750){\makebox(0,0)[r]{\strut{}}}%
      \put(355,1095){\makebox(0,0)[r]{\strut{}30}}%
      \put(355,1440){\makebox(0,0)[r]{\strut{}60}}%
      \put(355,1785){\makebox(0,0)[r]{\strut{}90}}%
      \put(355,2130){\makebox(0,0)[r]{\strut{}120}}%
      \put(355,2475){\makebox(0,0)[r]{\strut{}150}}%
      \put(-107,1785){\rotatebox{-270}{\makebox(0,0){\strut{}line of sight ($\zeta$)}}}%
      \put(3803,750){\makebox(0,0)[l]{\strut{} 0}}%
      \put(3803,1008){\makebox(0,0)[l]{\strut{}5e+01}}%
      \put(3803,1267){\makebox(0,0)[l]{\strut{}1e+02}}%
      \put(3803,1526){\makebox(0,0)[l]{\strut{}2e+02}}%
      \put(3803,1785){\makebox(0,0)[l]{\strut{}2e+02}}%
      \put(3803,2043){\makebox(0,0)[l]{\strut{}2e+02}}%
      \put(3803,2302){\makebox(0,0)[l]{\strut{}3e+02}}%
      \put(3803,2561){\makebox(0,0)[l]{\strut{}4e+02}}%
      \put(3803,2820){\makebox(0,0)[l]{\strut{}4e+02}}%
      \put(721,2613){\makebox(0,0)[l]{\strut{}5.11~GeV}}%
    }%
    \gplgaddtomacro\gplbacktext{%
    }%
    \gplgaddtomacro\gplfronttext{%
      \csname LTb\endcsname
      \put(4263,484){\makebox(0,0){\strut{}}}%
      \put(4983,484){\makebox(0,0){\strut{}}}%
      \put(5702,484){\makebox(0,0){\strut{}}}%
      \put(6421,484){\makebox(0,0){\strut{}}}%
      \put(7141,484){\makebox(0,0){\strut{}}}%
      \put(4041,750){\makebox(0,0)[r]{\strut{}}}%
      \put(4041,1095){\makebox(0,0)[r]{\strut{}}}%
      \put(4041,1440){\makebox(0,0)[r]{\strut{}}}%
      \put(4041,1785){\makebox(0,0)[r]{\strut{}}}%
      \put(4041,2130){\makebox(0,0)[r]{\strut{}}}%
      \put(4041,2475){\makebox(0,0)[r]{\strut{}}}%
      \put(7489,750){\makebox(0,0)[l]{\strut{} 0}}%
      \put(7489,980){\makebox(0,0)[l]{\strut{} 1}}%
      \put(7489,1210){\makebox(0,0)[l]{\strut{} 2}}%
      \put(7489,1440){\makebox(0,0)[l]{\strut{} 3}}%
      \put(7489,1670){\makebox(0,0)[l]{\strut{} 4}}%
      \put(7489,1900){\makebox(0,0)[l]{\strut{} 5}}%
      \put(7489,2130){\makebox(0,0)[l]{\strut{} 6}}%
      \put(7489,2360){\makebox(0,0)[l]{\strut{} 7}}%
      \put(7489,2590){\makebox(0,0)[l]{\strut{} 8}}%
      \put(7489,2820){\makebox(0,0)[l]{\strut{} 9}}%
      \put(4407,2613){\makebox(0,0)[l]{\strut{}16.1~GeV}}%
    }%
    \gplgaddtomacro\gplbacktext{%
    }%
    \gplgaddtomacro\gplfronttext{%
      \csname LTb\endcsname
      \put(7950,484){\makebox(0,0){\strut{}}}%
      \put(8670,484){\makebox(0,0){\strut{}}}%
      \put(9389,484){\makebox(0,0){\strut{}}}%
      \put(10108,484){\makebox(0,0){\strut{}}}%
      \put(10828,484){\makebox(0,0){\strut{}}}%
      \put(7728,750){\makebox(0,0)[r]{\strut{}}}%
      \put(7728,1095){\makebox(0,0)[r]{\strut{}}}%
      \put(7728,1440){\makebox(0,0)[r]{\strut{}}}%
      \put(7728,1785){\makebox(0,0)[r]{\strut{}}}%
      \put(7728,2130){\makebox(0,0)[r]{\strut{}}}%
      \put(7728,2475){\makebox(0,0)[r]{\strut{}}}%
      \put(11176,750){\makebox(0,0)[l]{\strut{} 0}}%
      \put(11176,1045){\makebox(0,0)[l]{\strut{}0.0001}}%
      \put(11176,1341){\makebox(0,0)[l]{\strut{}0.0002}}%
      \put(11176,1637){\makebox(0,0)[l]{\strut{}0.0003}}%
      \put(11176,1932){\makebox(0,0)[l]{\strut{}0.0004}}%
      \put(11176,2228){\makebox(0,0)[l]{\strut{}0.0005}}%
      \put(11176,2524){\makebox(0,0)[l]{\strut{}0.0006}}%
      \put(11176,2820){\makebox(0,0)[l]{\strut{}0.0007}}%
      \put(8094,2613){\makebox(0,0)[l]{\strut{}51.1~GeV}}%
    }%
    \gplbacktext
    \put(0,0){\includegraphics{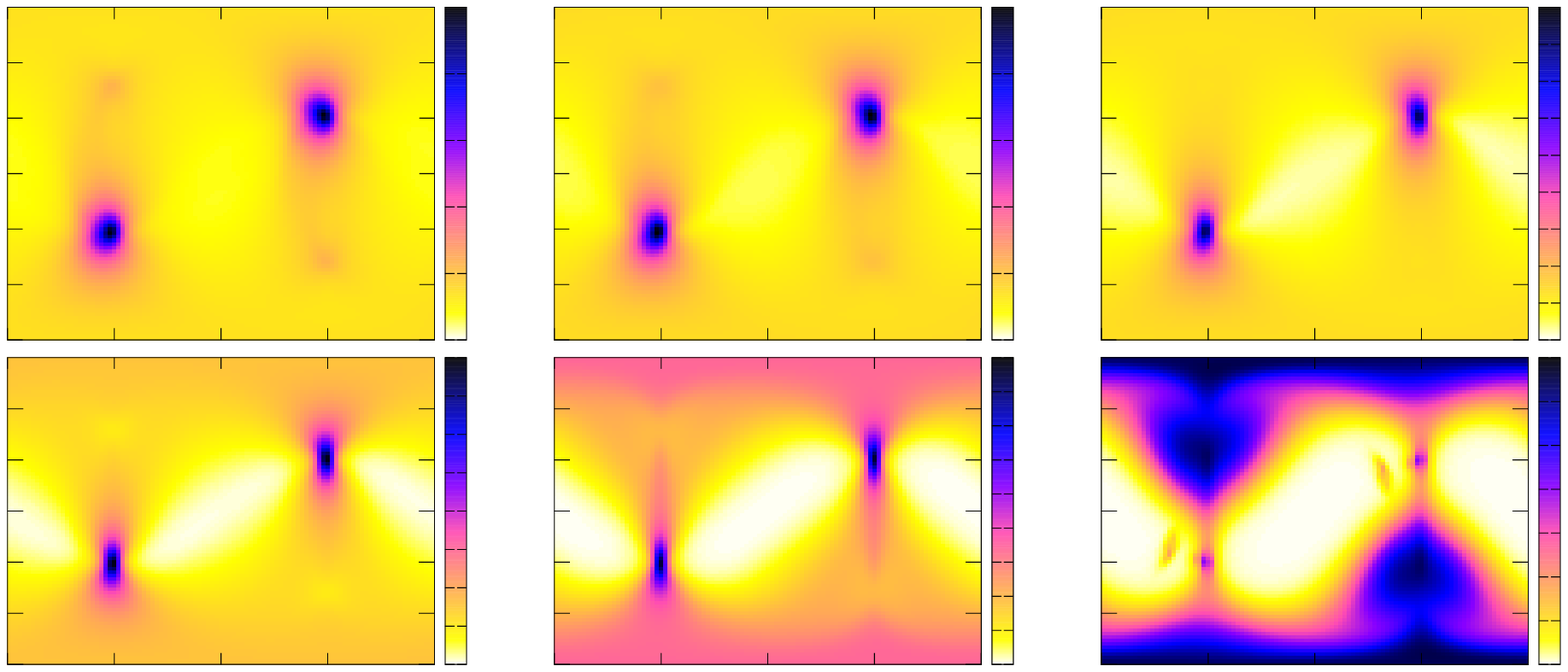}}%
    \gplfronttext
  \end{picture}%
\endgroup

%% file: skymaps_cl_r00025_b-3_c60_ri0.1_ro1_g10_d3.tex
\begingroup
  \makeatletter
  \providecommand\color[2][]{%
    \GenericError{(gnuplot) \space\space\space\@spaces}{%
      Package color not loaded in conjunction with
      terminal option `colourtext'%
    }{See the gnuplot documentation for explanation.%
    }{Either use 'blacktext' in gnuplot or load the package
      color.sty in LaTeX.}%
    \renewcommand\color[2][]{}%
  }%
  \providecommand\includegraphics[2][]{%
    \GenericError{(gnuplot) \space\space\space\@spaces}{%
      Package graphicx or graphics not loaded%
    }{See the gnuplot documentation for explanation.%
    }{The gnuplot epslatex terminal needs graphicx.sty or graphics.sty.}%
    \renewcommand\includegraphics[2][]{}%
  }%
  \providecommand\rotatebox[2]{#2}%
  \@ifundefined{ifGPcolor}{%
    \newif\ifGPcolor
    \GPcolortrue
  }{}%
  \@ifundefined{ifGPblacktext}{%
    \newif\ifGPblacktext
    \GPblacktextfalse
  }{}%
  \let\gplgaddtomacro\g@addto@macro
  \gdef\gplbacktext{}%
  \gdef\gplfronttext{}%
  \makeatother
  \ifGPblacktext
    \def\colorrgb#1{}%
    \def\colorgray#1{}%
  \else
    \ifGPcolor
      \def\colorrgb#1{\color[rgb]{#1}}%
      \def\colorgray#1{\color[gray]{#1}}%
      \expandafter\def\csname LTw\endcsname{\color{white}}%
      \expandafter\def\csname LTb\endcsname{\color{black}}%
      \expandafter\def\csname LTa\endcsname{\color{black}}%
      \expandafter\def\csname LT0\endcsname{\color[rgb]{1,0,0}}%
      \expandafter\def\csname LT1\endcsname{\color[rgb]{0,1,0}}%
      \expandafter\def\csname LT2\endcsname{\color[rgb]{0,0,1}}%
      \expandafter\def\csname LT3\endcsname{\color[rgb]{1,0,1}}%
      \expandafter\def\csname LT4\endcsname{\color[rgb]{0,1,1}}%
      \expandafter\def\csname LT5\endcsname{\color[rgb]{1,1,0}}%
      \expandafter\def\csname LT6\endcsname{\color[rgb]{0,0,0}}%
      \expandafter\def\csname LT7\endcsname{\color[rgb]{1,0.3,0}}%
      \expandafter\def\csname LT8\endcsname{\color[rgb]{0.5,0.5,0.5}}%
    \else
      \def\colorrgb#1{\color{black}}%
      \def\colorgray#1{\color[gray]{#1}}%
      \expandafter\def\csname LTw\endcsname{\color{white}}%
      \expandafter\def\csname LTb\endcsname{\color{black}}%
      \expandafter\def\csname LTa\endcsname{\color{black}}%
      \expandafter\def\csname LT0\endcsname{\color{black}}%
      \expandafter\def\csname LT1\endcsname{\color{black}}%
      \expandafter\def\csname LT2\endcsname{\color{black}}%
      \expandafter\def\csname LT3\endcsname{\color{black}}%
      \expandafter\def\csname LT4\endcsname{\color{black}}%
      \expandafter\def\csname LT5\endcsname{\color{black}}%
      \expandafter\def\csname LT6\endcsname{\color{black}}%
      \expandafter\def\csname LT7\endcsname{\color{black}}%
      \expandafter\def\csname LT8\endcsname{\color{black}}%
    \fi
  \fi
    \setlength{\unitlength}{0.0500bp}%
    \ifx\gptboxheight\undefined%
      \newlength{\gptboxheight}%
      \newlength{\gptboxwidth}%
      \newsavebox{\gptboxtext}%
    \fi%
    \setlength{\fboxrule}{0.5pt}%
    \setlength{\fboxsep}{1pt}%
\begin{picture}(11520.00,5760.00)%
    \gplgaddtomacro\gplbacktext{%
    }%
    \gplgaddtomacro\gplfronttext{%
      \csname LTb\endcsname
      \put(577,2662){\makebox(0,0){\strut{}}}%
      \put(1297,2662){\makebox(0,0){\strut{}}}%
      \put(2016,2662){\makebox(0,0){\strut{}}}%
      \put(2735,2662){\makebox(0,0){\strut{}}}%
      \put(3455,2662){\makebox(0,0){\strut{}}}%
      \put(364,2939){\makebox(0,0)[r]{\strut{}}}%
      \put(364,3313){\makebox(0,0)[r]{\strut{}30}}%
      \put(364,3687){\makebox(0,0)[r]{\strut{}60}}%
      \put(364,4061){\makebox(0,0)[r]{\strut{}90}}%
      \put(364,4435){\makebox(0,0)[r]{\strut{}120}}%
      \put(364,4809){\makebox(0,0)[r]{\strut{}150}}%
      \put(-98,4061){\rotatebox{-270}{\makebox(0,0){\strut{}line of sight ($\zeta$)}}}%
      \put(3803,2939){\makebox(0,0)[l]{\strut{} 0}}%
      \put(3803,3259){\makebox(0,0)[l]{\strut{} 5}}%
      \put(3803,3580){\makebox(0,0)[l]{\strut{}1e+01}}%
      \put(3803,3900){\makebox(0,0)[l]{\strut{}2e+01}}%
      \put(3803,4221){\makebox(0,0)[l]{\strut{}2e+01}}%
      \put(3803,4541){\makebox(0,0)[l]{\strut{}2e+01}}%
      \put(3803,4862){\makebox(0,0)[l]{\strut{}3e+01}}%
      \put(3803,5183){\makebox(0,0)[l]{\strut{}4e+01}}%
      \put(721,4959){\makebox(0,0)[l]{\strut{}161~MeV}}%
    }%
    \gplgaddtomacro\gplbacktext{%
    }%
    \gplgaddtomacro\gplfronttext{%
      \csname LTb\endcsname
      \put(4263,2662){\makebox(0,0){\strut{}}}%
      \put(4983,2662){\makebox(0,0){\strut{}}}%
      \put(5702,2662){\makebox(0,0){\strut{}}}%
      \put(6421,2662){\makebox(0,0){\strut{}}}%
      \put(7141,2662){\makebox(0,0){\strut{}}}%
      \put(4050,2939){\makebox(0,0)[r]{\strut{}}}%
      \put(4050,3313){\makebox(0,0)[r]{\strut{}}}%
      \put(4050,3687){\makebox(0,0)[r]{\strut{}}}%
      \put(4050,4061){\makebox(0,0)[r]{\strut{}}}%
      \put(4050,4435){\makebox(0,0)[r]{\strut{}}}%
      \put(4050,4809){\makebox(0,0)[r]{\strut{}}}%
      \put(7489,2939){\makebox(0,0)[l]{\strut{} 0}}%
      \put(7489,3219){\makebox(0,0)[l]{\strut{} 5}}%
      \put(7489,3500){\makebox(0,0)[l]{\strut{}1e+01}}%
      \put(7489,3780){\makebox(0,0)[l]{\strut{}2e+01}}%
      \put(7489,4061){\makebox(0,0)[l]{\strut{}2e+01}}%
      \put(7489,4341){\makebox(0,0)[l]{\strut{}2e+01}}%
      \put(7489,4622){\makebox(0,0)[l]{\strut{}3e+01}}%
      \put(7489,4902){\makebox(0,0)[l]{\strut{}4e+01}}%
      \put(7489,5183){\makebox(0,0)[l]{\strut{}4e+01}}%
      \put(4407,4959){\makebox(0,0)[l]{\strut{}511~MeV}}%
    }%
    \gplgaddtomacro\gplbacktext{%
    }%
    \gplgaddtomacro\gplfronttext{%
      \csname LTb\endcsname
      \put(7950,2662){\makebox(0,0){\strut{}}}%
      \put(8670,2662){\makebox(0,0){\strut{}}}%
      \put(9389,2662){\makebox(0,0){\strut{}}}%
      \put(10108,2662){\makebox(0,0){\strut{}}}%
      \put(10828,2662){\makebox(0,0){\strut{}}}%
      \put(7737,2939){\makebox(0,0)[r]{\strut{}}}%
      \put(7737,3313){\makebox(0,0)[r]{\strut{}}}%
      \put(7737,3687){\makebox(0,0)[r]{\strut{}}}%
      \put(7737,4061){\makebox(0,0)[r]{\strut{}}}%
      \put(7737,4435){\makebox(0,0)[r]{\strut{}}}%
      \put(7737,4809){\makebox(0,0)[r]{\strut{}}}%
      \put(11176,2939){\makebox(0,0)[l]{\strut{} 0}}%
      \put(11176,3219){\makebox(0,0)[l]{\strut{} 5}}%
      \put(11176,3500){\makebox(0,0)[l]{\strut{}1e+01}}%
      \put(11176,3780){\makebox(0,0)[l]{\strut{}2e+01}}%
      \put(11176,4061){\makebox(0,0)[l]{\strut{}2e+01}}%
      \put(11176,4341){\makebox(0,0)[l]{\strut{}2e+01}}%
      \put(11176,4622){\makebox(0,0)[l]{\strut{}3e+01}}%
      \put(11176,4902){\makebox(0,0)[l]{\strut{}4e+01}}%
      \put(11176,5183){\makebox(0,0)[l]{\strut{}4e+01}}%
      \put(8094,4959){\makebox(0,0)[l]{\strut{}1.61~GeV}}%
    }%
    \gplgaddtomacro\gplbacktext{%
    }%
    \gplgaddtomacro\gplfronttext{%
      \csname LTb\endcsname
      \put(577,484){\makebox(0,0){\strut{}}}%
      \put(1297,484){\makebox(0,0){\strut{}}}%
      \put(2016,484){\makebox(0,0){\strut{}}}%
      \put(2735,484){\makebox(0,0){\strut{}}}%
      \put(3455,484){\makebox(0,0){\strut{}}}%
      \put(355,750){\makebox(0,0)[r]{\strut{}}}%
      \put(355,1095){\makebox(0,0)[r]{\strut{}30}}%
      \put(355,1440){\makebox(0,0)[r]{\strut{}60}}%
      \put(355,1785){\makebox(0,0)[r]{\strut{}90}}%
      \put(355,2130){\makebox(0,0)[r]{\strut{}120}}%
      \put(355,2475){\makebox(0,0)[r]{\strut{}150}}%
      \put(-107,1785){\rotatebox{-270}{\makebox(0,0){\strut{}line of sight ($\zeta$)}}}%
      \put(3803,750){\makebox(0,0)[l]{\strut{} 0}}%
      \put(3803,980){\makebox(0,0)[l]{\strut{} 2}}%
      \put(3803,1210){\makebox(0,0)[l]{\strut{} 4}}%
      \put(3803,1440){\makebox(0,0)[l]{\strut{} 6}}%
      \put(3803,1670){\makebox(0,0)[l]{\strut{} 8}}%
      \put(3803,1900){\makebox(0,0)[l]{\strut{}1e+01}}%
      \put(3803,2130){\makebox(0,0)[l]{\strut{}1e+01}}%
      \put(3803,2360){\makebox(0,0)[l]{\strut{}1e+01}}%
      \put(3803,2590){\makebox(0,0)[l]{\strut{}2e+01}}%
      \put(3803,2820){\makebox(0,0)[l]{\strut{}2e+01}}%
      \put(721,2613){\makebox(0,0)[l]{\strut{}5.11~GeV}}%
    }%
    \gplgaddtomacro\gplbacktext{%
    }%
    \gplgaddtomacro\gplfronttext{%
      \csname LTb\endcsname
      \put(4263,484){\makebox(0,0){\strut{}}}%
      \put(4983,484){\makebox(0,0){\strut{}}}%
      \put(5702,484){\makebox(0,0){\strut{}}}%
      \put(6421,484){\makebox(0,0){\strut{}}}%
      \put(7141,484){\makebox(0,0){\strut{}}}%
      \put(4041,750){\makebox(0,0)[r]{\strut{}}}%
      \put(4041,1095){\makebox(0,0)[r]{\strut{}}}%
      \put(4041,1440){\makebox(0,0)[r]{\strut{}}}%
      \put(4041,1785){\makebox(0,0)[r]{\strut{}}}%
      \put(4041,2130){\makebox(0,0)[r]{\strut{}}}%
      \put(4041,2475){\makebox(0,0)[r]{\strut{}}}%
      \put(7489,750){\makebox(0,0)[l]{\strut{} 0}}%
      \put(7489,1164){\makebox(0,0)[l]{\strut{}0.5}}%
      \put(7489,1578){\makebox(0,0)[l]{\strut{} 1}}%
      \put(7489,1992){\makebox(0,0)[l]{\strut{} 2}}%
      \put(7489,2406){\makebox(0,0)[l]{\strut{} 2}}%
      \put(7489,2820){\makebox(0,0)[l]{\strut{} 2}}%
      \put(4407,2613){\makebox(0,0)[l]{\strut{}16.1~GeV}}%
    }%
    \gplgaddtomacro\gplbacktext{%
    }%
    \gplgaddtomacro\gplfronttext{%
      \csname LTb\endcsname
      \put(7950,484){\makebox(0,0){\strut{}}}%
      \put(8670,484){\makebox(0,0){\strut{}}}%
      \put(9389,484){\makebox(0,0){\strut{}}}%
      \put(10108,484){\makebox(0,0){\strut{}}}%
      \put(10828,484){\makebox(0,0){\strut{}}}%
      \put(7728,750){\makebox(0,0)[r]{\strut{}}}%
      \put(7728,1095){\makebox(0,0)[r]{\strut{}}}%
      \put(7728,1440){\makebox(0,0)[r]{\strut{}}}%
      \put(7728,1785){\makebox(0,0)[r]{\strut{}}}%
      \put(7728,2130){\makebox(0,0)[r]{\strut{}}}%
      \put(7728,2475){\makebox(0,0)[r]{\strut{}}}%
      \put(11176,750){\makebox(0,0)[l]{\strut{} 0}}%
      \put(11176,957){\makebox(0,0)[l]{\strut{}0.001}}%
      \put(11176,1164){\makebox(0,0)[l]{\strut{}0.002}}%
      \put(11176,1371){\makebox(0,0)[l]{\strut{}0.003}}%
      \put(11176,1578){\makebox(0,0)[l]{\strut{}0.004}}%
      \put(11176,1785){\makebox(0,0)[l]{\strut{}0.005}}%
      \put(11176,1992){\makebox(0,0)[l]{\strut{}0.006}}%
      \put(11176,2199){\makebox(0,0)[l]{\strut{}0.007}}%
      \put(11176,2406){\makebox(0,0)[l]{\strut{}0.008}}%
      \put(11176,2613){\makebox(0,0)[l]{\strut{}0.009}}%
      \put(11176,2820){\makebox(0,0)[l]{\strut{}0.01}}%
      \put(8094,2613){\makebox(0,0)[l]{\strut{}51.1~GeV}}%
    }%
    \gplbacktext
    \put(0,0){\includegraphics{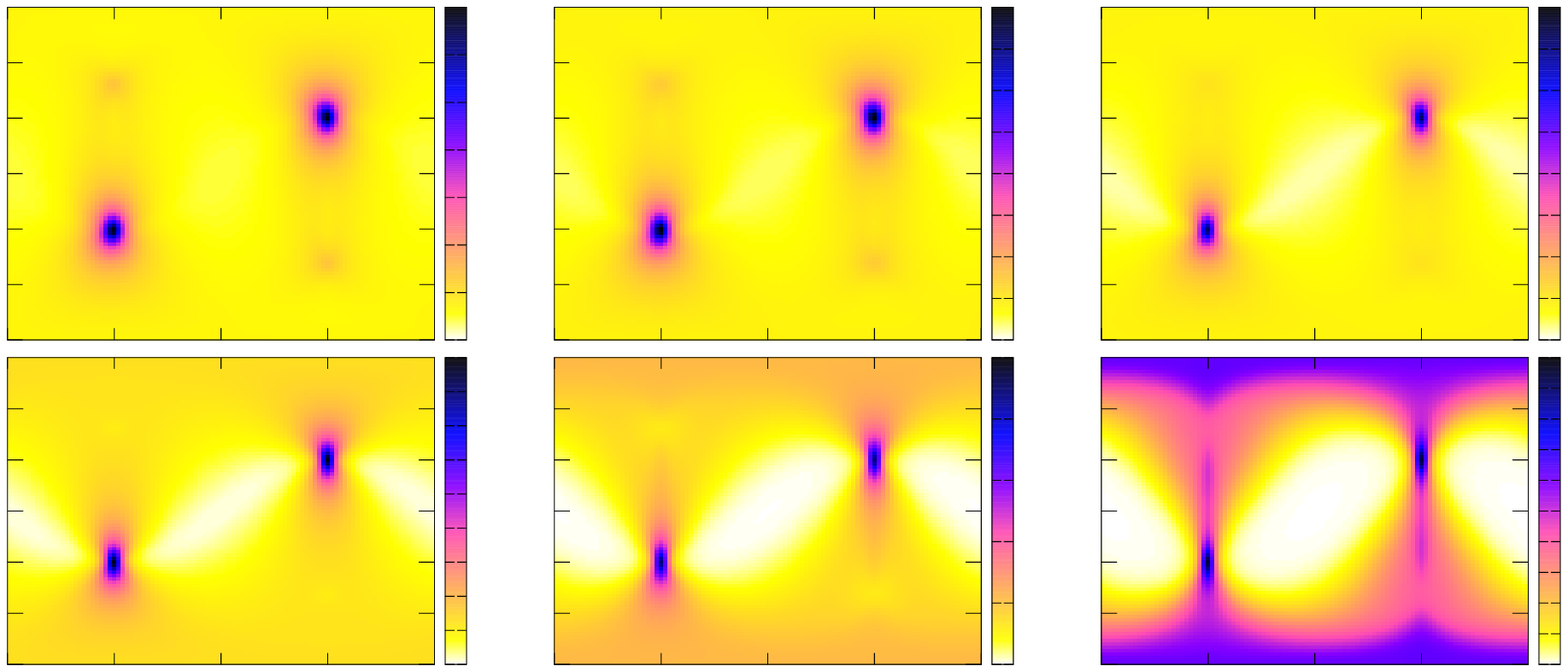}}%
    \gplfronttext
  \end{picture}%
\endgroup

%% file: skymaps_cl_r00025_b-3_c60_ri0.5_ro1_g10_d3.tex
\begingroup
  \makeatletter
  \providecommand\color[2][]{%
    \GenericError{(gnuplot) \space\space\space\@spaces}{%
      Package color not loaded in conjunction with
      terminal option `colourtext'%
    }{See the gnuplot documentation for explanation.%
    }{Either use 'blacktext' in gnuplot or load the package
      color.sty in LaTeX.}%
    \renewcommand\color[2][]{}%
  }%
  \providecommand\includegraphics[2][]{%
    \GenericError{(gnuplot) \space\space\space\@spaces}{%
      Package graphicx or graphics not loaded%
    }{See the gnuplot documentation for explanation.%
    }{The gnuplot epslatex terminal needs graphicx.sty or graphics.sty.}%
    \renewcommand\includegraphics[2][]{}%
  }%
  \providecommand\rotatebox[2]{#2}%
  \@ifundefined{ifGPcolor}{%
    \newif\ifGPcolor
    \GPcolortrue
  }{}%
  \@ifundefined{ifGPblacktext}{%
    \newif\ifGPblacktext
    \GPblacktextfalse
  }{}%
  \let\gplgaddtomacro\g@addto@macro
  \gdef\gplbacktext{}%
  \gdef\gplfronttext{}%
  \makeatother
  \ifGPblacktext
    \def\colorrgb#1{}%
    \def\colorgray#1{}%
  \else
    \ifGPcolor
      \def\colorrgb#1{\color[rgb]{#1}}%
      \def\colorgray#1{\color[gray]{#1}}%
      \expandafter\def\csname LTw\endcsname{\color{white}}%
      \expandafter\def\csname LTb\endcsname{\color{black}}%
      \expandafter\def\csname LTa\endcsname{\color{black}}%
      \expandafter\def\csname LT0\endcsname{\color[rgb]{1,0,0}}%
      \expandafter\def\csname LT1\endcsname{\color[rgb]{0,1,0}}%
      \expandafter\def\csname LT2\endcsname{\color[rgb]{0,0,1}}%
      \expandafter\def\csname LT3\endcsname{\color[rgb]{1,0,1}}%
      \expandafter\def\csname LT4\endcsname{\color[rgb]{0,1,1}}%
      \expandafter\def\csname LT5\endcsname{\color[rgb]{1,1,0}}%
      \expandafter\def\csname LT6\endcsname{\color[rgb]{0,0,0}}%
      \expandafter\def\csname LT7\endcsname{\color[rgb]{1,0.3,0}}%
      \expandafter\def\csname LT8\endcsname{\color[rgb]{0.5,0.5,0.5}}%
    \else
      \def\colorrgb#1{\color{black}}%
      \def\colorgray#1{\color[gray]{#1}}%
      \expandafter\def\csname LTw\endcsname{\color{white}}%
      \expandafter\def\csname LTb\endcsname{\color{black}}%
      \expandafter\def\csname LTa\endcsname{\color{black}}%
      \expandafter\def\csname LT0\endcsname{\color{black}}%
      \expandafter\def\csname LT1\endcsname{\color{black}}%
      \expandafter\def\csname LT2\endcsname{\color{black}}%
      \expandafter\def\csname LT3\endcsname{\color{black}}%
      \expandafter\def\csname LT4\endcsname{\color{black}}%
      \expandafter\def\csname LT5\endcsname{\color{black}}%
      \expandafter\def\csname LT6\endcsname{\color{black}}%
      \expandafter\def\csname LT7\endcsname{\color{black}}%
      \expandafter\def\csname LT8\endcsname{\color{black}}%
    \fi
  \fi
    \setlength{\unitlength}{0.0500bp}%
    \ifx\gptboxheight\undefined%
      \newlength{\gptboxheight}%
      \newlength{\gptboxwidth}%
      \newsavebox{\gptboxtext}%
    \fi%
    \setlength{\fboxrule}{0.5pt}%
    \setlength{\fboxsep}{1pt}%
\begin{picture}(11520.00,5760.00)%
    \gplgaddtomacro\gplbacktext{%
    }%
    \gplgaddtomacro\gplfronttext{%
      \csname LTb\endcsname
      \put(577,2662){\makebox(0,0){\strut{}}}%
      \put(1297,2662){\makebox(0,0){\strut{}}}%
      \put(2016,2662){\makebox(0,0){\strut{}}}%
      \put(2735,2662){\makebox(0,0){\strut{}}}%
      \put(3455,2662){\makebox(0,0){\strut{}}}%
      \put(364,2939){\makebox(0,0)[r]{\strut{}}}%
      \put(364,3313){\makebox(0,0)[r]{\strut{}30}}%
      \put(364,3687){\makebox(0,0)[r]{\strut{}60}}%
      \put(364,4061){\makebox(0,0)[r]{\strut{}90}}%
      \put(364,4435){\makebox(0,0)[r]{\strut{}120}}%
      \put(364,4809){\makebox(0,0)[r]{\strut{}150}}%
      \put(-98,4061){\rotatebox{-270}{\makebox(0,0){\strut{}line of sight ($\zeta$)}}}%
      \put(3803,2939){\makebox(0,0)[l]{\strut{} 0}}%
      \put(3803,3313){\makebox(0,0)[l]{\strut{} 1}}%
      \put(3803,3687){\makebox(0,0)[l]{\strut{} 2}}%
      \put(3803,4061){\makebox(0,0)[l]{\strut{} 3}}%
      \put(3803,4435){\makebox(0,0)[l]{\strut{} 4}}%
      \put(3803,4809){\makebox(0,0)[l]{\strut{} 5}}%
      \put(3803,5183){\makebox(0,0)[l]{\strut{} 6}}%
      \put(721,4959){\makebox(0,0)[l]{\strut{}161~MeV}}%
    }%
    \gplgaddtomacro\gplbacktext{%
    }%
    \gplgaddtomacro\gplfronttext{%
      \csname LTb\endcsname
      \put(4263,2662){\makebox(0,0){\strut{}}}%
      \put(4983,2662){\makebox(0,0){\strut{}}}%
      \put(5702,2662){\makebox(0,0){\strut{}}}%
      \put(6421,2662){\makebox(0,0){\strut{}}}%
      \put(7141,2662){\makebox(0,0){\strut{}}}%
      \put(4050,2939){\makebox(0,0)[r]{\strut{}}}%
      \put(4050,3313){\makebox(0,0)[r]{\strut{}}}%
      \put(4050,3687){\makebox(0,0)[r]{\strut{}}}%
      \put(4050,4061){\makebox(0,0)[r]{\strut{}}}%
      \put(4050,4435){\makebox(0,0)[r]{\strut{}}}%
      \put(4050,4809){\makebox(0,0)[r]{\strut{}}}%
      \put(7489,2939){\makebox(0,0)[l]{\strut{} 0}}%
      \put(7489,3313){\makebox(0,0)[l]{\strut{} 1}}%
      \put(7489,3687){\makebox(0,0)[l]{\strut{} 2}}%
      \put(7489,4061){\makebox(0,0)[l]{\strut{} 3}}%
      \put(7489,4435){\makebox(0,0)[l]{\strut{} 4}}%
      \put(7489,4809){\makebox(0,0)[l]{\strut{} 5}}%
      \put(7489,5183){\makebox(0,0)[l]{\strut{} 6}}%
      \put(4407,4959){\makebox(0,0)[l]{\strut{}511~MeV}}%
    }%
    \gplgaddtomacro\gplbacktext{%
    }%
    \gplgaddtomacro\gplfronttext{%
      \csname LTb\endcsname
      \put(7950,2662){\makebox(0,0){\strut{}}}%
      \put(8670,2662){\makebox(0,0){\strut{}}}%
      \put(9389,2662){\makebox(0,0){\strut{}}}%
      \put(10108,2662){\makebox(0,0){\strut{}}}%
      \put(10828,2662){\makebox(0,0){\strut{}}}%
      \put(7737,2939){\makebox(0,0)[r]{\strut{}}}%
      \put(7737,3313){\makebox(0,0)[r]{\strut{}}}%
      \put(7737,3687){\makebox(0,0)[r]{\strut{}}}%
      \put(7737,4061){\makebox(0,0)[r]{\strut{}}}%
      \put(7737,4435){\makebox(0,0)[r]{\strut{}}}%
      \put(7737,4809){\makebox(0,0)[r]{\strut{}}}%
      \put(11176,2939){\makebox(0,0)[l]{\strut{} 0}}%
      \put(11176,3313){\makebox(0,0)[l]{\strut{}0.5}}%
      \put(11176,3687){\makebox(0,0)[l]{\strut{} 1}}%
      \put(11176,4061){\makebox(0,0)[l]{\strut{} 2}}%
      \put(11176,4435){\makebox(0,0)[l]{\strut{} 2}}%
      \put(11176,4809){\makebox(0,0)[l]{\strut{} 2}}%
      \put(11176,5183){\makebox(0,0)[l]{\strut{} 3}}%
      \put(8094,4959){\makebox(0,0)[l]{\strut{}1.61~GeV}}%
    }%
    \gplgaddtomacro\gplbacktext{%
    }%
    \gplgaddtomacro\gplfronttext{%
      \csname LTb\endcsname
      \put(577,484){\makebox(0,0){\strut{}}}%
      \put(1297,484){\makebox(0,0){\strut{}}}%
      \put(2016,484){\makebox(0,0){\strut{}}}%
      \put(2735,484){\makebox(0,0){\strut{}}}%
      \put(3455,484){\makebox(0,0){\strut{}}}%
      \put(355,750){\makebox(0,0)[r]{\strut{}}}%
      \put(355,1095){\makebox(0,0)[r]{\strut{}30}}%
      \put(355,1440){\makebox(0,0)[r]{\strut{}60}}%
      \put(355,1785){\makebox(0,0)[r]{\strut{}90}}%
      \put(355,2130){\makebox(0,0)[r]{\strut{}120}}%
      \put(355,2475){\makebox(0,0)[r]{\strut{}150}}%
      \put(-107,1785){\rotatebox{-270}{\makebox(0,0){\strut{}line of sight ($\zeta$)}}}%
      \put(3803,750){\makebox(0,0)[l]{\strut{} 0}}%
      \put(3803,1164){\makebox(0,0)[l]{\strut{}0.05}}%
      \put(3803,1578){\makebox(0,0)[l]{\strut{}0.1}}%
      \put(3803,1992){\makebox(0,0)[l]{\strut{}0.2}}%
      \put(3803,2406){\makebox(0,0)[l]{\strut{}0.2}}%
      \put(3803,2820){\makebox(0,0)[l]{\strut{}0.2}}%
      \put(721,2613){\makebox(0,0)[l]{\strut{}5.11~GeV}}%
    }%
    \gplgaddtomacro\gplbacktext{%
    }%
    \gplgaddtomacro\gplfronttext{%
      \csname LTb\endcsname
      \put(4263,484){\makebox(0,0){\strut{}}}%
      \put(4983,484){\makebox(0,0){\strut{}}}%
      \put(5702,484){\makebox(0,0){\strut{}}}%
      \put(6421,484){\makebox(0,0){\strut{}}}%
      \put(7141,484){\makebox(0,0){\strut{}}}%
      \put(4041,750){\makebox(0,0)[r]{\strut{}}}%
      \put(4041,1095){\makebox(0,0)[r]{\strut{}}}%
      \put(4041,1440){\makebox(0,0)[r]{\strut{}}}%
      \put(4041,1785){\makebox(0,0)[r]{\strut{}}}%
      \put(4041,2130){\makebox(0,0)[r]{\strut{}}}%
      \put(4041,2475){\makebox(0,0)[r]{\strut{}}}%
      \put(7489,750){\makebox(0,0)[l]{\strut{} 0}}%
      \put(7489,980){\makebox(0,0)[l]{\strut{}0.0001}}%
      \put(7489,1210){\makebox(0,0)[l]{\strut{}0.0002}}%
      \put(7489,1440){\makebox(0,0)[l]{\strut{}0.0003}}%
      \put(7489,1670){\makebox(0,0)[l]{\strut{}0.0004}}%
      \put(7489,1899){\makebox(0,0)[l]{\strut{}0.0005}}%
      \put(7489,2130){\makebox(0,0)[l]{\strut{}0.0006}}%
      \put(7489,2360){\makebox(0,0)[l]{\strut{}0.0007}}%
      \put(7489,2590){\makebox(0,0)[l]{\strut{}0.0008}}%
      \put(7489,2820){\makebox(0,0)[l]{\strut{}0.0009}}%
      \put(4407,2613){\makebox(0,0)[l]{\strut{}16.1~GeV}}%
    }%
    \gplgaddtomacro\gplbacktext{%
    }%
    \gplgaddtomacro\gplfronttext{%
      \csname LTb\endcsname
      \put(7950,484){\makebox(0,0){\strut{}}}%
      \put(8670,484){\makebox(0,0){\strut{}}}%
      \put(9389,484){\makebox(0,0){\strut{}}}%
      \put(10108,484){\makebox(0,0){\strut{}}}%
      \put(10828,484){\makebox(0,0){\strut{}}}%
      \put(7728,750){\makebox(0,0)[r]{\strut{}}}%
      \put(7728,1095){\makebox(0,0)[r]{\strut{}}}%
      \put(7728,1440){\makebox(0,0)[r]{\strut{}}}%
      \put(7728,1785){\makebox(0,0)[r]{\strut{}}}%
      \put(7728,2130){\makebox(0,0)[r]{\strut{}}}%
      \put(7728,2475){\makebox(0,0)[r]{\strut{}}}%
      \put(11176,750){\makebox(0,0)[l]{\strut{} 0}}%
      \put(11176,980){\makebox(0,0)[l]{\strut{}1e-07}}%
      \put(11176,1210){\makebox(0,0)[l]{\strut{}2e-07}}%
      \put(11176,1440){\makebox(0,0)[l]{\strut{}3e-07}}%
      \put(11176,1670){\makebox(0,0)[l]{\strut{}4e-07}}%
      \put(11176,1900){\makebox(0,0)[l]{\strut{}5e-07}}%
      \put(11176,2130){\makebox(0,0)[l]{\strut{}6e-07}}%
      \put(11176,2360){\makebox(0,0)[l]{\strut{}7e-07}}%
      \put(11176,2590){\makebox(0,0)[l]{\strut{}8e-07}}%
      \put(11176,2820){\makebox(0,0)[l]{\strut{}9e-07}}%
      \put(8094,2613){\makebox(0,0)[l]{\strut{}51.1~GeV}}%
    }%
    \gplbacktext
    \put(0,0){\includegraphics{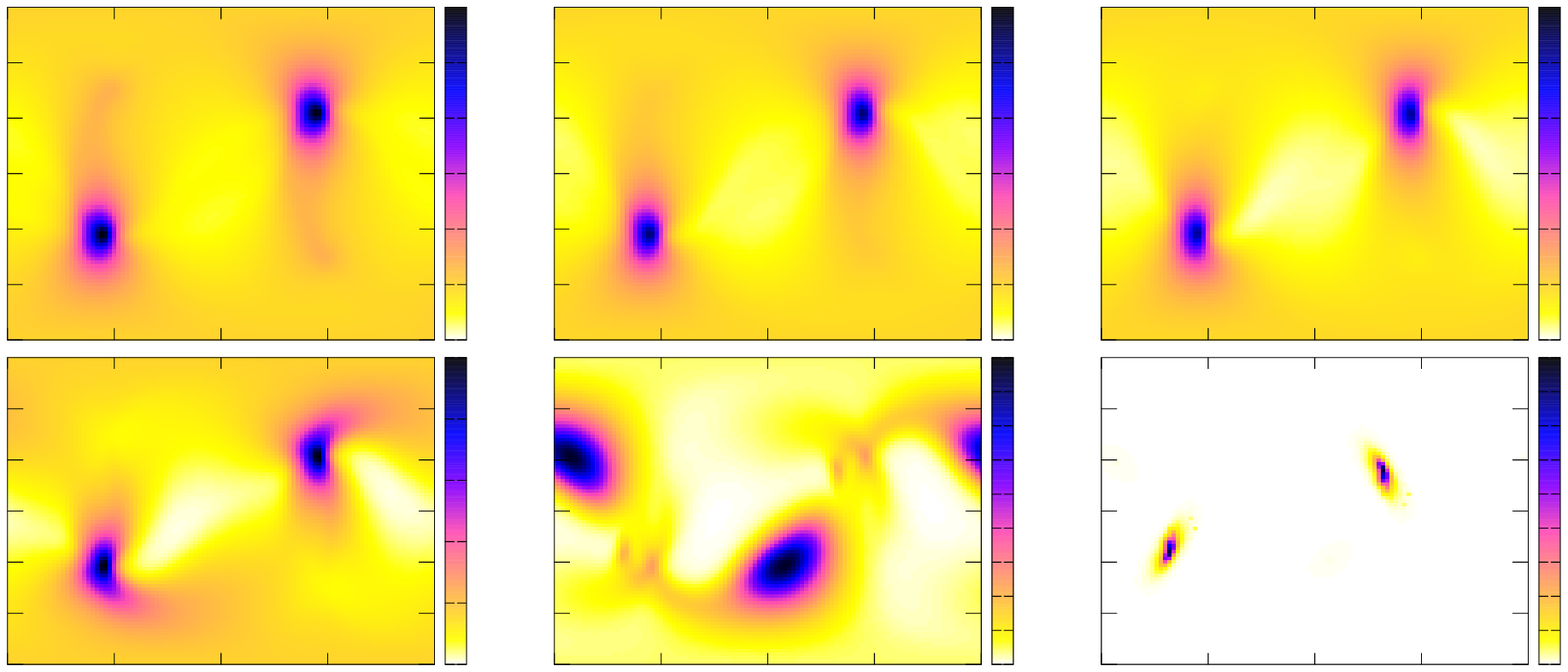}}%
    \gplfronttext
  \end{picture}%
\endgroup

%% file: skymaps_cl_r005_b-6_c60_ri0.2_ro1_g10_d3.tex
\begingroup
  \makeatletter
  \providecommand\color[2][]{%
    \GenericError{(gnuplot) \space\space\space\@spaces}{%
      Package color not loaded in conjunction with
      terminal option `colourtext'%
    }{See the gnuplot documentation for explanation.%
    }{Either use 'blacktext' in gnuplot or load the package
      color.sty in LaTeX.}%
    \renewcommand\color[2][]{}%
  }%
  \providecommand\includegraphics[2][]{%
    \GenericError{(gnuplot) \space\space\space\@spaces}{%
      Package graphicx or graphics not loaded%
    }{See the gnuplot documentation for explanation.%
    }{The gnuplot epslatex terminal needs graphicx.sty or graphics.sty.}%
    \renewcommand\includegraphics[2][]{}%
  }%
  \providecommand\rotatebox[2]{#2}%
  \@ifundefined{ifGPcolor}{%
    \newif\ifGPcolor
    \GPcolortrue
  }{}%
  \@ifundefined{ifGPblacktext}{%
    \newif\ifGPblacktext
    \GPblacktextfalse
  }{}%
  \let\gplgaddtomacro\g@addto@macro
  \gdef\gplbacktext{}%
  \gdef\gplfronttext{}%
  \makeatother
  \ifGPblacktext
    \def\colorrgb#1{}%
    \def\colorgray#1{}%
  \else
    \ifGPcolor
      \def\colorrgb#1{\color[rgb]{#1}}%
      \def\colorgray#1{\color[gray]{#1}}%
      \expandafter\def\csname LTw\endcsname{\color{white}}%
      \expandafter\def\csname LTb\endcsname{\color{black}}%
      \expandafter\def\csname LTa\endcsname{\color{black}}%
      \expandafter\def\csname LT0\endcsname{\color[rgb]{1,0,0}}%
      \expandafter\def\csname LT1\endcsname{\color[rgb]{0,1,0}}%
      \expandafter\def\csname LT2\endcsname{\color[rgb]{0,0,1}}%
      \expandafter\def\csname LT3\endcsname{\color[rgb]{1,0,1}}%
      \expandafter\def\csname LT4\endcsname{\color[rgb]{0,1,1}}%
      \expandafter\def\csname LT5\endcsname{\color[rgb]{1,1,0}}%
      \expandafter\def\csname LT6\endcsname{\color[rgb]{0,0,0}}%
      \expandafter\def\csname LT7\endcsname{\color[rgb]{1,0.3,0}}%
      \expandafter\def\csname LT8\endcsname{\color[rgb]{0.5,0.5,0.5}}%
    \else
      \def\colorrgb#1{\color{black}}%
      \def\colorgray#1{\color[gray]{#1}}%
      \expandafter\def\csname LTw\endcsname{\color{white}}%
      \expandafter\def\csname LTb\endcsname{\color{black}}%
      \expandafter\def\csname LTa\endcsname{\color{black}}%
      \expandafter\def\csname LT0\endcsname{\color{black}}%
      \expandafter\def\csname LT1\endcsname{\color{black}}%
      \expandafter\def\csname LT2\endcsname{\color{black}}%
      \expandafter\def\csname LT3\endcsname{\color{black}}%
      \expandafter\def\csname LT4\endcsname{\color{black}}%
      \expandafter\def\csname LT5\endcsname{\color{black}}%
      \expandafter\def\csname LT6\endcsname{\color{black}}%
      \expandafter\def\csname LT7\endcsname{\color{black}}%
      \expandafter\def\csname LT8\endcsname{\color{black}}%
    \fi
  \fi
    \setlength{\unitlength}{0.0500bp}%
    \ifx\gptboxheight\undefined%
      \newlength{\gptboxheight}%
      \newlength{\gptboxwidth}%
      \newsavebox{\gptboxtext}%
    \fi%
    \setlength{\fboxrule}{0.5pt}%
    \setlength{\fboxsep}{1pt}%
\begin{picture}(11520.00,5760.00)%
    \gplgaddtomacro\gplbacktext{%
    }%
    \gplgaddtomacro\gplfronttext{%
      \csname LTb\endcsname
      \put(577,2662){\makebox(0,0){\strut{}}}%
      \put(1297,2662){\makebox(0,0){\strut{}}}%
      \put(2016,2662){\makebox(0,0){\strut{}}}%
      \put(2735,2662){\makebox(0,0){\strut{}}}%
      \put(3455,2662){\makebox(0,0){\strut{}}}%
      \put(364,2939){\makebox(0,0)[r]{\strut{}}}%
      \put(364,3313){\makebox(0,0)[r]{\strut{}30}}%
      \put(364,3687){\makebox(0,0)[r]{\strut{}60}}%
      \put(364,4061){\makebox(0,0)[r]{\strut{}90}}%
      \put(364,4435){\makebox(0,0)[r]{\strut{}120}}%
      \put(364,4809){\makebox(0,0)[r]{\strut{}150}}%
      \put(-98,4061){\rotatebox{-270}{\makebox(0,0){\strut{}line of sight ($\zeta$)}}}%
      \put(3803,2939){\makebox(0,0)[l]{\strut{} 0}}%
      \put(3803,3163){\makebox(0,0)[l]{\strut{}5e+03}}%
      \put(3803,3387){\makebox(0,0)[l]{\strut{}1e+04}}%
      \put(3803,3612){\makebox(0,0)[l]{\strut{}2e+04}}%
      \put(3803,3836){\makebox(0,0)[l]{\strut{}2e+04}}%
      \put(3803,4061){\makebox(0,0)[l]{\strut{}2e+04}}%
      \put(3803,4285){\makebox(0,0)[l]{\strut{}3e+04}}%
      \put(3803,4509){\makebox(0,0)[l]{\strut{}4e+04}}%
      \put(3803,4734){\makebox(0,0)[l]{\strut{}4e+04}}%
      \put(3803,4958){\makebox(0,0)[l]{\strut{}4e+04}}%
      \put(3803,5183){\makebox(0,0)[l]{\strut{}5e+04}}%
      \put(721,4959){\makebox(0,0)[l]{\strut{}161~MeV}}%
    }%
    \gplgaddtomacro\gplbacktext{%
    }%
    \gplgaddtomacro\gplfronttext{%
      \csname LTb\endcsname
      \put(4263,2662){\makebox(0,0){\strut{}}}%
      \put(4983,2662){\makebox(0,0){\strut{}}}%
      \put(5702,2662){\makebox(0,0){\strut{}}}%
      \put(6421,2662){\makebox(0,0){\strut{}}}%
      \put(7141,2662){\makebox(0,0){\strut{}}}%
      \put(4050,2939){\makebox(0,0)[r]{\strut{}}}%
      \put(4050,3313){\makebox(0,0)[r]{\strut{}}}%
      \put(4050,3687){\makebox(0,0)[r]{\strut{}}}%
      \put(4050,4061){\makebox(0,0)[r]{\strut{}}}%
      \put(4050,4435){\makebox(0,0)[r]{\strut{}}}%
      \put(4050,4809){\makebox(0,0)[r]{\strut{}}}%
      \put(7489,2939){\makebox(0,0)[l]{\strut{} 0}}%
      \put(7489,3143){\makebox(0,0)[l]{\strut{}5e+03}}%
      \put(7489,3347){\makebox(0,0)[l]{\strut{}1e+04}}%
      \put(7489,3551){\makebox(0,0)[l]{\strut{}2e+04}}%
      \put(7489,3755){\makebox(0,0)[l]{\strut{}2e+04}}%
      \put(7489,3959){\makebox(0,0)[l]{\strut{}2e+04}}%
      \put(7489,4163){\makebox(0,0)[l]{\strut{}3e+04}}%
      \put(7489,4367){\makebox(0,0)[l]{\strut{}4e+04}}%
      \put(7489,4571){\makebox(0,0)[l]{\strut{}4e+04}}%
      \put(7489,4775){\makebox(0,0)[l]{\strut{}4e+04}}%
      \put(7489,4979){\makebox(0,0)[l]{\strut{}5e+04}}%
      \put(7489,5183){\makebox(0,0)[l]{\strut{}6e+04}}%
      \put(4407,4959){\makebox(0,0)[l]{\strut{}511~MeV}}%
    }%
    \gplgaddtomacro\gplbacktext{%
    }%
    \gplgaddtomacro\gplfronttext{%
      \csname LTb\endcsname
      \put(7950,2662){\makebox(0,0){\strut{}}}%
      \put(8670,2662){\makebox(0,0){\strut{}}}%
      \put(9389,2662){\makebox(0,0){\strut{}}}%
      \put(10108,2662){\makebox(0,0){\strut{}}}%
      \put(10828,2662){\makebox(0,0){\strut{}}}%
      \put(7737,2939){\makebox(0,0)[r]{\strut{}}}%
      \put(7737,3313){\makebox(0,0)[r]{\strut{}}}%
      \put(7737,3687){\makebox(0,0)[r]{\strut{}}}%
      \put(7737,4061){\makebox(0,0)[r]{\strut{}}}%
      \put(7737,4435){\makebox(0,0)[r]{\strut{}}}%
      \put(7737,4809){\makebox(0,0)[r]{\strut{}}}%
      \put(11176,2939){\makebox(0,0)[l]{\strut{} 0}}%
      \put(11176,3219){\makebox(0,0)[l]{\strut{}5e+03}}%
      \put(11176,3500){\makebox(0,0)[l]{\strut{}1e+04}}%
      \put(11176,3780){\makebox(0,0)[l]{\strut{}2e+04}}%
      \put(11176,4061){\makebox(0,0)[l]{\strut{}2e+04}}%
      \put(11176,4341){\makebox(0,0)[l]{\strut{}2e+04}}%
      \put(11176,4622){\makebox(0,0)[l]{\strut{}3e+04}}%
      \put(11176,4902){\makebox(0,0)[l]{\strut{}4e+04}}%
      \put(11176,5183){\makebox(0,0)[l]{\strut{}4e+04}}%
      \put(8094,4959){\makebox(0,0)[l]{\strut{}1.61~GeV}}%
    }%
    \gplgaddtomacro\gplbacktext{%
    }%
    \gplgaddtomacro\gplfronttext{%
      \csname LTb\endcsname
      \put(577,484){\makebox(0,0){\strut{}}}%
      \put(1297,484){\makebox(0,0){\strut{}}}%
      \put(2016,484){\makebox(0,0){\strut{}}}%
      \put(2735,484){\makebox(0,0){\strut{}}}%
      \put(3455,484){\makebox(0,0){\strut{}}}%
      \put(355,750){\makebox(0,0)[r]{\strut{}}}%
      \put(355,1095){\makebox(0,0)[r]{\strut{}30}}%
      \put(355,1440){\makebox(0,0)[r]{\strut{}60}}%
      \put(355,1785){\makebox(0,0)[r]{\strut{}90}}%
      \put(355,2130){\makebox(0,0)[r]{\strut{}120}}%
      \put(355,2475){\makebox(0,0)[r]{\strut{}150}}%
      \put(-107,1785){\rotatebox{-270}{\makebox(0,0){\strut{}line of sight ($\zeta$)}}}%
      \put(3803,750){\makebox(0,0)[l]{\strut{} 0}}%
      \put(3803,1095){\makebox(0,0)[l]{\strut{}2e+03}}%
      \put(3803,1440){\makebox(0,0)[l]{\strut{}4e+03}}%
      \put(3803,1785){\makebox(0,0)[l]{\strut{}6e+03}}%
      \put(3803,2130){\makebox(0,0)[l]{\strut{}8e+03}}%
      \put(3803,2475){\makebox(0,0)[l]{\strut{}1e+04}}%
      \put(3803,2820){\makebox(0,0)[l]{\strut{}1e+04}}%
      \put(721,2613){\makebox(0,0)[l]{\strut{}5.11~GeV}}%
    }%
    \gplgaddtomacro\gplbacktext{%
    }%
    \gplgaddtomacro\gplfronttext{%
      \csname LTb\endcsname
      \put(4263,484){\makebox(0,0){\strut{}}}%
      \put(4983,484){\makebox(0,0){\strut{}}}%
      \put(5702,484){\makebox(0,0){\strut{}}}%
      \put(6421,484){\makebox(0,0){\strut{}}}%
      \put(7141,484){\makebox(0,0){\strut{}}}%
      \put(4041,750){\makebox(0,0)[r]{\strut{}}}%
      \put(4041,1095){\makebox(0,0)[r]{\strut{}}}%
      \put(4041,1440){\makebox(0,0)[r]{\strut{}}}%
      \put(4041,1785){\makebox(0,0)[r]{\strut{}}}%
      \put(4041,2130){\makebox(0,0)[r]{\strut{}}}%
      \put(4041,2475){\makebox(0,0)[r]{\strut{}}}%
      \put(7489,750){\makebox(0,0)[l]{\strut{} 0}}%
      \put(7489,1095){\makebox(0,0)[l]{\strut{}5e+01}}%
      \put(7489,1440){\makebox(0,0)[l]{\strut{}1e+02}}%
      \put(7489,1785){\makebox(0,0)[l]{\strut{}2e+02}}%
      \put(7489,2130){\makebox(0,0)[l]{\strut{}2e+02}}%
      \put(7489,2475){\makebox(0,0)[l]{\strut{}2e+02}}%
      \put(7489,2820){\makebox(0,0)[l]{\strut{}3e+02}}%
      \put(4407,2613){\makebox(0,0)[l]{\strut{}16.1~GeV}}%
    }%
    \gplgaddtomacro\gplbacktext{%
    }%
    \gplgaddtomacro\gplfronttext{%
      \csname LTb\endcsname
      \put(7950,484){\makebox(0,0){\strut{}}}%
      \put(8670,484){\makebox(0,0){\strut{}}}%
      \put(9389,484){\makebox(0,0){\strut{}}}%
      \put(10108,484){\makebox(0,0){\strut{}}}%
      \put(10828,484){\makebox(0,0){\strut{}}}%
      \put(7728,750){\makebox(0,0)[r]{\strut{}}}%
      \put(7728,1095){\makebox(0,0)[r]{\strut{}}}%
      \put(7728,1440){\makebox(0,0)[r]{\strut{}}}%
      \put(7728,1785){\makebox(0,0)[r]{\strut{}}}%
      \put(7728,2130){\makebox(0,0)[r]{\strut{}}}%
      \put(7728,2475){\makebox(0,0)[r]{\strut{}}}%
      \put(11176,750){\makebox(0,0)[l]{\strut{} 0}}%
      \put(11176,1164){\makebox(0,0)[l]{\strut{}0.05}}%
      \put(11176,1578){\makebox(0,0)[l]{\strut{}0.1}}%
      \put(11176,1992){\makebox(0,0)[l]{\strut{}0.2}}%
      \put(11176,2406){\makebox(0,0)[l]{\strut{}0.2}}%
      \put(11176,2820){\makebox(0,0)[l]{\strut{}0.2}}%
      \put(8094,2613){\makebox(0,0)[l]{\strut{}51.1~GeV}}%
    }%
    \gplbacktext
    \put(0,0){\includegraphics{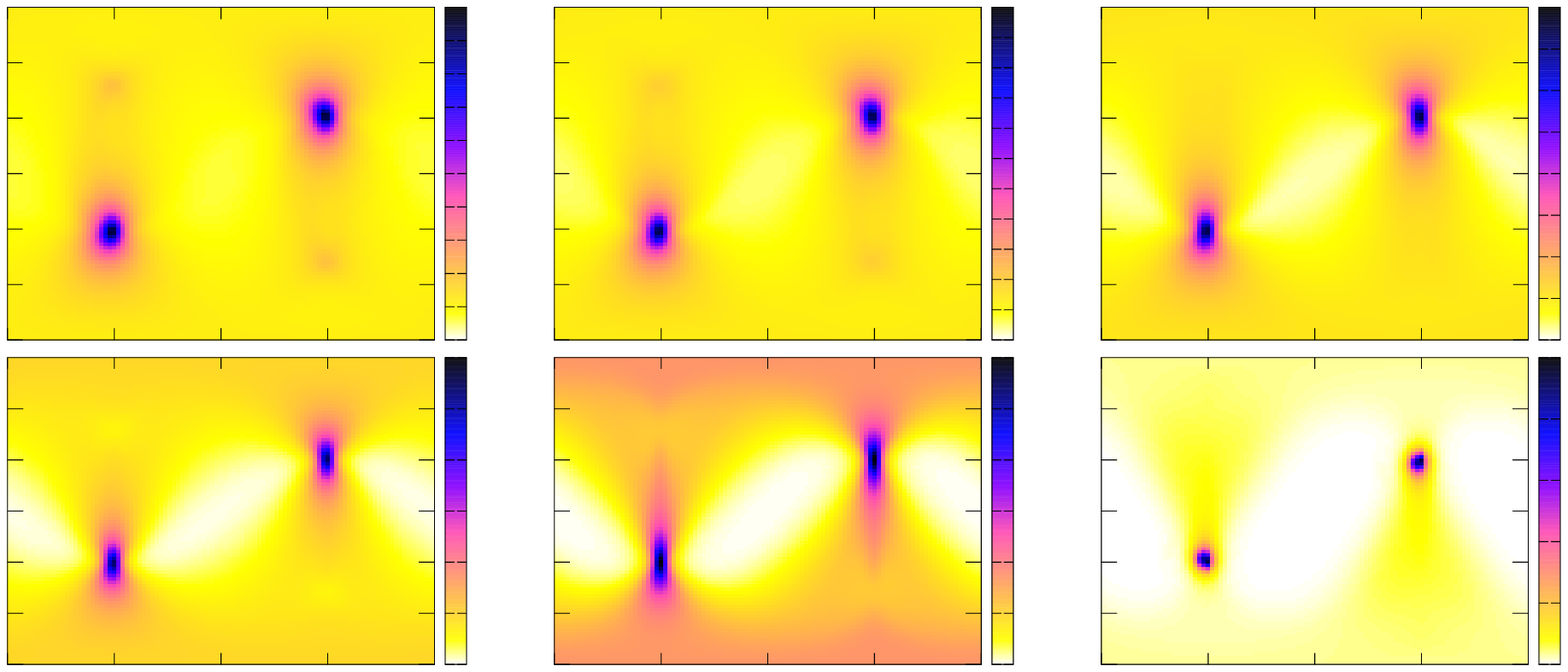}}%
    \gplfronttext
  \end{picture}%
\endgroup

%% file: skymaps_cl_r00025_b-3_c60_g10_d3_om6.tex
\begingroup
  \makeatletter
  \providecommand\color[2][]{%
    \GenericError{(gnuplot) \space\space\space\@spaces}{%
      Package color not loaded in conjunction with
      terminal option `colourtext'%
    }{See the gnuplot documentation for explanation.%
    }{Either use 'blacktext' in gnuplot or load the package
      color.sty in LaTeX.}%
    \renewcommand\color[2][]{}%
  }%
  \providecommand\includegraphics[2][]{%
    \GenericError{(gnuplot) \space\space\space\@spaces}{%
      Package graphicx or graphics not loaded%
    }{See the gnuplot documentation for explanation.%
    }{The gnuplot epslatex terminal needs graphicx.sty or graphics.sty.}%
    \renewcommand\includegraphics[2][]{}%
  }%
  \providecommand\rotatebox[2]{#2}%
  \@ifundefined{ifGPcolor}{%
    \newif\ifGPcolor
    \GPcolortrue
  }{}%
  \@ifundefined{ifGPblacktext}{%
    \newif\ifGPblacktext
    \GPblacktextfalse
  }{}%
  \let\gplgaddtomacro\g@addto@macro
  \gdef\gplbacktext{}%
  \gdef\gplfronttext{}%
  \makeatother
  \ifGPblacktext
    \def\colorrgb#1{}%
    \def\colorgray#1{}%
  \else
    \ifGPcolor
      \def\colorrgb#1{\color[rgb]{#1}}%
      \def\colorgray#1{\color[gray]{#1}}%
      \expandafter\def\csname LTw\endcsname{\color{white}}%
      \expandafter\def\csname LTb\endcsname{\color{black}}%
      \expandafter\def\csname LTa\endcsname{\color{black}}%
      \expandafter\def\csname LT0\endcsname{\color[rgb]{1,0,0}}%
      \expandafter\def\csname LT1\endcsname{\color[rgb]{0,1,0}}%
      \expandafter\def\csname LT2\endcsname{\color[rgb]{0,0,1}}%
      \expandafter\def\csname LT3\endcsname{\color[rgb]{1,0,1}}%
      \expandafter\def\csname LT4\endcsname{\color[rgb]{0,1,1}}%
      \expandafter\def\csname LT5\endcsname{\color[rgb]{1,1,0}}%
      \expandafter\def\csname LT6\endcsname{\color[rgb]{0,0,0}}%
      \expandafter\def\csname LT7\endcsname{\color[rgb]{1,0.3,0}}%
      \expandafter\def\csname LT8\endcsname{\color[rgb]{0.5,0.5,0.5}}%
    \else
      \def\colorrgb#1{\color{black}}%
      \def\colorgray#1{\color[gray]{#1}}%
      \expandafter\def\csname LTw\endcsname{\color{white}}%
      \expandafter\def\csname LTb\endcsname{\color{black}}%
      \expandafter\def\csname LTa\endcsname{\color{black}}%
      \expandafter\def\csname LT0\endcsname{\color{black}}%
      \expandafter\def\csname LT1\endcsname{\color{black}}%
      \expandafter\def\csname LT2\endcsname{\color{black}}%
      \expandafter\def\csname LT3\endcsname{\color{black}}%
      \expandafter\def\csname LT4\endcsname{\color{black}}%
      \expandafter\def\csname LT5\endcsname{\color{black}}%
      \expandafter\def\csname LT6\endcsname{\color{black}}%
      \expandafter\def\csname LT7\endcsname{\color{black}}%
      \expandafter\def\csname LT8\endcsname{\color{black}}%
    \fi
  \fi
    \setlength{\unitlength}{0.0500bp}%
    \ifx\gptboxheight\undefined%
      \newlength{\gptboxheight}%
      \newlength{\gptboxwidth}%
      \newsavebox{\gptboxtext}%
    \fi%
    \setlength{\fboxrule}{0.5pt}%
    \setlength{\fboxsep}{1pt}%
\begin{picture}(11520.00,8640.00)%
    \gplgaddtomacro\gplbacktext{%
    }%
    \gplgaddtomacro\gplfronttext{%
      \csname LTb\endcsname
      \put(577,5346){\makebox(0,0){\strut{}}}%
      \put(1297,5346){\makebox(0,0){\strut{}}}%
      \put(2016,5346){\makebox(0,0){\strut{}}}%
      \put(2735,5346){\makebox(0,0){\strut{}}}%
      \put(3455,5346){\makebox(0,0){\strut{}}}%
      \put(359,5617){\makebox(0,0)[r]{\strut{}}}%
      \put(359,5977){\makebox(0,0)[r]{\strut{}30}}%
      \put(359,6337){\makebox(0,0)[r]{\strut{}60}}%
      \put(359,6696){\makebox(0,0)[r]{\strut{}90}}%
      \put(359,7055){\makebox(0,0)[r]{\strut{}120}}%
      \put(359,7415){\makebox(0,0)[r]{\strut{}150}}%
      \put(-103,6696){\rotatebox{-270}{\makebox(0,0){\strut{}line of sight ($\zeta$)}}}%
      \put(3803,5617){\makebox(0,0)[l]{\strut{} 0}}%
      \put(3803,5856){\makebox(0,0)[l]{\strut{} 5}}%
      \put(3803,6096){\makebox(0,0)[l]{\strut{}1e+01}}%
      \put(3803,6336){\makebox(0,0)[l]{\strut{}2e+01}}%
      \put(3803,6576){\makebox(0,0)[l]{\strut{}2e+01}}%
      \put(3803,6815){\makebox(0,0)[l]{\strut{}2e+01}}%
      \put(3803,7055){\makebox(0,0)[l]{\strut{}3e+01}}%
      \put(3803,7295){\makebox(0,0)[l]{\strut{}4e+01}}%
      \put(3803,7535){\makebox(0,0)[l]{\strut{}4e+01}}%
      \put(3803,7775){\makebox(0,0)[l]{\strut{}4e+01}}%
      \put(721,7559){\makebox(0,0)[l]{\strut{}0.1-0.2}}%
    }%
    \gplgaddtomacro\gplbacktext{%
    }%
    \gplgaddtomacro\gplfronttext{%
      \csname LTb\endcsname
      \put(4263,5346){\makebox(0,0){\strut{}}}%
      \put(4983,5346){\makebox(0,0){\strut{}}}%
      \put(5702,5346){\makebox(0,0){\strut{}}}%
      \put(6421,5346){\makebox(0,0){\strut{}}}%
      \put(7141,5346){\makebox(0,0){\strut{}}}%
      \put(4045,5617){\makebox(0,0)[r]{\strut{}}}%
      \put(4045,5977){\makebox(0,0)[r]{\strut{}}}%
      \put(4045,6337){\makebox(0,0)[r]{\strut{}}}%
      \put(4045,6696){\makebox(0,0)[r]{\strut{}}}%
      \put(4045,7055){\makebox(0,0)[r]{\strut{}}}%
      \put(4045,7415){\makebox(0,0)[r]{\strut{}}}%
      \put(7489,5617){\makebox(0,0)[l]{\strut{} 0}}%
      \put(7489,6048){\makebox(0,0)[l]{\strut{} 5}}%
      \put(7489,6480){\makebox(0,0)[l]{\strut{}1e+01}}%
      \put(7489,6911){\makebox(0,0)[l]{\strut{}2e+01}}%
      \put(7489,7343){\makebox(0,0)[l]{\strut{}2e+01}}%
      \put(7489,7775){\makebox(0,0)[l]{\strut{}2e+01}}%
      \put(4407,7559){\makebox(0,0)[l]{\strut{}0.2-0.3}}%
    }%
    \gplgaddtomacro\gplbacktext{%
    }%
    \gplgaddtomacro\gplfronttext{%
      \csname LTb\endcsname
      \put(7950,5346){\makebox(0,0){\strut{}}}%
      \put(8670,5346){\makebox(0,0){\strut{}}}%
      \put(9389,5346){\makebox(0,0){\strut{}}}%
      \put(10108,5346){\makebox(0,0){\strut{}}}%
      \put(10828,5346){\makebox(0,0){\strut{}}}%
      \put(7732,5617){\makebox(0,0)[r]{\strut{}}}%
      \put(7732,5977){\makebox(0,0)[r]{\strut{}}}%
      \put(7732,6337){\makebox(0,0)[r]{\strut{}}}%
      \put(7732,6696){\makebox(0,0)[r]{\strut{}}}%
      \put(7732,7055){\makebox(0,0)[r]{\strut{}}}%
      \put(7732,7415){\makebox(0,0)[r]{\strut{}}}%
      \put(11176,5617){\makebox(0,0)[l]{\strut{} 0}}%
      \put(11176,5886){\makebox(0,0)[l]{\strut{} 2}}%
      \put(11176,6156){\makebox(0,0)[l]{\strut{} 4}}%
      \put(11176,6426){\makebox(0,0)[l]{\strut{} 6}}%
      \put(11176,6696){\makebox(0,0)[l]{\strut{} 8}}%
      \put(11176,6965){\makebox(0,0)[l]{\strut{}1e+01}}%
      \put(11176,7235){\makebox(0,0)[l]{\strut{}1e+01}}%
      \put(11176,7505){\makebox(0,0)[l]{\strut{}1e+01}}%
      \put(11176,7775){\makebox(0,0)[l]{\strut{}2e+01}}%
      \put(8094,7559){\makebox(0,0)[l]{\strut{}0.3-0.4}}%
    }%
    \gplgaddtomacro\gplbacktext{%
    }%
    \gplgaddtomacro\gplfronttext{%
      \csname LTb\endcsname
      \put(577,3100){\makebox(0,0){\strut{}}}%
      \put(1297,3100){\makebox(0,0){\strut{}}}%
      \put(2016,3100){\makebox(0,0){\strut{}}}%
      \put(2735,3100){\makebox(0,0){\strut{}}}%
      \put(3455,3100){\makebox(0,0){\strut{}}}%
      \put(359,3371){\makebox(0,0)[r]{\strut{}}}%
      \put(359,3731){\makebox(0,0)[r]{\strut{}30}}%
      \put(359,4091){\makebox(0,0)[r]{\strut{}60}}%
      \put(359,4450){\makebox(0,0)[r]{\strut{}90}}%
      \put(359,4809){\makebox(0,0)[r]{\strut{}120}}%
      \put(359,5169){\makebox(0,0)[r]{\strut{}150}}%
      \put(-103,4450){\rotatebox{-270}{\makebox(0,0){\strut{}line of sight ($\zeta$)}}}%
      \put(3803,3371){\makebox(0,0)[l]{\strut{} 0}}%
      \put(3803,3730){\makebox(0,0)[l]{\strut{} 2}}%
      \put(3803,4090){\makebox(0,0)[l]{\strut{} 4}}%
      \put(3803,4450){\makebox(0,0)[l]{\strut{} 6}}%
      \put(3803,4809){\makebox(0,0)[l]{\strut{} 8}}%
      \put(3803,5169){\makebox(0,0)[l]{\strut{}1e+01}}%
      \put(3803,5529){\makebox(0,0)[l]{\strut{}1e+01}}%
      \put(721,5313){\makebox(0,0)[l]{\strut{}0.4-0.5}}%
    }%
    \gplgaddtomacro\gplbacktext{%
    }%
    \gplgaddtomacro\gplfronttext{%
      \csname LTb\endcsname
      \put(4263,3100){\makebox(0,0){\strut{}}}%
      \put(4983,3100){\makebox(0,0){\strut{}}}%
      \put(5702,3100){\makebox(0,0){\strut{}}}%
      \put(6421,3100){\makebox(0,0){\strut{}}}%
      \put(7141,3100){\makebox(0,0){\strut{}}}%
      \put(4045,3371){\makebox(0,0)[r]{\strut{}}}%
      \put(4045,3731){\makebox(0,0)[r]{\strut{}}}%
      \put(4045,4091){\makebox(0,0)[r]{\strut{}}}%
      \put(4045,4450){\makebox(0,0)[r]{\strut{}}}%
      \put(4045,4809){\makebox(0,0)[r]{\strut{}}}%
      \put(4045,5169){\makebox(0,0)[r]{\strut{}}}%
      \put(7489,3371){\makebox(0,0)[l]{\strut{} 0}}%
      \put(7489,3640){\makebox(0,0)[l]{\strut{} 1}}%
      \put(7489,3910){\makebox(0,0)[l]{\strut{} 2}}%
      \put(7489,4180){\makebox(0,0)[l]{\strut{} 3}}%
      \put(7489,4450){\makebox(0,0)[l]{\strut{} 4}}%
      \put(7489,4719){\makebox(0,0)[l]{\strut{} 5}}%
      \put(7489,4989){\makebox(0,0)[l]{\strut{} 6}}%
      \put(7489,5259){\makebox(0,0)[l]{\strut{} 7}}%
      \put(7489,5529){\makebox(0,0)[l]{\strut{} 8}}%
      \put(4407,5313){\makebox(0,0)[l]{\strut{}0.5-0.6}}%
    }%
    \gplgaddtomacro\gplbacktext{%
    }%
    \gplgaddtomacro\gplfronttext{%
      \csname LTb\endcsname
      \put(7950,3100){\makebox(0,0){\strut{}}}%
      \put(8670,3100){\makebox(0,0){\strut{}}}%
      \put(9389,3100){\makebox(0,0){\strut{}}}%
      \put(10108,3100){\makebox(0,0){\strut{}}}%
      \put(10828,3100){\makebox(0,0){\strut{}}}%
      \put(7732,3371){\makebox(0,0)[r]{\strut{}}}%
      \put(7732,3731){\makebox(0,0)[r]{\strut{}}}%
      \put(7732,4091){\makebox(0,0)[r]{\strut{}}}%
      \put(7732,4450){\makebox(0,0)[r]{\strut{}}}%
      \put(7732,4809){\makebox(0,0)[r]{\strut{}}}%
      \put(7732,5169){\makebox(0,0)[r]{\strut{}}}%
      \put(11176,3371){\makebox(0,0)[l]{\strut{} 0}}%
      \put(11176,3679){\makebox(0,0)[l]{\strut{} 1}}%
      \put(11176,3987){\makebox(0,0)[l]{\strut{} 2}}%
      \put(11176,4295){\makebox(0,0)[l]{\strut{} 3}}%
      \put(11176,4604){\makebox(0,0)[l]{\strut{} 4}}%
      \put(11176,4912){\makebox(0,0)[l]{\strut{} 5}}%
      \put(11176,5220){\makebox(0,0)[l]{\strut{} 6}}%
      \put(11176,5529){\makebox(0,0)[l]{\strut{} 7}}%
      \put(8094,5313){\makebox(0,0)[l]{\strut{}0.6-0.7}}%
    }%
    \gplgaddtomacro\gplbacktext{%
    }%
    \gplgaddtomacro\gplfronttext{%
      \csname LTb\endcsname
      \put(577,853){\makebox(0,0){\strut{}0}}%
      \put(1297,853){\makebox(0,0){\strut{}0.25}}%
      \put(2016,853){\makebox(0,0){\strut{}0.5}}%
      \put(2735,853){\makebox(0,0){\strut{}0.75}}%
      \put(3455,853){\makebox(0,0){\strut{}1}}%
      \put(2016,523){\makebox(0,0){\strut{}phase ($\phi$)}}%
      \put(359,1124){\makebox(0,0)[r]{\strut{}}}%
      \put(359,1484){\makebox(0,0)[r]{\strut{}30}}%
      \put(359,1844){\makebox(0,0)[r]{\strut{}60}}%
      \put(359,2203){\makebox(0,0)[r]{\strut{}90}}%
      \put(359,2562){\makebox(0,0)[r]{\strut{}120}}%
      \put(359,2922){\makebox(0,0)[r]{\strut{}150}}%
      \put(-103,2203){\rotatebox{-270}{\makebox(0,0){\strut{}line of sight ($\zeta$)}}}%
      \put(3803,1124){\makebox(0,0)[l]{\strut{} 0}}%
      \put(3803,1363){\makebox(0,0)[l]{\strut{}0.5}}%
      \put(3803,1603){\makebox(0,0)[l]{\strut{} 1}}%
      \put(3803,1843){\makebox(0,0)[l]{\strut{} 2}}%
      \put(3803,2083){\makebox(0,0)[l]{\strut{} 2}}%
      \put(3803,2322){\makebox(0,0)[l]{\strut{} 2}}%
      \put(3803,2562){\makebox(0,0)[l]{\strut{} 3}}%
      \put(3803,2802){\makebox(0,0)[l]{\strut{} 4}}%
      \put(3803,3042){\makebox(0,0)[l]{\strut{} 4}}%
      \put(3803,3282){\makebox(0,0)[l]{\strut{} 4}}%
      \put(721,3066){\makebox(0,0)[l]{\strut{}0.7-0.8}}%
    }%
    \gplgaddtomacro\gplbacktext{%
    }%
    \gplgaddtomacro\gplfronttext{%
      \csname LTb\endcsname
      \put(4263,853){\makebox(0,0){\strut{}0}}%
      \put(4983,853){\makebox(0,0){\strut{}0.25}}%
      \put(5702,853){\makebox(0,0){\strut{}0.5}}%
      \put(6421,853){\makebox(0,0){\strut{}0.75}}%
      \put(7141,853){\makebox(0,0){\strut{}1}}%
      \put(5702,523){\makebox(0,0){\strut{}phase ($\phi$)}}%
      \put(4045,1124){\makebox(0,0)[r]{\strut{}}}%
      \put(4045,1484){\makebox(0,0)[r]{\strut{}}}%
      \put(4045,1844){\makebox(0,0)[r]{\strut{}}}%
      \put(4045,2203){\makebox(0,0)[r]{\strut{}}}%
      \put(4045,2562){\makebox(0,0)[r]{\strut{}}}%
      \put(4045,2922){\makebox(0,0)[r]{\strut{}}}%
      \put(7489,1124){\makebox(0,0)[l]{\strut{} 0}}%
      \put(7489,1432){\makebox(0,0)[l]{\strut{}0.5}}%
      \put(7489,1740){\makebox(0,0)[l]{\strut{} 1}}%
      \put(7489,2048){\makebox(0,0)[l]{\strut{} 2}}%
      \put(7489,2357){\makebox(0,0)[l]{\strut{} 2}}%
      \put(7489,2665){\makebox(0,0)[l]{\strut{} 2}}%
      \put(7489,2973){\makebox(0,0)[l]{\strut{} 3}}%
      \put(7489,3282){\makebox(0,0)[l]{\strut{} 4}}%
      \put(4407,3066){\makebox(0,0)[l]{\strut{}0.8-0.9}}%
    }%
    \gplgaddtomacro\gplbacktext{%
    }%
    \gplgaddtomacro\gplfronttext{%
      \csname LTb\endcsname
      \put(7950,853){\makebox(0,0){\strut{}0}}%
      \put(8670,853){\makebox(0,0){\strut{}0.25}}%
      \put(9389,853){\makebox(0,0){\strut{}0.5}}%
      \put(10108,853){\makebox(0,0){\strut{}0.75}}%
      \put(10828,853){\makebox(0,0){\strut{}1}}%
      \put(9389,523){\makebox(0,0){\strut{}phase ($\phi$)}}%
      \put(7732,1124){\makebox(0,0)[r]{\strut{}}}%
      \put(7732,1484){\makebox(0,0)[r]{\strut{}}}%
      \put(7732,1844){\makebox(0,0)[r]{\strut{}}}%
      \put(7732,2203){\makebox(0,0)[r]{\strut{}}}%
      \put(7732,2562){\makebox(0,0)[r]{\strut{}}}%
      \put(7732,2922){\makebox(0,0)[r]{\strut{}}}%
      \put(11176,1124){\makebox(0,0)[l]{\strut{} 0}}%
      \put(11176,1483){\makebox(0,0)[l]{\strut{}0.5}}%
      \put(11176,1843){\makebox(0,0)[l]{\strut{} 1}}%
      \put(11176,2203){\makebox(0,0)[l]{\strut{} 2}}%
      \put(11176,2562){\makebox(0,0)[l]{\strut{} 2}}%
      \put(11176,2922){\makebox(0,0)[l]{\strut{} 2}}%
      \put(11176,3282){\makebox(0,0)[l]{\strut{} 3}}%
      \put(8094,3066){\makebox(0,0)[l]{\strut{}0.9-1.0}}%
    }%
    \gplbacktext
    \put(0,0){\includegraphics{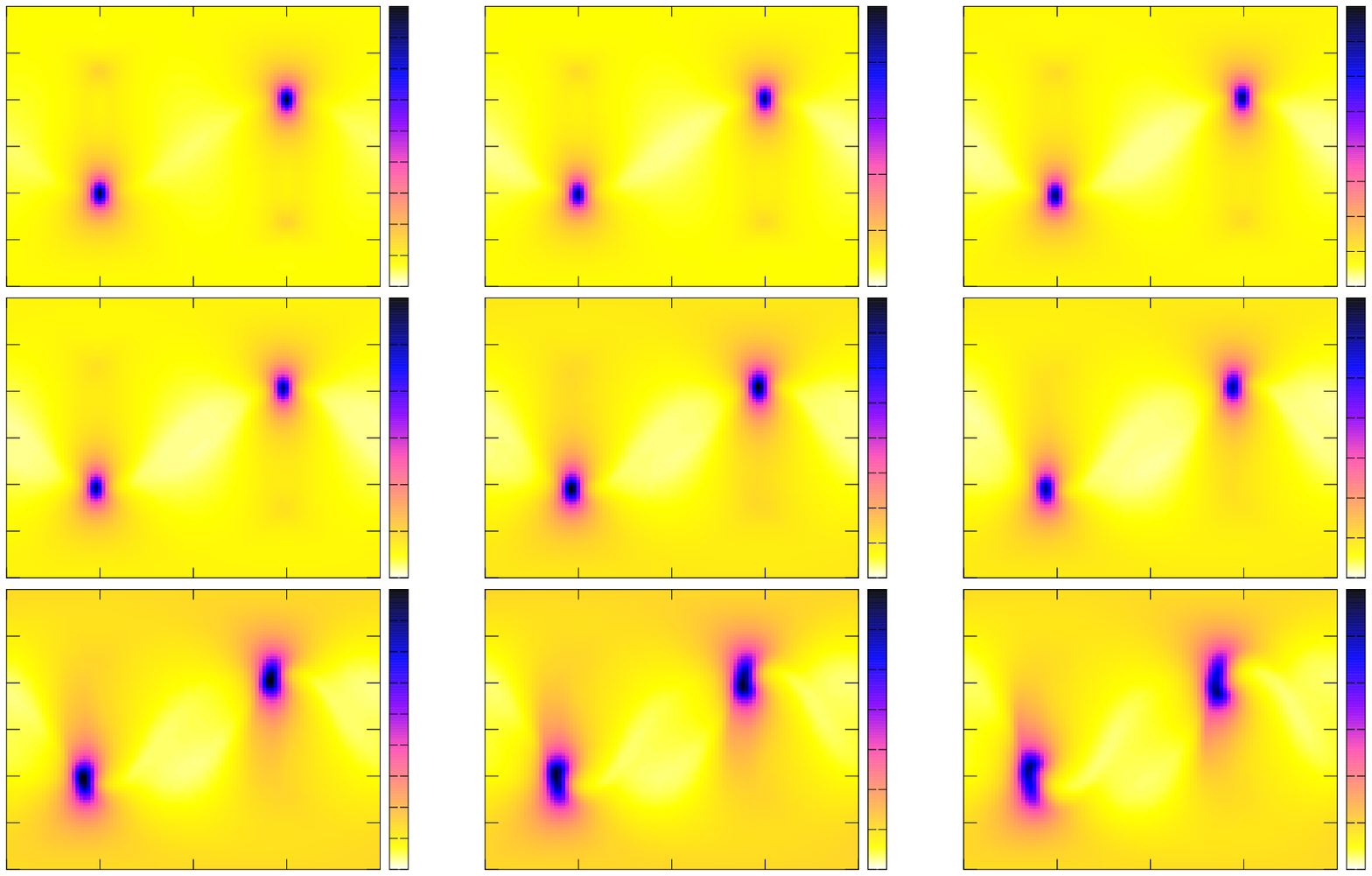}}%
    \gplfronttext
  \end{picture}%
\endgroup

%% file: skymaps_cl_r00025_b-3_c60_g10_d3_om8.tex
\begingroup
  \makeatletter
  \providecommand\color[2][]{%
    \GenericError{(gnuplot) \space\space\space\@spaces}{%
      Package color not loaded in conjunction with
      terminal option `colourtext'%
    }{See the gnuplot documentation for explanation.%
    }{Either use 'blacktext' in gnuplot or load the package
      color.sty in LaTeX.}%
    \renewcommand\color[2][]{}%
  }%
  \providecommand\includegraphics[2][]{%
    \GenericError{(gnuplot) \space\space\space\@spaces}{%
      Package graphicx or graphics not loaded%
    }{See the gnuplot documentation for explanation.%
    }{The gnuplot epslatex terminal needs graphicx.sty or graphics.sty.}%
    \renewcommand\includegraphics[2][]{}%
  }%
  \providecommand\rotatebox[2]{#2}%
  \@ifundefined{ifGPcolor}{%
    \newif\ifGPcolor
    \GPcolortrue
  }{}%
  \@ifundefined{ifGPblacktext}{%
    \newif\ifGPblacktext
    \GPblacktextfalse
  }{}%
  \let\gplgaddtomacro\g@addto@macro
  \gdef\gplbacktext{}%
  \gdef\gplfronttext{}%
  \makeatother
  \ifGPblacktext
    \def\colorrgb#1{}%
    \def\colorgray#1{}%
  \else
    \ifGPcolor
      \def\colorrgb#1{\color[rgb]{#1}}%
      \def\colorgray#1{\color[gray]{#1}}%
      \expandafter\def\csname LTw\endcsname{\color{white}}%
      \expandafter\def\csname LTb\endcsname{\color{black}}%
      \expandafter\def\csname LTa\endcsname{\color{black}}%
      \expandafter\def\csname LT0\endcsname{\color[rgb]{1,0,0}}%
      \expandafter\def\csname LT1\endcsname{\color[rgb]{0,1,0}}%
      \expandafter\def\csname LT2\endcsname{\color[rgb]{0,0,1}}%
      \expandafter\def\csname LT3\endcsname{\color[rgb]{1,0,1}}%
      \expandafter\def\csname LT4\endcsname{\color[rgb]{0,1,1}}%
      \expandafter\def\csname LT5\endcsname{\color[rgb]{1,1,0}}%
      \expandafter\def\csname LT6\endcsname{\color[rgb]{0,0,0}}%
      \expandafter\def\csname LT7\endcsname{\color[rgb]{1,0.3,0}}%
      \expandafter\def\csname LT8\endcsname{\color[rgb]{0.5,0.5,0.5}}%
    \else
      \def\colorrgb#1{\color{black}}%
      \def\colorgray#1{\color[gray]{#1}}%
      \expandafter\def\csname LTw\endcsname{\color{white}}%
      \expandafter\def\csname LTb\endcsname{\color{black}}%
      \expandafter\def\csname LTa\endcsname{\color{black}}%
      \expandafter\def\csname LT0\endcsname{\color{black}}%
      \expandafter\def\csname LT1\endcsname{\color{black}}%
      \expandafter\def\csname LT2\endcsname{\color{black}}%
      \expandafter\def\csname LT3\endcsname{\color{black}}%
      \expandafter\def\csname LT4\endcsname{\color{black}}%
      \expandafter\def\csname LT5\endcsname{\color{black}}%
      \expandafter\def\csname LT6\endcsname{\color{black}}%
      \expandafter\def\csname LT7\endcsname{\color{black}}%
      \expandafter\def\csname LT8\endcsname{\color{black}}%
    \fi
  \fi
    \setlength{\unitlength}{0.0500bp}%
    \ifx\gptboxheight\undefined%
      \newlength{\gptboxheight}%
      \newlength{\gptboxwidth}%
      \newsavebox{\gptboxtext}%
    \fi%
    \setlength{\fboxrule}{0.5pt}%
    \setlength{\fboxsep}{1pt}%
\begin{picture}(11520.00,8640.00)%
    \gplgaddtomacro\gplbacktext{%
    }%
    \gplgaddtomacro\gplfronttext{%
      \csname LTb\endcsname
      \put(577,5346){\makebox(0,0){\strut{}}}%
      \put(1297,5346){\makebox(0,0){\strut{}}}%
      \put(2016,5346){\makebox(0,0){\strut{}}}%
      \put(2735,5346){\makebox(0,0){\strut{}}}%
      \put(3455,5346){\makebox(0,0){\strut{}}}%
      \put(359,5617){\makebox(0,0)[r]{\strut{}}}%
      \put(359,5977){\makebox(0,0)[r]{\strut{}30}}%
      \put(359,6337){\makebox(0,0)[r]{\strut{}60}}%
      \put(359,6696){\makebox(0,0)[r]{\strut{}90}}%
      \put(359,7055){\makebox(0,0)[r]{\strut{}120}}%
      \put(359,7415){\makebox(0,0)[r]{\strut{}150}}%
      \put(-103,6696){\rotatebox{-270}{\makebox(0,0){\strut{}line of sight ($\zeta$)}}}%
      \put(3803,5617){\makebox(0,0)[l]{\strut{} 0}}%
      \put(3803,5976){\makebox(0,0)[l]{\strut{} 5}}%
      \put(3803,6336){\makebox(0,0)[l]{\strut{}1e+01}}%
      \put(3803,6696){\makebox(0,0)[l]{\strut{}2e+01}}%
      \put(3803,7055){\makebox(0,0)[l]{\strut{}2e+01}}%
      \put(3803,7415){\makebox(0,0)[l]{\strut{}2e+01}}%
      \put(3803,7775){\makebox(0,0)[l]{\strut{}3e+01}}%
      \put(721,7559){\makebox(0,0)[l]{\strut{}0.1-0.2}}%
    }%
    \gplgaddtomacro\gplbacktext{%
    }%
    \gplgaddtomacro\gplfronttext{%
      \csname LTb\endcsname
      \put(4263,5346){\makebox(0,0){\strut{}}}%
      \put(4983,5346){\makebox(0,0){\strut{}}}%
      \put(5702,5346){\makebox(0,0){\strut{}}}%
      \put(6421,5346){\makebox(0,0){\strut{}}}%
      \put(7141,5346){\makebox(0,0){\strut{}}}%
      \put(4045,5617){\makebox(0,0)[r]{\strut{}}}%
      \put(4045,5977){\makebox(0,0)[r]{\strut{}}}%
      \put(4045,6337){\makebox(0,0)[r]{\strut{}}}%
      \put(4045,6696){\makebox(0,0)[r]{\strut{}}}%
      \put(4045,7055){\makebox(0,0)[r]{\strut{}}}%
      \put(4045,7415){\makebox(0,0)[r]{\strut{}}}%
      \put(7489,5617){\makebox(0,0)[l]{\strut{} 0}}%
      \put(7489,5925){\makebox(0,0)[l]{\strut{} 1}}%
      \put(7489,6233){\makebox(0,0)[l]{\strut{} 2}}%
      \put(7489,6541){\makebox(0,0)[l]{\strut{} 3}}%
      \put(7489,6850){\makebox(0,0)[l]{\strut{} 4}}%
      \put(7489,7158){\makebox(0,0)[l]{\strut{} 5}}%
      \put(7489,7466){\makebox(0,0)[l]{\strut{} 6}}%
      \put(7489,7775){\makebox(0,0)[l]{\strut{} 7}}%
      \put(4407,7559){\makebox(0,0)[l]{\strut{}0.2-0.3}}%
    }%
    \gplgaddtomacro\gplbacktext{%
    }%
    \gplgaddtomacro\gplfronttext{%
      \csname LTb\endcsname
      \put(7950,5346){\makebox(0,0){\strut{}}}%
      \put(8670,5346){\makebox(0,0){\strut{}}}%
      \put(9389,5346){\makebox(0,0){\strut{}}}%
      \put(10108,5346){\makebox(0,0){\strut{}}}%
      \put(10828,5346){\makebox(0,0){\strut{}}}%
      \put(7732,5617){\makebox(0,0)[r]{\strut{}}}%
      \put(7732,5977){\makebox(0,0)[r]{\strut{}}}%
      \put(7732,6337){\makebox(0,0)[r]{\strut{}}}%
      \put(7732,6696){\makebox(0,0)[r]{\strut{}}}%
      \put(7732,7055){\makebox(0,0)[r]{\strut{}}}%
      \put(7732,7415){\makebox(0,0)[r]{\strut{}}}%
      \put(11176,5617){\makebox(0,0)[l]{\strut{} 0}}%
      \put(11176,6048){\makebox(0,0)[l]{\strut{}0.5}}%
      \put(11176,6480){\makebox(0,0)[l]{\strut{} 1}}%
      \put(11176,6911){\makebox(0,0)[l]{\strut{} 2}}%
      \put(11176,7343){\makebox(0,0)[l]{\strut{} 2}}%
      \put(11176,7775){\makebox(0,0)[l]{\strut{} 2}}%
      \put(8094,7559){\makebox(0,0)[l]{\strut{}0.3-0.4}}%
    }%
    \gplgaddtomacro\gplbacktext{%
    }%
    \gplgaddtomacro\gplfronttext{%
      \csname LTb\endcsname
      \put(577,3100){\makebox(0,0){\strut{}}}%
      \put(1297,3100){\makebox(0,0){\strut{}}}%
      \put(2016,3100){\makebox(0,0){\strut{}}}%
      \put(2735,3100){\makebox(0,0){\strut{}}}%
      \put(3455,3100){\makebox(0,0){\strut{}}}%
      \put(359,3371){\makebox(0,0)[r]{\strut{}}}%
      \put(359,3731){\makebox(0,0)[r]{\strut{}30}}%
      \put(359,4091){\makebox(0,0)[r]{\strut{}60}}%
      \put(359,4450){\makebox(0,0)[r]{\strut{}90}}%
      \put(359,4809){\makebox(0,0)[r]{\strut{}120}}%
      \put(359,5169){\makebox(0,0)[r]{\strut{}150}}%
      \put(-103,4450){\rotatebox{-270}{\makebox(0,0){\strut{}line of sight ($\zeta$)}}}%
      \put(3803,3371){\makebox(0,0)[l]{\strut{} 0}}%
      \put(3803,3586){\makebox(0,0)[l]{\strut{}0.1}}%
      \put(3803,3802){\makebox(0,0)[l]{\strut{}0.2}}%
      \put(3803,4018){\makebox(0,0)[l]{\strut{}0.3}}%
      \put(3803,4234){\makebox(0,0)[l]{\strut{}0.4}}%
      \put(3803,4450){\makebox(0,0)[l]{\strut{}0.5}}%
      \put(3803,4665){\makebox(0,0)[l]{\strut{}0.6}}%
      \put(3803,4881){\makebox(0,0)[l]{\strut{}0.7}}%
      \put(3803,5097){\makebox(0,0)[l]{\strut{}0.8}}%
      \put(3803,5313){\makebox(0,0)[l]{\strut{}0.9}}%
      \put(3803,5529){\makebox(0,0)[l]{\strut{} 1}}%
      \put(721,5313){\makebox(0,0)[l]{\strut{}0.4-0.5}}%
    }%
    \gplgaddtomacro\gplbacktext{%
    }%
    \gplgaddtomacro\gplfronttext{%
      \csname LTb\endcsname
      \put(4263,3100){\makebox(0,0){\strut{}}}%
      \put(4983,3100){\makebox(0,0){\strut{}}}%
      \put(5702,3100){\makebox(0,0){\strut{}}}%
      \put(6421,3100){\makebox(0,0){\strut{}}}%
      \put(7141,3100){\makebox(0,0){\strut{}}}%
      \put(4045,3371){\makebox(0,0)[r]{\strut{}}}%
      \put(4045,3731){\makebox(0,0)[r]{\strut{}}}%
      \put(4045,4091){\makebox(0,0)[r]{\strut{}}}%
      \put(4045,4450){\makebox(0,0)[r]{\strut{}}}%
      \put(4045,4809){\makebox(0,0)[r]{\strut{}}}%
      \put(4045,5169){\makebox(0,0)[r]{\strut{}}}%
      \put(7489,3371){\makebox(0,0)[l]{\strut{} 0}}%
      \put(7489,3586){\makebox(0,0)[l]{\strut{}0.05}}%
      \put(7489,3802){\makebox(0,0)[l]{\strut{}0.1}}%
      \put(7489,4018){\makebox(0,0)[l]{\strut{}0.2}}%
      \put(7489,4234){\makebox(0,0)[l]{\strut{}0.2}}%
      \put(7489,4450){\makebox(0,0)[l]{\strut{}0.2}}%
      \put(7489,4665){\makebox(0,0)[l]{\strut{}0.3}}%
      \put(7489,4881){\makebox(0,0)[l]{\strut{}0.3}}%
      \put(7489,5097){\makebox(0,0)[l]{\strut{}0.4}}%
      \put(7489,5313){\makebox(0,0)[l]{\strut{}0.4}}%
      \put(7489,5529){\makebox(0,0)[l]{\strut{}0.5}}%
      \put(4407,5313){\makebox(0,0)[l]{\strut{}0.5-0.6}}%
    }%
    \gplgaddtomacro\gplbacktext{%
    }%
    \gplgaddtomacro\gplfronttext{%
      \csname LTb\endcsname
      \put(7950,3100){\makebox(0,0){\strut{}}}%
      \put(8670,3100){\makebox(0,0){\strut{}}}%
      \put(9389,3100){\makebox(0,0){\strut{}}}%
      \put(10108,3100){\makebox(0,0){\strut{}}}%
      \put(10828,3100){\makebox(0,0){\strut{}}}%
      \put(7732,3371){\makebox(0,0)[r]{\strut{}}}%
      \put(7732,3731){\makebox(0,0)[r]{\strut{}}}%
      \put(7732,4091){\makebox(0,0)[r]{\strut{}}}%
      \put(7732,4450){\makebox(0,0)[r]{\strut{}}}%
      \put(7732,4809){\makebox(0,0)[r]{\strut{}}}%
      \put(7732,5169){\makebox(0,0)[r]{\strut{}}}%
      \put(11176,3371){\makebox(0,0)[l]{\strut{} 0}}%
      \put(11176,3730){\makebox(0,0)[l]{\strut{}0.05}}%
      \put(11176,4090){\makebox(0,0)[l]{\strut{}0.1}}%
      \put(11176,4450){\makebox(0,0)[l]{\strut{}0.2}}%
      \put(11176,4809){\makebox(0,0)[l]{\strut{}0.2}}%
      \put(11176,5169){\makebox(0,0)[l]{\strut{}0.2}}%
      \put(11176,5529){\makebox(0,0)[l]{\strut{}0.3}}%
      \put(8094,5313){\makebox(0,0)[l]{\strut{}0.6-0.7}}%
    }%
    \gplgaddtomacro\gplbacktext{%
    }%
    \gplgaddtomacro\gplfronttext{%
      \csname LTb\endcsname
      \put(577,853){\makebox(0,0){\strut{}0}}%
      \put(1297,853){\makebox(0,0){\strut{}0.25}}%
      \put(2016,853){\makebox(0,0){\strut{}0.5}}%
      \put(2735,853){\makebox(0,0){\strut{}0.75}}%
      \put(3455,853){\makebox(0,0){\strut{}1}}%
      \put(2016,523){\makebox(0,0){\strut{}phase ($\phi$)}}%
      \put(359,1124){\makebox(0,0)[r]{\strut{}}}%
      \put(359,1484){\makebox(0,0)[r]{\strut{}30}}%
      \put(359,1844){\makebox(0,0)[r]{\strut{}60}}%
      \put(359,2203){\makebox(0,0)[r]{\strut{}90}}%
      \put(359,2562){\makebox(0,0)[r]{\strut{}120}}%
      \put(359,2922){\makebox(0,0)[r]{\strut{}150}}%
      \put(-103,2203){\rotatebox{-270}{\makebox(0,0){\strut{}line of sight ($\zeta$)}}}%
      \put(3803,1124){\makebox(0,0)[l]{\strut{} 0}}%
      \put(3803,1363){\makebox(0,0)[l]{\strut{}0.02}}%
      \put(3803,1603){\makebox(0,0)[l]{\strut{}0.04}}%
      \put(3803,1843){\makebox(0,0)[l]{\strut{}0.06}}%
      \put(3803,2083){\makebox(0,0)[l]{\strut{}0.08}}%
      \put(3803,2322){\makebox(0,0)[l]{\strut{}0.1}}%
      \put(3803,2562){\makebox(0,0)[l]{\strut{}0.1}}%
      \put(3803,2802){\makebox(0,0)[l]{\strut{}0.1}}%
      \put(3803,3042){\makebox(0,0)[l]{\strut{}0.2}}%
      \put(3803,3282){\makebox(0,0)[l]{\strut{}0.2}}%
      \put(721,3066){\makebox(0,0)[l]{\strut{}0.7-0.8}}%
    }%
    \gplgaddtomacro\gplbacktext{%
    }%
    \gplgaddtomacro\gplfronttext{%
      \csname LTb\endcsname
      \put(4263,853){\makebox(0,0){\strut{}0}}%
      \put(4983,853){\makebox(0,0){\strut{}0.25}}%
      \put(5702,853){\makebox(0,0){\strut{}0.5}}%
      \put(6421,853){\makebox(0,0){\strut{}0.75}}%
      \put(7141,853){\makebox(0,0){\strut{}1}}%
      \put(5702,523){\makebox(0,0){\strut{}phase ($\phi$)}}%
      \put(4045,1124){\makebox(0,0)[r]{\strut{}}}%
      \put(4045,1484){\makebox(0,0)[r]{\strut{}}}%
      \put(4045,1844){\makebox(0,0)[r]{\strut{}}}%
      \put(4045,2203){\makebox(0,0)[r]{\strut{}}}%
      \put(4045,2562){\makebox(0,0)[r]{\strut{}}}%
      \put(4045,2922){\makebox(0,0)[r]{\strut{}}}%
      \put(7489,1124){\makebox(0,0)[l]{\strut{} 0}}%
      \put(7489,1432){\makebox(0,0)[l]{\strut{}0.02}}%
      \put(7489,1740){\makebox(0,0)[l]{\strut{}0.04}}%
      \put(7489,2048){\makebox(0,0)[l]{\strut{}0.06}}%
      \put(7489,2357){\makebox(0,0)[l]{\strut{}0.08}}%
      \put(7489,2665){\makebox(0,0)[l]{\strut{}0.1}}%
      \put(7489,2973){\makebox(0,0)[l]{\strut{}0.1}}%
      \put(7489,3282){\makebox(0,0)[l]{\strut{}0.1}}%
      \put(4407,3066){\makebox(0,0)[l]{\strut{}0.8-0.9}}%
    }%
    \gplgaddtomacro\gplbacktext{%
    }%
    \gplgaddtomacro\gplfronttext{%
      \csname LTb\endcsname
      \put(7950,853){\makebox(0,0){\strut{}0}}%
      \put(8670,853){\makebox(0,0){\strut{}0.25}}%
      \put(9389,853){\makebox(0,0){\strut{}0.5}}%
      \put(10108,853){\makebox(0,0){\strut{}0.75}}%
      \put(10828,853){\makebox(0,0){\strut{}1}}%
      \put(9389,523){\makebox(0,0){\strut{}phase ($\phi$)}}%
      \put(7732,1124){\makebox(0,0)[r]{\strut{}}}%
      \put(7732,1484){\makebox(0,0)[r]{\strut{}}}%
      \put(7732,1844){\makebox(0,0)[r]{\strut{}}}%
      \put(7732,2203){\makebox(0,0)[r]{\strut{}}}%
      \put(7732,2562){\makebox(0,0)[r]{\strut{}}}%
      \put(7732,2922){\makebox(0,0)[r]{\strut{}}}%
      \put(11176,1124){\makebox(0,0)[l]{\strut{} 0}}%
      \put(11176,1363){\makebox(0,0)[l]{\strut{}0.01}}%
      \put(11176,1603){\makebox(0,0)[l]{\strut{}0.02}}%
      \put(11176,1843){\makebox(0,0)[l]{\strut{}0.03}}%
      \put(11176,2083){\makebox(0,0)[l]{\strut{}0.04}}%
      \put(11176,2322){\makebox(0,0)[l]{\strut{}0.05}}%
      \put(11176,2562){\makebox(0,0)[l]{\strut{}0.06}}%
      \put(11176,2802){\makebox(0,0)[l]{\strut{}0.07}}%
      \put(11176,3042){\makebox(0,0)[l]{\strut{}0.08}}%
      \put(11176,3282){\makebox(0,0)[l]{\strut{}0.09}}%
      \put(8094,3066){\makebox(0,0)[l]{\strut{}0.9-1.0}}%
    }%
    \gplbacktext
    \put(0,0){\includegraphics{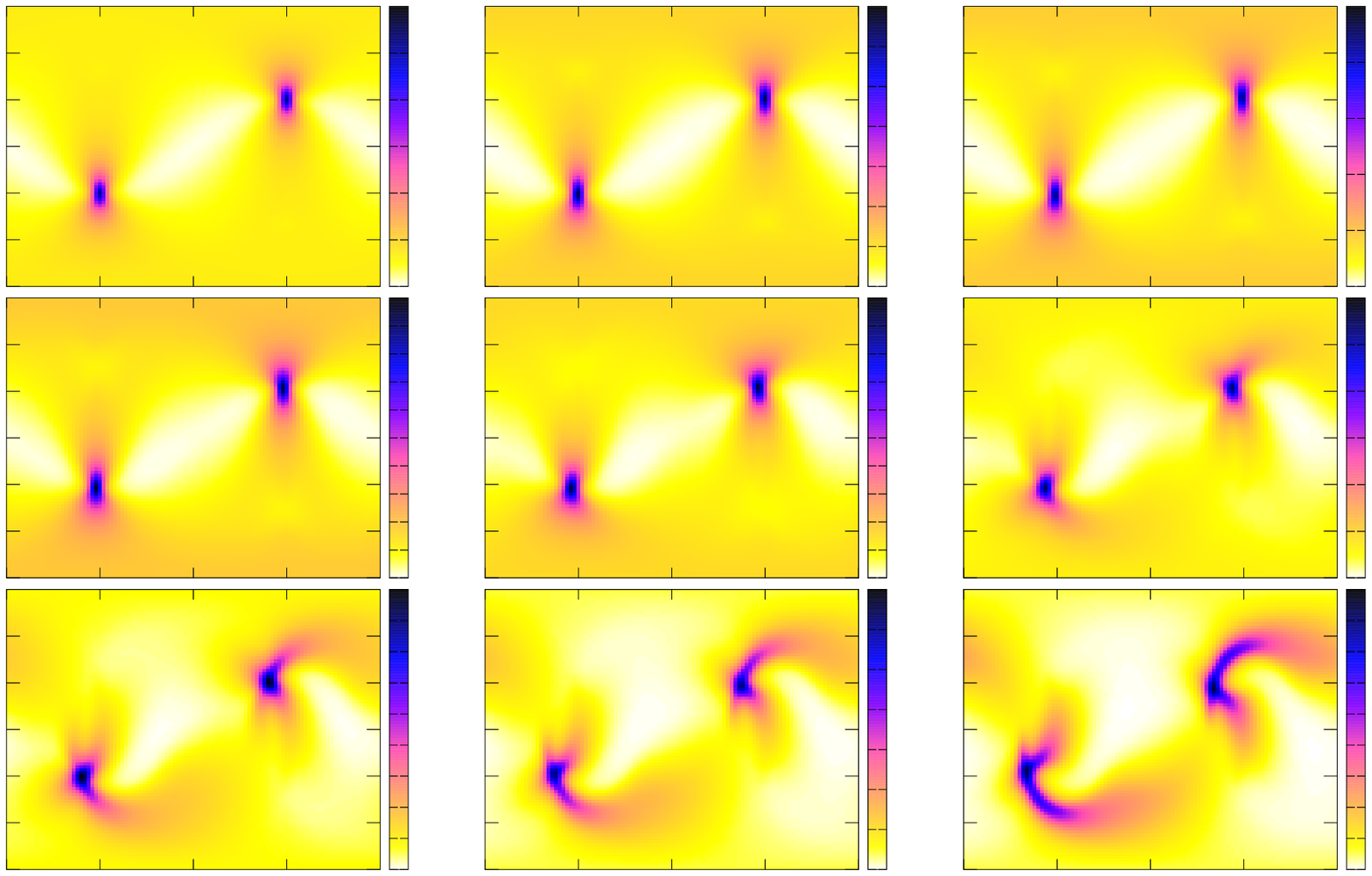}}%
    \gplfronttext
  \end{picture}%
\endgroup

%% file: skymaps_cl_r00025_b-3_c60_g10_d3_om10.tex
\begingroup
  \makeatletter
  \providecommand\color[2][]{%
    \GenericError{(gnuplot) \space\space\space\@spaces}{%
      Package color not loaded in conjunction with
      terminal option `colourtext'%
    }{See the gnuplot documentation for explanation.%
    }{Either use 'blacktext' in gnuplot or load the package
      color.sty in LaTeX.}%
    \renewcommand\color[2][]{}%
  }%
  \providecommand\includegraphics[2][]{%
    \GenericError{(gnuplot) \space\space\space\@spaces}{%
      Package graphicx or graphics not loaded%
    }{See the gnuplot documentation for explanation.%
    }{The gnuplot epslatex terminal needs graphicx.sty or graphics.sty.}%
    \renewcommand\includegraphics[2][]{}%
  }%
  \providecommand\rotatebox[2]{#2}%
  \@ifundefined{ifGPcolor}{%
    \newif\ifGPcolor
    \GPcolortrue
  }{}%
  \@ifundefined{ifGPblacktext}{%
    \newif\ifGPblacktext
    \GPblacktextfalse
  }{}%
  \let\gplgaddtomacro\g@addto@macro
  \gdef\gplbacktext{}%
  \gdef\gplfronttext{}%
  \makeatother
  \ifGPblacktext
    \def\colorrgb#1{}%
    \def\colorgray#1{}%
  \else
    \ifGPcolor
      \def\colorrgb#1{\color[rgb]{#1}}%
      \def\colorgray#1{\color[gray]{#1}}%
      \expandafter\def\csname LTw\endcsname{\color{white}}%
      \expandafter\def\csname LTb\endcsname{\color{black}}%
      \expandafter\def\csname LTa\endcsname{\color{black}}%
      \expandafter\def\csname LT0\endcsname{\color[rgb]{1,0,0}}%
      \expandafter\def\csname LT1\endcsname{\color[rgb]{0,1,0}}%
      \expandafter\def\csname LT2\endcsname{\color[rgb]{0,0,1}}%
      \expandafter\def\csname LT3\endcsname{\color[rgb]{1,0,1}}%
      \expandafter\def\csname LT4\endcsname{\color[rgb]{0,1,1}}%
      \expandafter\def\csname LT5\endcsname{\color[rgb]{1,1,0}}%
      \expandafter\def\csname LT6\endcsname{\color[rgb]{0,0,0}}%
      \expandafter\def\csname LT7\endcsname{\color[rgb]{1,0.3,0}}%
      \expandafter\def\csname LT8\endcsname{\color[rgb]{0.5,0.5,0.5}}%
    \else
      \def\colorrgb#1{\color{black}}%
      \def\colorgray#1{\color[gray]{#1}}%
      \expandafter\def\csname LTw\endcsname{\color{white}}%
      \expandafter\def\csname LTb\endcsname{\color{black}}%
      \expandafter\def\csname LTa\endcsname{\color{black}}%
      \expandafter\def\csname LT0\endcsname{\color{black}}%
      \expandafter\def\csname LT1\endcsname{\color{black}}%
      \expandafter\def\csname LT2\endcsname{\color{black}}%
      \expandafter\def\csname LT3\endcsname{\color{black}}%
      \expandafter\def\csname LT4\endcsname{\color{black}}%
      \expandafter\def\csname LT5\endcsname{\color{black}}%
      \expandafter\def\csname LT6\endcsname{\color{black}}%
      \expandafter\def\csname LT7\endcsname{\color{black}}%
      \expandafter\def\csname LT8\endcsname{\color{black}}%
    \fi
  \fi
    \setlength{\unitlength}{0.0500bp}%
    \ifx\gptboxheight\undefined%
      \newlength{\gptboxheight}%
      \newlength{\gptboxwidth}%
      \newsavebox{\gptboxtext}%
    \fi%
    \setlength{\fboxrule}{0.5pt}%
    \setlength{\fboxsep}{1pt}%
\begin{picture}(11520.00,8640.00)%
    \gplgaddtomacro\gplbacktext{%
    }%
    \gplgaddtomacro\gplfronttext{%
      \csname LTb\endcsname
      \put(577,5346){\makebox(0,0){\strut{}}}%
      \put(1297,5346){\makebox(0,0){\strut{}}}%
      \put(2016,5346){\makebox(0,0){\strut{}}}%
      \put(2735,5346){\makebox(0,0){\strut{}}}%
      \put(3455,5346){\makebox(0,0){\strut{}}}%
      \put(359,5617){\makebox(0,0)[r]{\strut{}}}%
      \put(359,5977){\makebox(0,0)[r]{\strut{}30}}%
      \put(359,6337){\makebox(0,0)[r]{\strut{}60}}%
      \put(359,6696){\makebox(0,0)[r]{\strut{}90}}%
      \put(359,7055){\makebox(0,0)[r]{\strut{}120}}%
      \put(359,7415){\makebox(0,0)[r]{\strut{}150}}%
      \put(-103,6696){\rotatebox{-270}{\makebox(0,0){\strut{}line of sight ($\zeta$)}}}%
      \put(3803,5617){\makebox(0,0)[l]{\strut{} 0}}%
      \put(3803,5856){\makebox(0,0)[l]{\strut{}0.002}}%
      \put(3803,6096){\makebox(0,0)[l]{\strut{}0.004}}%
      \put(3803,6336){\makebox(0,0)[l]{\strut{}0.006}}%
      \put(3803,6576){\makebox(0,0)[l]{\strut{}0.008}}%
      \put(3803,6815){\makebox(0,0)[l]{\strut{}0.01}}%
      \put(3803,7055){\makebox(0,0)[l]{\strut{}0.01}}%
      \put(3803,7295){\makebox(0,0)[l]{\strut{}0.01}}%
      \put(3803,7535){\makebox(0,0)[l]{\strut{}0.02}}%
      \put(3803,7775){\makebox(0,0)[l]{\strut{}0.02}}%
      \put(721,7559){\makebox(0,0)[l]{\strut{}0.1-0.2}}%
    }%
    \gplgaddtomacro\gplbacktext{%
    }%
    \gplgaddtomacro\gplfronttext{%
      \csname LTb\endcsname
      \put(4263,5346){\makebox(0,0){\strut{}}}%
      \put(4983,5346){\makebox(0,0){\strut{}}}%
      \put(5702,5346){\makebox(0,0){\strut{}}}%
      \put(6421,5346){\makebox(0,0){\strut{}}}%
      \put(7141,5346){\makebox(0,0){\strut{}}}%
      \put(4045,5617){\makebox(0,0)[r]{\strut{}}}%
      \put(4045,5977){\makebox(0,0)[r]{\strut{}}}%
      \put(4045,6337){\makebox(0,0)[r]{\strut{}}}%
      \put(4045,6696){\makebox(0,0)[r]{\strut{}}}%
      \put(4045,7055){\makebox(0,0)[r]{\strut{}}}%
      \put(4045,7415){\makebox(0,0)[r]{\strut{}}}%
      \put(7489,5617){\makebox(0,0)[l]{\strut{} 0}}%
      \put(7489,5856){\makebox(0,0)[l]{\strut{}2e-06}}%
      \put(7489,6096){\makebox(0,0)[l]{\strut{}4e-06}}%
      \put(7489,6336){\makebox(0,0)[l]{\strut{}6e-06}}%
      \put(7489,6576){\makebox(0,0)[l]{\strut{}8e-06}}%
      \put(7489,6815){\makebox(0,0)[l]{\strut{}1e-05}}%
      \put(7489,7055){\makebox(0,0)[l]{\strut{}1e-05}}%
      \put(7489,7295){\makebox(0,0)[l]{\strut{}1e-05}}%
      \put(7489,7535){\makebox(0,0)[l]{\strut{}2e-05}}%
      \put(7489,7775){\makebox(0,0)[l]{\strut{}2e-05}}%
      \put(4407,7559){\makebox(0,0)[l]{\strut{}0.2-0.3}}%
    }%
    \gplgaddtomacro\gplbacktext{%
    }%
    \gplgaddtomacro\gplfronttext{%
      \csname LTb\endcsname
      \put(7950,5346){\makebox(0,0){\strut{}}}%
      \put(8670,5346){\makebox(0,0){\strut{}}}%
      \put(9389,5346){\makebox(0,0){\strut{}}}%
      \put(10108,5346){\makebox(0,0){\strut{}}}%
      \put(10828,5346){\makebox(0,0){\strut{}}}%
      \put(7732,5617){\makebox(0,0)[r]{\strut{}}}%
      \put(7732,5977){\makebox(0,0)[r]{\strut{}}}%
      \put(7732,6337){\makebox(0,0)[r]{\strut{}}}%
      \put(7732,6696){\makebox(0,0)[r]{\strut{}}}%
      \put(7732,7055){\makebox(0,0)[r]{\strut{}}}%
      \put(7732,7415){\makebox(0,0)[r]{\strut{}}}%
      \put(11176,5617){\makebox(0,0)[l]{\strut{} 0}}%
      \put(11176,5886){\makebox(0,0)[l]{\strut{}2e-08}}%
      \put(11176,6156){\makebox(0,0)[l]{\strut{}4e-08}}%
      \put(11176,6426){\makebox(0,0)[l]{\strut{}6e-08}}%
      \put(11176,6696){\makebox(0,0)[l]{\strut{}8e-08}}%
      \put(11176,6965){\makebox(0,0)[l]{\strut{}1e-07}}%
      \put(11176,7235){\makebox(0,0)[l]{\strut{}1e-07}}%
      \put(11176,7505){\makebox(0,0)[l]{\strut{}1e-07}}%
      \put(11176,7775){\makebox(0,0)[l]{\strut{}2e-07}}%
      \put(8094,7559){\makebox(0,0)[l]{\strut{}0.3-0.4}}%
    }%
    \gplgaddtomacro\gplbacktext{%
    }%
    \gplgaddtomacro\gplfronttext{%
      \csname LTb\endcsname
      \put(577,3100){\makebox(0,0){\strut{}}}%
      \put(1297,3100){\makebox(0,0){\strut{}}}%
      \put(2016,3100){\makebox(0,0){\strut{}}}%
      \put(2735,3100){\makebox(0,0){\strut{}}}%
      \put(3455,3100){\makebox(0,0){\strut{}}}%
      \put(359,3371){\makebox(0,0)[r]{\strut{}}}%
      \put(359,3731){\makebox(0,0)[r]{\strut{}30}}%
      \put(359,4091){\makebox(0,0)[r]{\strut{}60}}%
      \put(359,4450){\makebox(0,0)[r]{\strut{}90}}%
      \put(359,4809){\makebox(0,0)[r]{\strut{}120}}%
      \put(359,5169){\makebox(0,0)[r]{\strut{}150}}%
      \put(-103,4450){\rotatebox{-270}{\makebox(0,0){\strut{}line of sight ($\zeta$)}}}%
      \put(3803,3371){\makebox(0,0)[l]{\strut{} 0}}%
      \put(3803,3730){\makebox(0,0)[l]{\strut{}1e-08}}%
      \put(3803,4090){\makebox(0,0)[l]{\strut{}2e-08}}%
      \put(3803,4450){\makebox(0,0)[l]{\strut{}3e-08}}%
      \put(3803,4809){\makebox(0,0)[l]{\strut{}4e-08}}%
      \put(3803,5169){\makebox(0,0)[l]{\strut{}5e-08}}%
      \put(3803,5529){\makebox(0,0)[l]{\strut{}6e-08}}%
      \put(721,5313){\makebox(0,0)[l]{\strut{}0.4-0.5}}%
    }%
    \gplgaddtomacro\gplbacktext{%
    }%
    \gplgaddtomacro\gplfronttext{%
      \csname LTb\endcsname
      \put(4263,3100){\makebox(0,0){\strut{}}}%
      \put(4983,3100){\makebox(0,0){\strut{}}}%
      \put(5702,3100){\makebox(0,0){\strut{}}}%
      \put(6421,3100){\makebox(0,0){\strut{}}}%
      \put(7141,3100){\makebox(0,0){\strut{}}}%
      \put(4045,3371){\makebox(0,0)[r]{\strut{}}}%
      \put(4045,3731){\makebox(0,0)[r]{\strut{}}}%
      \put(4045,4091){\makebox(0,0)[r]{\strut{}}}%
      \put(4045,4450){\makebox(0,0)[r]{\strut{}}}%
      \put(4045,4809){\makebox(0,0)[r]{\strut{}}}%
      \put(4045,5169){\makebox(0,0)[r]{\strut{}}}%
      \put(7489,3371){\makebox(0,0)[l]{\strut{} 0}}%
      \put(7489,3610){\makebox(0,0)[l]{\strut{}5e-09}}%
      \put(7489,3850){\makebox(0,0)[l]{\strut{}1e-08}}%
      \put(7489,4090){\makebox(0,0)[l]{\strut{}2e-08}}%
      \put(7489,4330){\makebox(0,0)[l]{\strut{}2e-08}}%
      \put(7489,4569){\makebox(0,0)[l]{\strut{}2e-08}}%
      \put(7489,4809){\makebox(0,0)[l]{\strut{}3e-08}}%
      \put(7489,5049){\makebox(0,0)[l]{\strut{}3e-08}}%
      \put(7489,5289){\makebox(0,0)[l]{\strut{}4e-08}}%
      \put(7489,5529){\makebox(0,0)[l]{\strut{}4e-08}}%
      \put(4407,5313){\makebox(0,0)[l]{\strut{}0.5-0.6}}%
    }%
    \gplgaddtomacro\gplbacktext{%
    }%
    \gplgaddtomacro\gplfronttext{%
      \csname LTb\endcsname
      \put(7950,3100){\makebox(0,0){\strut{}}}%
      \put(8670,3100){\makebox(0,0){\strut{}}}%
      \put(9389,3100){\makebox(0,0){\strut{}}}%
      \put(10108,3100){\makebox(0,0){\strut{}}}%
      \put(10828,3100){\makebox(0,0){\strut{}}}%
      \put(7732,3371){\makebox(0,0)[r]{\strut{}}}%
      \put(7732,3731){\makebox(0,0)[r]{\strut{}}}%
      \put(7732,4091){\makebox(0,0)[r]{\strut{}}}%
      \put(7732,4450){\makebox(0,0)[r]{\strut{}}}%
      \put(7732,4809){\makebox(0,0)[r]{\strut{}}}%
      \put(7732,5169){\makebox(0,0)[r]{\strut{}}}%
      \put(11176,3371){\makebox(0,0)[l]{\strut{} 0}}%
      \put(11176,3679){\makebox(0,0)[l]{\strut{}5e-09}}%
      \put(11176,3987){\makebox(0,0)[l]{\strut{}1e-08}}%
      \put(11176,4295){\makebox(0,0)[l]{\strut{}2e-08}}%
      \put(11176,4604){\makebox(0,0)[l]{\strut{}2e-08}}%
      \put(11176,4912){\makebox(0,0)[l]{\strut{}2e-08}}%
      \put(11176,5220){\makebox(0,0)[l]{\strut{}3e-08}}%
      \put(11176,5529){\makebox(0,0)[l]{\strut{}3e-08}}%
      \put(8094,5313){\makebox(0,0)[l]{\strut{}0.6-0.7}}%
    }%
    \gplgaddtomacro\gplbacktext{%
    }%
    \gplgaddtomacro\gplfronttext{%
      \csname LTb\endcsname
      \put(577,853){\makebox(0,0){\strut{}0}}%
      \put(1297,853){\makebox(0,0){\strut{}0.25}}%
      \put(2016,853){\makebox(0,0){\strut{}0.5}}%
      \put(2735,853){\makebox(0,0){\strut{}0.75}}%
      \put(3455,853){\makebox(0,0){\strut{}1}}%
      \put(2016,523){\makebox(0,0){\strut{}phase ($\phi$)}}%
      \put(359,1124){\makebox(0,0)[r]{\strut{}}}%
      \put(359,1484){\makebox(0,0)[r]{\strut{}30}}%
      \put(359,1844){\makebox(0,0)[r]{\strut{}60}}%
      \put(359,2203){\makebox(0,0)[r]{\strut{}90}}%
      \put(359,2562){\makebox(0,0)[r]{\strut{}120}}%
      \put(359,2922){\makebox(0,0)[r]{\strut{}150}}%
      \put(-103,2203){\rotatebox{-270}{\makebox(0,0){\strut{}line of sight ($\zeta$)}}}%
      \put(3803,1124){\makebox(0,0)[l]{\strut{} 0}}%
      \put(3803,1363){\makebox(0,0)[l]{\strut{}2e-09}}%
      \put(3803,1603){\makebox(0,0)[l]{\strut{}4e-09}}%
      \put(3803,1843){\makebox(0,0)[l]{\strut{}6e-09}}%
      \put(3803,2083){\makebox(0,0)[l]{\strut{}8e-09}}%
      \put(3803,2322){\makebox(0,0)[l]{\strut{}1e-08}}%
      \put(3803,2562){\makebox(0,0)[l]{\strut{}1e-08}}%
      \put(3803,2802){\makebox(0,0)[l]{\strut{}1e-08}}%
      \put(3803,3042){\makebox(0,0)[l]{\strut{}2e-08}}%
      \put(3803,3282){\makebox(0,0)[l]{\strut{}2e-08}}%
      \put(721,3066){\makebox(0,0)[l]{\strut{}0.7-0.8}}%
    }%
    \gplgaddtomacro\gplbacktext{%
    }%
    \gplgaddtomacro\gplfronttext{%
      \csname LTb\endcsname
      \put(4263,853){\makebox(0,0){\strut{}0}}%
      \put(4983,853){\makebox(0,0){\strut{}0.25}}%
      \put(5702,853){\makebox(0,0){\strut{}0.5}}%
      \put(6421,853){\makebox(0,0){\strut{}0.75}}%
      \put(7141,853){\makebox(0,0){\strut{}1}}%
      \put(5702,523){\makebox(0,0){\strut{}phase ($\phi$)}}%
      \put(4045,1124){\makebox(0,0)[r]{\strut{}}}%
      \put(4045,1484){\makebox(0,0)[r]{\strut{}}}%
      \put(4045,1844){\makebox(0,0)[r]{\strut{}}}%
      \put(4045,2203){\makebox(0,0)[r]{\strut{}}}%
      \put(4045,2562){\makebox(0,0)[r]{\strut{}}}%
      \put(4045,2922){\makebox(0,0)[r]{\strut{}}}%
      \put(7489,1124){\makebox(0,0)[l]{\strut{} 0}}%
      \put(7489,1432){\makebox(0,0)[l]{\strut{}2e-09}}%
      \put(7489,1740){\makebox(0,0)[l]{\strut{}4e-09}}%
      \put(7489,2048){\makebox(0,0)[l]{\strut{}6e-09}}%
      \put(7489,2357){\makebox(0,0)[l]{\strut{}8e-09}}%
      \put(7489,2665){\makebox(0,0)[l]{\strut{}1e-08}}%
      \put(7489,2973){\makebox(0,0)[l]{\strut{}1e-08}}%
      \put(7489,3281){\makebox(0,0)[l]{\strut{}1e-08}}%
      \put(4407,3066){\makebox(0,0)[l]{\strut{}0.8-0.9}}%
    }%
    \gplgaddtomacro\gplbacktext{%
    }%
    \gplgaddtomacro\gplfronttext{%
      \csname LTb\endcsname
      \put(7950,853){\makebox(0,0){\strut{}0}}%
      \put(8670,853){\makebox(0,0){\strut{}0.25}}%
      \put(9389,853){\makebox(0,0){\strut{}0.5}}%
      \put(10108,853){\makebox(0,0){\strut{}0.75}}%
      \put(10828,853){\makebox(0,0){\strut{}1}}%
      \put(9389,523){\makebox(0,0){\strut{}phase ($\phi$)}}%
      \put(7732,1124){\makebox(0,0)[r]{\strut{}}}%
      \put(7732,1484){\makebox(0,0)[r]{\strut{}}}%
      \put(7732,1844){\makebox(0,0)[r]{\strut{}}}%
      \put(7732,2203){\makebox(0,0)[r]{\strut{}}}%
      \put(7732,2562){\makebox(0,0)[r]{\strut{}}}%
      \put(7732,2922){\makebox(0,0)[r]{\strut{}}}%
      \put(11176,1124){\makebox(0,0)[l]{\strut{} 0}}%
      \put(11176,1363){\makebox(0,0)[l]{\strut{}5e-07}}%
      \put(11176,1603){\makebox(0,0)[l]{\strut{}1e-06}}%
      \put(11176,1843){\makebox(0,0)[l]{\strut{}2e-06}}%
      \put(11176,2083){\makebox(0,0)[l]{\strut{}2e-06}}%
      \put(11176,2322){\makebox(0,0)[l]{\strut{}2e-06}}%
      \put(11176,2562){\makebox(0,0)[l]{\strut{}3e-06}}%
      \put(11176,2802){\makebox(0,0)[l]{\strut{}3e-06}}%
      \put(11176,3042){\makebox(0,0)[l]{\strut{}4e-06}}%
      \put(11176,3282){\makebox(0,0)[l]{\strut{}5e-06}}%
      \put(8094,3066){\makebox(0,0)[l]{\strut{}0.9-1.0}}%
    }%
    \gplbacktext
    \put(0,0){\includegraphics{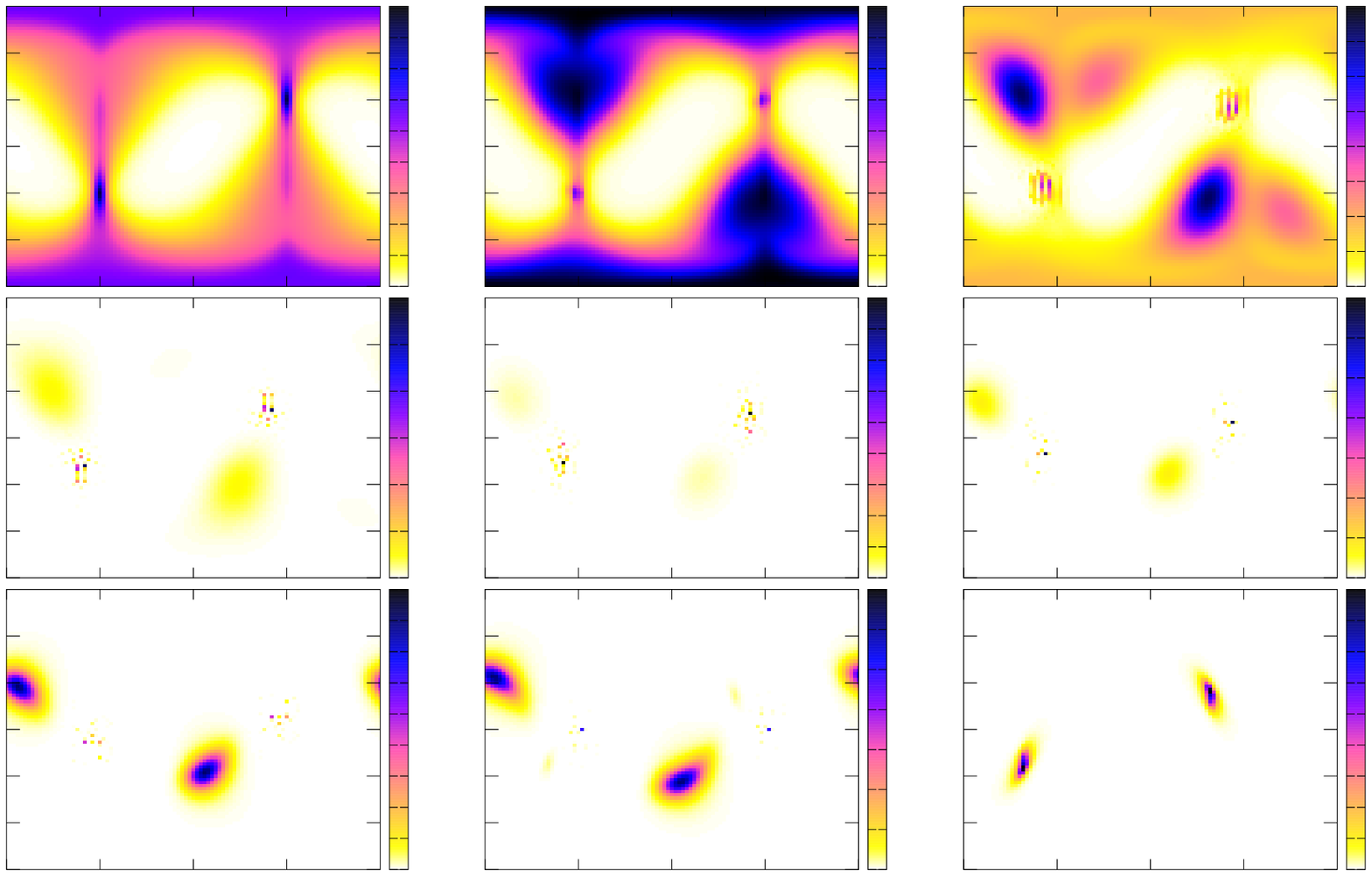}}%
    \gplfronttext
  \end{picture}%
\endgroup

%% file: skymaps_cl_r00025_b-3_ri0.2_ro1_g10_d3_om8.tex
\begingroup
  \makeatletter
  \providecommand\color[2][]{%
    \GenericError{(gnuplot) \space\space\space\@spaces}{%
      Package color not loaded in conjunction with
      terminal option `colourtext'%
    }{See the gnuplot documentation for explanation.%
    }{Either use 'blacktext' in gnuplot or load the package
      color.sty in LaTeX.}%
    \renewcommand\color[2][]{}%
  }%
  \providecommand\includegraphics[2][]{%
    \GenericError{(gnuplot) \space\space\space\@spaces}{%
      Package graphicx or graphics not loaded%
    }{See the gnuplot documentation for explanation.%
    }{The gnuplot epslatex terminal needs graphicx.sty or graphics.sty.}%
    \renewcommand\includegraphics[2][]{}%
  }%
  \providecommand\rotatebox[2]{#2}%
  \@ifundefined{ifGPcolor}{%
    \newif\ifGPcolor
    \GPcolortrue
  }{}%
  \@ifundefined{ifGPblacktext}{%
    \newif\ifGPblacktext
    \GPblacktextfalse
  }{}%
  \let\gplgaddtomacro\g@addto@macro
  \gdef\gplbacktext{}%
  \gdef\gplfronttext{}%
  \makeatother
  \ifGPblacktext
    \def\colorrgb#1{}%
    \def\colorgray#1{}%
  \else
    \ifGPcolor
      \def\colorrgb#1{\color[rgb]{#1}}%
      \def\colorgray#1{\color[gray]{#1}}%
      \expandafter\def\csname LTw\endcsname{\color{white}}%
      \expandafter\def\csname LTb\endcsname{\color{black}}%
      \expandafter\def\csname LTa\endcsname{\color{black}}%
      \expandafter\def\csname LT0\endcsname{\color[rgb]{1,0,0}}%
      \expandafter\def\csname LT1\endcsname{\color[rgb]{0,1,0}}%
      \expandafter\def\csname LT2\endcsname{\color[rgb]{0,0,1}}%
      \expandafter\def\csname LT3\endcsname{\color[rgb]{1,0,1}}%
      \expandafter\def\csname LT4\endcsname{\color[rgb]{0,1,1}}%
      \expandafter\def\csname LT5\endcsname{\color[rgb]{1,1,0}}%
      \expandafter\def\csname LT6\endcsname{\color[rgb]{0,0,0}}%
      \expandafter\def\csname LT7\endcsname{\color[rgb]{1,0.3,0}}%
      \expandafter\def\csname LT8\endcsname{\color[rgb]{0.5,0.5,0.5}}%
    \else
      \def\colorrgb#1{\color{black}}%
      \def\colorgray#1{\color[gray]{#1}}%
      \expandafter\def\csname LTw\endcsname{\color{white}}%
      \expandafter\def\csname LTb\endcsname{\color{black}}%
      \expandafter\def\csname LTa\endcsname{\color{black}}%
      \expandafter\def\csname LT0\endcsname{\color{black}}%
      \expandafter\def\csname LT1\endcsname{\color{black}}%
      \expandafter\def\csname LT2\endcsname{\color{black}}%
      \expandafter\def\csname LT3\endcsname{\color{black}}%
      \expandafter\def\csname LT4\endcsname{\color{black}}%
      \expandafter\def\csname LT5\endcsname{\color{black}}%
      \expandafter\def\csname LT6\endcsname{\color{black}}%
      \expandafter\def\csname LT7\endcsname{\color{black}}%
      \expandafter\def\csname LT8\endcsname{\color{black}}%
    \fi
  \fi
    \setlength{\unitlength}{0.0500bp}%
    \ifx\gptboxheight\undefined%
      \newlength{\gptboxheight}%
      \newlength{\gptboxwidth}%
      \newsavebox{\gptboxtext}%
    \fi%
    \setlength{\fboxrule}{0.5pt}%
    \setlength{\fboxsep}{1pt}%
\begin{picture}(10080.00,4320.00)%
    \gplgaddtomacro\gplbacktext{%
    }%
    \gplgaddtomacro\gplfronttext{%
      \csname LTb\endcsname
      \put(8064,2535){\makebox(0,0){\strut{}}}%
      \put(8417,2535){\makebox(0,0){\strut{}}}%
      \put(8769,2535){\makebox(0,0){\strut{}}}%
      \put(9121,2535){\makebox(0,0){\strut{}}}%
      \put(9474,2535){\makebox(0,0){\strut{}}}%
      \put(7849,2809){\makebox(0,0)[r]{\strut{}}}%
      \put(7849,2989){\makebox(0,0)[r]{\strut{}}}%
      \put(7849,3169){\makebox(0,0)[r]{\strut{}}}%
      \put(7849,3348){\makebox(0,0)[r]{\strut{}}}%
      \put(7849,3527){\makebox(0,0)[r]{\strut{}}}%
      \put(7849,3707){\makebox(0,0)[r]{\strut{}}}%
      \put(9712,2809){\makebox(0,0)[l]{\strut{} 0}}%
      \put(9712,2963){\makebox(0,0)[l]{\strut{}0.2}}%
      \put(9712,3117){\makebox(0,0)[l]{\strut{}0.4}}%
      \put(9712,3271){\makebox(0,0)[l]{\strut{}0.6}}%
      \put(9712,3425){\makebox(0,0)[l]{\strut{}0.8}}%
      \put(9712,3579){\makebox(0,0)[l]{\strut{} 1}}%
      \put(9712,3733){\makebox(0,0)[l]{\strut{} 1}}%
      \put(9712,3887){\makebox(0,0)[l]{\strut{} 1}}%
    }%
    \gplgaddtomacro\gplbacktext{%
      \csname LTb\endcsname
      \put(372,2808){\makebox(0,0)[r]{\strut{}}}%
      \put(504,2588){\makebox(0,0){\strut{}}}%
      \put(857,2588){\makebox(0,0){\strut{}}}%
      \put(1210,2588){\makebox(0,0){\strut{}}}%
      \put(1562,2588){\makebox(0,0){\strut{}}}%
      \put(1915,2588){\makebox(0,0){\strut{}}}%
    }%
    \gplgaddtomacro\gplfronttext{%
      \csname LTb\endcsname
      \put(152,3347){\rotatebox{-270}{\makebox(0,0){\strut{}$\chi=30$\degr}}}%
      \put(1209,4217){\makebox(0,0){\strut{}$\zeta=18$\degr}}%
    }%
    \gplgaddtomacro\gplbacktext{%
      \csname LTb\endcsname
      \put(1884,2808){\makebox(0,0)[r]{\strut{}}}%
      \put(2016,2588){\makebox(0,0){\strut{}}}%
      \put(2369,2588){\makebox(0,0){\strut{}}}%
      \put(2721,2588){\makebox(0,0){\strut{}}}%
      \put(3074,2588){\makebox(0,0){\strut{}}}%
      \put(3426,2588){\makebox(0,0){\strut{}}}%
    }%
    \gplgaddtomacro\gplfronttext{%
      \csname LTb\endcsname
      \put(2721,4217){\makebox(0,0){\strut{}$\zeta=36$\degr}}%
    }%
    \gplgaddtomacro\gplbacktext{%
      \csname LTb\endcsname
      \put(3396,2808){\makebox(0,0)[r]{\strut{}}}%
      \put(3528,2588){\makebox(0,0){\strut{}}}%
      \put(3881,2588){\makebox(0,0){\strut{}}}%
      \put(4233,2588){\makebox(0,0){\strut{}}}%
      \put(4586,2588){\makebox(0,0){\strut{}}}%
      \put(4938,2588){\makebox(0,0){\strut{}}}%
    }%
    \gplgaddtomacro\gplfronttext{%
      \csname LTb\endcsname
      \put(4233,4217){\makebox(0,0){\strut{}$\zeta=54$\degr}}%
    }%
    \gplgaddtomacro\gplbacktext{%
      \csname LTb\endcsname
      \put(4908,2808){\makebox(0,0)[r]{\strut{}}}%
      \put(5040,2588){\makebox(0,0){\strut{}}}%
      \put(5393,2588){\makebox(0,0){\strut{}}}%
      \put(5745,2588){\makebox(0,0){\strut{}}}%
      \put(6098,2588){\makebox(0,0){\strut{}}}%
      \put(6450,2588){\makebox(0,0){\strut{}}}%
    }%
    \gplgaddtomacro\gplfronttext{%
      \csname LTb\endcsname
      \put(5745,4217){\makebox(0,0){\strut{}$\zeta=72$\degr}}%
    }%
    \gplgaddtomacro\gplbacktext{%
      \csname LTb\endcsname
      \put(6420,2808){\makebox(0,0)[r]{\strut{}}}%
      \put(6552,2588){\makebox(0,0){\strut{}}}%
      \put(6905,2588){\makebox(0,0){\strut{}}}%
      \put(7257,2588){\makebox(0,0){\strut{}}}%
      \put(7610,2588){\makebox(0,0){\strut{}}}%
      \put(7962,2588){\makebox(0,0){\strut{}}}%
    }%
    \gplgaddtomacro\gplfronttext{%
      \csname LTb\endcsname
      \put(7257,4217){\makebox(0,0){\strut{}$\zeta=90$\degr}}%
    }%
    \gplgaddtomacro\gplbacktext{%
    }%
    \gplgaddtomacro\gplfronttext{%
      \csname LTb\endcsname
      \put(8064,1239){\makebox(0,0){\strut{}}}%
      \put(8417,1239){\makebox(0,0){\strut{}}}%
      \put(8769,1239){\makebox(0,0){\strut{}}}%
      \put(9121,1239){\makebox(0,0){\strut{}}}%
      \put(9474,1239){\makebox(0,0){\strut{}}}%
      \put(7849,1513){\makebox(0,0)[r]{\strut{}}}%
      \put(7849,1693){\makebox(0,0)[r]{\strut{}}}%
      \put(7849,1873){\makebox(0,0)[r]{\strut{}}}%
      \put(7849,2052){\makebox(0,0)[r]{\strut{}}}%
      \put(7849,2231){\makebox(0,0)[r]{\strut{}}}%
      \put(7849,2411){\makebox(0,0)[r]{\strut{}}}%
      \put(9712,1513){\makebox(0,0)[l]{\strut{} 0}}%
      \put(9712,1647){\makebox(0,0)[l]{\strut{}0.5}}%
      \put(9712,1782){\makebox(0,0)[l]{\strut{} 1}}%
      \put(9712,1917){\makebox(0,0)[l]{\strut{} 2}}%
      \put(9712,2052){\makebox(0,0)[l]{\strut{} 2}}%
      \put(9712,2186){\makebox(0,0)[l]{\strut{} 2}}%
      \put(9712,2321){\makebox(0,0)[l]{\strut{} 3}}%
      \put(9712,2456){\makebox(0,0)[l]{\strut{} 4}}%
      \put(9712,2591){\makebox(0,0)[l]{\strut{} 4}}%
    }%
    \gplgaddtomacro\gplbacktext{%
      \csname LTb\endcsname
      \put(372,1512){\makebox(0,0)[r]{\strut{}}}%
      \put(504,1292){\makebox(0,0){\strut{}}}%
      \put(857,1292){\makebox(0,0){\strut{}}}%
      \put(1210,1292){\makebox(0,0){\strut{}}}%
      \put(1562,1292){\makebox(0,0){\strut{}}}%
      \put(1915,1292){\makebox(0,0){\strut{}}}%
    }%
    \gplgaddtomacro\gplfronttext{%
      \csname LTb\endcsname
      \put(152,2051){\rotatebox{-270}{\makebox(0,0){\strut{}$\chi=60$\degr}}}%
    }%
    \gplgaddtomacro\gplbacktext{%
      \csname LTb\endcsname
      \put(1884,1512){\makebox(0,0)[r]{\strut{}}}%
      \put(2016,1292){\makebox(0,0){\strut{}}}%
      \put(2369,1292){\makebox(0,0){\strut{}}}%
      \put(2721,1292){\makebox(0,0){\strut{}}}%
      \put(3074,1292){\makebox(0,0){\strut{}}}%
      \put(3426,1292){\makebox(0,0){\strut{}}}%
    }%
    \gplgaddtomacro\gplfronttext{%
    }%
    \gplgaddtomacro\gplbacktext{%
      \csname LTb\endcsname
      \put(3396,1512){\makebox(0,0)[r]{\strut{}}}%
      \put(3528,1292){\makebox(0,0){\strut{}}}%
      \put(3881,1292){\makebox(0,0){\strut{}}}%
      \put(4233,1292){\makebox(0,0){\strut{}}}%
      \put(4586,1292){\makebox(0,0){\strut{}}}%
      \put(4938,1292){\makebox(0,0){\strut{}}}%
    }%
    \gplgaddtomacro\gplfronttext{%
    }%
    \gplgaddtomacro\gplbacktext{%
      \csname LTb\endcsname
      \put(4908,1512){\makebox(0,0)[r]{\strut{}}}%
      \put(5040,1292){\makebox(0,0){\strut{}}}%
      \put(5393,1292){\makebox(0,0){\strut{}}}%
      \put(5745,1292){\makebox(0,0){\strut{}}}%
      \put(6098,1292){\makebox(0,0){\strut{}}}%
      \put(6450,1292){\makebox(0,0){\strut{}}}%
    }%
    \gplgaddtomacro\gplfronttext{%
    }%
    \gplgaddtomacro\gplbacktext{%
      \csname LTb\endcsname
      \put(6420,1512){\makebox(0,0)[r]{\strut{}}}%
      \put(6552,1292){\makebox(0,0){\strut{}}}%
      \put(6905,1292){\makebox(0,0){\strut{}}}%
      \put(7257,1292){\makebox(0,0){\strut{}}}%
      \put(7610,1292){\makebox(0,0){\strut{}}}%
      \put(7962,1292){\makebox(0,0){\strut{}}}%
    }%
    \gplgaddtomacro\gplfronttext{%
    }%
    \gplgaddtomacro\gplbacktext{%
    }%
    \gplgaddtomacro\gplfronttext{%
      \csname LTb\endcsname
      \put(8064,-57){\makebox(0,0){\strut{}0}}%
      \put(8417,-57){\makebox(0,0){\strut{}0.25}}%
      \put(8769,-57){\makebox(0,0){\strut{}0.5}}%
      \put(9121,-57){\makebox(0,0){\strut{}0.75}}%
      \put(9474,-57){\makebox(0,0){\strut{}1}}%
      \put(8769,-387){\makebox(0,0){\strut{}phase ($\phi$)}}%
      \put(7849,217){\makebox(0,0)[r]{\strut{}}}%
      \put(7849,397){\makebox(0,0)[r]{\strut{}}}%
      \put(7849,577){\makebox(0,0)[r]{\strut{}}}%
      \put(7849,756){\makebox(0,0)[r]{\strut{}}}%
      \put(7849,935){\makebox(0,0)[r]{\strut{}}}%
      \put(7849,1115){\makebox(0,0)[r]{\strut{}}}%
      \put(9712,217){\makebox(0,0)[l]{\strut{} 0}}%
      \put(9712,351){\makebox(0,0)[l]{\strut{}0.5}}%
      \put(9712,486){\makebox(0,0)[l]{\strut{} 1}}%
      \put(9712,621){\makebox(0,0)[l]{\strut{} 2}}%
      \put(9712,756){\makebox(0,0)[l]{\strut{} 2}}%
      \put(9712,890){\makebox(0,0)[l]{\strut{} 2}}%
      \put(9712,1025){\makebox(0,0)[l]{\strut{} 3}}%
      \put(9712,1160){\makebox(0,0)[l]{\strut{} 4}}%
      \put(9712,1295){\makebox(0,0)[l]{\strut{} 4}}%
    }%
    \gplgaddtomacro\gplbacktext{%
      \csname LTb\endcsname
      \put(372,216){\makebox(0,0)[r]{\strut{}}}%
      \put(504,-4){\makebox(0,0){\strut{}0}}%
      \put(857,-4){\makebox(0,0){\strut{}0.25}}%
      \put(1210,-4){\makebox(0,0){\strut{}0.5}}%
      \put(1562,-4){\makebox(0,0){\strut{}0.75}}%
      \put(1915,-4){\makebox(0,0){\strut{}1}}%
    }%
    \gplgaddtomacro\gplfronttext{%
      \csname LTb\endcsname
      \put(152,755){\rotatebox{-270}{\makebox(0,0){\strut{}$\chi=90$\degr}}}%
      \put(1209,-334){\makebox(0,0){\strut{}phase ($\phi$)}}%
    }%
    \gplgaddtomacro\gplbacktext{%
      \csname LTb\endcsname
      \put(1884,216){\makebox(0,0)[r]{\strut{}}}%
      \put(2016,-4){\makebox(0,0){\strut{}0}}%
      \put(2369,-4){\makebox(0,0){\strut{}0.25}}%
      \put(2721,-4){\makebox(0,0){\strut{}0.5}}%
      \put(3074,-4){\makebox(0,0){\strut{}0.75}}%
      \put(3426,-4){\makebox(0,0){\strut{}1}}%
    }%
    \gplgaddtomacro\gplfronttext{%
      \csname LTb\endcsname
      \put(2721,-334){\makebox(0,0){\strut{}phase ($\phi$)}}%
    }%
    \gplgaddtomacro\gplbacktext{%
      \csname LTb\endcsname
      \put(3396,216){\makebox(0,0)[r]{\strut{}}}%
      \put(3528,-4){\makebox(0,0){\strut{}0}}%
      \put(3881,-4){\makebox(0,0){\strut{}0.25}}%
      \put(4233,-4){\makebox(0,0){\strut{}0.5}}%
      \put(4586,-4){\makebox(0,0){\strut{}0.75}}%
      \put(4938,-4){\makebox(0,0){\strut{}1}}%
    }%
    \gplgaddtomacro\gplfronttext{%
      \csname LTb\endcsname
      \put(4233,-334){\makebox(0,0){\strut{}phase ($\phi$)}}%
    }%
    \gplgaddtomacro\gplbacktext{%
      \csname LTb\endcsname
      \put(4908,216){\makebox(0,0)[r]{\strut{}}}%
      \put(5040,-4){\makebox(0,0){\strut{}0}}%
      \put(5393,-4){\makebox(0,0){\strut{}0.25}}%
      \put(5745,-4){\makebox(0,0){\strut{}0.5}}%
      \put(6098,-4){\makebox(0,0){\strut{}0.75}}%
      \put(6450,-4){\makebox(0,0){\strut{}1}}%
    }%
    \gplgaddtomacro\gplfronttext{%
      \csname LTb\endcsname
      \put(5745,-334){\makebox(0,0){\strut{}phase ($\phi$)}}%
    }%
    \gplgaddtomacro\gplbacktext{%
      \csname LTb\endcsname
      \put(6420,216){\makebox(0,0)[r]{\strut{}}}%
      \put(6552,-4){\makebox(0,0){\strut{}0}}%
      \put(6905,-4){\makebox(0,0){\strut{}0.25}}%
      \put(7257,-4){\makebox(0,0){\strut{}0.5}}%
      \put(7610,-4){\makebox(0,0){\strut{}0.75}}%
      \put(7962,-4){\makebox(0,0){\strut{}1}}%
    }%
    \gplgaddtomacro\gplfronttext{%
      \csname LTb\endcsname
      \put(7257,-334){\makebox(0,0){\strut{}phase ($\phi$)}}%
    }%
    \gplbacktext
    \put(0,0){\includegraphics{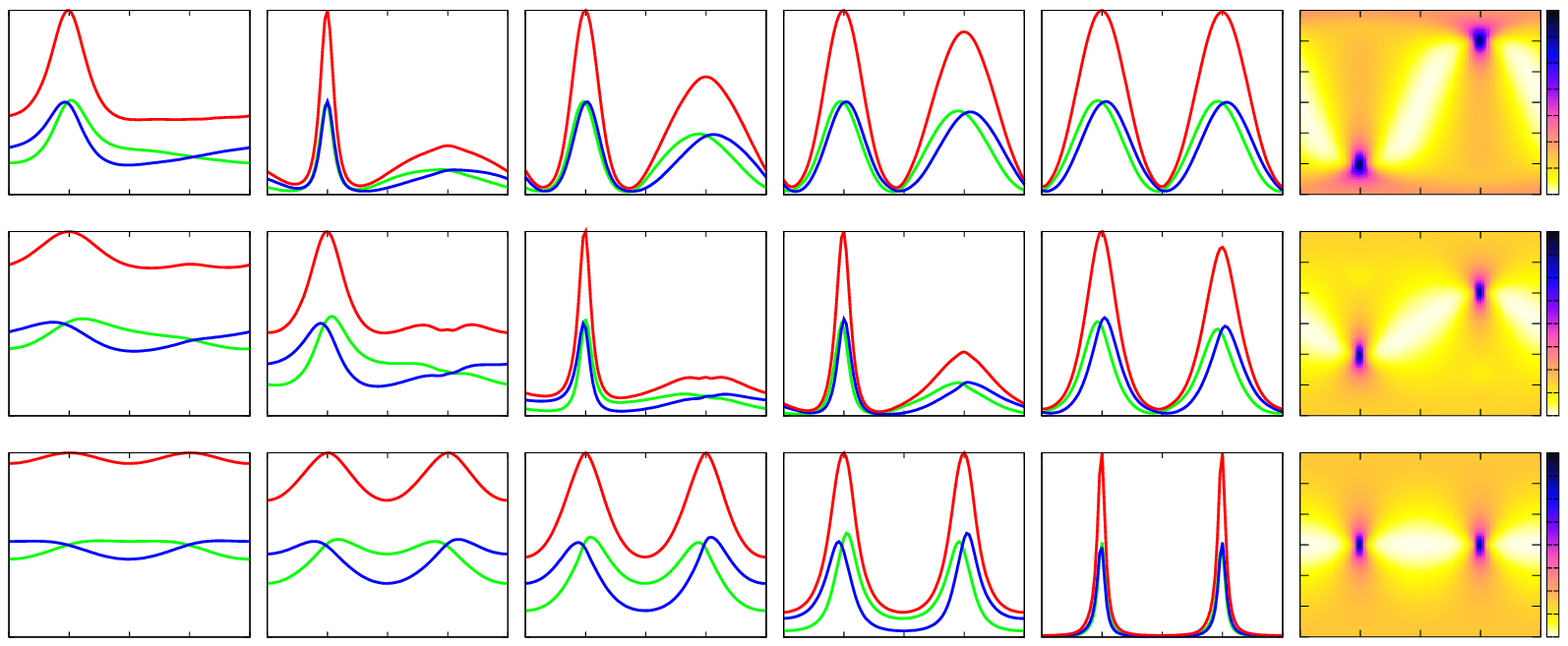}}%
    \gplfronttext
  \end{picture}%
\endgroup

%% file: skymaps_cl_r005_b-6_ri0.2_ro1_g10_d3_om8.tex
\begingroup
  \makeatletter
  \providecommand\color[2][]{%
    \GenericError{(gnuplot) \space\space\space\@spaces}{%
      Package color not loaded in conjunction with
      terminal option `colourtext'%
    }{See the gnuplot documentation for explanation.%
    }{Either use 'blacktext' in gnuplot or load the package
      color.sty in LaTeX.}%
    \renewcommand\color[2][]{}%
  }%
  \providecommand\includegraphics[2][]{%
    \GenericError{(gnuplot) \space\space\space\@spaces}{%
      Package graphicx or graphics not loaded%
    }{See the gnuplot documentation for explanation.%
    }{The gnuplot epslatex terminal needs graphicx.sty or graphics.sty.}%
    \renewcommand\includegraphics[2][]{}%
  }%
  \providecommand\rotatebox[2]{#2}%
  \@ifundefined{ifGPcolor}{%
    \newif\ifGPcolor
    \GPcolortrue
  }{}%
  \@ifundefined{ifGPblacktext}{%
    \newif\ifGPblacktext
    \GPblacktextfalse
  }{}%
  \let\gplgaddtomacro\g@addto@macro
  \gdef\gplbacktext{}%
  \gdef\gplfronttext{}%
  \makeatother
  \ifGPblacktext
    \def\colorrgb#1{}%
    \def\colorgray#1{}%
  \else
    \ifGPcolor
      \def\colorrgb#1{\color[rgb]{#1}}%
      \def\colorgray#1{\color[gray]{#1}}%
      \expandafter\def\csname LTw\endcsname{\color{white}}%
      \expandafter\def\csname LTb\endcsname{\color{black}}%
      \expandafter\def\csname LTa\endcsname{\color{black}}%
      \expandafter\def\csname LT0\endcsname{\color[rgb]{1,0,0}}%
      \expandafter\def\csname LT1\endcsname{\color[rgb]{0,1,0}}%
      \expandafter\def\csname LT2\endcsname{\color[rgb]{0,0,1}}%
      \expandafter\def\csname LT3\endcsname{\color[rgb]{1,0,1}}%
      \expandafter\def\csname LT4\endcsname{\color[rgb]{0,1,1}}%
      \expandafter\def\csname LT5\endcsname{\color[rgb]{1,1,0}}%
      \expandafter\def\csname LT6\endcsname{\color[rgb]{0,0,0}}%
      \expandafter\def\csname LT7\endcsname{\color[rgb]{1,0.3,0}}%
      \expandafter\def\csname LT8\endcsname{\color[rgb]{0.5,0.5,0.5}}%
    \else
      \def\colorrgb#1{\color{black}}%
      \def\colorgray#1{\color[gray]{#1}}%
      \expandafter\def\csname LTw\endcsname{\color{white}}%
      \expandafter\def\csname LTb\endcsname{\color{black}}%
      \expandafter\def\csname LTa\endcsname{\color{black}}%
      \expandafter\def\csname LT0\endcsname{\color{black}}%
      \expandafter\def\csname LT1\endcsname{\color{black}}%
      \expandafter\def\csname LT2\endcsname{\color{black}}%
      \expandafter\def\csname LT3\endcsname{\color{black}}%
      \expandafter\def\csname LT4\endcsname{\color{black}}%
      \expandafter\def\csname LT5\endcsname{\color{black}}%
      \expandafter\def\csname LT6\endcsname{\color{black}}%
      \expandafter\def\csname LT7\endcsname{\color{black}}%
      \expandafter\def\csname LT8\endcsname{\color{black}}%
    \fi
  \fi
    \setlength{\unitlength}{0.0500bp}%
    \ifx\gptboxheight\undefined%
      \newlength{\gptboxheight}%
      \newlength{\gptboxwidth}%
      \newsavebox{\gptboxtext}%
    \fi%
    \setlength{\fboxrule}{0.5pt}%
    \setlength{\fboxsep}{1pt}%
\begin{picture}(10080.00,4320.00)%
    \gplgaddtomacro\gplbacktext{%
    }%
    \gplgaddtomacro\gplfronttext{%
      \csname LTb\endcsname
      \put(8064,2535){\makebox(0,0){\strut{}}}%
      \put(8417,2535){\makebox(0,0){\strut{}}}%
      \put(8769,2535){\makebox(0,0){\strut{}}}%
      \put(9121,2535){\makebox(0,0){\strut{}}}%
      \put(9474,2535){\makebox(0,0){\strut{}}}%
      \put(7849,2809){\makebox(0,0)[r]{\strut{}}}%
      \put(7849,2989){\makebox(0,0)[r]{\strut{}}}%
      \put(7849,3169){\makebox(0,0)[r]{\strut{}}}%
      \put(7849,3348){\makebox(0,0)[r]{\strut{}}}%
      \put(7849,3527){\makebox(0,0)[r]{\strut{}}}%
      \put(7849,3707){\makebox(0,0)[r]{\strut{}}}%
      \put(9712,2809){\makebox(0,0)[l]{\strut{} 0}}%
      \put(9712,2928){\makebox(0,0)[l]{\strut{}5e+02}}%
      \put(9712,3048){\makebox(0,0)[l]{\strut{}1e+03}}%
      \put(9712,3168){\makebox(0,0)[l]{\strut{}2e+03}}%
      \put(9712,3288){\makebox(0,0)[l]{\strut{}2e+03}}%
      \put(9712,3407){\makebox(0,0)[l]{\strut{}2e+03}}%
      \put(9712,3527){\makebox(0,0)[l]{\strut{}3e+03}}%
      \put(9712,3647){\makebox(0,0)[l]{\strut{}4e+03}}%
      \put(9712,3767){\makebox(0,0)[l]{\strut{}4e+03}}%
      \put(9712,3887){\makebox(0,0)[l]{\strut{}4e+03}}%
    }%
    \gplgaddtomacro\gplbacktext{%
      \csname LTb\endcsname
      \put(372,2808){\makebox(0,0)[r]{\strut{}}}%
      \put(372,2816){\makebox(0,0)[r]{\strut{}}}%
      \put(372,2824){\makebox(0,0)[r]{\strut{}}}%
      \put(372,2832){\makebox(0,0)[r]{\strut{}}}%
      \put(372,2839){\makebox(0,0)[r]{\strut{}}}%
      \put(372,2847){\makebox(0,0)[r]{\strut{}}}%
      \put(504,2588){\makebox(0,0){\strut{}}}%
      \put(857,2588){\makebox(0,0){\strut{}}}%
      \put(1210,2588){\makebox(0,0){\strut{}}}%
      \put(1562,2588){\makebox(0,0){\strut{}}}%
      \put(1915,2588){\makebox(0,0){\strut{}}}%
    }%
    \gplgaddtomacro\gplfronttext{%
      \csname LTb\endcsname
      \put(152,3347){\rotatebox{-270}{\makebox(0,0){\strut{}$\chi=30$\degr}}}%
      \put(1209,4217){\makebox(0,0){\strut{}$\zeta=18$\degr}}%
    }%
    \gplgaddtomacro\gplbacktext{%
      \csname LTb\endcsname
      \put(1884,2808){\makebox(0,0)[r]{\strut{}}}%
      \put(1884,2813){\makebox(0,0)[r]{\strut{}}}%
      \put(1884,2817){\makebox(0,0)[r]{\strut{}}}%
      \put(1884,2822){\makebox(0,0)[r]{\strut{}}}%
      \put(1884,2827){\makebox(0,0)[r]{\strut{}}}%
      \put(1884,2831){\makebox(0,0)[r]{\strut{}}}%
      \put(2016,2588){\makebox(0,0){\strut{}}}%
      \put(2369,2588){\makebox(0,0){\strut{}}}%
      \put(2721,2588){\makebox(0,0){\strut{}}}%
      \put(3074,2588){\makebox(0,0){\strut{}}}%
      \put(3426,2588){\makebox(0,0){\strut{}}}%
    }%
    \gplgaddtomacro\gplfronttext{%
      \csname LTb\endcsname
      \put(2721,4217){\makebox(0,0){\strut{}$\zeta=36$\degr}}%
    }%
    \gplgaddtomacro\gplbacktext{%
      \csname LTb\endcsname
      \put(3396,2808){\makebox(0,0)[r]{\strut{}}}%
      \put(3396,2819){\makebox(0,0)[r]{\strut{}}}%
      \put(3396,2829){\makebox(0,0)[r]{\strut{}}}%
      \put(3396,2840){\makebox(0,0)[r]{\strut{}}}%
      \put(3396,2851){\makebox(0,0)[r]{\strut{}}}%
      \put(3396,2862){\makebox(0,0)[r]{\strut{}}}%
      \put(3528,2588){\makebox(0,0){\strut{}}}%
      \put(3881,2588){\makebox(0,0){\strut{}}}%
      \put(4233,2588){\makebox(0,0){\strut{}}}%
      \put(4586,2588){\makebox(0,0){\strut{}}}%
      \put(4938,2588){\makebox(0,0){\strut{}}}%
    }%
    \gplgaddtomacro\gplfronttext{%
      \csname LTb\endcsname
      \put(4233,4217){\makebox(0,0){\strut{}$\zeta=54$\degr}}%
    }%
    \gplgaddtomacro\gplbacktext{%
      \csname LTb\endcsname
      \put(4908,2808){\makebox(0,0)[r]{\strut{}}}%
      \put(4908,2823){\makebox(0,0)[r]{\strut{}}}%
      \put(4908,2838){\makebox(0,0)[r]{\strut{}}}%
      \put(4908,2853){\makebox(0,0)[r]{\strut{}}}%
      \put(4908,2868){\makebox(0,0)[r]{\strut{}}}%
      \put(4908,2883){\makebox(0,0)[r]{\strut{}}}%
      \put(5040,2588){\makebox(0,0){\strut{}}}%
      \put(5393,2588){\makebox(0,0){\strut{}}}%
      \put(5745,2588){\makebox(0,0){\strut{}}}%
      \put(6098,2588){\makebox(0,0){\strut{}}}%
      \put(6450,2588){\makebox(0,0){\strut{}}}%
    }%
    \gplgaddtomacro\gplfronttext{%
      \csname LTb\endcsname
      \put(5745,4217){\makebox(0,0){\strut{}$\zeta=72$\degr}}%
    }%
    \gplgaddtomacro\gplbacktext{%
      \csname LTb\endcsname
      \put(6420,2808){\makebox(0,0)[r]{\strut{}}}%
      \put(6420,2825){\makebox(0,0)[r]{\strut{}}}%
      \put(6420,2843){\makebox(0,0)[r]{\strut{}}}%
      \put(6420,2860){\makebox(0,0)[r]{\strut{}}}%
      \put(6420,2877){\makebox(0,0)[r]{\strut{}}}%
      \put(6420,2895){\makebox(0,0)[r]{\strut{}}}%
      \put(6552,2588){\makebox(0,0){\strut{}}}%
      \put(6905,2588){\makebox(0,0){\strut{}}}%
      \put(7257,2588){\makebox(0,0){\strut{}}}%
      \put(7610,2588){\makebox(0,0){\strut{}}}%
      \put(7962,2588){\makebox(0,0){\strut{}}}%
    }%
    \gplgaddtomacro\gplfronttext{%
      \csname LTb\endcsname
      \put(7257,4217){\makebox(0,0){\strut{}$\zeta=90$\degr}}%
    }%
    \gplgaddtomacro\gplbacktext{%
    }%
    \gplgaddtomacro\gplfronttext{%
      \csname LTb\endcsname
      \put(8064,1239){\makebox(0,0){\strut{}}}%
      \put(8417,1239){\makebox(0,0){\strut{}}}%
      \put(8769,1239){\makebox(0,0){\strut{}}}%
      \put(9121,1239){\makebox(0,0){\strut{}}}%
      \put(9474,1239){\makebox(0,0){\strut{}}}%
      \put(7849,1513){\makebox(0,0)[r]{\strut{}}}%
      \put(7849,1693){\makebox(0,0)[r]{\strut{}}}%
      \put(7849,1873){\makebox(0,0)[r]{\strut{}}}%
      \put(7849,2052){\makebox(0,0)[r]{\strut{}}}%
      \put(7849,2231){\makebox(0,0)[r]{\strut{}}}%
      \put(7849,2411){\makebox(0,0)[r]{\strut{}}}%
      \put(9712,1513){\makebox(0,0)[l]{\strut{} 0}}%
      \put(9712,1692){\makebox(0,0)[l]{\strut{}2e+03}}%
      \put(9712,1872){\makebox(0,0)[l]{\strut{}4e+03}}%
      \put(9712,2052){\makebox(0,0)[l]{\strut{}6e+03}}%
      \put(9712,2231){\makebox(0,0)[l]{\strut{}8e+03}}%
      \put(9712,2411){\makebox(0,0)[l]{\strut{}1e+04}}%
      \put(9712,2591){\makebox(0,0)[l]{\strut{}1e+04}}%
    }%
    \gplgaddtomacro\gplbacktext{%
      \csname LTb\endcsname
      \put(372,1512){\makebox(0,0)[r]{\strut{}}}%
      \put(372,1519){\makebox(0,0)[r]{\strut{}}}%
      \put(372,1527){\makebox(0,0)[r]{\strut{}}}%
      \put(372,1534){\makebox(0,0)[r]{\strut{}}}%
      \put(372,1542){\makebox(0,0)[r]{\strut{}}}%
      \put(372,1549){\makebox(0,0)[r]{\strut{}}}%
      \put(504,1292){\makebox(0,0){\strut{}}}%
      \put(857,1292){\makebox(0,0){\strut{}}}%
      \put(1210,1292){\makebox(0,0){\strut{}}}%
      \put(1562,1292){\makebox(0,0){\strut{}}}%
      \put(1915,1292){\makebox(0,0){\strut{}}}%
    }%
    \gplgaddtomacro\gplfronttext{%
      \csname LTb\endcsname
      \put(152,2051){\rotatebox{-270}{\makebox(0,0){\strut{}$\chi=60$\degr}}}%
    }%
    \gplgaddtomacro\gplbacktext{%
      \csname LTb\endcsname
      \put(1884,1512){\makebox(0,0)[r]{\strut{}}}%
      \put(1884,1517){\makebox(0,0)[r]{\strut{}}}%
      \put(1884,1522){\makebox(0,0)[r]{\strut{}}}%
      \put(1884,1527){\makebox(0,0)[r]{\strut{}}}%
      \put(1884,1532){\makebox(0,0)[r]{\strut{}}}%
      \put(1884,1537){\makebox(0,0)[r]{\strut{}}}%
      \put(2016,1292){\makebox(0,0){\strut{}}}%
      \put(2369,1292){\makebox(0,0){\strut{}}}%
      \put(2721,1292){\makebox(0,0){\strut{}}}%
      \put(3074,1292){\makebox(0,0){\strut{}}}%
      \put(3426,1292){\makebox(0,0){\strut{}}}%
    }%
    \gplgaddtomacro\gplfronttext{%
    }%
    \gplgaddtomacro\gplbacktext{%
      \csname LTb\endcsname
      \put(3396,1512){\makebox(0,0)[r]{\strut{}}}%
      \put(3396,1514){\makebox(0,0)[r]{\strut{}}}%
      \put(3396,1516){\makebox(0,0)[r]{\strut{}}}%
      \put(3396,1518){\makebox(0,0)[r]{\strut{}}}%
      \put(3396,1521){\makebox(0,0)[r]{\strut{}}}%
      \put(3396,1523){\makebox(0,0)[r]{\strut{}}}%
      \put(3528,1292){\makebox(0,0){\strut{}}}%
      \put(3881,1292){\makebox(0,0){\strut{}}}%
      \put(4233,1292){\makebox(0,0){\strut{}}}%
      \put(4586,1292){\makebox(0,0){\strut{}}}%
      \put(4938,1292){\makebox(0,0){\strut{}}}%
    }%
    \gplgaddtomacro\gplfronttext{%
    }%
    \gplgaddtomacro\gplbacktext{%
      \csname LTb\endcsname
      \put(4908,1512){\makebox(0,0)[r]{\strut{}}}%
      \put(4908,1515){\makebox(0,0)[r]{\strut{}}}%
      \put(4908,1517){\makebox(0,0)[r]{\strut{}}}%
      \put(4908,1520){\makebox(0,0)[r]{\strut{}}}%
      \put(4908,1523){\makebox(0,0)[r]{\strut{}}}%
      \put(4908,1525){\makebox(0,0)[r]{\strut{}}}%
      \put(5040,1292){\makebox(0,0){\strut{}}}%
      \put(5393,1292){\makebox(0,0){\strut{}}}%
      \put(5745,1292){\makebox(0,0){\strut{}}}%
      \put(6098,1292){\makebox(0,0){\strut{}}}%
      \put(6450,1292){\makebox(0,0){\strut{}}}%
    }%
    \gplgaddtomacro\gplfronttext{%
    }%
    \gplgaddtomacro\gplbacktext{%
      \csname LTb\endcsname
      \put(6420,1512){\makebox(0,0)[r]{\strut{}}}%
      \put(6420,1518){\makebox(0,0)[r]{\strut{}}}%
      \put(6420,1523){\makebox(0,0)[r]{\strut{}}}%
      \put(6420,1529){\makebox(0,0)[r]{\strut{}}}%
      \put(6420,1534){\makebox(0,0)[r]{\strut{}}}%
      \put(6420,1540){\makebox(0,0)[r]{\strut{}}}%
      \put(6552,1292){\makebox(0,0){\strut{}}}%
      \put(6905,1292){\makebox(0,0){\strut{}}}%
      \put(7257,1292){\makebox(0,0){\strut{}}}%
      \put(7610,1292){\makebox(0,0){\strut{}}}%
      \put(7962,1292){\makebox(0,0){\strut{}}}%
    }%
    \gplgaddtomacro\gplfronttext{%
    }%
    \gplgaddtomacro\gplbacktext{%
    }%
    \gplgaddtomacro\gplfronttext{%
      \csname LTb\endcsname
      \put(8064,-57){\makebox(0,0){\strut{}0}}%
      \put(8417,-57){\makebox(0,0){\strut{}0.25}}%
      \put(8769,-57){\makebox(0,0){\strut{}0.5}}%
      \put(9121,-57){\makebox(0,0){\strut{}0.75}}%
      \put(9474,-57){\makebox(0,0){\strut{}1}}%
      \put(8769,-387){\makebox(0,0){\strut{}phase ($\phi$)}}%
      \put(7849,217){\makebox(0,0)[r]{\strut{}}}%
      \put(7849,397){\makebox(0,0)[r]{\strut{}}}%
      \put(7849,577){\makebox(0,0)[r]{\strut{}}}%
      \put(7849,756){\makebox(0,0)[r]{\strut{}}}%
      \put(7849,935){\makebox(0,0)[r]{\strut{}}}%
      \put(7849,1115){\makebox(0,0)[r]{\strut{}}}%
      \put(9712,217){\makebox(0,0)[l]{\strut{} 0}}%
      \put(9712,396){\makebox(0,0)[l]{\strut{}2e+03}}%
      \put(9712,576){\makebox(0,0)[l]{\strut{}4e+03}}%
      \put(9712,756){\makebox(0,0)[l]{\strut{}6e+03}}%
      \put(9712,935){\makebox(0,0)[l]{\strut{}8e+03}}%
      \put(9712,1115){\makebox(0,0)[l]{\strut{}1e+04}}%
      \put(9712,1295){\makebox(0,0)[l]{\strut{}1e+04}}%
    }%
    \gplgaddtomacro\gplbacktext{%
      \csname LTb\endcsname
      \put(372,216){\makebox(0,0)[r]{\strut{}}}%
      \put(372,223){\makebox(0,0)[r]{\strut{}}}%
      \put(372,231){\makebox(0,0)[r]{\strut{}}}%
      \put(372,238){\makebox(0,0)[r]{\strut{}}}%
      \put(372,246){\makebox(0,0)[r]{\strut{}}}%
      \put(372,253){\makebox(0,0)[r]{\strut{}}}%
      \put(504,-4){\makebox(0,0){\strut{}0}}%
      \put(857,-4){\makebox(0,0){\strut{}0.25}}%
      \put(1210,-4){\makebox(0,0){\strut{}0.5}}%
      \put(1562,-4){\makebox(0,0){\strut{}0.75}}%
      \put(1915,-4){\makebox(0,0){\strut{}1}}%
    }%
    \gplgaddtomacro\gplfronttext{%
      \csname LTb\endcsname
      \put(152,755){\rotatebox{-270}{\makebox(0,0){\strut{}$\chi=90$\degr}}}%
      \put(1209,-334){\makebox(0,0){\strut{}phase ($\phi$)}}%
    }%
    \gplgaddtomacro\gplbacktext{%
      \csname LTb\endcsname
      \put(1884,216){\makebox(0,0)[r]{\strut{}}}%
      \put(1884,223){\makebox(0,0)[r]{\strut{}}}%
      \put(1884,230){\makebox(0,0)[r]{\strut{}}}%
      \put(1884,236){\makebox(0,0)[r]{\strut{}}}%
      \put(1884,243){\makebox(0,0)[r]{\strut{}}}%
      \put(1884,250){\makebox(0,0)[r]{\strut{}}}%
      \put(2016,-4){\makebox(0,0){\strut{}0}}%
      \put(2369,-4){\makebox(0,0){\strut{}0.25}}%
      \put(2721,-4){\makebox(0,0){\strut{}0.5}}%
      \put(3074,-4){\makebox(0,0){\strut{}0.75}}%
      \put(3426,-4){\makebox(0,0){\strut{}1}}%
    }%
    \gplgaddtomacro\gplfronttext{%
      \csname LTb\endcsname
      \put(2721,-334){\makebox(0,0){\strut{}phase ($\phi$)}}%
    }%
    \gplgaddtomacro\gplbacktext{%
      \csname LTb\endcsname
      \put(3396,216){\makebox(0,0)[r]{\strut{}}}%
      \put(3396,222){\makebox(0,0)[r]{\strut{}}}%
      \put(3396,227){\makebox(0,0)[r]{\strut{}}}%
      \put(3396,233){\makebox(0,0)[r]{\strut{}}}%
      \put(3396,239){\makebox(0,0)[r]{\strut{}}}%
      \put(3396,244){\makebox(0,0)[r]{\strut{}}}%
      \put(3528,-4){\makebox(0,0){\strut{}0}}%
      \put(3881,-4){\makebox(0,0){\strut{}0.25}}%
      \put(4233,-4){\makebox(0,0){\strut{}0.5}}%
      \put(4586,-4){\makebox(0,0){\strut{}0.75}}%
      \put(4938,-4){\makebox(0,0){\strut{}1}}%
    }%
    \gplgaddtomacro\gplfronttext{%
      \csname LTb\endcsname
      \put(4233,-334){\makebox(0,0){\strut{}phase ($\phi$)}}%
    }%
    \gplgaddtomacro\gplbacktext{%
      \csname LTb\endcsname
      \put(4908,216){\makebox(0,0)[r]{\strut{}}}%
      \put(4908,220){\makebox(0,0)[r]{\strut{}}}%
      \put(4908,224){\makebox(0,0)[r]{\strut{}}}%
      \put(4908,228){\makebox(0,0)[r]{\strut{}}}%
      \put(4908,232){\makebox(0,0)[r]{\strut{}}}%
      \put(4908,236){\makebox(0,0)[r]{\strut{}}}%
      \put(5040,-4){\makebox(0,0){\strut{}0}}%
      \put(5393,-4){\makebox(0,0){\strut{}0.25}}%
      \put(5745,-4){\makebox(0,0){\strut{}0.5}}%
      \put(6098,-4){\makebox(0,0){\strut{}0.75}}%
      \put(6450,-4){\makebox(0,0){\strut{}1}}%
    }%
    \gplgaddtomacro\gplfronttext{%
      \csname LTb\endcsname
      \put(5745,-334){\makebox(0,0){\strut{}phase ($\phi$)}}%
    }%
    \gplgaddtomacro\gplbacktext{%
      \csname LTb\endcsname
      \put(6420,216){\makebox(0,0)[r]{\strut{}}}%
      \put(6420,218){\makebox(0,0)[r]{\strut{}}}%
      \put(6420,219){\makebox(0,0)[r]{\strut{}}}%
      \put(6420,221){\makebox(0,0)[r]{\strut{}}}%
      \put(6420,222){\makebox(0,0)[r]{\strut{}}}%
      \put(6420,224){\makebox(0,0)[r]{\strut{}}}%
      \put(6552,-4){\makebox(0,0){\strut{}0}}%
      \put(6905,-4){\makebox(0,0){\strut{}0.25}}%
      \put(7257,-4){\makebox(0,0){\strut{}0.5}}%
      \put(7610,-4){\makebox(0,0){\strut{}0.75}}%
      \put(7962,-4){\makebox(0,0){\strut{}1}}%
    }%
    \gplgaddtomacro\gplfronttext{%
      \csname LTb\endcsname
      \put(7257,-334){\makebox(0,0){\strut{}phase ($\phi$)}}%
    }%
    \gplbacktext
    \put(0,0){\includegraphics{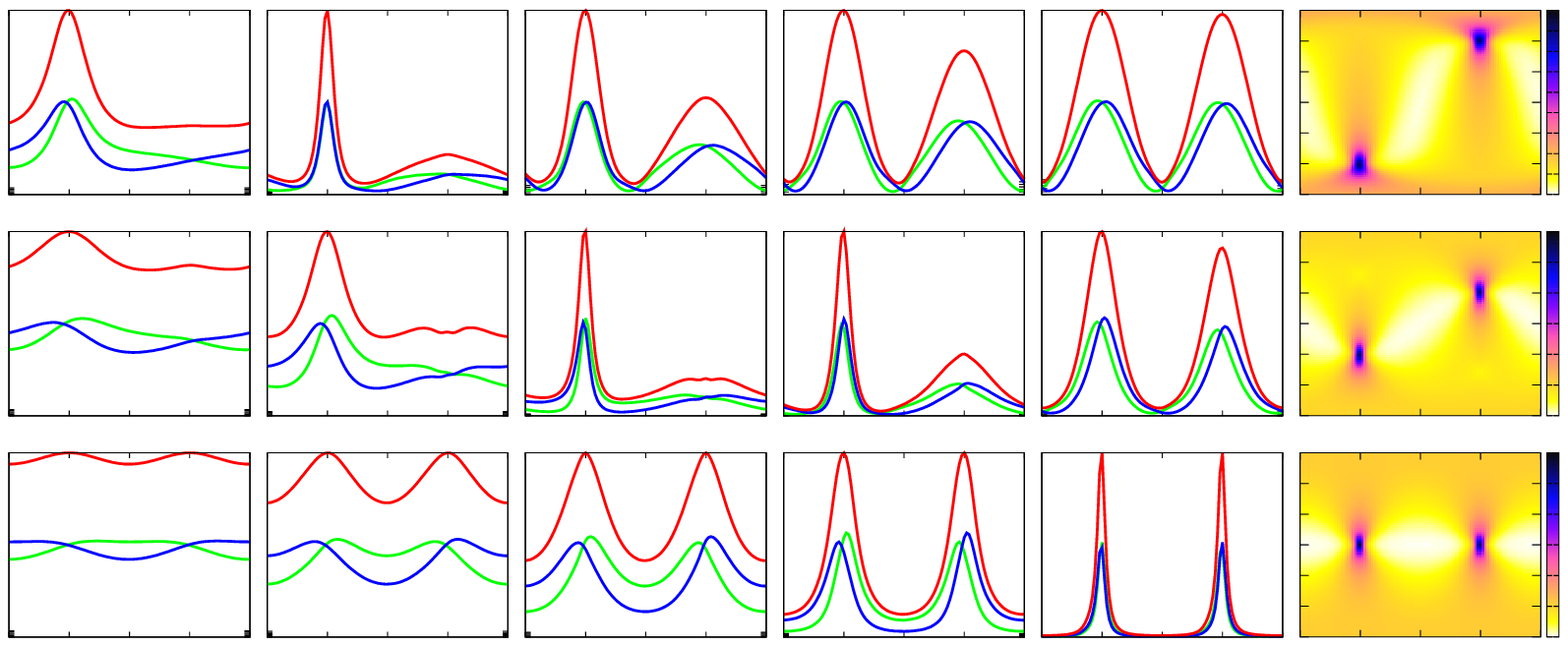}}%
    \gplfronttext
  \end{picture}%
\endgroup

%% file: skymaps_cl_r00025_b-3_ri1_ro5_g10_d3_om8.tex
\begingroup
  \makeatletter
  \providecommand\color[2][]{%
    \GenericError{(gnuplot) \space\space\space\@spaces}{%
      Package color not loaded in conjunction with
      terminal option `colourtext'%
    }{See the gnuplot documentation for explanation.%
    }{Either use 'blacktext' in gnuplot or load the package
      color.sty in LaTeX.}%
    \renewcommand\color[2][]{}%
  }%
  \providecommand\includegraphics[2][]{%
    \GenericError{(gnuplot) \space\space\space\@spaces}{%
      Package graphicx or graphics not loaded%
    }{See the gnuplot documentation for explanation.%
    }{The gnuplot epslatex terminal needs graphicx.sty or graphics.sty.}%
    \renewcommand\includegraphics[2][]{}%
  }%
  \providecommand\rotatebox[2]{#2}%
  \@ifundefined{ifGPcolor}{%
    \newif\ifGPcolor
    \GPcolortrue
  }{}%
  \@ifundefined{ifGPblacktext}{%
    \newif\ifGPblacktext
    \GPblacktextfalse
  }{}%
  \let\gplgaddtomacro\g@addto@macro
  \gdef\gplbacktext{}%
  \gdef\gplfronttext{}%
  \makeatother
  \ifGPblacktext
    \def\colorrgb#1{}%
    \def\colorgray#1{}%
  \else
    \ifGPcolor
      \def\colorrgb#1{\color[rgb]{#1}}%
      \def\colorgray#1{\color[gray]{#1}}%
      \expandafter\def\csname LTw\endcsname{\color{white}}%
      \expandafter\def\csname LTb\endcsname{\color{black}}%
      \expandafter\def\csname LTa\endcsname{\color{black}}%
      \expandafter\def\csname LT0\endcsname{\color[rgb]{1,0,0}}%
      \expandafter\def\csname LT1\endcsname{\color[rgb]{0,1,0}}%
      \expandafter\def\csname LT2\endcsname{\color[rgb]{0,0,1}}%
      \expandafter\def\csname LT3\endcsname{\color[rgb]{1,0,1}}%
      \expandafter\def\csname LT4\endcsname{\color[rgb]{0,1,1}}%
      \expandafter\def\csname LT5\endcsname{\color[rgb]{1,1,0}}%
      \expandafter\def\csname LT6\endcsname{\color[rgb]{0,0,0}}%
      \expandafter\def\csname LT7\endcsname{\color[rgb]{1,0.3,0}}%
      \expandafter\def\csname LT8\endcsname{\color[rgb]{0.5,0.5,0.5}}%
    \else
      \def\colorrgb#1{\color{black}}%
      \def\colorgray#1{\color[gray]{#1}}%
      \expandafter\def\csname LTw\endcsname{\color{white}}%
      \expandafter\def\csname LTb\endcsname{\color{black}}%
      \expandafter\def\csname LTa\endcsname{\color{black}}%
      \expandafter\def\csname LT0\endcsname{\color{black}}%
      \expandafter\def\csname LT1\endcsname{\color{black}}%
      \expandafter\def\csname LT2\endcsname{\color{black}}%
      \expandafter\def\csname LT3\endcsname{\color{black}}%
      \expandafter\def\csname LT4\endcsname{\color{black}}%
      \expandafter\def\csname LT5\endcsname{\color{black}}%
      \expandafter\def\csname LT6\endcsname{\color{black}}%
      \expandafter\def\csname LT7\endcsname{\color{black}}%
      \expandafter\def\csname LT8\endcsname{\color{black}}%
    \fi
  \fi
    \setlength{\unitlength}{0.0500bp}%
    \ifx\gptboxheight\undefined%
      \newlength{\gptboxheight}%
      \newlength{\gptboxwidth}%
      \newsavebox{\gptboxtext}%
    \fi%
    \setlength{\fboxrule}{0.5pt}%
    \setlength{\fboxsep}{1pt}%
\begin{picture}(10080.00,4320.00)%
    \gplgaddtomacro\gplbacktext{%
    }%
    \gplgaddtomacro\gplfronttext{%
      \csname LTb\endcsname
      \put(8064,2535){\makebox(0,0){\strut{}}}%
      \put(8417,2535){\makebox(0,0){\strut{}}}%
      \put(8769,2535){\makebox(0,0){\strut{}}}%
      \put(9121,2535){\makebox(0,0){\strut{}}}%
      \put(9474,2535){\makebox(0,0){\strut{}}}%
      \put(7849,2809){\makebox(0,0)[r]{\strut{}}}%
      \put(7849,2989){\makebox(0,0)[r]{\strut{}}}%
      \put(7849,3169){\makebox(0,0)[r]{\strut{}}}%
      \put(7849,3348){\makebox(0,0)[r]{\strut{}}}%
      \put(7849,3527){\makebox(0,0)[r]{\strut{}}}%
      \put(7849,3707){\makebox(0,0)[r]{\strut{}}}%
      \put(9712,2809){\makebox(0,0)[l]{\strut{} 0}}%
      \put(9712,2943){\makebox(0,0)[l]{\strut{}0.0002}}%
      \put(9712,3078){\makebox(0,0)[l]{\strut{}0.0004}}%
      \put(9712,3213){\makebox(0,0)[l]{\strut{}0.0006}}%
      \put(9712,3348){\makebox(0,0)[l]{\strut{}0.0008}}%
      \put(9712,3482){\makebox(0,0)[l]{\strut{}0.001}}%
      \put(9712,3617){\makebox(0,0)[l]{\strut{}0.001}}%
      \put(9712,3752){\makebox(0,0)[l]{\strut{}0.001}}%
      \put(9712,3887){\makebox(0,0)[l]{\strut{}0.002}}%
    }%
    \gplgaddtomacro\gplbacktext{%
      \csname LTb\endcsname
      \put(372,2808){\makebox(0,0)[r]{\strut{}}}%
      \put(504,2588){\makebox(0,0){\strut{}}}%
      \put(857,2588){\makebox(0,0){\strut{}}}%
      \put(1210,2588){\makebox(0,0){\strut{}}}%
      \put(1562,2588){\makebox(0,0){\strut{}}}%
      \put(1915,2588){\makebox(0,0){\strut{}}}%
    }%
    \gplgaddtomacro\gplfronttext{%
      \csname LTb\endcsname
      \put(152,3347){\rotatebox{-270}{\makebox(0,0){\strut{}$\chi=30$\degr}}}%
      \put(1209,4217){\makebox(0,0){\strut{}$\zeta=18$\degr}}%
    }%
    \gplgaddtomacro\gplbacktext{%
      \csname LTb\endcsname
      \put(1884,2808){\makebox(0,0)[r]{\strut{}}}%
      \put(2016,2588){\makebox(0,0){\strut{}}}%
      \put(2369,2588){\makebox(0,0){\strut{}}}%
      \put(2721,2588){\makebox(0,0){\strut{}}}%
      \put(3074,2588){\makebox(0,0){\strut{}}}%
      \put(3426,2588){\makebox(0,0){\strut{}}}%
    }%
    \gplgaddtomacro\gplfronttext{%
      \csname LTb\endcsname
      \put(2721,4217){\makebox(0,0){\strut{}$\zeta=36$\degr}}%
    }%
    \gplgaddtomacro\gplbacktext{%
      \csname LTb\endcsname
      \put(3396,2808){\makebox(0,0)[r]{\strut{}}}%
      \put(3528,2588){\makebox(0,0){\strut{}}}%
      \put(3881,2588){\makebox(0,0){\strut{}}}%
      \put(4233,2588){\makebox(0,0){\strut{}}}%
      \put(4586,2588){\makebox(0,0){\strut{}}}%
      \put(4938,2588){\makebox(0,0){\strut{}}}%
    }%
    \gplgaddtomacro\gplfronttext{%
      \csname LTb\endcsname
      \put(4233,4217){\makebox(0,0){\strut{}$\zeta=54$\degr}}%
    }%
    \gplgaddtomacro\gplbacktext{%
      \csname LTb\endcsname
      \put(4908,2808){\makebox(0,0)[r]{\strut{}}}%
      \put(5040,2588){\makebox(0,0){\strut{}}}%
      \put(5393,2588){\makebox(0,0){\strut{}}}%
      \put(5745,2588){\makebox(0,0){\strut{}}}%
      \put(6098,2588){\makebox(0,0){\strut{}}}%
      \put(6450,2588){\makebox(0,0){\strut{}}}%
    }%
    \gplgaddtomacro\gplfronttext{%
      \csname LTb\endcsname
      \put(5745,4217){\makebox(0,0){\strut{}$\zeta=72$\degr}}%
    }%
    \gplgaddtomacro\gplbacktext{%
      \csname LTb\endcsname
      \put(6420,2808){\makebox(0,0)[r]{\strut{}}}%
      \put(6552,2588){\makebox(0,0){\strut{}}}%
      \put(6905,2588){\makebox(0,0){\strut{}}}%
      \put(7257,2588){\makebox(0,0){\strut{}}}%
      \put(7610,2588){\makebox(0,0){\strut{}}}%
      \put(7962,2588){\makebox(0,0){\strut{}}}%
    }%
    \gplgaddtomacro\gplfronttext{%
      \csname LTb\endcsname
      \put(7257,4217){\makebox(0,0){\strut{}$\zeta=90$\degr}}%
    }%
    \gplgaddtomacro\gplbacktext{%
    }%
    \gplgaddtomacro\gplfronttext{%
      \csname LTb\endcsname
      \put(8064,1239){\makebox(0,0){\strut{}}}%
      \put(8417,1239){\makebox(0,0){\strut{}}}%
      \put(8769,1239){\makebox(0,0){\strut{}}}%
      \put(9121,1239){\makebox(0,0){\strut{}}}%
      \put(9474,1239){\makebox(0,0){\strut{}}}%
      \put(7849,1513){\makebox(0,0)[r]{\strut{}}}%
      \put(7849,1693){\makebox(0,0)[r]{\strut{}}}%
      \put(7849,1873){\makebox(0,0)[r]{\strut{}}}%
      \put(7849,2052){\makebox(0,0)[r]{\strut{}}}%
      \put(7849,2231){\makebox(0,0)[r]{\strut{}}}%
      \put(7849,2411){\makebox(0,0)[r]{\strut{}}}%
      \put(9712,1513){\makebox(0,0)[l]{\strut{} 0}}%
      \put(9712,1632){\makebox(0,0)[l]{\strut{}0.002}}%
      \put(9712,1752){\makebox(0,0)[l]{\strut{}0.004}}%
      \put(9712,1872){\makebox(0,0)[l]{\strut{}0.006}}%
      \put(9712,1992){\makebox(0,0)[l]{\strut{}0.008}}%
      \put(9712,2111){\makebox(0,0)[l]{\strut{}0.01}}%
      \put(9712,2231){\makebox(0,0)[l]{\strut{}0.01}}%
      \put(9712,2351){\makebox(0,0)[l]{\strut{}0.01}}%
      \put(9712,2471){\makebox(0,0)[l]{\strut{}0.02}}%
      \put(9712,2591){\makebox(0,0)[l]{\strut{}0.02}}%
    }%
    \gplgaddtomacro\gplbacktext{%
      \csname LTb\endcsname
      \put(372,1512){\makebox(0,0)[r]{\strut{}}}%
      \put(504,1292){\makebox(0,0){\strut{}}}%
      \put(857,1292){\makebox(0,0){\strut{}}}%
      \put(1210,1292){\makebox(0,0){\strut{}}}%
      \put(1562,1292){\makebox(0,0){\strut{}}}%
      \put(1915,1292){\makebox(0,0){\strut{}}}%
    }%
    \gplgaddtomacro\gplfronttext{%
      \csname LTb\endcsname
      \put(152,2051){\rotatebox{-270}{\makebox(0,0){\strut{}$\chi=60$\degr}}}%
    }%
    \gplgaddtomacro\gplbacktext{%
      \csname LTb\endcsname
      \put(1884,1512){\makebox(0,0)[r]{\strut{}}}%
      \put(2016,1292){\makebox(0,0){\strut{}}}%
      \put(2369,1292){\makebox(0,0){\strut{}}}%
      \put(2721,1292){\makebox(0,0){\strut{}}}%
      \put(3074,1292){\makebox(0,0){\strut{}}}%
      \put(3426,1292){\makebox(0,0){\strut{}}}%
    }%
    \gplgaddtomacro\gplfronttext{%
    }%
    \gplgaddtomacro\gplbacktext{%
      \csname LTb\endcsname
      \put(3396,1512){\makebox(0,0)[r]{\strut{}}}%
      \put(3528,1292){\makebox(0,0){\strut{}}}%
      \put(3881,1292){\makebox(0,0){\strut{}}}%
      \put(4233,1292){\makebox(0,0){\strut{}}}%
      \put(4586,1292){\makebox(0,0){\strut{}}}%
      \put(4938,1292){\makebox(0,0){\strut{}}}%
    }%
    \gplgaddtomacro\gplfronttext{%
    }%
    \gplgaddtomacro\gplbacktext{%
      \csname LTb\endcsname
      \put(4908,1512){\makebox(0,0)[r]{\strut{}}}%
      \put(5040,1292){\makebox(0,0){\strut{}}}%
      \put(5393,1292){\makebox(0,0){\strut{}}}%
      \put(5745,1292){\makebox(0,0){\strut{}}}%
      \put(6098,1292){\makebox(0,0){\strut{}}}%
      \put(6450,1292){\makebox(0,0){\strut{}}}%
    }%
    \gplgaddtomacro\gplfronttext{%
    }%
    \gplgaddtomacro\gplbacktext{%
      \csname LTb\endcsname
      \put(6420,1512){\makebox(0,0)[r]{\strut{}}}%
      \put(6552,1292){\makebox(0,0){\strut{}}}%
      \put(6905,1292){\makebox(0,0){\strut{}}}%
      \put(7257,1292){\makebox(0,0){\strut{}}}%
      \put(7610,1292){\makebox(0,0){\strut{}}}%
      \put(7962,1292){\makebox(0,0){\strut{}}}%
    }%
    \gplgaddtomacro\gplfronttext{%
    }%
    \gplgaddtomacro\gplbacktext{%
    }%
    \gplgaddtomacro\gplfronttext{%
      \csname LTb\endcsname
      \put(8064,-57){\makebox(0,0){\strut{}0}}%
      \put(8417,-57){\makebox(0,0){\strut{}0.25}}%
      \put(8769,-57){\makebox(0,0){\strut{}0.5}}%
      \put(9121,-57){\makebox(0,0){\strut{}0.75}}%
      \put(9474,-57){\makebox(0,0){\strut{}1}}%
      \put(8769,-387){\makebox(0,0){\strut{}phase ($\phi$)}}%
      \put(7849,217){\makebox(0,0)[r]{\strut{}}}%
      \put(7849,397){\makebox(0,0)[r]{\strut{}}}%
      \put(7849,577){\makebox(0,0)[r]{\strut{}}}%
      \put(7849,756){\makebox(0,0)[r]{\strut{}}}%
      \put(7849,935){\makebox(0,0)[r]{\strut{}}}%
      \put(7849,1115){\makebox(0,0)[r]{\strut{}}}%
      \put(9712,217){\makebox(0,0)[l]{\strut{} 0}}%
      \put(9712,432){\makebox(0,0)[l]{\strut{}0.005}}%
      \put(9712,648){\makebox(0,0)[l]{\strut{}0.01}}%
      \put(9712,863){\makebox(0,0)[l]{\strut{}0.01}}%
      \put(9712,1079){\makebox(0,0)[l]{\strut{}0.02}}%
      \put(9712,1295){\makebox(0,0)[l]{\strut{}0.03}}%
    }%
    \gplgaddtomacro\gplbacktext{%
      \csname LTb\endcsname
      \put(372,216){\makebox(0,0)[r]{\strut{}}}%
      \put(504,-4){\makebox(0,0){\strut{}0}}%
      \put(857,-4){\makebox(0,0){\strut{}0.25}}%
      \put(1210,-4){\makebox(0,0){\strut{}0.5}}%
      \put(1562,-4){\makebox(0,0){\strut{}0.75}}%
      \put(1915,-4){\makebox(0,0){\strut{}1}}%
    }%
    \gplgaddtomacro\gplfronttext{%
      \csname LTb\endcsname
      \put(152,755){\rotatebox{-270}{\makebox(0,0){\strut{}$\chi=90$\degr}}}%
      \put(1209,-334){\makebox(0,0){\strut{}phase ($\phi$)}}%
    }%
    \gplgaddtomacro\gplbacktext{%
      \csname LTb\endcsname
      \put(1884,216){\makebox(0,0)[r]{\strut{}}}%
      \put(2016,-4){\makebox(0,0){\strut{}0}}%
      \put(2369,-4){\makebox(0,0){\strut{}0.25}}%
      \put(2721,-4){\makebox(0,0){\strut{}0.5}}%
      \put(3074,-4){\makebox(0,0){\strut{}0.75}}%
      \put(3426,-4){\makebox(0,0){\strut{}1}}%
    }%
    \gplgaddtomacro\gplfronttext{%
      \csname LTb\endcsname
      \put(2721,-334){\makebox(0,0){\strut{}phase ($\phi$)}}%
    }%
    \gplgaddtomacro\gplbacktext{%
      \csname LTb\endcsname
      \put(3396,216){\makebox(0,0)[r]{\strut{}}}%
      \put(3528,-4){\makebox(0,0){\strut{}0}}%
      \put(3881,-4){\makebox(0,0){\strut{}0.25}}%
      \put(4233,-4){\makebox(0,0){\strut{}0.5}}%
      \put(4586,-4){\makebox(0,0){\strut{}0.75}}%
      \put(4938,-4){\makebox(0,0){\strut{}1}}%
    }%
    \gplgaddtomacro\gplfronttext{%
      \csname LTb\endcsname
      \put(4233,-334){\makebox(0,0){\strut{}phase ($\phi$)}}%
    }%
    \gplgaddtomacro\gplbacktext{%
      \csname LTb\endcsname
      \put(4908,216){\makebox(0,0)[r]{\strut{}}}%
      \put(5040,-4){\makebox(0,0){\strut{}0}}%
      \put(5393,-4){\makebox(0,0){\strut{}0.25}}%
      \put(5745,-4){\makebox(0,0){\strut{}0.5}}%
      \put(6098,-4){\makebox(0,0){\strut{}0.75}}%
      \put(6450,-4){\makebox(0,0){\strut{}1}}%
    }%
    \gplgaddtomacro\gplfronttext{%
      \csname LTb\endcsname
      \put(5745,-334){\makebox(0,0){\strut{}phase ($\phi$)}}%
    }%
    \gplgaddtomacro\gplbacktext{%
      \csname LTb\endcsname
      \put(6420,216){\makebox(0,0)[r]{\strut{}}}%
      \put(6552,-4){\makebox(0,0){\strut{}0}}%
      \put(6905,-4){\makebox(0,0){\strut{}0.25}}%
      \put(7257,-4){\makebox(0,0){\strut{}0.5}}%
      \put(7610,-4){\makebox(0,0){\strut{}0.75}}%
      \put(7962,-4){\makebox(0,0){\strut{}1}}%
    }%
    \gplgaddtomacro\gplfronttext{%
      \csname LTb\endcsname
      \put(7257,-334){\makebox(0,0){\strut{}phase ($\phi$)}}%
    }%
    \gplbacktext
    \put(0,0){\includegraphics{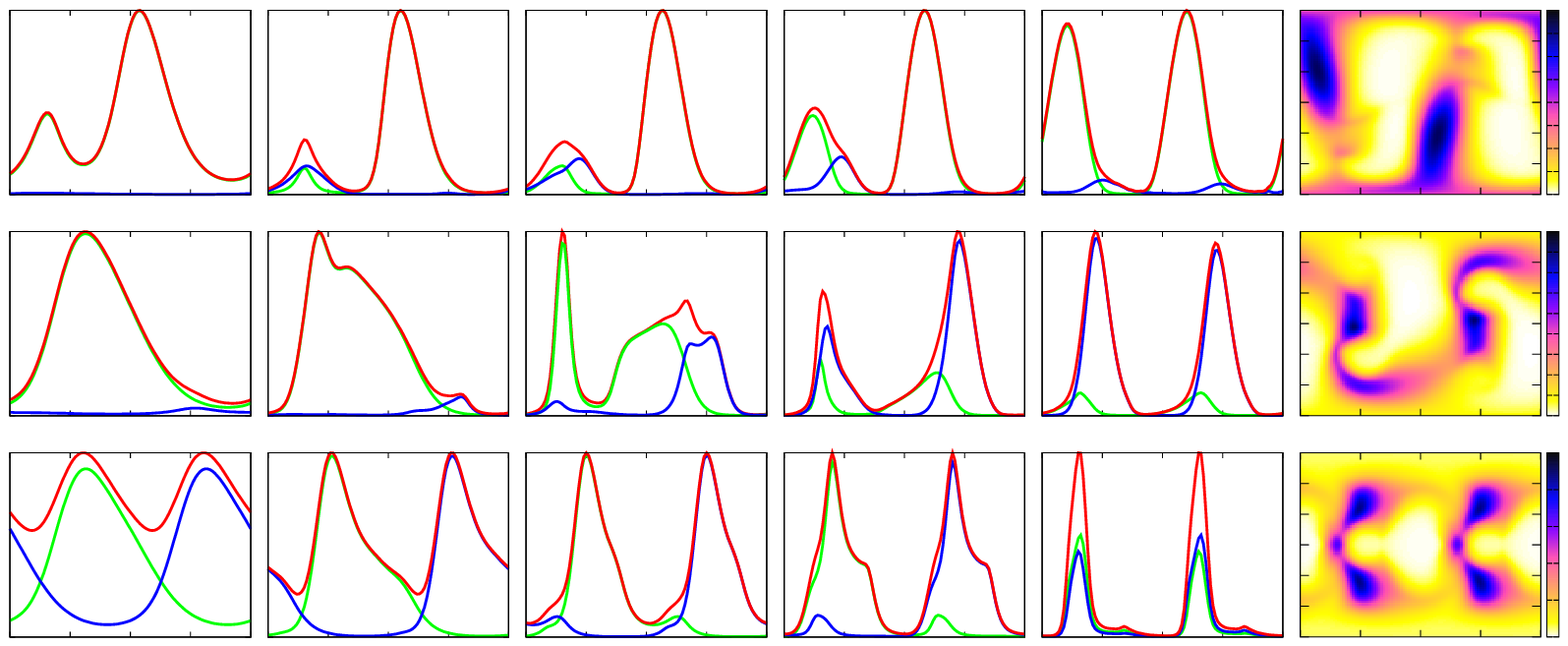}}%
    \gplfronttext
  \end{picture}%
\endgroup

%% file: skymaps_cl_r00025_b-3_ri1_ro5_g10_d3_om6.tex
\begingroup
  \makeatletter
  \providecommand\color[2][]{%
    \GenericError{(gnuplot) \space\space\space\@spaces}{%
      Package color not loaded in conjunction with
      terminal option `colourtext'%
    }{See the gnuplot documentation for explanation.%
    }{Either use 'blacktext' in gnuplot or load the package
      color.sty in LaTeX.}%
    \renewcommand\color[2][]{}%
  }%
  \providecommand\includegraphics[2][]{%
    \GenericError{(gnuplot) \space\space\space\@spaces}{%
      Package graphicx or graphics not loaded%
    }{See the gnuplot documentation for explanation.%
    }{The gnuplot epslatex terminal needs graphicx.sty or graphics.sty.}%
    \renewcommand\includegraphics[2][]{}%
  }%
  \providecommand\rotatebox[2]{#2}%
  \@ifundefined{ifGPcolor}{%
    \newif\ifGPcolor
    \GPcolortrue
  }{}%
  \@ifundefined{ifGPblacktext}{%
    \newif\ifGPblacktext
    \GPblacktextfalse
  }{}%
  \let\gplgaddtomacro\g@addto@macro
  \gdef\gplbacktext{}%
  \gdef\gplfronttext{}%
  \makeatother
  \ifGPblacktext
    \def\colorrgb#1{}%
    \def\colorgray#1{}%
  \else
    \ifGPcolor
      \def\colorrgb#1{\color[rgb]{#1}}%
      \def\colorgray#1{\color[gray]{#1}}%
      \expandafter\def\csname LTw\endcsname{\color{white}}%
      \expandafter\def\csname LTb\endcsname{\color{black}}%
      \expandafter\def\csname LTa\endcsname{\color{black}}%
      \expandafter\def\csname LT0\endcsname{\color[rgb]{1,0,0}}%
      \expandafter\def\csname LT1\endcsname{\color[rgb]{0,1,0}}%
      \expandafter\def\csname LT2\endcsname{\color[rgb]{0,0,1}}%
      \expandafter\def\csname LT3\endcsname{\color[rgb]{1,0,1}}%
      \expandafter\def\csname LT4\endcsname{\color[rgb]{0,1,1}}%
      \expandafter\def\csname LT5\endcsname{\color[rgb]{1,1,0}}%
      \expandafter\def\csname LT6\endcsname{\color[rgb]{0,0,0}}%
      \expandafter\def\csname LT7\endcsname{\color[rgb]{1,0.3,0}}%
      \expandafter\def\csname LT8\endcsname{\color[rgb]{0.5,0.5,0.5}}%
    \else
      \def\colorrgb#1{\color{black}}%
      \def\colorgray#1{\color[gray]{#1}}%
      \expandafter\def\csname LTw\endcsname{\color{white}}%
      \expandafter\def\csname LTb\endcsname{\color{black}}%
      \expandafter\def\csname LTa\endcsname{\color{black}}%
      \expandafter\def\csname LT0\endcsname{\color{black}}%
      \expandafter\def\csname LT1\endcsname{\color{black}}%
      \expandafter\def\csname LT2\endcsname{\color{black}}%
      \expandafter\def\csname LT3\endcsname{\color{black}}%
      \expandafter\def\csname LT4\endcsname{\color{black}}%
      \expandafter\def\csname LT5\endcsname{\color{black}}%
      \expandafter\def\csname LT6\endcsname{\color{black}}%
      \expandafter\def\csname LT7\endcsname{\color{black}}%
      \expandafter\def\csname LT8\endcsname{\color{black}}%
    \fi
  \fi
    \setlength{\unitlength}{0.0500bp}%
    \ifx\gptboxheight\undefined%
      \newlength{\gptboxheight}%
      \newlength{\gptboxwidth}%
      \newsavebox{\gptboxtext}%
    \fi%
    \setlength{\fboxrule}{0.5pt}%
    \setlength{\fboxsep}{1pt}%
\begin{picture}(10080.00,4320.00)%
    \gplgaddtomacro\gplbacktext{%
    }%
    \gplgaddtomacro\gplfronttext{%
      \csname LTb\endcsname
      \put(8064,2535){\makebox(0,0){\strut{}}}%
      \put(8417,2535){\makebox(0,0){\strut{}}}%
      \put(8769,2535){\makebox(0,0){\strut{}}}%
      \put(9121,2535){\makebox(0,0){\strut{}}}%
      \put(9474,2535){\makebox(0,0){\strut{}}}%
      \put(7849,2809){\makebox(0,0)[r]{\strut{}}}%
      \put(7849,2989){\makebox(0,0)[r]{\strut{}}}%
      \put(7849,3169){\makebox(0,0)[r]{\strut{}}}%
      \put(7849,3348){\makebox(0,0)[r]{\strut{}}}%
      \put(7849,3527){\makebox(0,0)[r]{\strut{}}}%
      \put(7849,3707){\makebox(0,0)[r]{\strut{}}}%
      \put(9712,2809){\makebox(0,0)[l]{\strut{} 0}}%
      \put(9712,2963){\makebox(0,0)[l]{\strut{}0.1}}%
      \put(9712,3117){\makebox(0,0)[l]{\strut{}0.2}}%
      \put(9712,3271){\makebox(0,0)[l]{\strut{}0.3}}%
      \put(9712,3425){\makebox(0,0)[l]{\strut{}0.4}}%
      \put(9712,3579){\makebox(0,0)[l]{\strut{}0.5}}%
      \put(9712,3733){\makebox(0,0)[l]{\strut{}0.6}}%
      \put(9712,3887){\makebox(0,0)[l]{\strut{}0.7}}%
    }%
    \gplgaddtomacro\gplbacktext{%
      \csname LTb\endcsname
      \put(372,2808){\makebox(0,0)[r]{\strut{}}}%
      \put(504,2588){\makebox(0,0){\strut{}}}%
      \put(857,2588){\makebox(0,0){\strut{}}}%
      \put(1210,2588){\makebox(0,0){\strut{}}}%
      \put(1562,2588){\makebox(0,0){\strut{}}}%
      \put(1915,2588){\makebox(0,0){\strut{}}}%
    }%
    \gplgaddtomacro\gplfronttext{%
      \csname LTb\endcsname
      \put(152,3347){\rotatebox{-270}{\makebox(0,0){\strut{}$\chi=30$\degr}}}%
      \put(1209,4217){\makebox(0,0){\strut{}$\zeta=18$\degr}}%
    }%
    \gplgaddtomacro\gplbacktext{%
      \csname LTb\endcsname
      \put(1884,2808){\makebox(0,0)[r]{\strut{}}}%
      \put(2016,2588){\makebox(0,0){\strut{}}}%
      \put(2369,2588){\makebox(0,0){\strut{}}}%
      \put(2721,2588){\makebox(0,0){\strut{}}}%
      \put(3074,2588){\makebox(0,0){\strut{}}}%
      \put(3426,2588){\makebox(0,0){\strut{}}}%
    }%
    \gplgaddtomacro\gplfronttext{%
      \csname LTb\endcsname
      \put(2721,4217){\makebox(0,0){\strut{}$\zeta=36$\degr}}%
    }%
    \gplgaddtomacro\gplbacktext{%
      \csname LTb\endcsname
      \put(3396,2808){\makebox(0,0)[r]{\strut{}}}%
      \put(3528,2588){\makebox(0,0){\strut{}}}%
      \put(3881,2588){\makebox(0,0){\strut{}}}%
      \put(4233,2588){\makebox(0,0){\strut{}}}%
      \put(4586,2588){\makebox(0,0){\strut{}}}%
      \put(4938,2588){\makebox(0,0){\strut{}}}%
    }%
    \gplgaddtomacro\gplfronttext{%
      \csname LTb\endcsname
      \put(4233,4217){\makebox(0,0){\strut{}$\zeta=54$\degr}}%
    }%
    \gplgaddtomacro\gplbacktext{%
      \csname LTb\endcsname
      \put(4908,2808){\makebox(0,0)[r]{\strut{}}}%
      \put(5040,2588){\makebox(0,0){\strut{}}}%
      \put(5393,2588){\makebox(0,0){\strut{}}}%
      \put(5745,2588){\makebox(0,0){\strut{}}}%
      \put(6098,2588){\makebox(0,0){\strut{}}}%
      \put(6450,2588){\makebox(0,0){\strut{}}}%
    }%
    \gplgaddtomacro\gplfronttext{%
      \csname LTb\endcsname
      \put(5745,4217){\makebox(0,0){\strut{}$\zeta=72$\degr}}%
    }%
    \gplgaddtomacro\gplbacktext{%
      \csname LTb\endcsname
      \put(6420,2808){\makebox(0,0)[r]{\strut{}}}%
      \put(6552,2588){\makebox(0,0){\strut{}}}%
      \put(6905,2588){\makebox(0,0){\strut{}}}%
      \put(7257,2588){\makebox(0,0){\strut{}}}%
      \put(7610,2588){\makebox(0,0){\strut{}}}%
      \put(7962,2588){\makebox(0,0){\strut{}}}%
    }%
    \gplgaddtomacro\gplfronttext{%
      \csname LTb\endcsname
      \put(7257,4217){\makebox(0,0){\strut{}$\zeta=90$\degr}}%
    }%
    \gplgaddtomacro\gplbacktext{%
    }%
    \gplgaddtomacro\gplfronttext{%
      \csname LTb\endcsname
      \put(8064,1239){\makebox(0,0){\strut{}}}%
      \put(8417,1239){\makebox(0,0){\strut{}}}%
      \put(8769,1239){\makebox(0,0){\strut{}}}%
      \put(9121,1239){\makebox(0,0){\strut{}}}%
      \put(9474,1239){\makebox(0,0){\strut{}}}%
      \put(7849,1513){\makebox(0,0)[r]{\strut{}}}%
      \put(7849,1693){\makebox(0,0)[r]{\strut{}}}%
      \put(7849,1873){\makebox(0,0)[r]{\strut{}}}%
      \put(7849,2052){\makebox(0,0)[r]{\strut{}}}%
      \put(7849,2231){\makebox(0,0)[r]{\strut{}}}%
      \put(7849,2411){\makebox(0,0)[r]{\strut{}}}%
      \put(9712,1513){\makebox(0,0)[l]{\strut{} 0}}%
      \put(9712,1647){\makebox(0,0)[l]{\strut{}0.1}}%
      \put(9712,1782){\makebox(0,0)[l]{\strut{}0.2}}%
      \put(9712,1917){\makebox(0,0)[l]{\strut{}0.3}}%
      \put(9712,2052){\makebox(0,0)[l]{\strut{}0.4}}%
      \put(9712,2186){\makebox(0,0)[l]{\strut{}0.5}}%
      \put(9712,2321){\makebox(0,0)[l]{\strut{}0.6}}%
      \put(9712,2456){\makebox(0,0)[l]{\strut{}0.7}}%
      \put(9712,2591){\makebox(0,0)[l]{\strut{}0.8}}%
    }%
    \gplgaddtomacro\gplbacktext{%
      \csname LTb\endcsname
      \put(372,1512){\makebox(0,0)[r]{\strut{}}}%
      \put(504,1292){\makebox(0,0){\strut{}}}%
      \put(857,1292){\makebox(0,0){\strut{}}}%
      \put(1210,1292){\makebox(0,0){\strut{}}}%
      \put(1562,1292){\makebox(0,0){\strut{}}}%
      \put(1915,1292){\makebox(0,0){\strut{}}}%
    }%
    \gplgaddtomacro\gplfronttext{%
      \csname LTb\endcsname
      \put(152,2051){\rotatebox{-270}{\makebox(0,0){\strut{}$\chi=60$\degr}}}%
    }%
    \gplgaddtomacro\gplbacktext{%
      \csname LTb\endcsname
      \put(1884,1512){\makebox(0,0)[r]{\strut{}}}%
      \put(2016,1292){\makebox(0,0){\strut{}}}%
      \put(2369,1292){\makebox(0,0){\strut{}}}%
      \put(2721,1292){\makebox(0,0){\strut{}}}%
      \put(3074,1292){\makebox(0,0){\strut{}}}%
      \put(3426,1292){\makebox(0,0){\strut{}}}%
    }%
    \gplgaddtomacro\gplfronttext{%
    }%
    \gplgaddtomacro\gplbacktext{%
      \csname LTb\endcsname
      \put(3396,1512){\makebox(0,0)[r]{\strut{}}}%
      \put(3528,1292){\makebox(0,0){\strut{}}}%
      \put(3881,1292){\makebox(0,0){\strut{}}}%
      \put(4233,1292){\makebox(0,0){\strut{}}}%
      \put(4586,1292){\makebox(0,0){\strut{}}}%
      \put(4938,1292){\makebox(0,0){\strut{}}}%
    }%
    \gplgaddtomacro\gplfronttext{%
    }%
    \gplgaddtomacro\gplbacktext{%
      \csname LTb\endcsname
      \put(4908,1512){\makebox(0,0)[r]{\strut{}}}%
      \put(5040,1292){\makebox(0,0){\strut{}}}%
      \put(5393,1292){\makebox(0,0){\strut{}}}%
      \put(5745,1292){\makebox(0,0){\strut{}}}%
      \put(6098,1292){\makebox(0,0){\strut{}}}%
      \put(6450,1292){\makebox(0,0){\strut{}}}%
    }%
    \gplgaddtomacro\gplfronttext{%
    }%
    \gplgaddtomacro\gplbacktext{%
      \csname LTb\endcsname
      \put(6420,1512){\makebox(0,0)[r]{\strut{}}}%
      \put(6552,1292){\makebox(0,0){\strut{}}}%
      \put(6905,1292){\makebox(0,0){\strut{}}}%
      \put(7257,1292){\makebox(0,0){\strut{}}}%
      \put(7610,1292){\makebox(0,0){\strut{}}}%
      \put(7962,1292){\makebox(0,0){\strut{}}}%
    }%
    \gplgaddtomacro\gplfronttext{%
    }%
    \gplgaddtomacro\gplbacktext{%
    }%
    \gplgaddtomacro\gplfronttext{%
      \csname LTb\endcsname
      \put(8064,-57){\makebox(0,0){\strut{}0}}%
      \put(8417,-57){\makebox(0,0){\strut{}0.25}}%
      \put(8769,-57){\makebox(0,0){\strut{}0.5}}%
      \put(9121,-57){\makebox(0,0){\strut{}0.75}}%
      \put(9474,-57){\makebox(0,0){\strut{}1}}%
      \put(8769,-387){\makebox(0,0){\strut{}phase ($\phi$)}}%
      \put(7849,217){\makebox(0,0)[r]{\strut{}}}%
      \put(7849,397){\makebox(0,0)[r]{\strut{}}}%
      \put(7849,577){\makebox(0,0)[r]{\strut{}}}%
      \put(7849,756){\makebox(0,0)[r]{\strut{}}}%
      \put(7849,935){\makebox(0,0)[r]{\strut{}}}%
      \put(7849,1115){\makebox(0,0)[r]{\strut{}}}%
      \put(9712,217){\makebox(0,0)[l]{\strut{} 0}}%
      \put(9712,371){\makebox(0,0)[l]{\strut{}0.1}}%
      \put(9712,525){\makebox(0,0)[l]{\strut{}0.2}}%
      \put(9712,679){\makebox(0,0)[l]{\strut{}0.3}}%
      \put(9712,833){\makebox(0,0)[l]{\strut{}0.4}}%
      \put(9712,986){\makebox(0,0)[l]{\strut{}0.5}}%
      \put(9712,1140){\makebox(0,0)[l]{\strut{}0.6}}%
      \put(9712,1294){\makebox(0,0)[l]{\strut{}0.7}}%
    }%
    \gplgaddtomacro\gplbacktext{%
      \csname LTb\endcsname
      \put(372,216){\makebox(0,0)[r]{\strut{}}}%
      \put(504,-4){\makebox(0,0){\strut{}0}}%
      \put(857,-4){\makebox(0,0){\strut{}0.25}}%
      \put(1210,-4){\makebox(0,0){\strut{}0.5}}%
      \put(1562,-4){\makebox(0,0){\strut{}0.75}}%
      \put(1915,-4){\makebox(0,0){\strut{}1}}%
    }%
    \gplgaddtomacro\gplfronttext{%
      \csname LTb\endcsname
      \put(152,755){\rotatebox{-270}{\makebox(0,0){\strut{}$\chi=90$\degr}}}%
      \put(1209,-334){\makebox(0,0){\strut{}phase ($\phi$)}}%
    }%
    \gplgaddtomacro\gplbacktext{%
      \csname LTb\endcsname
      \put(1884,216){\makebox(0,0)[r]{\strut{}}}%
      \put(2016,-4){\makebox(0,0){\strut{}0}}%
      \put(2369,-4){\makebox(0,0){\strut{}0.25}}%
      \put(2721,-4){\makebox(0,0){\strut{}0.5}}%
      \put(3074,-4){\makebox(0,0){\strut{}0.75}}%
      \put(3426,-4){\makebox(0,0){\strut{}1}}%
    }%
    \gplgaddtomacro\gplfronttext{%
      \csname LTb\endcsname
      \put(2721,-334){\makebox(0,0){\strut{}phase ($\phi$)}}%
    }%
    \gplgaddtomacro\gplbacktext{%
      \csname LTb\endcsname
      \put(3396,216){\makebox(0,0)[r]{\strut{}}}%
      \put(3528,-4){\makebox(0,0){\strut{}0}}%
      \put(3881,-4){\makebox(0,0){\strut{}0.25}}%
      \put(4233,-4){\makebox(0,0){\strut{}0.5}}%
      \put(4586,-4){\makebox(0,0){\strut{}0.75}}%
      \put(4938,-4){\makebox(0,0){\strut{}1}}%
    }%
    \gplgaddtomacro\gplfronttext{%
      \csname LTb\endcsname
      \put(4233,-334){\makebox(0,0){\strut{}phase ($\phi$)}}%
    }%
    \gplgaddtomacro\gplbacktext{%
      \csname LTb\endcsname
      \put(4908,216){\makebox(0,0)[r]{\strut{}}}%
      \put(5040,-4){\makebox(0,0){\strut{}0}}%
      \put(5393,-4){\makebox(0,0){\strut{}0.25}}%
      \put(5745,-4){\makebox(0,0){\strut{}0.5}}%
      \put(6098,-4){\makebox(0,0){\strut{}0.75}}%
      \put(6450,-4){\makebox(0,0){\strut{}1}}%
    }%
    \gplgaddtomacro\gplfronttext{%
      \csname LTb\endcsname
      \put(5745,-334){\makebox(0,0){\strut{}phase ($\phi$)}}%
    }%
    \gplgaddtomacro\gplbacktext{%
      \csname LTb\endcsname
      \put(6420,216){\makebox(0,0)[r]{\strut{}}}%
      \put(6552,-4){\makebox(0,0){\strut{}0}}%
      \put(6905,-4){\makebox(0,0){\strut{}0.25}}%
      \put(7257,-4){\makebox(0,0){\strut{}0.5}}%
      \put(7610,-4){\makebox(0,0){\strut{}0.75}}%
      \put(7962,-4){\makebox(0,0){\strut{}1}}%
    }%
    \gplgaddtomacro\gplfronttext{%
      \csname LTb\endcsname
      \put(7257,-334){\makebox(0,0){\strut{}phase ($\phi$)}}%
    }%
    \gplbacktext
    \put(0,0){\includegraphics{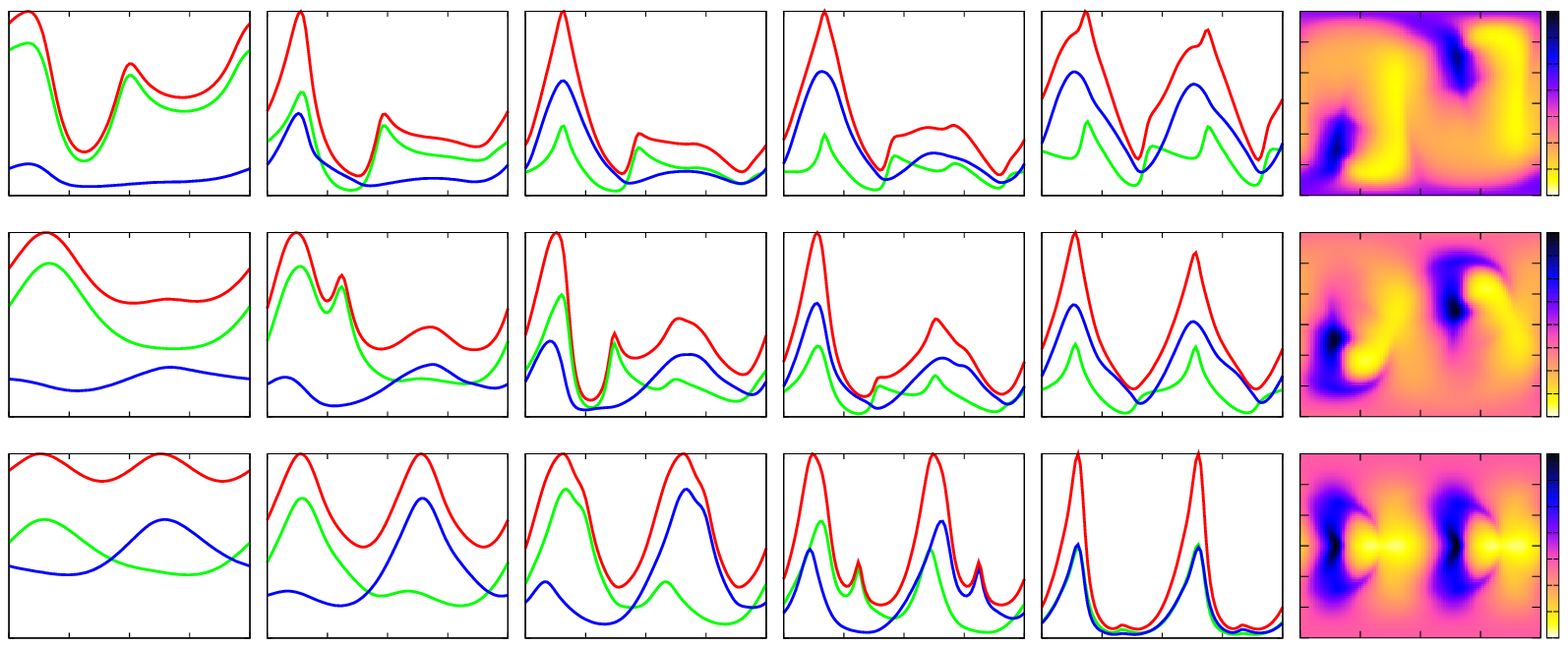}}%
    \gplfronttext
  \end{picture}%
\endgroup

%% file: courbe_lumiere_r00025_b-3_c60_ri0.2_ro1_g10_dx_z60.tex
\begingroup
  \makeatletter
  \providecommand\color[2][]{%
    \GenericError{(gnuplot) \space\space\space\@spaces}{%
      Package color not loaded in conjunction with
      terminal option `colourtext'%
    }{See the gnuplot documentation for explanation.%
    }{Either use 'blacktext' in gnuplot or load the package
      color.sty in LaTeX.}%
    \renewcommand\color[2][]{}%
  }%
  \providecommand\includegraphics[2][]{%
    \GenericError{(gnuplot) \space\space\space\@spaces}{%
      Package graphicx or graphics not loaded%
    }{See the gnuplot documentation for explanation.%
    }{The gnuplot epslatex terminal needs graphicx.sty or graphics.sty.}%
    \renewcommand\includegraphics[2][]{}%
  }%
  \providecommand\rotatebox[2]{#2}%
  \@ifundefined{ifGPcolor}{%
    \newif\ifGPcolor
    \GPcolortrue
  }{}%
  \@ifundefined{ifGPblacktext}{%
    \newif\ifGPblacktext
    \GPblacktextfalse
  }{}%
  \let\gplgaddtomacro\g@addto@macro
  \gdef\gplbacktext{}%
  \gdef\gplfronttext{}%
  \makeatother
  \ifGPblacktext
    \def\colorrgb#1{}%
    \def\colorgray#1{}%
  \else
    \ifGPcolor
      \def\colorrgb#1{\color[rgb]{#1}}%
      \def\colorgray#1{\color[gray]{#1}}%
      \expandafter\def\csname LTw\endcsname{\color{white}}%
      \expandafter\def\csname LTb\endcsname{\color{black}}%
      \expandafter\def\csname LTa\endcsname{\color{black}}%
      \expandafter\def\csname LT0\endcsname{\color[rgb]{1,0,0}}%
      \expandafter\def\csname LT1\endcsname{\color[rgb]{0,1,0}}%
      \expandafter\def\csname LT2\endcsname{\color[rgb]{0,0,1}}%
      \expandafter\def\csname LT3\endcsname{\color[rgb]{1,0,1}}%
      \expandafter\def\csname LT4\endcsname{\color[rgb]{0,1,1}}%
      \expandafter\def\csname LT5\endcsname{\color[rgb]{1,1,0}}%
      \expandafter\def\csname LT6\endcsname{\color[rgb]{0,0,0}}%
      \expandafter\def\csname LT7\endcsname{\color[rgb]{1,0.3,0}}%
      \expandafter\def\csname LT8\endcsname{\color[rgb]{0.5,0.5,0.5}}%
    \else
      \def\colorrgb#1{\color{black}}%
      \def\colorgray#1{\color[gray]{#1}}%
      \expandafter\def\csname LTw\endcsname{\color{white}}%
      \expandafter\def\csname LTb\endcsname{\color{black}}%
      \expandafter\def\csname LTa\endcsname{\color{black}}%
      \expandafter\def\csname LT0\endcsname{\color{black}}%
      \expandafter\def\csname LT1\endcsname{\color{black}}%
      \expandafter\def\csname LT2\endcsname{\color{black}}%
      \expandafter\def\csname LT3\endcsname{\color{black}}%
      \expandafter\def\csname LT4\endcsname{\color{black}}%
      \expandafter\def\csname LT5\endcsname{\color{black}}%
      \expandafter\def\csname LT6\endcsname{\color{black}}%
      \expandafter\def\csname LT7\endcsname{\color{black}}%
      \expandafter\def\csname LT8\endcsname{\color{black}}%
    \fi
  \fi
    \setlength{\unitlength}{0.0500bp}%
    \ifx\gptboxheight\undefined%
      \newlength{\gptboxheight}%
      \newlength{\gptboxwidth}%
      \newsavebox{\gptboxtext}%
    \fi%
    \setlength{\fboxrule}{0.5pt}%
    \setlength{\fboxsep}{1pt}%
\begin{picture}(8640.00,8640.00)%
    \gplgaddtomacro\gplbacktext{%
      \csname LTb\endcsname
      \put(264,7310){\makebox(0,0)[r]{\strut{} 0}}%
      \put(264,7513){\makebox(0,0)[r]{\strut{} 1}}%
      \put(264,7716){\makebox(0,0)[r]{\strut{} 2}}%
      \put(264,7920){\makebox(0,0)[r]{\strut{} 3}}%
      \put(264,8123){\makebox(0,0)[r]{\strut{} 4}}%
      \put(264,8326){\makebox(0,0)[r]{\strut{} 5}}%
      \put(264,8529){\makebox(0,0)[r]{\strut{} 6}}%
      \put(396,7090){\makebox(0,0){\strut{}}}%
      \put(1278,7090){\makebox(0,0){\strut{}}}%
      \put(2160,7090){\makebox(0,0){\strut{}}}%
      \put(3041,7090){\makebox(0,0){\strut{}}}%
      \put(3923,7090){\makebox(0,0){\strut{}}}%
    }%
    \gplgaddtomacro\gplfronttext{%
      \csname LTb\endcsname
      \put(2159,8859){\makebox(0,0){\strut{}$q=3$}}%
      \csname LTb\endcsname
      \put(2512,8285){\makebox(0,0)[l]{\strut{}5~MeV}}%
    }%
    \gplgaddtomacro\gplbacktext{%
      \csname LTb\endcsname
      \put(264,5870){\makebox(0,0)[r]{\strut{} 0}}%
      \put(264,6073){\makebox(0,0)[r]{\strut{} 2}}%
      \put(264,6276){\makebox(0,0)[r]{\strut{} 4}}%
      \put(264,6480){\makebox(0,0)[r]{\strut{} 6}}%
      \put(264,6683){\makebox(0,0)[r]{\strut{} 8}}%
      \put(264,6886){\makebox(0,0)[r]{\strut{}1e+01}}%
      \put(264,7089){\makebox(0,0)[r]{\strut{}1e+01}}%
      \put(396,5650){\makebox(0,0){\strut{}}}%
      \put(1278,5650){\makebox(0,0){\strut{}}}%
      \put(2160,5650){\makebox(0,0){\strut{}}}%
      \put(3041,5650){\makebox(0,0){\strut{}}}%
      \put(3923,5650){\makebox(0,0){\strut{}}}%
    }%
    \gplgaddtomacro\gplfronttext{%
      \csname LTb\endcsname
      \put(2512,6845){\makebox(0,0)[l]{\strut{}51~MeV}}%
    }%
    \gplgaddtomacro\gplbacktext{%
      \csname LTb\endcsname
      \put(264,4430){\makebox(0,0)[r]{\strut{} 0}}%
      \put(264,4582){\makebox(0,0)[r]{\strut{} 2}}%
      \put(264,4735){\makebox(0,0)[r]{\strut{} 4}}%
      \put(264,4887){\makebox(0,0)[r]{\strut{} 6}}%
      \put(264,5040){\makebox(0,0)[r]{\strut{} 8}}%
      \put(264,5192){\makebox(0,0)[r]{\strut{}1e+01}}%
      \put(264,5344){\makebox(0,0)[r]{\strut{}1e+01}}%
      \put(264,5497){\makebox(0,0)[r]{\strut{}1e+01}}%
      \put(264,5649){\makebox(0,0)[r]{\strut{}2e+01}}%
      \put(396,4210){\makebox(0,0){\strut{}}}%
      \put(1278,4210){\makebox(0,0){\strut{}}}%
      \put(2160,4210){\makebox(0,0){\strut{}}}%
      \put(3041,4210){\makebox(0,0){\strut{}}}%
      \put(3923,4210){\makebox(0,0){\strut{}}}%
    }%
    \gplgaddtomacro\gplfronttext{%
      \csname LTb\endcsname
      \put(2512,5405){\makebox(0,0)[l]{\strut{}511~MeV}}%
    }%
    \gplgaddtomacro\gplbacktext{%
      \csname LTb\endcsname
      \put(264,2990){\makebox(0,0)[r]{\strut{} 0}}%
      \put(264,3164){\makebox(0,0)[r]{\strut{}0.5}}%
      \put(264,3339){\makebox(0,0)[r]{\strut{} 1}}%
      \put(264,3513){\makebox(0,0)[r]{\strut{} 2}}%
      \put(264,3687){\makebox(0,0)[r]{\strut{} 2}}%
      \put(264,3861){\makebox(0,0)[r]{\strut{} 2}}%
      \put(264,4036){\makebox(0,0)[r]{\strut{} 3}}%
      \put(264,4210){\makebox(0,0)[r]{\strut{} 4}}%
      \put(396,2770){\makebox(0,0){\strut{}}}%
      \put(1278,2770){\makebox(0,0){\strut{}}}%
      \put(2160,2770){\makebox(0,0){\strut{}}}%
      \put(3041,2770){\makebox(0,0){\strut{}}}%
      \put(3923,2770){\makebox(0,0){\strut{}}}%
    }%
    \gplgaddtomacro\gplfronttext{%
      \csname LTb\endcsname
      \put(2512,3966){\makebox(0,0)[l]{\strut{}5~GeV}}%
    }%
    \gplgaddtomacro\gplbacktext{%
      \csname LTb\endcsname
      \put(264,1550){\makebox(0,0)[r]{\strut{} 0}}%
      \put(264,1703){\makebox(0,0)[r]{\strut{}1e-06}}%
      \put(264,1855){\makebox(0,0)[r]{\strut{}2e-06}}%
      \put(264,2008){\makebox(0,0)[r]{\strut{}3e-06}}%
      \put(264,2160){\makebox(0,0)[r]{\strut{}4e-06}}%
      \put(264,2313){\makebox(0,0)[r]{\strut{}5e-06}}%
      \put(264,2465){\makebox(0,0)[r]{\strut{}6e-06}}%
      \put(264,2618){\makebox(0,0)[r]{\strut{}7e-06}}%
      \put(264,2770){\makebox(0,0)[r]{\strut{}8e-06}}%
      \put(396,1330){\makebox(0,0){\strut{}}}%
      \put(1278,1330){\makebox(0,0){\strut{}}}%
      \put(2160,1330){\makebox(0,0){\strut{}}}%
      \put(3041,1330){\makebox(0,0){\strut{}}}%
      \put(3923,1330){\makebox(0,0){\strut{}}}%
    }%
    \gplgaddtomacro\gplfronttext{%
      \csname LTb\endcsname
      \put(2512,2526){\makebox(0,0)[l]{\strut{}51~GeV}}%
    }%
    \gplgaddtomacro\gplbacktext{%
      \csname LTb\endcsname
      \put(264,110){\makebox(0,0)[r]{\strut{} 0}}%
      \put(264,284){\makebox(0,0)[r]{\strut{}5e-14}}%
      \put(264,459){\makebox(0,0)[r]{\strut{}1e-13}}%
      \put(264,633){\makebox(0,0)[r]{\strut{}2e-13}}%
      \put(264,807){\makebox(0,0)[r]{\strut{}2e-13}}%
      \put(264,981){\makebox(0,0)[r]{\strut{}2e-13}}%
      \put(264,1156){\makebox(0,0)[r]{\strut{}3e-13}}%
      \put(264,1330){\makebox(0,0)[r]{\strut{}3e-13}}%
      \put(396,-110){\makebox(0,0){\strut{}0}}%
      \put(1278,-110){\makebox(0,0){\strut{}0.25}}%
      \put(2160,-110){\makebox(0,0){\strut{}0.5}}%
      \put(3041,-110){\makebox(0,0){\strut{}0.75}}%
      \put(3923,-110){\makebox(0,0){\strut{}1}}%
    }%
    \gplgaddtomacro\gplfronttext{%
      \csname LTb\endcsname
      \put(2512,1086){\makebox(0,0)[l]{\strut{}511~GeV}}%
    }%
    \gplgaddtomacro\gplbacktext{%
      \csname LTb\endcsname
      \put(4584,7310){\makebox(0,0)[r]{\strut{} 0}}%
      \put(4584,7462){\makebox(0,0)[r]{\strut{}2e+04}}%
      \put(4584,7615){\makebox(0,0)[r]{\strut{}4e+04}}%
      \put(4584,7767){\makebox(0,0)[r]{\strut{}6e+04}}%
      \put(4584,7920){\makebox(0,0)[r]{\strut{}8e+04}}%
      \put(4584,8072){\makebox(0,0)[r]{\strut{}1e+05}}%
      \put(4584,8224){\makebox(0,0)[r]{\strut{}1e+05}}%
      \put(4584,8377){\makebox(0,0)[r]{\strut{}1e+05}}%
      \put(4584,8529){\makebox(0,0)[r]{\strut{}2e+05}}%
      \put(4716,7090){\makebox(0,0){\strut{}}}%
      \put(5598,7090){\makebox(0,0){\strut{}}}%
      \put(6480,7090){\makebox(0,0){\strut{}}}%
      \put(7361,7090){\makebox(0,0){\strut{}}}%
      \put(8243,7090){\makebox(0,0){\strut{}}}%
    }%
    \gplgaddtomacro\gplfronttext{%
      \csname LTb\endcsname
      \put(6479,8859){\makebox(0,0){\strut{}$q=1$}}%
      \csname LTb\endcsname
      \put(6832,8285){\makebox(0,0)[l]{\strut{}5~MeV}}%
    }%
    \gplgaddtomacro\gplbacktext{%
      \csname LTb\endcsname
      \put(4584,5870){\makebox(0,0)[r]{\strut{} 0}}%
      \put(4584,6073){\makebox(0,0)[r]{\strut{}5e+04}}%
      \put(4584,6276){\makebox(0,0)[r]{\strut{}1e+05}}%
      \put(4584,6480){\makebox(0,0)[r]{\strut{}2e+05}}%
      \put(4584,6683){\makebox(0,0)[r]{\strut{}2e+05}}%
      \put(4584,6886){\makebox(0,0)[r]{\strut{}2e+05}}%
      \put(4584,7089){\makebox(0,0)[r]{\strut{}3e+05}}%
      \put(4716,5650){\makebox(0,0){\strut{}}}%
      \put(5598,5650){\makebox(0,0){\strut{}}}%
      \put(6480,5650){\makebox(0,0){\strut{}}}%
      \put(7361,5650){\makebox(0,0){\strut{}}}%
      \put(8243,5650){\makebox(0,0){\strut{}}}%
    }%
    \gplgaddtomacro\gplfronttext{%
      \csname LTb\endcsname
      \put(6832,6845){\makebox(0,0)[l]{\strut{}51~MeV}}%
    }%
    \gplgaddtomacro\gplbacktext{%
      \csname LTb\endcsname
      \put(4584,4430){\makebox(0,0)[r]{\strut{} 0}}%
      \put(4584,4582){\makebox(0,0)[r]{\strut{}5e+04}}%
      \put(4584,4735){\makebox(0,0)[r]{\strut{}1e+05}}%
      \put(4584,4887){\makebox(0,0)[r]{\strut{}2e+05}}%
      \put(4584,5040){\makebox(0,0)[r]{\strut{}2e+05}}%
      \put(4584,5192){\makebox(0,0)[r]{\strut{}2e+05}}%
      \put(4584,5344){\makebox(0,0)[r]{\strut{}3e+05}}%
      \put(4584,5497){\makebox(0,0)[r]{\strut{}4e+05}}%
      \put(4584,5649){\makebox(0,0)[r]{\strut{}4e+05}}%
      \put(4716,4210){\makebox(0,0){\strut{}}}%
      \put(5598,4210){\makebox(0,0){\strut{}}}%
      \put(6480,4210){\makebox(0,0){\strut{}}}%
      \put(7361,4210){\makebox(0,0){\strut{}}}%
      \put(8243,4210){\makebox(0,0){\strut{}}}%
    }%
    \gplgaddtomacro\gplfronttext{%
      \csname LTb\endcsname
      \put(6832,5405){\makebox(0,0)[l]{\strut{}511~MeV}}%
    }%
    \gplgaddtomacro\gplbacktext{%
      \csname LTb\endcsname
      \put(4584,2990){\makebox(0,0)[r]{\strut{} 0}}%
      \put(4584,3126){\makebox(0,0)[r]{\strut{}5e+03}}%
      \put(4584,3261){\makebox(0,0)[r]{\strut{}1e+04}}%
      \put(4584,3397){\makebox(0,0)[r]{\strut{}2e+04}}%
      \put(4584,3532){\makebox(0,0)[r]{\strut{}2e+04}}%
      \put(4584,3668){\makebox(0,0)[r]{\strut{}2e+04}}%
      \put(4584,3803){\makebox(0,0)[r]{\strut{}3e+04}}%
      \put(4584,3939){\makebox(0,0)[r]{\strut{}4e+04}}%
      \put(4584,4074){\makebox(0,0)[r]{\strut{}4e+04}}%
      \put(4584,4210){\makebox(0,0)[r]{\strut{}4e+04}}%
      \put(4716,2770){\makebox(0,0){\strut{}}}%
      \put(5598,2770){\makebox(0,0){\strut{}}}%
      \put(6480,2770){\makebox(0,0){\strut{}}}%
      \put(7361,2770){\makebox(0,0){\strut{}}}%
      \put(8243,2770){\makebox(0,0){\strut{}}}%
    }%
    \gplgaddtomacro\gplfronttext{%
      \csname LTb\endcsname
      \put(6832,3966){\makebox(0,0)[l]{\strut{}5~GeV}}%
    }%
    \gplgaddtomacro\gplbacktext{%
      \csname LTb\endcsname
      \put(4584,1550){\makebox(0,0)[r]{\strut{} 0}}%
      \put(4584,1753){\makebox(0,0)[r]{\strut{}0.01}}%
      \put(4584,1957){\makebox(0,0)[r]{\strut{}0.02}}%
      \put(4584,2160){\makebox(0,0)[r]{\strut{}0.03}}%
      \put(4584,2363){\makebox(0,0)[r]{\strut{}0.04}}%
      \put(4584,2567){\makebox(0,0)[r]{\strut{}0.05}}%
      \put(4584,2770){\makebox(0,0)[r]{\strut{}0.06}}%
      \put(4716,1330){\makebox(0,0){\strut{}}}%
      \put(5598,1330){\makebox(0,0){\strut{}}}%
      \put(6480,1330){\makebox(0,0){\strut{}}}%
      \put(7361,1330){\makebox(0,0){\strut{}}}%
      \put(8243,1330){\makebox(0,0){\strut{}}}%
    }%
    \gplgaddtomacro\gplfronttext{%
      \csname LTb\endcsname
      \put(6832,2526){\makebox(0,0)[l]{\strut{}51~GeV}}%
    }%
    \gplgaddtomacro\gplbacktext{%
      \csname LTb\endcsname
      \put(4584,110){\makebox(0,0)[r]{\strut{} 0}}%
      \put(4584,313){\makebox(0,0)[r]{\strut{}1e-09}}%
      \put(4584,517){\makebox(0,0)[r]{\strut{}2e-09}}%
      \put(4584,720){\makebox(0,0)[r]{\strut{}3e-09}}%
      \put(4584,923){\makebox(0,0)[r]{\strut{}4e-09}}%
      \put(4584,1127){\makebox(0,0)[r]{\strut{}5e-09}}%
      \put(4584,1330){\makebox(0,0)[r]{\strut{}6e-09}}%
      \put(4716,-110){\makebox(0,0){\strut{}0}}%
      \put(5598,-110){\makebox(0,0){\strut{}0.25}}%
      \put(6480,-110){\makebox(0,0){\strut{}0.5}}%
      \put(7361,-110){\makebox(0,0){\strut{}0.75}}%
      \put(8243,-110){\makebox(0,0){\strut{}1}}%
    }%
    \gplgaddtomacro\gplfronttext{%
      \csname LTb\endcsname
      \put(6832,1086){\makebox(0,0)[l]{\strut{}511~GeV}}%
    }%
    \gplbacktext
    \put(0,0){\includegraphics{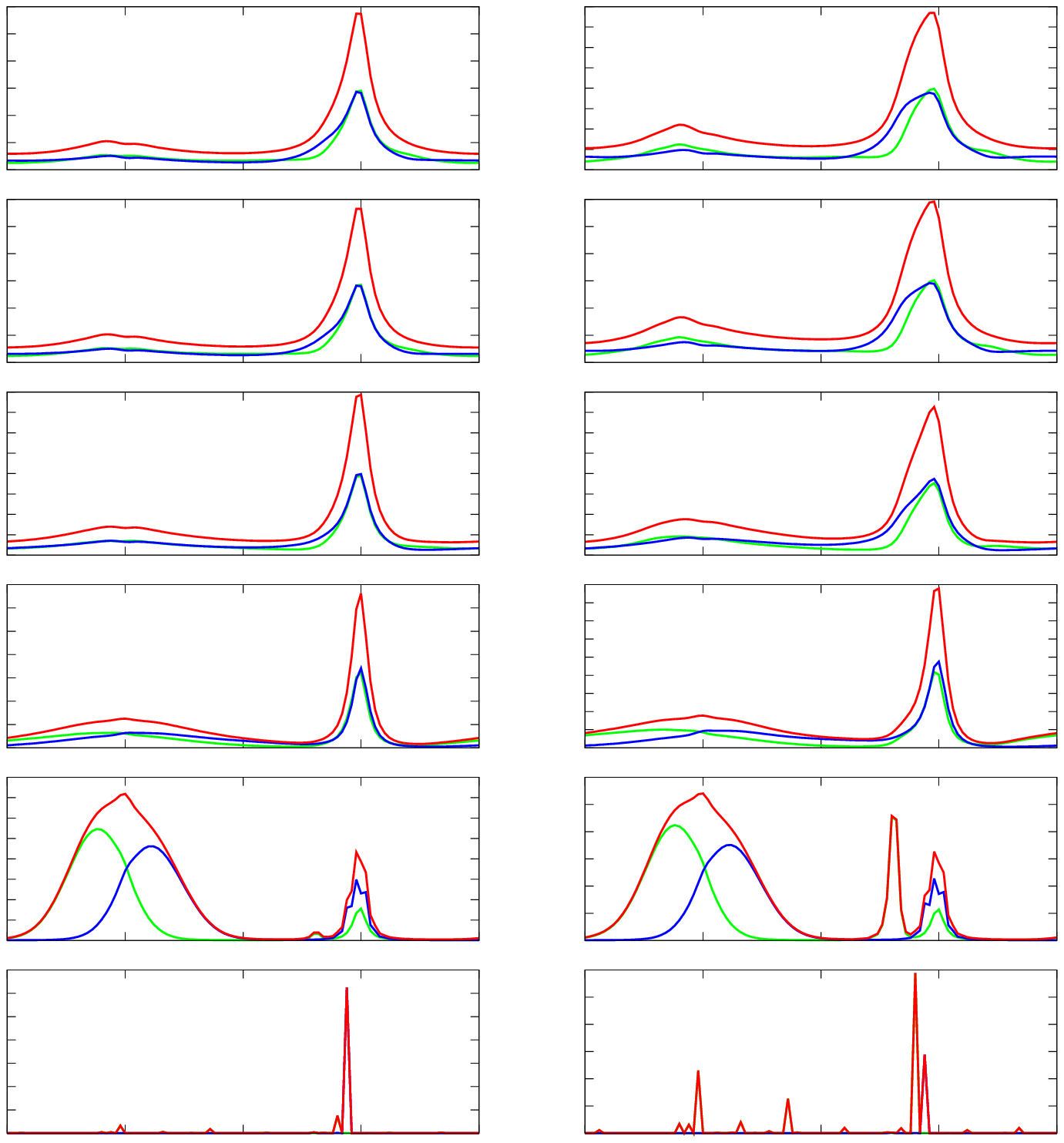}}%
    \gplfronttext
  \end{picture}%
\endgroup

%% file: courbe_lumiere_r005_b-6_c60_ri0.2_ro1_g10_dx_z60.tex
\begingroup
  \makeatletter
  \providecommand\color[2][]{%
    \GenericError{(gnuplot) \space\space\space\@spaces}{%
      Package color not loaded in conjunction with
      terminal option `colourtext'%
    }{See the gnuplot documentation for explanation.%
    }{Either use 'blacktext' in gnuplot or load the package
      color.sty in LaTeX.}%
    \renewcommand\color[2][]{}%
  }%
  \providecommand\includegraphics[2][]{%
    \GenericError{(gnuplot) \space\space\space\@spaces}{%
      Package graphicx or graphics not loaded%
    }{See the gnuplot documentation for explanation.%
    }{The gnuplot epslatex terminal needs graphicx.sty or graphics.sty.}%
    \renewcommand\includegraphics[2][]{}%
  }%
  \providecommand\rotatebox[2]{#2}%
  \@ifundefined{ifGPcolor}{%
    \newif\ifGPcolor
    \GPcolortrue
  }{}%
  \@ifundefined{ifGPblacktext}{%
    \newif\ifGPblacktext
    \GPblacktextfalse
  }{}%
  \let\gplgaddtomacro\g@addto@macro
  \gdef\gplbacktext{}%
  \gdef\gplfronttext{}%
  \makeatother
  \ifGPblacktext
    \def\colorrgb#1{}%
    \def\colorgray#1{}%
  \else
    \ifGPcolor
      \def\colorrgb#1{\color[rgb]{#1}}%
      \def\colorgray#1{\color[gray]{#1}}%
      \expandafter\def\csname LTw\endcsname{\color{white}}%
      \expandafter\def\csname LTb\endcsname{\color{black}}%
      \expandafter\def\csname LTa\endcsname{\color{black}}%
      \expandafter\def\csname LT0\endcsname{\color[rgb]{1,0,0}}%
      \expandafter\def\csname LT1\endcsname{\color[rgb]{0,1,0}}%
      \expandafter\def\csname LT2\endcsname{\color[rgb]{0,0,1}}%
      \expandafter\def\csname LT3\endcsname{\color[rgb]{1,0,1}}%
      \expandafter\def\csname LT4\endcsname{\color[rgb]{0,1,1}}%
      \expandafter\def\csname LT5\endcsname{\color[rgb]{1,1,0}}%
      \expandafter\def\csname LT6\endcsname{\color[rgb]{0,0,0}}%
      \expandafter\def\csname LT7\endcsname{\color[rgb]{1,0.3,0}}%
      \expandafter\def\csname LT8\endcsname{\color[rgb]{0.5,0.5,0.5}}%
    \else
      \def\colorrgb#1{\color{black}}%
      \def\colorgray#1{\color[gray]{#1}}%
      \expandafter\def\csname LTw\endcsname{\color{white}}%
      \expandafter\def\csname LTb\endcsname{\color{black}}%
      \expandafter\def\csname LTa\endcsname{\color{black}}%
      \expandafter\def\csname LT0\endcsname{\color{black}}%
      \expandafter\def\csname LT1\endcsname{\color{black}}%
      \expandafter\def\csname LT2\endcsname{\color{black}}%
      \expandafter\def\csname LT3\endcsname{\color{black}}%
      \expandafter\def\csname LT4\endcsname{\color{black}}%
      \expandafter\def\csname LT5\endcsname{\color{black}}%
      \expandafter\def\csname LT6\endcsname{\color{black}}%
      \expandafter\def\csname LT7\endcsname{\color{black}}%
      \expandafter\def\csname LT8\endcsname{\color{black}}%
    \fi
  \fi
    \setlength{\unitlength}{0.0500bp}%
    \ifx\gptboxheight\undefined%
      \newlength{\gptboxheight}%
      \newlength{\gptboxwidth}%
      \newsavebox{\gptboxtext}%
    \fi%
    \setlength{\fboxrule}{0.5pt}%
    \setlength{\fboxsep}{1pt}%
\begin{picture}(8640.00,8640.00)%
    \gplgaddtomacro\gplbacktext{%
      \csname LTb\endcsname
      \put(264,7310){\makebox(0,0)[r]{\strut{} 0}}%
      \put(264,7445){\makebox(0,0)[r]{\strut{}2e+03}}%
      \put(264,7581){\makebox(0,0)[r]{\strut{}4e+03}}%
      \put(264,7716){\makebox(0,0)[r]{\strut{}6e+03}}%
      \put(264,7852){\makebox(0,0)[r]{\strut{}8e+03}}%
      \put(264,7987){\makebox(0,0)[r]{\strut{}1e+04}}%
      \put(264,8123){\makebox(0,0)[r]{\strut{}1e+04}}%
      \put(264,8258){\makebox(0,0)[r]{\strut{}1e+04}}%
      \put(264,8394){\makebox(0,0)[r]{\strut{}2e+04}}%
      \put(264,8529){\makebox(0,0)[r]{\strut{}2e+04}}%
      \put(396,7090){\makebox(0,0){\strut{}}}%
      \put(1278,7090){\makebox(0,0){\strut{}}}%
      \put(2160,7090){\makebox(0,0){\strut{}}}%
      \put(3041,7090){\makebox(0,0){\strut{}}}%
      \put(3923,7090){\makebox(0,0){\strut{}}}%
    }%
    \gplgaddtomacro\gplfronttext{%
      \csname LTb\endcsname
      \put(2159,8859){\makebox(0,0){\strut{}$q=3$}}%
      \csname LTb\endcsname
      \put(2512,8285){\makebox(0,0)[l]{\strut{}5~MeV}}%
    }%
    \gplgaddtomacro\gplbacktext{%
      \csname LTb\endcsname
      \put(264,5870){\makebox(0,0)[r]{\strut{} 0}}%
      \put(264,6044){\makebox(0,0)[r]{\strut{}5e+03}}%
      \put(264,6218){\makebox(0,0)[r]{\strut{}1e+04}}%
      \put(264,6392){\makebox(0,0)[r]{\strut{}2e+04}}%
      \put(264,6567){\makebox(0,0)[r]{\strut{}2e+04}}%
      \put(264,6741){\makebox(0,0)[r]{\strut{}2e+04}}%
      \put(264,6915){\makebox(0,0)[r]{\strut{}3e+04}}%
      \put(264,7089){\makebox(0,0)[r]{\strut{}4e+04}}%
      \put(396,5650){\makebox(0,0){\strut{}}}%
      \put(1278,5650){\makebox(0,0){\strut{}}}%
      \put(2160,5650){\makebox(0,0){\strut{}}}%
      \put(3041,5650){\makebox(0,0){\strut{}}}%
      \put(3923,5650){\makebox(0,0){\strut{}}}%
    }%
    \gplgaddtomacro\gplfronttext{%
      \csname LTb\endcsname
      \put(2512,6845){\makebox(0,0)[l]{\strut{}51~MeV}}%
    }%
    \gplgaddtomacro\gplbacktext{%
      \csname LTb\endcsname
      \put(264,4430){\makebox(0,0)[r]{\strut{} 0}}%
      \put(264,4565){\makebox(0,0)[r]{\strut{}5e+03}}%
      \put(264,4701){\makebox(0,0)[r]{\strut{}1e+04}}%
      \put(264,4836){\makebox(0,0)[r]{\strut{}2e+04}}%
      \put(264,4972){\makebox(0,0)[r]{\strut{}2e+04}}%
      \put(264,5107){\makebox(0,0)[r]{\strut{}2e+04}}%
      \put(264,5243){\makebox(0,0)[r]{\strut{}3e+04}}%
      \put(264,5378){\makebox(0,0)[r]{\strut{}4e+04}}%
      \put(264,5514){\makebox(0,0)[r]{\strut{}4e+04}}%
      \put(264,5649){\makebox(0,0)[r]{\strut{}4e+04}}%
      \put(396,4210){\makebox(0,0){\strut{}}}%
      \put(1278,4210){\makebox(0,0){\strut{}}}%
      \put(2160,4210){\makebox(0,0){\strut{}}}%
      \put(3041,4210){\makebox(0,0){\strut{}}}%
      \put(3923,4210){\makebox(0,0){\strut{}}}%
    }%
    \gplgaddtomacro\gplfronttext{%
      \csname LTb\endcsname
      \put(2512,5405){\makebox(0,0)[l]{\strut{}511~MeV}}%
    }%
    \gplgaddtomacro\gplbacktext{%
      \csname LTb\endcsname
      \put(264,2990){\makebox(0,0)[r]{\strut{} 0}}%
      \put(264,3101){\makebox(0,0)[r]{\strut{}1e+03}}%
      \put(264,3212){\makebox(0,0)[r]{\strut{}2e+03}}%
      \put(264,3323){\makebox(0,0)[r]{\strut{}3e+03}}%
      \put(264,3434){\makebox(0,0)[r]{\strut{}4e+03}}%
      \put(264,3545){\makebox(0,0)[r]{\strut{}5e+03}}%
      \put(264,3655){\makebox(0,0)[r]{\strut{}6e+03}}%
      \put(264,3766){\makebox(0,0)[r]{\strut{}7e+03}}%
      \put(264,3877){\makebox(0,0)[r]{\strut{}8e+03}}%
      \put(264,3988){\makebox(0,0)[r]{\strut{}9e+03}}%
      \put(264,4099){\makebox(0,0)[r]{\strut{}1e+04}}%
      \put(264,4210){\makebox(0,0)[r]{\strut{}1e+04}}%
      \put(396,2770){\makebox(0,0){\strut{}}}%
      \put(1278,2770){\makebox(0,0){\strut{}}}%
      \put(2160,2770){\makebox(0,0){\strut{}}}%
      \put(3041,2770){\makebox(0,0){\strut{}}}%
      \put(3923,2770){\makebox(0,0){\strut{}}}%
    }%
    \gplgaddtomacro\gplfronttext{%
      \csname LTb\endcsname
      \put(2512,3966){\makebox(0,0)[l]{\strut{}5~GeV}}%
    }%
    \gplgaddtomacro\gplbacktext{%
      \csname LTb\endcsname
      \put(264,1550){\makebox(0,0)[r]{\strut{} 0}}%
      \put(264,1794){\makebox(0,0)[r]{\strut{}0.05}}%
      \put(264,2038){\makebox(0,0)[r]{\strut{}0.1}}%
      \put(264,2282){\makebox(0,0)[r]{\strut{}0.2}}%
      \put(264,2526){\makebox(0,0)[r]{\strut{}0.2}}%
      \put(264,2770){\makebox(0,0)[r]{\strut{}0.2}}%
      \put(396,1330){\makebox(0,0){\strut{}}}%
      \put(1278,1330){\makebox(0,0){\strut{}}}%
      \put(2160,1330){\makebox(0,0){\strut{}}}%
      \put(3041,1330){\makebox(0,0){\strut{}}}%
      \put(3923,1330){\makebox(0,0){\strut{}}}%
    }%
    \gplgaddtomacro\gplfronttext{%
      \csname LTb\endcsname
      \put(2512,2526){\makebox(0,0)[l]{\strut{}51~GeV}}%
    }%
    \gplgaddtomacro\gplbacktext{%
      \csname LTb\endcsname
      \put(264,110){\makebox(0,0)[r]{\strut{} 0}}%
      \put(264,284){\makebox(0,0)[r]{\strut{}1e-07}}%
      \put(264,459){\makebox(0,0)[r]{\strut{}2e-07}}%
      \put(264,633){\makebox(0,0)[r]{\strut{}3e-07}}%
      \put(264,807){\makebox(0,0)[r]{\strut{}4e-07}}%
      \put(264,981){\makebox(0,0)[r]{\strut{}5e-07}}%
      \put(264,1156){\makebox(0,0)[r]{\strut{}6e-07}}%
      \put(264,1330){\makebox(0,0)[r]{\strut{}7e-07}}%
      \put(396,-110){\makebox(0,0){\strut{}0}}%
      \put(1278,-110){\makebox(0,0){\strut{}0.25}}%
      \put(2160,-110){\makebox(0,0){\strut{}0.5}}%
      \put(3041,-110){\makebox(0,0){\strut{}0.75}}%
      \put(3923,-110){\makebox(0,0){\strut{}1}}%
    }%
    \gplgaddtomacro\gplfronttext{%
      \csname LTb\endcsname
      \put(2512,1086){\makebox(0,0)[l]{\strut{}511~GeV}}%
    }%
    \gplgaddtomacro\gplbacktext{%
      \csname LTb\endcsname
      \put(4584,7310){\makebox(0,0)[r]{\strut{} 0}}%
      \put(4584,7513){\makebox(0,0)[r]{\strut{}2e+05}}%
      \put(4584,7716){\makebox(0,0)[r]{\strut{}4e+05}}%
      \put(4584,7920){\makebox(0,0)[r]{\strut{}6e+05}}%
      \put(4584,8123){\makebox(0,0)[r]{\strut{}8e+05}}%
      \put(4584,8326){\makebox(0,0)[r]{\strut{}1e+06}}%
      \put(4584,8529){\makebox(0,0)[r]{\strut{}1e+06}}%
      \put(4716,7090){\makebox(0,0){\strut{}}}%
      \put(5598,7090){\makebox(0,0){\strut{}}}%
      \put(6480,7090){\makebox(0,0){\strut{}}}%
      \put(7361,7090){\makebox(0,0){\strut{}}}%
      \put(8243,7090){\makebox(0,0){\strut{}}}%
    }%
    \gplgaddtomacro\gplfronttext{%
      \csname LTb\endcsname
      \put(6479,8859){\makebox(0,0){\strut{}$q=1$}}%
      \csname LTb\endcsname
      \put(6832,8285){\makebox(0,0)[l]{\strut{}5~MeV}}%
    }%
    \gplgaddtomacro\gplbacktext{%
      \csname LTb\endcsname
      \put(4584,5870){\makebox(0,0)[r]{\strut{} 0}}%
      \put(4584,6114){\makebox(0,0)[r]{\strut{}5e+05}}%
      \put(4584,6358){\makebox(0,0)[r]{\strut{}1e+06}}%
      \put(4584,6601){\makebox(0,0)[r]{\strut{}2e+06}}%
      \put(4584,6845){\makebox(0,0)[r]{\strut{}2e+06}}%
      \put(4584,7089){\makebox(0,0)[r]{\strut{}2e+06}}%
      \put(4716,5650){\makebox(0,0){\strut{}}}%
      \put(5598,5650){\makebox(0,0){\strut{}}}%
      \put(6480,5650){\makebox(0,0){\strut{}}}%
      \put(7361,5650){\makebox(0,0){\strut{}}}%
      \put(8243,5650){\makebox(0,0){\strut{}}}%
    }%
    \gplgaddtomacro\gplfronttext{%
      \csname LTb\endcsname
      \put(6832,6845){\makebox(0,0)[l]{\strut{}51~MeV}}%
    }%
    \gplgaddtomacro\gplbacktext{%
      \csname LTb\endcsname
      \put(4584,4430){\makebox(0,0)[r]{\strut{} 0}}%
      \put(4584,4633){\makebox(0,0)[r]{\strut{}5e+05}}%
      \put(4584,4836){\makebox(0,0)[r]{\strut{}1e+06}}%
      \put(4584,5040){\makebox(0,0)[r]{\strut{}2e+06}}%
      \put(4584,5243){\makebox(0,0)[r]{\strut{}2e+06}}%
      \put(4584,5446){\makebox(0,0)[r]{\strut{}2e+06}}%
      \put(4584,5649){\makebox(0,0)[r]{\strut{}3e+06}}%
      \put(4716,4210){\makebox(0,0){\strut{}}}%
      \put(5598,4210){\makebox(0,0){\strut{}}}%
      \put(6480,4210){\makebox(0,0){\strut{}}}%
      \put(7361,4210){\makebox(0,0){\strut{}}}%
      \put(8243,4210){\makebox(0,0){\strut{}}}%
    }%
    \gplgaddtomacro\gplfronttext{%
      \csname LTb\endcsname
      \put(6832,5405){\makebox(0,0)[l]{\strut{}511~MeV}}%
    }%
    \gplgaddtomacro\gplbacktext{%
      \csname LTb\endcsname
      \put(4584,2990){\makebox(0,0)[r]{\strut{} 0}}%
      \put(4584,3164){\makebox(0,0)[r]{\strut{}5e+04}}%
      \put(4584,3339){\makebox(0,0)[r]{\strut{}1e+05}}%
      \put(4584,3513){\makebox(0,0)[r]{\strut{}2e+05}}%
      \put(4584,3687){\makebox(0,0)[r]{\strut{}2e+05}}%
      \put(4584,3861){\makebox(0,0)[r]{\strut{}2e+05}}%
      \put(4584,4036){\makebox(0,0)[r]{\strut{}3e+05}}%
      \put(4584,4210){\makebox(0,0)[r]{\strut{}4e+05}}%
      \put(4716,2770){\makebox(0,0){\strut{}}}%
      \put(5598,2770){\makebox(0,0){\strut{}}}%
      \put(6480,2770){\makebox(0,0){\strut{}}}%
      \put(7361,2770){\makebox(0,0){\strut{}}}%
      \put(8243,2770){\makebox(0,0){\strut{}}}%
    }%
    \gplgaddtomacro\gplfronttext{%
      \csname LTb\endcsname
      \put(6832,3966){\makebox(0,0)[l]{\strut{}5~GeV}}%
    }%
    \gplgaddtomacro\gplbacktext{%
      \csname LTb\endcsname
      \put(4584,1550){\makebox(0,0)[r]{\strut{} 0}}%
      \put(4584,1686){\makebox(0,0)[r]{\strut{}0.5}}%
      \put(4584,1821){\makebox(0,0)[r]{\strut{} 1}}%
      \put(4584,1957){\makebox(0,0)[r]{\strut{} 2}}%
      \put(4584,2092){\makebox(0,0)[r]{\strut{} 2}}%
      \put(4584,2228){\makebox(0,0)[r]{\strut{} 2}}%
      \put(4584,2363){\makebox(0,0)[r]{\strut{} 3}}%
      \put(4584,2499){\makebox(0,0)[r]{\strut{} 4}}%
      \put(4584,2634){\makebox(0,0)[r]{\strut{} 4}}%
      \put(4584,2770){\makebox(0,0)[r]{\strut{} 4}}%
      \put(4716,1330){\makebox(0,0){\strut{}}}%
      \put(5598,1330){\makebox(0,0){\strut{}}}%
      \put(6480,1330){\makebox(0,0){\strut{}}}%
      \put(7361,1330){\makebox(0,0){\strut{}}}%
      \put(8243,1330){\makebox(0,0){\strut{}}}%
    }%
    \gplgaddtomacro\gplfronttext{%
      \csname LTb\endcsname
      \put(6832,2526){\makebox(0,0)[l]{\strut{}51~GeV}}%
    }%
    \gplgaddtomacro\gplbacktext{%
      \csname LTb\endcsname
      \put(4584,110){\makebox(0,0)[r]{\strut{} 0}}%
      \put(4584,263){\makebox(0,0)[r]{\strut{}2e-06}}%
      \put(4584,415){\makebox(0,0)[r]{\strut{}4e-06}}%
      \put(4584,568){\makebox(0,0)[r]{\strut{}6e-06}}%
      \put(4584,720){\makebox(0,0)[r]{\strut{}8e-06}}%
      \put(4584,873){\makebox(0,0)[r]{\strut{}1e-05}}%
      \put(4584,1025){\makebox(0,0)[r]{\strut{}1e-05}}%
      \put(4584,1178){\makebox(0,0)[r]{\strut{}1e-05}}%
      \put(4584,1330){\makebox(0,0)[r]{\strut{}2e-05}}%
      \put(4716,-110){\makebox(0,0){\strut{}0}}%
      \put(5598,-110){\makebox(0,0){\strut{}0.25}}%
      \put(6480,-110){\makebox(0,0){\strut{}0.5}}%
      \put(7361,-110){\makebox(0,0){\strut{}0.75}}%
      \put(8243,-110){\makebox(0,0){\strut{}1}}%
    }%
    \gplgaddtomacro\gplfronttext{%
      \csname LTb\endcsname
      \put(6832,1086){\makebox(0,0)[l]{\strut{}511~GeV}}%
    }%
    \gplbacktext
    \put(0,0){\includegraphics{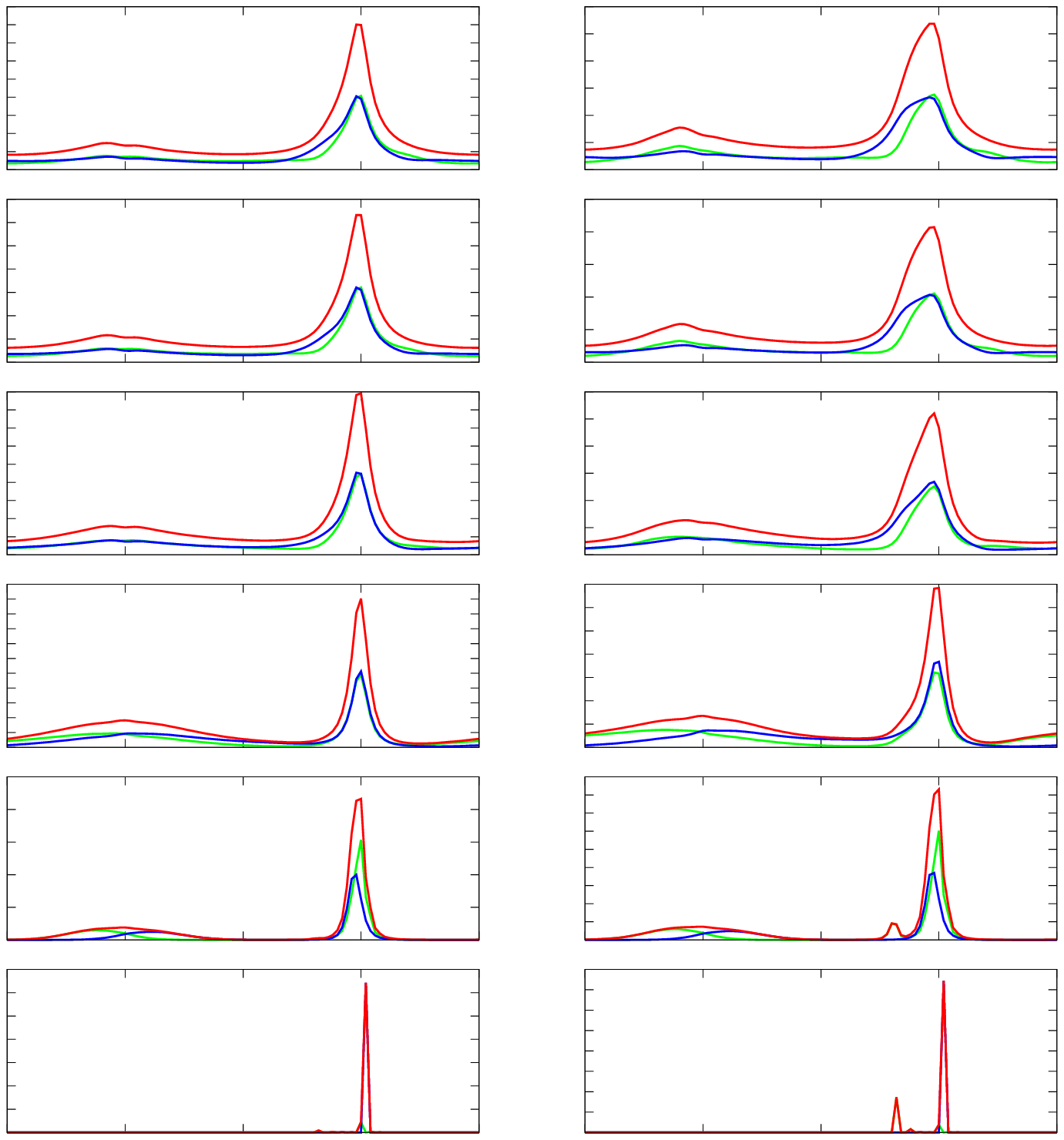}}%
    \gplfronttext
  \end{picture}%
\endgroup

%% file: Ecut_r0.0025_c60_q0_g10_nr64.tex
\begingroup
  \makeatletter
  \providecommand\color[2][]{%
    \GenericError{(gnuplot) \space\space\space\@spaces}{%
      Package color not loaded in conjunction with
      terminal option `colourtext'%
    }{See the gnuplot documentation for explanation.%
    }{Either use 'blacktext' in gnuplot or load the package
      color.sty in LaTeX.}%
    \renewcommand\color[2][]{}%
  }%
  \providecommand\includegraphics[2][]{%
    \GenericError{(gnuplot) \space\space\space\@spaces}{%
      Package graphicx or graphics not loaded%
    }{See the gnuplot documentation for explanation.%
    }{The gnuplot epslatex terminal needs graphicx.sty or graphics.sty.}%
    \renewcommand\includegraphics[2][]{}%
  }%
  \providecommand\rotatebox[2]{#2}%
  \@ifundefined{ifGPcolor}{%
    \newif\ifGPcolor
    \GPcolortrue
  }{}%
  \@ifundefined{ifGPblacktext}{%
    \newif\ifGPblacktext
    \GPblacktextfalse
  }{}%
  \let\gplgaddtomacro\g@addto@macro
  \gdef\gplbacktext{}%
  \gdef\gplfronttext{}%
  \makeatother
  \ifGPblacktext
    \def\colorrgb#1{}%
    \def\colorgray#1{}%
  \else
    \ifGPcolor
      \def\colorrgb#1{\color[rgb]{#1}}%
      \def\colorgray#1{\color[gray]{#1}}%
      \expandafter\def\csname LTw\endcsname{\color{white}}%
      \expandafter\def\csname LTb\endcsname{\color{black}}%
      \expandafter\def\csname LTa\endcsname{\color{black}}%
      \expandafter\def\csname LT0\endcsname{\color[rgb]{1,0,0}}%
      \expandafter\def\csname LT1\endcsname{\color[rgb]{0,1,0}}%
      \expandafter\def\csname LT2\endcsname{\color[rgb]{0,0,1}}%
      \expandafter\def\csname LT3\endcsname{\color[rgb]{1,0,1}}%
      \expandafter\def\csname LT4\endcsname{\color[rgb]{0,1,1}}%
      \expandafter\def\csname LT5\endcsname{\color[rgb]{1,1,0}}%
      \expandafter\def\csname LT6\endcsname{\color[rgb]{0,0,0}}%
      \expandafter\def\csname LT7\endcsname{\color[rgb]{1,0.3,0}}%
      \expandafter\def\csname LT8\endcsname{\color[rgb]{0.5,0.5,0.5}}%
    \else
      \def\colorrgb#1{\color{black}}%
      \def\colorgray#1{\color[gray]{#1}}%
      \expandafter\def\csname LTw\endcsname{\color{white}}%
      \expandafter\def\csname LTb\endcsname{\color{black}}%
      \expandafter\def\csname LTa\endcsname{\color{black}}%
      \expandafter\def\csname LT0\endcsname{\color{black}}%
      \expandafter\def\csname LT1\endcsname{\color{black}}%
      \expandafter\def\csname LT2\endcsname{\color{black}}%
      \expandafter\def\csname LT3\endcsname{\color{black}}%
      \expandafter\def\csname LT4\endcsname{\color{black}}%
      \expandafter\def\csname LT5\endcsname{\color{black}}%
      \expandafter\def\csname LT6\endcsname{\color{black}}%
      \expandafter\def\csname LT7\endcsname{\color{black}}%
      \expandafter\def\csname LT8\endcsname{\color{black}}%
    \fi
  \fi
    \setlength{\unitlength}{0.0500bp}%
    \ifx\gptboxheight\undefined%
      \newlength{\gptboxheight}%
      \newlength{\gptboxwidth}%
      \newsavebox{\gptboxtext}%
    \fi%
    \setlength{\fboxrule}{0.5pt}%
    \setlength{\fboxsep}{1pt}%
\begin{picture}(5760.00,4320.00)%
    \gplgaddtomacro\gplbacktext{%
      \csname LTb\endcsname
      \put(946,704){\makebox(0,0)[r]{\strut{}$-2.5$}}%
      \put(946,1044){\makebox(0,0)[r]{\strut{}$-2$}}%
      \put(946,1383){\makebox(0,0)[r]{\strut{}$-1.5$}}%
      \put(946,1723){\makebox(0,0)[r]{\strut{}$-1$}}%
      \put(946,2062){\makebox(0,0)[r]{\strut{}$-0.5$}}%
      \put(946,2402){\makebox(0,0)[r]{\strut{}$0$}}%
      \put(946,2741){\makebox(0,0)[r]{\strut{}$0.5$}}%
      \put(946,3081){\makebox(0,0)[r]{\strut{}$1$}}%
      \put(946,3420){\makebox(0,0)[r]{\strut{}$1.5$}}%
      \put(946,3760){\makebox(0,0)[r]{\strut{}$2$}}%
      \put(946,4099){\makebox(0,0)[r]{\strut{}$2.5$}}%
      \put(1435,484){\makebox(0,0){\strut{}$-6$}}%
      \put(2149,484){\makebox(0,0){\strut{}$-5$}}%
      \put(2863,484){\makebox(0,0){\strut{}$-4$}}%
      \put(3578,484){\makebox(0,0){\strut{}$-3$}}%
      \put(4292,484){\makebox(0,0){\strut{}$-2$}}%
      \put(5006,484){\makebox(0,0){\strut{}$-1$}}%
    }%
    \gplgaddtomacro\gplfronttext{%
      \csname LTb\endcsname
      \put(198,2401){\rotatebox{-270}{\makebox(0,0){\strut{}$\log E_{\rm cut} ~ (\textrm{GeV})$}}}%
      \put(3220,154){\makebox(0,0){\strut{}$\log(B/B_{\rm qed})$}}%
      \csname LTb\endcsname
      \put(2398,3926){\makebox(0,0)[r]{\strut{}$r_{\rm in}=0.1$}}%
      \csname LTb\endcsname
      \put(2398,3706){\makebox(0,0)[r]{\strut{}0.2}}%
      \csname LTb\endcsname
      \put(2398,3486){\makebox(0,0)[r]{\strut{}0.5}}%
      \csname LTb\endcsname
      \put(2398,3266){\makebox(0,0)[r]{\strut{}$B^{3/4}$}}%
    }%
    \gplbacktext
    \put(0,0){\includegraphics{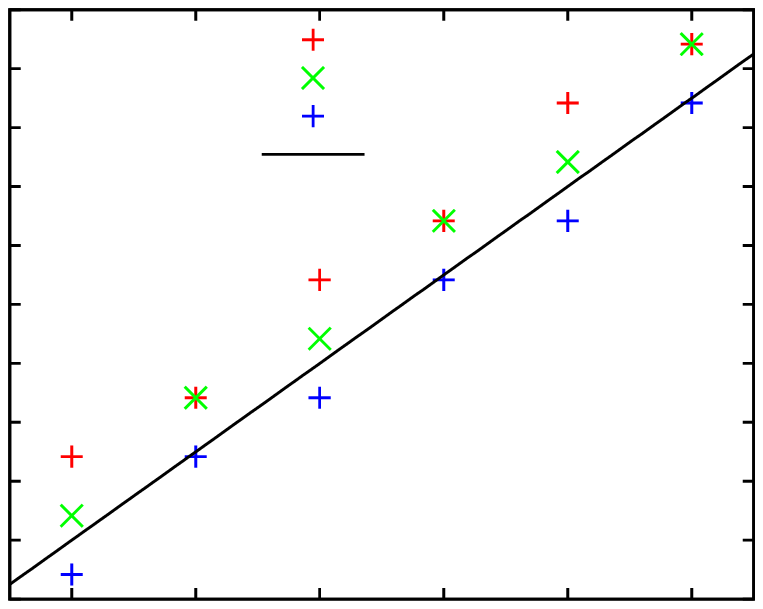}}%
    \gplfronttext
  \end{picture}%
\endgroup

%% file: Ecut_r0.05_c60_q0_g10_nr64.tex
\begingroup
  \makeatletter
  \providecommand\color[2][]{%
    \GenericError{(gnuplot) \space\space\space\@spaces}{%
      Package color not loaded in conjunction with
      terminal option `colourtext'%
    }{See the gnuplot documentation for explanation.%
    }{Either use 'blacktext' in gnuplot or load the package
      color.sty in LaTeX.}%
    \renewcommand\color[2][]{}%
  }%
  \providecommand\includegraphics[2][]{%
    \GenericError{(gnuplot) \space\space\space\@spaces}{%
      Package graphicx or graphics not loaded%
    }{See the gnuplot documentation for explanation.%
    }{The gnuplot epslatex terminal needs graphicx.sty or graphics.sty.}%
    \renewcommand\includegraphics[2][]{}%
  }%
  \providecommand\rotatebox[2]{#2}%
  \@ifundefined{ifGPcolor}{%
    \newif\ifGPcolor
    \GPcolortrue
  }{}%
  \@ifundefined{ifGPblacktext}{%
    \newif\ifGPblacktext
    \GPblacktextfalse
  }{}%
  \let\gplgaddtomacro\g@addto@macro
  \gdef\gplbacktext{}%
  \gdef\gplfronttext{}%
  \makeatother
  \ifGPblacktext
    \def\colorrgb#1{}%
    \def\colorgray#1{}%
  \else
    \ifGPcolor
      \def\colorrgb#1{\color[rgb]{#1}}%
      \def\colorgray#1{\color[gray]{#1}}%
      \expandafter\def\csname LTw\endcsname{\color{white}}%
      \expandafter\def\csname LTb\endcsname{\color{black}}%
      \expandafter\def\csname LTa\endcsname{\color{black}}%
      \expandafter\def\csname LT0\endcsname{\color[rgb]{1,0,0}}%
      \expandafter\def\csname LT1\endcsname{\color[rgb]{0,1,0}}%
      \expandafter\def\csname LT2\endcsname{\color[rgb]{0,0,1}}%
      \expandafter\def\csname LT3\endcsname{\color[rgb]{1,0,1}}%
      \expandafter\def\csname LT4\endcsname{\color[rgb]{0,1,1}}%
      \expandafter\def\csname LT5\endcsname{\color[rgb]{1,1,0}}%
      \expandafter\def\csname LT6\endcsname{\color[rgb]{0,0,0}}%
      \expandafter\def\csname LT7\endcsname{\color[rgb]{1,0.3,0}}%
      \expandafter\def\csname LT8\endcsname{\color[rgb]{0.5,0.5,0.5}}%
    \else
      \def\colorrgb#1{\color{black}}%
      \def\colorgray#1{\color[gray]{#1}}%
      \expandafter\def\csname LTw\endcsname{\color{white}}%
      \expandafter\def\csname LTb\endcsname{\color{black}}%
      \expandafter\def\csname LTa\endcsname{\color{black}}%
      \expandafter\def\csname LT0\endcsname{\color{black}}%
      \expandafter\def\csname LT1\endcsname{\color{black}}%
      \expandafter\def\csname LT2\endcsname{\color{black}}%
      \expandafter\def\csname LT3\endcsname{\color{black}}%
      \expandafter\def\csname LT4\endcsname{\color{black}}%
      \expandafter\def\csname LT5\endcsname{\color{black}}%
      \expandafter\def\csname LT6\endcsname{\color{black}}%
      \expandafter\def\csname LT7\endcsname{\color{black}}%
      \expandafter\def\csname LT8\endcsname{\color{black}}%
    \fi
  \fi
    \setlength{\unitlength}{0.0500bp}%
    \ifx\gptboxheight\undefined%
      \newlength{\gptboxheight}%
      \newlength{\gptboxwidth}%
      \newsavebox{\gptboxtext}%
    \fi%
    \setlength{\fboxrule}{0.5pt}%
    \setlength{\fboxsep}{1pt}%
\begin{picture}(5760.00,4320.00)%
    \gplgaddtomacro\gplbacktext{%
      \csname LTb\endcsname
      \put(814,704){\makebox(0,0)[r]{\strut{}$0$}}%
      \put(814,1044){\makebox(0,0)[r]{\strut{}$0.5$}}%
      \put(814,1383){\makebox(0,0)[r]{\strut{}$1$}}%
      \put(814,1723){\makebox(0,0)[r]{\strut{}$1.5$}}%
      \put(814,2062){\makebox(0,0)[r]{\strut{}$2$}}%
      \put(814,2402){\makebox(0,0)[r]{\strut{}$2.5$}}%
      \put(814,2741){\makebox(0,0)[r]{\strut{}$3$}}%
      \put(814,3081){\makebox(0,0)[r]{\strut{}$3.5$}}%
      \put(814,3420){\makebox(0,0)[r]{\strut{}$4$}}%
      \put(814,3760){\makebox(0,0)[r]{\strut{}$4.5$}}%
      \put(814,4099){\makebox(0,0)[r]{\strut{}$5$}}%
      \put(1314,484){\makebox(0,0){\strut{}$-6$}}%
      \put(2050,484){\makebox(0,0){\strut{}$-5$}}%
      \put(2786,484){\makebox(0,0){\strut{}$-4$}}%
      \put(3523,484){\makebox(0,0){\strut{}$-3$}}%
      \put(4259,484){\makebox(0,0){\strut{}$-2$}}%
      \put(4995,484){\makebox(0,0){\strut{}$-1$}}%
    }%
    \gplgaddtomacro\gplfronttext{%
      \csname LTb\endcsname
      \put(198,2401){\rotatebox{-270}{\makebox(0,0){\strut{}$\log E_{\rm cut} ~ (\textrm{GeV})$}}}%
      \put(3154,154){\makebox(0,0){\strut{}$\log(B/B_{\rm qed})$}}%
      \csname LTb\endcsname
      \put(2266,3926){\makebox(0,0)[r]{\strut{}$r_{\rm in}=0.1$}}%
      \csname LTb\endcsname
      \put(2266,3706){\makebox(0,0)[r]{\strut{}0.2}}%
      \csname LTb\endcsname
      \put(2266,3486){\makebox(0,0)[r]{\strut{}0.5}}%
      \csname LTb\endcsname
      \put(2266,3266){\makebox(0,0)[r]{\strut{}$B^{3/4}$}}%
    }%
    \gplbacktext
    \put(0,0){\includegraphics{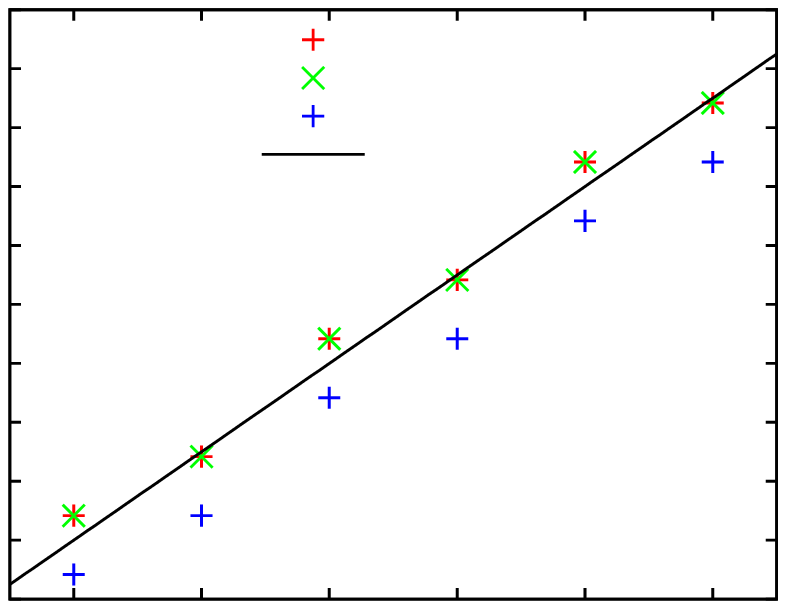}}%
    \gplfronttext
  \end{picture}%
\endgroup

%% file: courbure_s_ms_c60.tex
\begingroup
  \makeatletter
  \providecommand\color[2][]{%
    \GenericError{(gnuplot) \space\space\space\@spaces}{%
      Package color not loaded in conjunction with
      terminal option `colourtext'%
    }{See the gnuplot documentation for explanation.%
    }{Either use 'blacktext' in gnuplot or load the package
      color.sty in LaTeX.}%
    \renewcommand\color[2][]{}%
  }%
  \providecommand\includegraphics[2][]{%
    \GenericError{(gnuplot) \space\space\space\@spaces}{%
      Package graphicx or graphics not loaded%
    }{See the gnuplot documentation for explanation.%
    }{The gnuplot epslatex terminal needs graphicx.sty or graphics.sty.}%
    \renewcommand\includegraphics[2][]{}%
  }%
  \providecommand\rotatebox[2]{#2}%
  \@ifundefined{ifGPcolor}{%
    \newif\ifGPcolor
    \GPcolortrue
  }{}%
  \@ifundefined{ifGPblacktext}{%
    \newif\ifGPblacktext
    \GPblacktextfalse
  }{}%
  \let\gplgaddtomacro\g@addto@macro
  \gdef\gplbacktext{}%
  \gdef\gplfronttext{}%
  \makeatother
  \ifGPblacktext
    \def\colorrgb#1{}%
    \def\colorgray#1{}%
  \else
    \ifGPcolor
      \def\colorrgb#1{\color[rgb]{#1}}%
      \def\colorgray#1{\color[gray]{#1}}%
      \expandafter\def\csname LTw\endcsname{\color{white}}%
      \expandafter\def\csname LTb\endcsname{\color{black}}%
      \expandafter\def\csname LTa\endcsname{\color{black}}%
      \expandafter\def\csname LT0\endcsname{\color[rgb]{1,0,0}}%
      \expandafter\def\csname LT1\endcsname{\color[rgb]{0,1,0}}%
      \expandafter\def\csname LT2\endcsname{\color[rgb]{0,0,1}}%
      \expandafter\def\csname LT3\endcsname{\color[rgb]{1,0,1}}%
      \expandafter\def\csname LT4\endcsname{\color[rgb]{0,1,1}}%
      \expandafter\def\csname LT5\endcsname{\color[rgb]{1,1,0}}%
      \expandafter\def\csname LT6\endcsname{\color[rgb]{0,0,0}}%
      \expandafter\def\csname LT7\endcsname{\color[rgb]{1,0.3,0}}%
      \expandafter\def\csname LT8\endcsname{\color[rgb]{0.5,0.5,0.5}}%
    \else
      \def\colorrgb#1{\color{black}}%
      \def\colorgray#1{\color[gray]{#1}}%
      \expandafter\def\csname LTw\endcsname{\color{white}}%
      \expandafter\def\csname LTb\endcsname{\color{black}}%
      \expandafter\def\csname LTa\endcsname{\color{black}}%
      \expandafter\def\csname LT0\endcsname{\color{black}}%
      \expandafter\def\csname LT1\endcsname{\color{black}}%
      \expandafter\def\csname LT2\endcsname{\color{black}}%
      \expandafter\def\csname LT3\endcsname{\color{black}}%
      \expandafter\def\csname LT4\endcsname{\color{black}}%
      \expandafter\def\csname LT5\endcsname{\color{black}}%
      \expandafter\def\csname LT6\endcsname{\color{black}}%
      \expandafter\def\csname LT7\endcsname{\color{black}}%
      \expandafter\def\csname LT8\endcsname{\color{black}}%
    \fi
  \fi
    \setlength{\unitlength}{0.0500bp}%
    \ifx\gptboxheight\undefined%
      \newlength{\gptboxheight}%
      \newlength{\gptboxwidth}%
      \newsavebox{\gptboxtext}%
    \fi%
    \setlength{\fboxrule}{0.5pt}%
    \setlength{\fboxsep}{1pt}%
\begin{picture}(5760.00,8640.00)%
    \gplgaddtomacro\gplbacktext{%
      \csname LTb\endcsname%
      \put(444,4664){\makebox(0,0)[r]{\strut{}-2}}%
      \put(444,5053){\makebox(0,0)[r]{\strut{}}}%
      \put(444,5442){\makebox(0,0)[r]{\strut{}-1}}%
      \put(444,5831){\makebox(0,0)[r]{\strut{}}}%
      \put(444,6220){\makebox(0,0)[r]{\strut{}0}}%
      \put(444,6609){\makebox(0,0)[r]{\strut{}}}%
      \put(444,6998){\makebox(0,0)[r]{\strut{}1}}%
      \put(444,7387){\makebox(0,0)[r]{\strut{}}}%
      \put(444,7776){\makebox(0,0)[r]{\strut{}2}}%
      \put(964,4445){\makebox(0,0){\strut{}}}%
      \put(1353,4445){\makebox(0,0){\strut{}-1}}%
      \put(1742,4445){\makebox(0,0){\strut{}}}%
      \put(2131,4445){\makebox(0,0){\strut{}0}}%
      \put(2520,4445){\makebox(0,0){\strut{}}}%
      \put(2909,4445){\makebox(0,0){\strut{}1}}%
      \put(3298,4445){\makebox(0,0){\strut{}}}%
    }%
    \gplgaddtomacro\gplfronttext{%
      \csname LTb\endcsname%
      \put(2131,8105){\makebox(0,0){\strut{}$\log \rho_c(e^-)$ for 0.1~s PSR}}%
      \csname LTb\endcsname%
      \put(4051,4665){\makebox(0,0)[l]{\strut{}$-2$}}%
      \put(4051,4947){\makebox(0,0)[l]{\strut{}$-1.5$}}%
      \put(4051,5230){\makebox(0,0)[l]{\strut{}$-1$}}%
      \put(4051,5513){\makebox(0,0)[l]{\strut{}$-0.5$}}%
      \put(4051,5795){\makebox(0,0)[l]{\strut{}$0$}}%
      \put(4051,6078){\makebox(0,0)[l]{\strut{}$0.5$}}%
      \put(4051,6361){\makebox(0,0)[l]{\strut{}$1$}}%
      \put(4051,6644){\makebox(0,0)[l]{\strut{}$1.5$}}%
      \put(4051,6926){\makebox(0,0)[l]{\strut{}$2$}}%
      \put(4051,7209){\makebox(0,0)[l]{\strut{}$2.5$}}%
      \put(4051,7492){\makebox(0,0)[l]{\strut{}$3$}}%
      \put(4051,7775){\makebox(0,0)[l]{\strut{}$3.5$}}%
    }%
    \gplgaddtomacro\gplbacktext{%
      \csname LTb\endcsname%
      \put(444,863){\makebox(0,0)[r]{\strut{}}}%
      \put(444,1252){\makebox(0,0)[r]{\strut{}}}%
      \put(444,1641){\makebox(0,0)[r]{\strut{}}}%
      \put(444,2030){\makebox(0,0)[r]{\strut{}}}%
      \put(444,2419){\makebox(0,0)[r]{\strut{}}}%
      \put(444,2807){\makebox(0,0)[r]{\strut{}}}%
      \put(444,3196){\makebox(0,0)[r]{\strut{}}}%
      \put(444,3585){\makebox(0,0)[r]{\strut{}}}%
      \put(444,3974){\makebox(0,0)[r]{\strut{}}}%
      \put(964,644){\makebox(0,0){\strut{}}}%
      \put(1353,644){\makebox(0,0){\strut{}-1}}%
      \put(1742,644){\makebox(0,0){\strut{}}}%
      \put(2131,644){\makebox(0,0){\strut{}0}}%
      \put(2519,644){\makebox(0,0){\strut{}}}%
      \put(2908,644){\makebox(0,0){\strut{}1}}%
      \put(3297,644){\makebox(0,0){\strut{}}}%
    }%
    \gplgaddtomacro\gplfronttext{%
      \csname LTb\endcsname%
      \put(2130,4303){\makebox(0,0){\strut{}$\log \rho_c(e^-)$ for 5~ms PSR}}%
      \csname LTb\endcsname%
      \put(4050,864){\makebox(0,0)[l]{\strut{}$-2$}}%
      \put(4050,1146){\makebox(0,0)[l]{\strut{}$-1.5$}}%
      \put(4050,1429){\makebox(0,0)[l]{\strut{}$-1$}}%
      \put(4050,1711){\makebox(0,0)[l]{\strut{}$-0.5$}}%
      \put(4050,1994){\makebox(0,0)[l]{\strut{}$0$}}%
      \put(4050,2277){\makebox(0,0)[l]{\strut{}$0.5$}}%
      \put(4050,2559){\makebox(0,0)[l]{\strut{}$1$}}%
      \put(4050,2842){\makebox(0,0)[l]{\strut{}$1.5$}}%
      \put(4050,3125){\makebox(0,0)[l]{\strut{}$2$}}%
      \put(4050,3407){\makebox(0,0)[l]{\strut{}$2.5$}}%
      \put(4050,3690){\makebox(0,0)[l]{\strut{}$3$}}%
      \put(4050,3973){\makebox(0,0)[l]{\strut{}$3.5$}}%
    }%
    \gplbacktext
    \put(0,0){\includegraphics{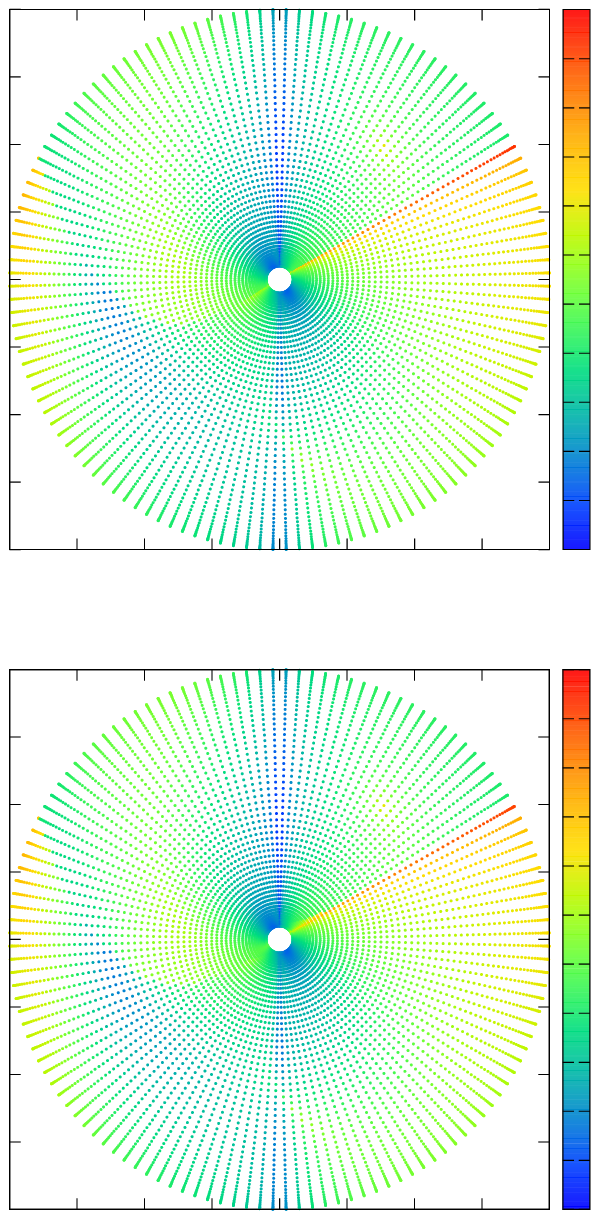}}%
    \gplfronttext
  \end{picture}%
\endgroup

%% file: luminosite_rx_c60_g10_nr128.tex
\begingroup
  \makeatletter
  \providecommand\color[2][]{%
    \GenericError{(gnuplot) \space\space\space\@spaces}{%
      Package color not loaded in conjunction with
      terminal option `colourtext'%
    }{See the gnuplot documentation for explanation.%
    }{Either use 'blacktext' in gnuplot or load the package
      color.sty in LaTeX.}%
    \renewcommand\color[2][]{}%
  }%
  \providecommand\includegraphics[2][]{%
    \GenericError{(gnuplot) \space\space\space\@spaces}{%
      Package graphicx or graphics not loaded%
    }{See the gnuplot documentation for explanation.%
    }{The gnuplot epslatex terminal needs graphicx.sty or graphics.sty.}%
    \renewcommand\includegraphics[2][]{}%
  }%
  \providecommand\rotatebox[2]{#2}%
  \@ifundefined{ifGPcolor}{%
    \newif\ifGPcolor
    \GPcolortrue
  }{}%
  \@ifundefined{ifGPblacktext}{%
    \newif\ifGPblacktext
    \GPblacktextfalse
  }{}%
  \let\gplgaddtomacro\g@addto@macro
  \gdef\gplbacktext{}%
  \gdef\gplfronttext{}%
  \makeatother
  \ifGPblacktext
    \def\colorrgb#1{}%
    \def\colorgray#1{}%
  \else
    \ifGPcolor
      \def\colorrgb#1{\color[rgb]{#1}}%
      \def\colorgray#1{\color[gray]{#1}}%
      \expandafter\def\csname LTw\endcsname{\color{white}}%
      \expandafter\def\csname LTb\endcsname{\color{black}}%
      \expandafter\def\csname LTa\endcsname{\color{black}}%
      \expandafter\def\csname LT0\endcsname{\color[rgb]{1,0,0}}%
      \expandafter\def\csname LT1\endcsname{\color[rgb]{0,1,0}}%
      \expandafter\def\csname LT2\endcsname{\color[rgb]{0,0,1}}%
      \expandafter\def\csname LT3\endcsname{\color[rgb]{1,0,1}}%
      \expandafter\def\csname LT4\endcsname{\color[rgb]{0,1,1}}%
      \expandafter\def\csname LT5\endcsname{\color[rgb]{1,1,0}}%
      \expandafter\def\csname LT6\endcsname{\color[rgb]{0,0,0}}%
      \expandafter\def\csname LT7\endcsname{\color[rgb]{1,0.3,0}}%
      \expandafter\def\csname LT8\endcsname{\color[rgb]{0.5,0.5,0.5}}%
    \else
      \def\colorrgb#1{\color{black}}%
      \def\colorgray#1{\color[gray]{#1}}%
      \expandafter\def\csname LTw\endcsname{\color{white}}%
      \expandafter\def\csname LTb\endcsname{\color{black}}%
      \expandafter\def\csname LTa\endcsname{\color{black}}%
      \expandafter\def\csname LT0\endcsname{\color{black}}%
      \expandafter\def\csname LT1\endcsname{\color{black}}%
      \expandafter\def\csname LT2\endcsname{\color{black}}%
      \expandafter\def\csname LT3\endcsname{\color{black}}%
      \expandafter\def\csname LT4\endcsname{\color{black}}%
      \expandafter\def\csname LT5\endcsname{\color{black}}%
      \expandafter\def\csname LT6\endcsname{\color{black}}%
      \expandafter\def\csname LT7\endcsname{\color{black}}%
      \expandafter\def\csname LT8\endcsname{\color{black}}%
    \fi
  \fi
    \setlength{\unitlength}{0.0500bp}%
    \ifx\gptboxheight\undefined%
      \newlength{\gptboxheight}%
      \newlength{\gptboxwidth}%
      \newsavebox{\gptboxtext}%
    \fi%
    \setlength{\fboxrule}{0.5pt}%
    \setlength{\fboxsep}{1pt}%
\begin{picture}(5760.00,4320.00)%
    \gplgaddtomacro\gplbacktext{%
      \csname LTb\endcsname
      \put(682,704){\makebox(0,0)[r]{\strut{}$18$}}%
      \put(682,1270){\makebox(0,0)[r]{\strut{}$20$}}%
      \put(682,1836){\makebox(0,0)[r]{\strut{}$22$}}%
      \put(682,2402){\makebox(0,0)[r]{\strut{}$24$}}%
      \put(682,2967){\makebox(0,0)[r]{\strut{}$26$}}%
      \put(682,3533){\makebox(0,0)[r]{\strut{}$28$}}%
      \put(682,4099){\makebox(0,0)[r]{\strut{}$30$}}%
      \put(1193,484){\makebox(0,0){\strut{}$-6$}}%
      \put(1951,484){\makebox(0,0){\strut{}$-5$}}%
      \put(2709,484){\makebox(0,0){\strut{}$-4$}}%
      \put(3468,484){\makebox(0,0){\strut{}$-3$}}%
      \put(4226,484){\makebox(0,0){\strut{}$-2$}}%
      \put(4984,484){\makebox(0,0){\strut{}$-1$}}%
    }%
    \gplgaddtomacro\gplfronttext{%
      \csname LTb\endcsname
      \put(198,2401){\rotatebox{-270}{\makebox(0,0){\strut{}$L_\gamma ~ (\textrm{W})$}}}%
      \put(3088,154){\makebox(0,0){\strut{}$\log(B/\BQ)$}}%
      \csname LTb\endcsname
      \put(4376,1757){\makebox(0,0)[r]{\strut{}$r_{\rm  in}=0.1$}}%
      \csname LTb\endcsname
      \put(4376,1537){\makebox(0,0)[r]{\strut{}0.2}}%
      \csname LTb\endcsname
      \put(4376,1317){\makebox(0,0)[r]{\strut{}0.5}}%
      \csname LTb\endcsname
      \put(4376,1097){\makebox(0,0)[r]{\strut{}100~ms}}%
      \csname LTb\endcsname
      \put(4376,877){\makebox(0,0)[r]{\strut{}5~ms}}%
      \csname LTb\endcsname
      \put(3468,3703){\rotatebox{30}{\makebox(0,0){\strut{}$L_\gamma = 10^{34} \, b^2$}}}%
      \put(3468,2571){\rotatebox{30}{\makebox(0,0){\strut{}$L_\gamma = 10^{30} \, b^2$}}}%
    }%
    \gplbacktext
    \put(0,0){\includegraphics{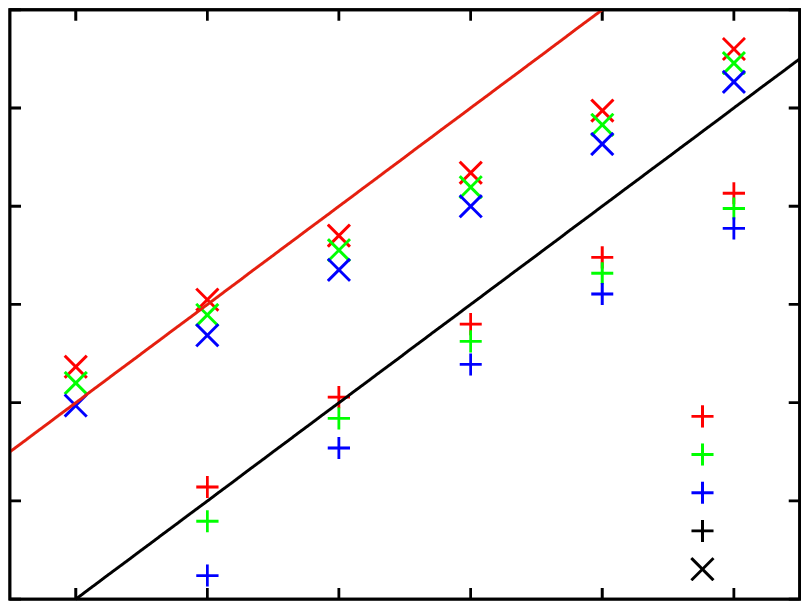}}%
    \gplfronttext
  \end{picture}%
\endgroup

%% file: luminosite_Edot_c60_g10_d3_nr128.tex
\begingroup
  \makeatletter
  \providecommand\color[2][]{%
    \GenericError{(gnuplot) \space\space\space\@spaces}{%
      Package color not loaded in conjunction with
      terminal option `colourtext'%
    }{See the gnuplot documentation for explanation.%
    }{Either use 'blacktext' in gnuplot or load the package
      color.sty in LaTeX.}%
    \renewcommand\color[2][]{}%
  }%
  \providecommand\includegraphics[2][]{%
    \GenericError{(gnuplot) \space\space\space\@spaces}{%
      Package graphicx or graphics not loaded%
    }{See the gnuplot documentation for explanation.%
    }{The gnuplot epslatex terminal needs graphicx.sty or graphics.sty.}%
    \renewcommand\includegraphics[2][]{}%
  }%
  \providecommand\rotatebox[2]{#2}%
  \@ifundefined{ifGPcolor}{%
    \newif\ifGPcolor
    \GPcolortrue
  }{}%
  \@ifundefined{ifGPblacktext}{%
    \newif\ifGPblacktext
    \GPblacktextfalse
  }{}%
  \let\gplgaddtomacro\g@addto@macro
  \gdef\gplbacktext{}%
  \gdef\gplfronttext{}%
  \makeatother
  \ifGPblacktext
    \def\colorrgb#1{}%
    \def\colorgray#1{}%
  \else
    \ifGPcolor
      \def\colorrgb#1{\color[rgb]{#1}}%
      \def\colorgray#1{\color[gray]{#1}}%
      \expandafter\def\csname LTw\endcsname{\color{white}}%
      \expandafter\def\csname LTb\endcsname{\color{black}}%
      \expandafter\def\csname LTa\endcsname{\color{black}}%
      \expandafter\def\csname LT0\endcsname{\color[rgb]{1,0,0}}%
      \expandafter\def\csname LT1\endcsname{\color[rgb]{0,1,0}}%
      \expandafter\def\csname LT2\endcsname{\color[rgb]{0,0,1}}%
      \expandafter\def\csname LT3\endcsname{\color[rgb]{1,0,1}}%
      \expandafter\def\csname LT4\endcsname{\color[rgb]{0,1,1}}%
      \expandafter\def\csname LT5\endcsname{\color[rgb]{1,1,0}}%
      \expandafter\def\csname LT6\endcsname{\color[rgb]{0,0,0}}%
      \expandafter\def\csname LT7\endcsname{\color[rgb]{1,0.3,0}}%
      \expandafter\def\csname LT8\endcsname{\color[rgb]{0.5,0.5,0.5}}%
    \else
      \def\colorrgb#1{\color{black}}%
      \def\colorgray#1{\color[gray]{#1}}%
      \expandafter\def\csname LTw\endcsname{\color{white}}%
      \expandafter\def\csname LTb\endcsname{\color{black}}%
      \expandafter\def\csname LTa\endcsname{\color{black}}%
      \expandafter\def\csname LT0\endcsname{\color{black}}%
      \expandafter\def\csname LT1\endcsname{\color{black}}%
      \expandafter\def\csname LT2\endcsname{\color{black}}%
      \expandafter\def\csname LT3\endcsname{\color{black}}%
      \expandafter\def\csname LT4\endcsname{\color{black}}%
      \expandafter\def\csname LT5\endcsname{\color{black}}%
      \expandafter\def\csname LT6\endcsname{\color{black}}%
      \expandafter\def\csname LT7\endcsname{\color{black}}%
      \expandafter\def\csname LT8\endcsname{\color{black}}%
    \fi
  \fi
    \setlength{\unitlength}{0.0500bp}%
    \ifx\gptboxheight\undefined%
      \newlength{\gptboxheight}%
      \newlength{\gptboxwidth}%
      \newsavebox{\gptboxtext}%
    \fi%
    \setlength{\fboxrule}{0.5pt}%
    \setlength{\fboxsep}{1pt}%
\begin{picture}(5760.00,4320.00)%
    \gplgaddtomacro\gplbacktext{%
      \csname LTb\endcsname
      \put(682,704){\makebox(0,0)[r]{\strut{}$21$}}%
      \put(682,1189){\makebox(0,0)[r]{\strut{}$22$}}%
      \put(682,1674){\makebox(0,0)[r]{\strut{}$23$}}%
      \put(682,2159){\makebox(0,0)[r]{\strut{}$24$}}%
      \put(682,2644){\makebox(0,0)[r]{\strut{}$25$}}%
      \put(682,3129){\makebox(0,0)[r]{\strut{}$26$}}%
      \put(682,3614){\makebox(0,0)[r]{\strut{}$27$}}%
      \put(682,4099){\makebox(0,0)[r]{\strut{}$28$}}%
      \put(1139,484){\makebox(0,0){\strut{}$25$}}%
      \put(1789,484){\makebox(0,0){\strut{}$26$}}%
      \put(2439,484){\makebox(0,0){\strut{}$27$}}%
      \put(3089,484){\makebox(0,0){\strut{}$28$}}%
      \put(3738,484){\makebox(0,0){\strut{}$29$}}%
      \put(4388,484){\makebox(0,0){\strut{}$30$}}%
      \put(5038,484){\makebox(0,0){\strut{}$31$}}%
    }%
    \gplgaddtomacro\gplfronttext{%
      \csname LTb\endcsname
      \put(198,2401){\rotatebox{-270}{\makebox(0,0){\strut{}$\log L_\gamma ~ (\textrm{W})$}}}%
      \put(3088,154){\makebox(0,0){\strut{}$\log \dot E ~ (\textrm{W})$}}%
      \csname LTb\endcsname
      \put(4376,1757){\makebox(0,0)[r]{\strut{}$r_{\rm  in}=0.1$}}%
      \csname LTb\endcsname
      \put(4376,1537){\makebox(0,0)[r]{\strut{}0.2}}%
      \csname LTb\endcsname
      \put(4376,1317){\makebox(0,0)[r]{\strut{}0.5}}%
      \csname LTb\endcsname
      \put(4376,1097){\makebox(0,0)[r]{\strut{}100~ms}}%
      \csname LTb\endcsname
      \put(4376,877){\makebox(0,0)[r]{\strut{}5~ms}}%
      \csname LTb\endcsname
      \put(1789,3275){\rotatebox{35}{\makebox(0,0){\strut{}$L_\gamma = \dot E$}}}%
      \put(1789,2305){\rotatebox{35}{\makebox(0,0){\strut{}$L_\gamma = \dot E/10^2$}}}%
      \put(1789,1335){\rotatebox{35}{\makebox(0,0){\strut{}$L_\gamma = \dot E/10^4$}}}%
    }%
    \gplbacktext
    \put(0,0){\includegraphics{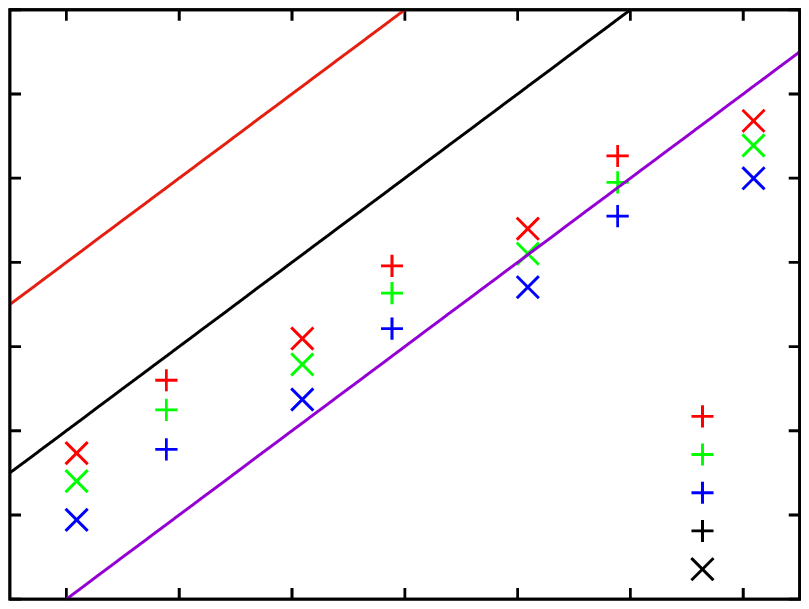}}%
    \gplfronttext
  \end{picture}%
\endgroup

%% file: fermi_sensitivity.tex
\begingroup
  \makeatletter
  \providecommand\color[2][]{%
    \GenericError{(gnuplot) \space\space\space\@spaces}{%
      Package color not loaded in conjunction with
      terminal option `colourtext'%
    }{See the gnuplot documentation for explanation.%
    }{Either use 'blacktext' in gnuplot or load the package
      color.sty in LaTeX.}%
    \renewcommand\color[2][]{}%
  }%
  \providecommand\includegraphics[2][]{%
    \GenericError{(gnuplot) \space\space\space\@spaces}{%
      Package graphicx or graphics not loaded%
    }{See the gnuplot documentation for explanation.%
    }{The gnuplot epslatex terminal needs graphicx.sty or graphics.sty.}%
    \renewcommand\includegraphics[2][]{}%
  }%
  \providecommand\rotatebox[2]{#2}%
  \@ifundefined{ifGPcolor}{%
    \newif\ifGPcolor
    \GPcolortrue
  }{}%
  \@ifundefined{ifGPblacktext}{%
    \newif\ifGPblacktext
    \GPblacktextfalse
  }{}%
  \let\gplgaddtomacro\g@addto@macro
  \gdef\gplbacktext{}%
  \gdef\gplfronttext{}%
  \makeatother
  \ifGPblacktext
    \def\colorrgb#1{}%
    \def\colorgray#1{}%
  \else
    \ifGPcolor
      \def\colorrgb#1{\color[rgb]{#1}}%
      \def\colorgray#1{\color[gray]{#1}}%
      \expandafter\def\csname LTw\endcsname{\color{white}}%
      \expandafter\def\csname LTb\endcsname{\color{black}}%
      \expandafter\def\csname LTa\endcsname{\color{black}}%
      \expandafter\def\csname LT0\endcsname{\color[rgb]{1,0,0}}%
      \expandafter\def\csname LT1\endcsname{\color[rgb]{0,1,0}}%
      \expandafter\def\csname LT2\endcsname{\color[rgb]{0,0,1}}%
      \expandafter\def\csname LT3\endcsname{\color[rgb]{1,0,1}}%
      \expandafter\def\csname LT4\endcsname{\color[rgb]{0,1,1}}%
      \expandafter\def\csname LT5\endcsname{\color[rgb]{1,1,0}}%
      \expandafter\def\csname LT6\endcsname{\color[rgb]{0,0,0}}%
      \expandafter\def\csname LT7\endcsname{\color[rgb]{1,0.3,0}}%
      \expandafter\def\csname LT8\endcsname{\color[rgb]{0.5,0.5,0.5}}%
    \else
      \def\colorrgb#1{\color{black}}%
      \def\colorgray#1{\color[gray]{#1}}%
      \expandafter\def\csname LTw\endcsname{\color{white}}%
      \expandafter\def\csname LTb\endcsname{\color{black}}%
      \expandafter\def\csname LTa\endcsname{\color{black}}%
      \expandafter\def\csname LT0\endcsname{\color{black}}%
      \expandafter\def\csname LT1\endcsname{\color{black}}%
      \expandafter\def\csname LT2\endcsname{\color{black}}%
      \expandafter\def\csname LT3\endcsname{\color{black}}%
      \expandafter\def\csname LT4\endcsname{\color{black}}%
      \expandafter\def\csname LT5\endcsname{\color{black}}%
      \expandafter\def\csname LT6\endcsname{\color{black}}%
      \expandafter\def\csname LT7\endcsname{\color{black}}%
      \expandafter\def\csname LT8\endcsname{\color{black}}%
    \fi
  \fi
    \setlength{\unitlength}{0.0500bp}%
    \ifx\gptboxheight\undefined%
      \newlength{\gptboxheight}%
      \newlength{\gptboxwidth}%
      \newsavebox{\gptboxtext}%
    \fi%
    \setlength{\fboxrule}{0.5pt}%
    \setlength{\fboxsep}{1pt}%
\begin{picture}(5760.00,4320.00)%
    \gplgaddtomacro\gplbacktext{%
      \csname LTb\endcsname
      \put(1078,704){\makebox(0,0)[r]{\strut{}$-17$}}%
      \put(1078,1044){\makebox(0,0)[r]{\strut{}$-16.5$}}%
      \put(1078,1383){\makebox(0,0)[r]{\strut{}$-16$}}%
      \put(1078,1723){\makebox(0,0)[r]{\strut{}$-15.5$}}%
      \put(1078,2062){\makebox(0,0)[r]{\strut{}$-15$}}%
      \put(1078,2402){\makebox(0,0)[r]{\strut{}$-14.5$}}%
      \put(1078,2741){\makebox(0,0)[r]{\strut{}$-14$}}%
      \put(1078,3081){\makebox(0,0)[r]{\strut{}$-13.5$}}%
      \put(1078,3420){\makebox(0,0)[r]{\strut{}$-13$}}%
      \put(1078,3760){\makebox(0,0)[r]{\strut{}$-12.5$}}%
      \put(1078,4099){\makebox(0,0)[r]{\strut{}$-12$}}%
      \put(1210,484){\makebox(0,0){\strut{}$1$}}%
      \put(1803,484){\makebox(0,0){\strut{}$2$}}%
      \put(2397,484){\makebox(0,0){\strut{}$3$}}%
      \put(2990,484){\makebox(0,0){\strut{}$4$}}%
      \put(3583,484){\makebox(0,0){\strut{}$5$}}%
      \put(4176,484){\makebox(0,0){\strut{}$6$}}%
      \put(4770,484){\makebox(0,0){\strut{}$7$}}%
      \put(5363,484){\makebox(0,0){\strut{}$8$}}%
    }%
    \gplgaddtomacro\gplfronttext{%
      \csname LTb\endcsname
      \put(198,2401){\rotatebox{-270}{\makebox(0,0){\strut{}$\log(E^2 \, dN/dE\,dt) ~ (W/m^2)$}}}%
      \put(3286,154){\makebox(0,0){\strut{}$\log(E/\textrm{MeV})$}}%
      \csname LTb\endcsname
      \put(4376,3926){\makebox(0,0)[r]{\strut{}(-2,m)}}%
      \csname LTb\endcsname
      \put(4376,3706){\makebox(0,0)[r]{\strut{}(-3,m)}}%
      \csname LTb\endcsname
      \put(4376,3486){\makebox(0,0)[r]{\strut{}(-2,w)}}%
      \csname LTb\endcsname
      \put(4376,3266){\makebox(0,0)[r]{\strut{}(-3,w)}}%
      \csname LTb\endcsname
      \put(4376,3046){\makebox(0,0)[r]{\strut{}(-5,m)}}%
      \csname LTb\endcsname
      \put(4376,2826){\makebox(0,0)[r]{\strut{}(-6,m)}}%
      \csname LTb\endcsname
      \put(4376,2606){\makebox(0,0)[r]{\strut{}(-5,w)}}%
      \csname LTb\endcsname
      \put(4376,2386){\makebox(0,0)[r]{\strut{}(-6,w)}}%
      \csname LTb\endcsname
      \put(4376,2166){\makebox(0,0)[r]{\strut{}Fermi}}%
      \csname LTb\endcsname
      \put(4376,1946){\makebox(0,0)[r]{\strut{}CTA 50h}}%
    }%
    \gplbacktext
    \put(0,0){\includegraphics{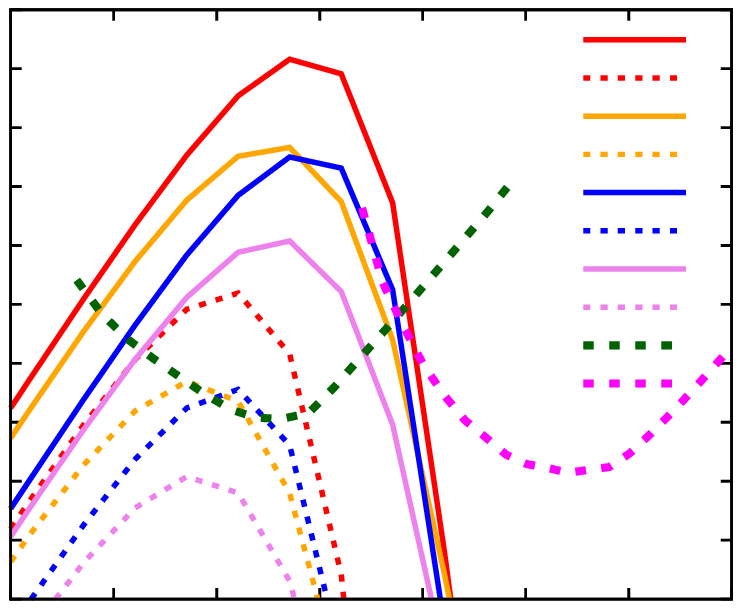}}%
    \gplfronttext
  \end{picture}%
\endgroup

%% file: cavite_cx.tex
\begingroup
  \makeatletter
  \providecommand\color[2][]{%
    \GenericError{(gnuplot) \space\space\space\@spaces}{%
      Package color not loaded in conjunction with
      terminal option `colourtext'%
    }{See the gnuplot documentation for explanation.%
    }{Either use 'blacktext' in gnuplot or load the package
      color.sty in LaTeX.}%
    \renewcommand\color[2][]{}%
  }%
  \providecommand\includegraphics[2][]{%
    \GenericError{(gnuplot) \space\space\space\@spaces}{%
      Package graphicx or graphics not loaded%
    }{See the gnuplot documentation for explanation.%
    }{The gnuplot epslatex terminal needs graphicx.sty or graphics.sty.}%
    \renewcommand\includegraphics[2][]{}%
  }%
  \providecommand\rotatebox[2]{#2}%
  \@ifundefined{ifGPcolor}{%
    \newif\ifGPcolor
    \GPcolortrue
  }{}%
  \@ifundefined{ifGPblacktext}{%
    \newif\ifGPblacktext
    \GPblacktextfalse
  }{}%
  \let\gplgaddtomacro\g@addto@macro
  \gdef\gplbacktext{}%
  \gdef\gplfronttext{}%
  \makeatother
  \ifGPblacktext
    \def\colorrgb#1{}%
    \def\colorgray#1{}%
  \else
    \ifGPcolor
      \def\colorrgb#1{\color[rgb]{#1}}%
      \def\colorgray#1{\color[gray]{#1}}%
      \expandafter\def\csname LTw\endcsname{\color{white}}%
      \expandafter\def\csname LTb\endcsname{\color{black}}%
      \expandafter\def\csname LTa\endcsname{\color{black}}%
      \expandafter\def\csname LT0\endcsname{\color[rgb]{1,0,0}}%
      \expandafter\def\csname LT1\endcsname{\color[rgb]{0,1,0}}%
      \expandafter\def\csname LT2\endcsname{\color[rgb]{0,0,1}}%
      \expandafter\def\csname LT3\endcsname{\color[rgb]{1,0,1}}%
      \expandafter\def\csname LT4\endcsname{\color[rgb]{0,1,1}}%
      \expandafter\def\csname LT5\endcsname{\color[rgb]{1,1,0}}%
      \expandafter\def\csname LT6\endcsname{\color[rgb]{0,0,0}}%
      \expandafter\def\csname LT7\endcsname{\color[rgb]{1,0.3,0}}%
      \expandafter\def\csname LT8\endcsname{\color[rgb]{0.5,0.5,0.5}}%
    \else
      \def\colorrgb#1{\color{black}}%
      \def\colorgray#1{\color[gray]{#1}}%
      \expandafter\def\csname LTw\endcsname{\color{white}}%
      \expandafter\def\csname LTb\endcsname{\color{black}}%
      \expandafter\def\csname LTa\endcsname{\color{black}}%
      \expandafter\def\csname LT0\endcsname{\color{black}}%
      \expandafter\def\csname LT1\endcsname{\color{black}}%
      \expandafter\def\csname LT2\endcsname{\color{black}}%
      \expandafter\def\csname LT3\endcsname{\color{black}}%
      \expandafter\def\csname LT4\endcsname{\color{black}}%
      \expandafter\def\csname LT5\endcsname{\color{black}}%
      \expandafter\def\csname LT6\endcsname{\color{black}}%
      \expandafter\def\csname LT7\endcsname{\color{black}}%
      \expandafter\def\csname LT8\endcsname{\color{black}}%
    \fi
  \fi
    \setlength{\unitlength}{0.0500bp}%
    \ifx\gptboxheight\undefined%
      \newlength{\gptboxheight}%
      \newlength{\gptboxwidth}%
      \newsavebox{\gptboxtext}%
    \fi%
    \setlength{\fboxrule}{0.5pt}%
    \setlength{\fboxsep}{1pt}%
\begin{picture}(5760.00,8640.00)%
    \gplgaddtomacro\gplbacktext{%
      \csname LTb\endcsname
      \put(88,5980){\makebox(0,0)[r]{\strut{}-2}}%
      \put(88,6285){\makebox(0,0)[r]{\strut{}}}%
      \put(88,6590){\makebox(0,0)[r]{\strut{}-1}}%
      \put(88,6895){\makebox(0,0)[r]{\strut{}}}%
      \put(88,7200){\makebox(0,0)[r]{\strut{}0}}%
      \put(88,7504){\makebox(0,0)[r]{\strut{}}}%
      \put(88,7809){\makebox(0,0)[r]{\strut{}1}}%
      \put(88,8114){\makebox(0,0)[r]{\strut{}}}%
      \put(88,8419){\makebox(0,0)[r]{\strut{}2}}%
      \put(220,5760){\makebox(0,0){\strut{}}}%
      \put(525,5760){\makebox(0,0){\strut{}}}%
      \put(830,5760){\makebox(0,0){\strut{}}}%
      \put(1135,5760){\makebox(0,0){\strut{}}}%
      \put(1440,5760){\makebox(0,0){\strut{}}}%
      \put(1744,5760){\makebox(0,0){\strut{}}}%
      \put(2049,5760){\makebox(0,0){\strut{}}}%
      \put(2354,5760){\makebox(0,0){\strut{}}}%
      \put(2659,5760){\makebox(0,0){\strut{}}}%
      \put(2415,8175){\makebox(0,0)[l]{\strut{}$e^-$}}%
    }%
    \gplgaddtomacro\gplfronttext{%
      \csname LTb\endcsname
      \put(1439,8749){\makebox(0,0){\strut{}$\chi=0$\degr}}%
    }%
    \gplgaddtomacro\gplbacktext{%
      \csname LTb\endcsname
      \put(2968,5980){\makebox(0,0)[r]{\strut{}}}%
      \put(2968,6285){\makebox(0,0)[r]{\strut{}}}%
      \put(2968,6590){\makebox(0,0)[r]{\strut{}}}%
      \put(2968,6895){\makebox(0,0)[r]{\strut{}}}%
      \put(2968,7200){\makebox(0,0)[r]{\strut{}}}%
      \put(2968,7504){\makebox(0,0)[r]{\strut{}}}%
      \put(2968,7809){\makebox(0,0)[r]{\strut{}}}%
      \put(2968,8114){\makebox(0,0)[r]{\strut{}}}%
      \put(2968,8419){\makebox(0,0)[r]{\strut{}}}%
      \put(3100,5760){\makebox(0,0){\strut{}}}%
      \put(3405,5760){\makebox(0,0){\strut{}}}%
      \put(3710,5760){\makebox(0,0){\strut{}}}%
      \put(4015,5760){\makebox(0,0){\strut{}}}%
      \put(4320,5760){\makebox(0,0){\strut{}}}%
      \put(4624,5760){\makebox(0,0){\strut{}}}%
      \put(4929,5760){\makebox(0,0){\strut{}}}%
      \put(5234,5760){\makebox(0,0){\strut{}}}%
      \put(5539,5760){\makebox(0,0){\strut{}}}%
      \put(5295,8175){\makebox(0,0)[l]{\strut{}$e^+$}}%
    }%
    \gplgaddtomacro\gplfronttext{%
      \csname LTb\endcsname
      \put(4319,8749){\makebox(0,0){\strut{}$\chi=0$\degr}}%
    }%
    \gplgaddtomacro\gplbacktext{%
      \csname LTb\endcsname
      \put(88,3100){\makebox(0,0)[r]{\strut{}-2}}%
      \put(88,3405){\makebox(0,0)[r]{\strut{}}}%
      \put(88,3710){\makebox(0,0)[r]{\strut{}-1}}%
      \put(88,4015){\makebox(0,0)[r]{\strut{}}}%
      \put(88,4320){\makebox(0,0)[r]{\strut{}0}}%
      \put(88,4624){\makebox(0,0)[r]{\strut{}}}%
      \put(88,4929){\makebox(0,0)[r]{\strut{}1}}%
      \put(88,5234){\makebox(0,0)[r]{\strut{}}}%
      \put(88,5539){\makebox(0,0)[r]{\strut{}2}}%
      \put(220,2880){\makebox(0,0){\strut{}}}%
      \put(525,2880){\makebox(0,0){\strut{}}}%
      \put(830,2880){\makebox(0,0){\strut{}}}%
      \put(1135,2880){\makebox(0,0){\strut{}}}%
      \put(1440,2880){\makebox(0,0){\strut{}}}%
      \put(1744,2880){\makebox(0,0){\strut{}}}%
      \put(2049,2880){\makebox(0,0){\strut{}}}%
      \put(2354,2880){\makebox(0,0){\strut{}}}%
      \put(2659,2880){\makebox(0,0){\strut{}}}%
      \put(2415,5295){\makebox(0,0)[l]{\strut{}$e^-$}}%
    }%
    \gplgaddtomacro\gplfronttext{%
      \csname LTb\endcsname
      \put(1439,5869){\makebox(0,0){\strut{}$\chi=60$\degr}}%
    }%
    \gplgaddtomacro\gplbacktext{%
      \csname LTb\endcsname
      \put(2968,3100){\makebox(0,0)[r]{\strut{}}}%
      \put(2968,3405){\makebox(0,0)[r]{\strut{}}}%
      \put(2968,3710){\makebox(0,0)[r]{\strut{}}}%
      \put(2968,4015){\makebox(0,0)[r]{\strut{}}}%
      \put(2968,4320){\makebox(0,0)[r]{\strut{}}}%
      \put(2968,4624){\makebox(0,0)[r]{\strut{}}}%
      \put(2968,4929){\makebox(0,0)[r]{\strut{}}}%
      \put(2968,5234){\makebox(0,0)[r]{\strut{}}}%
      \put(2968,5539){\makebox(0,0)[r]{\strut{}}}%
      \put(3100,2880){\makebox(0,0){\strut{}}}%
      \put(3405,2880){\makebox(0,0){\strut{}}}%
      \put(3710,2880){\makebox(0,0){\strut{}}}%
      \put(4015,2880){\makebox(0,0){\strut{}}}%
      \put(4320,2880){\makebox(0,0){\strut{}}}%
      \put(4624,2880){\makebox(0,0){\strut{}}}%
      \put(4929,2880){\makebox(0,0){\strut{}}}%
      \put(5234,2880){\makebox(0,0){\strut{}}}%
      \put(5539,2880){\makebox(0,0){\strut{}}}%
      \put(5295,5295){\makebox(0,0)[l]{\strut{}$e^+$}}%
    }%
    \gplgaddtomacro\gplfronttext{%
      \csname LTb\endcsname
      \put(4319,5869){\makebox(0,0){\strut{}$\chi=60$\degr}}%
    }%
    \gplgaddtomacro\gplbacktext{%
      \csname LTb\endcsname
      \put(87,220){\makebox(0,0)[r]{\strut{}-2}}%
      \put(87,525){\makebox(0,0)[r]{\strut{}}}%
      \put(87,830){\makebox(0,0)[r]{\strut{}-1}}%
      \put(87,1135){\makebox(0,0)[r]{\strut{}}}%
      \put(87,1440){\makebox(0,0)[r]{\strut{}0}}%
      \put(87,1745){\makebox(0,0)[r]{\strut{}}}%
      \put(87,2050){\makebox(0,0)[r]{\strut{}1}}%
      \put(87,2355){\makebox(0,0)[r]{\strut{}}}%
      \put(87,2660){\makebox(0,0)[r]{\strut{}2}}%
      \put(219,0){\makebox(0,0){\strut{}-2}}%
      \put(524,0){\makebox(0,0){\strut{}}}%
      \put(829,0){\makebox(0,0){\strut{}-1}}%
      \put(1134,0){\makebox(0,0){\strut{}}}%
      \put(1440,0){\makebox(0,0){\strut{}0}}%
      \put(1745,0){\makebox(0,0){\strut{}}}%
      \put(2050,0){\makebox(0,0){\strut{}1}}%
      \put(2355,0){\makebox(0,0){\strut{}}}%
      \put(2660,0){\makebox(0,0){\strut{}2}}%
      \put(2416,2416){\makebox(0,0)[l]{\strut{}$e^-$}}%
    }%
    \gplgaddtomacro\gplfronttext{%
      \csname LTb\endcsname
      \put(1439,2990){\makebox(0,0){\strut{}$\chi=90$\degr}}%
    }%
    \gplgaddtomacro\gplbacktext{%
      \csname LTb\endcsname
      \put(2967,220){\makebox(0,0)[r]{\strut{}}}%
      \put(2967,525){\makebox(0,0)[r]{\strut{}}}%
      \put(2967,830){\makebox(0,0)[r]{\strut{}}}%
      \put(2967,1135){\makebox(0,0)[r]{\strut{}}}%
      \put(2967,1440){\makebox(0,0)[r]{\strut{}}}%
      \put(2967,1745){\makebox(0,0)[r]{\strut{}}}%
      \put(2967,2050){\makebox(0,0)[r]{\strut{}}}%
      \put(2967,2355){\makebox(0,0)[r]{\strut{}}}%
      \put(2967,2660){\makebox(0,0)[r]{\strut{}}}%
      \put(3099,0){\makebox(0,0){\strut{}-2}}%
      \put(3404,0){\makebox(0,0){\strut{}}}%
      \put(3709,0){\makebox(0,0){\strut{}-1}}%
      \put(4014,0){\makebox(0,0){\strut{}}}%
      \put(4320,0){\makebox(0,0){\strut{}0}}%
      \put(4625,0){\makebox(0,0){\strut{}}}%
      \put(4930,0){\makebox(0,0){\strut{}1}}%
      \put(5235,0){\makebox(0,0){\strut{}}}%
      \put(5540,0){\makebox(0,0){\strut{}2}}%
      \put(5296,2416){\makebox(0,0)[l]{\strut{}$e^+$}}%
    }%
    \gplgaddtomacro\gplfronttext{%
      \csname LTb\endcsname
      \put(4319,2990){\makebox(0,0){\strut{}$\chi=90$\degr}}%
    }%
    \gplbacktext
    \put(0,0){\includegraphics{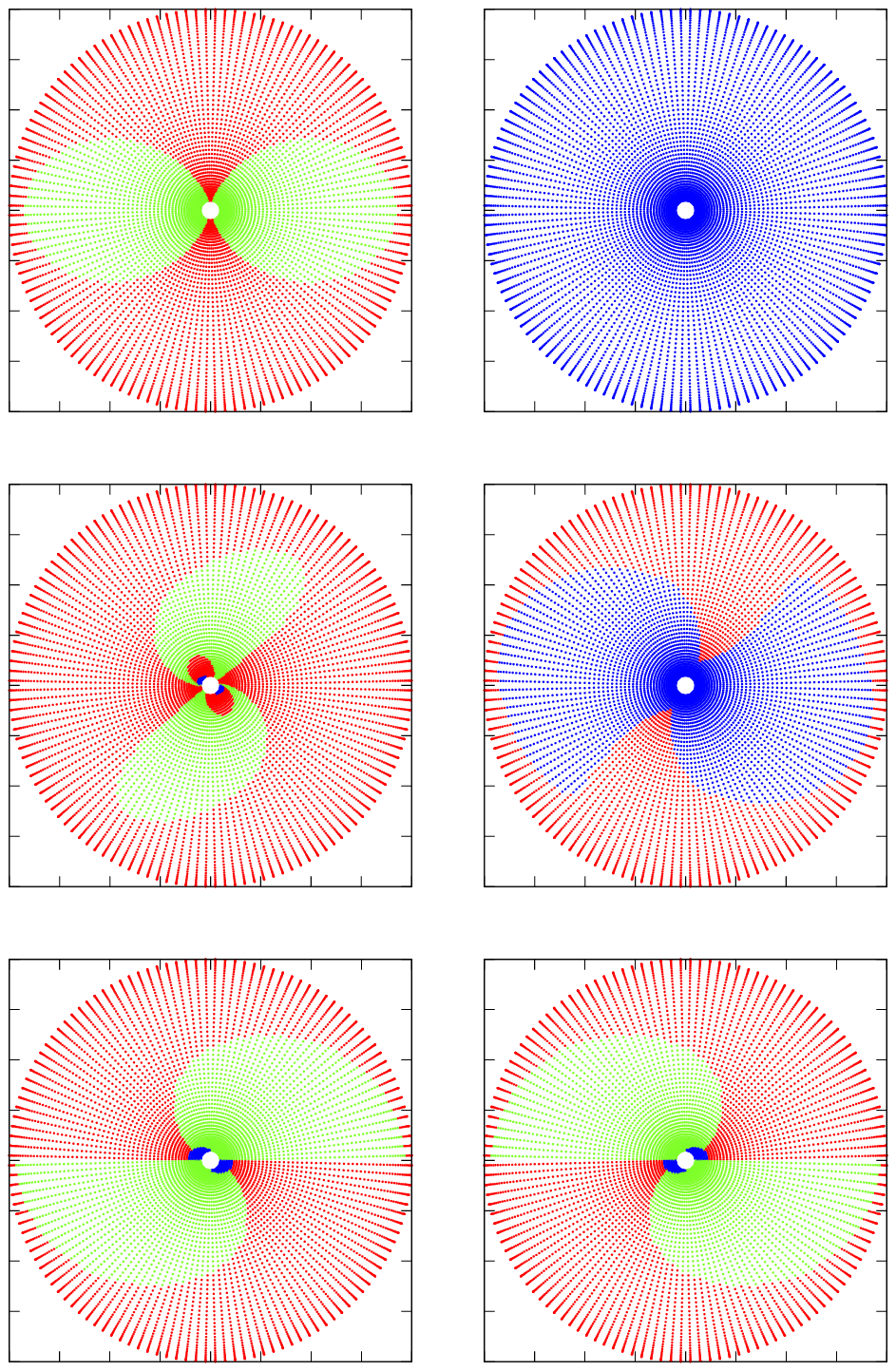}}%
    \gplfronttext
  \end{picture}%
\endgroup

%% file: carte_fuite_cx.tex
\begingroup
  \makeatletter
  \providecommand\color[2][]{%
    \GenericError{(gnuplot) \space\space\space\@spaces}{%
      Package color not loaded in conjunction with
      terminal option `colourtext'%
    }{See the gnuplot documentation for explanation.%
    }{Either use 'blacktext' in gnuplot or load the package
      color.sty in LaTeX.}%
    \renewcommand\color[2][]{}%
  }%
  \providecommand\includegraphics[2][]{%
    \GenericError{(gnuplot) \space\space\space\@spaces}{%
      Package graphicx or graphics not loaded%
    }{See the gnuplot documentation for explanation.%
    }{The gnuplot epslatex terminal needs graphicx.sty or graphics.sty.}%
    \renewcommand\includegraphics[2][]{}%
  }%
  \providecommand\rotatebox[2]{#2}%
  \@ifundefined{ifGPcolor}{%
    \newif\ifGPcolor
    \GPcolortrue
  }{}%
  \@ifundefined{ifGPblacktext}{%
    \newif\ifGPblacktext
    \GPblacktextfalse
  }{}%
  \let\gplgaddtomacro\g@addto@macro
  \gdef\gplbacktext{}%
  \gdef\gplfronttext{}%
  \makeatother
  \ifGPblacktext
    \def\colorrgb#1{}%
    \def\colorgray#1{}%
  \else
    \ifGPcolor
      \def\colorrgb#1{\color[rgb]{#1}}%
      \def\colorgray#1{\color[gray]{#1}}%
      \expandafter\def\csname LTw\endcsname{\color{white}}%
      \expandafter\def\csname LTb\endcsname{\color{black}}%
      \expandafter\def\csname LTa\endcsname{\color{black}}%
      \expandafter\def\csname LT0\endcsname{\color[rgb]{1,0,0}}%
      \expandafter\def\csname LT1\endcsname{\color[rgb]{0,1,0}}%
      \expandafter\def\csname LT2\endcsname{\color[rgb]{0,0,1}}%
      \expandafter\def\csname LT3\endcsname{\color[rgb]{1,0,1}}%
      \expandafter\def\csname LT4\endcsname{\color[rgb]{0,1,1}}%
      \expandafter\def\csname LT5\endcsname{\color[rgb]{1,1,0}}%
      \expandafter\def\csname LT6\endcsname{\color[rgb]{0,0,0}}%
      \expandafter\def\csname LT7\endcsname{\color[rgb]{1,0.3,0}}%
      \expandafter\def\csname LT8\endcsname{\color[rgb]{0.5,0.5,0.5}}%
    \else
      \def\colorrgb#1{\color{black}}%
      \def\colorgray#1{\color[gray]{#1}}%
      \expandafter\def\csname LTw\endcsname{\color{white}}%
      \expandafter\def\csname LTb\endcsname{\color{black}}%
      \expandafter\def\csname LTa\endcsname{\color{black}}%
      \expandafter\def\csname LT0\endcsname{\color{black}}%
      \expandafter\def\csname LT1\endcsname{\color{black}}%
      \expandafter\def\csname LT2\endcsname{\color{black}}%
      \expandafter\def\csname LT3\endcsname{\color{black}}%
      \expandafter\def\csname LT4\endcsname{\color{black}}%
      \expandafter\def\csname LT5\endcsname{\color{black}}%
      \expandafter\def\csname LT6\endcsname{\color{black}}%
      \expandafter\def\csname LT7\endcsname{\color{black}}%
      \expandafter\def\csname LT8\endcsname{\color{black}}%
    \fi
  \fi
    \setlength{\unitlength}{0.0500bp}%
    \ifx\gptboxheight\undefined%
      \newlength{\gptboxheight}%
      \newlength{\gptboxwidth}%
      \newsavebox{\gptboxtext}%
    \fi%
    \setlength{\fboxrule}{0.5pt}%
    \setlength{\fboxsep}{1pt}%
\begin{picture}(8640.00,5760.00)%
    \gplgaddtomacro\gplbacktext{%
      \csname LTb\endcsname
      \put(0,852){\makebox(0,0)[r]{\strut{}$0$}}%
      \put(0,1528){\makebox(0,0)[r]{\strut{}$30$}}%
      \put(0,2204){\makebox(0,0)[r]{\strut{}$60$}}%
      \put(0,2880){\makebox(0,0)[r]{\strut{}$90$}}%
      \put(0,3555){\makebox(0,0)[r]{\strut{}$120$}}%
      \put(0,4231){\makebox(0,0)[r]{\strut{}$150$}}%
      \put(0,4907){\makebox(0,0)[r]{\strut{}$180$}}%
      \put(132,632){\makebox(0,0){\strut{}$0$}}%
      \put(808,632){\makebox(0,0){\strut{}$60$}}%
      \put(1484,632){\makebox(0,0){\strut{}$120$}}%
      \put(2160,632){\makebox(0,0){\strut{}$180$}}%
      \put(2835,632){\makebox(0,0){\strut{}$240$}}%
      \put(3511,632){\makebox(0,0){\strut{}$300$}}%
      \put(4187,632){\makebox(0,0){\strut{}$360$}}%
    }%
    \gplgaddtomacro\gplfronttext{%
      \csname LTb\endcsname
      \put(-616,2879){\rotatebox{-270}{\makebox(0,0){\strut{}$\theta$}}}%
      \put(2159,302){\makebox(0,0){\strut{}$\phi$}}%
      \put(2159,5237){\makebox(0,0){\strut{}$\chi=60$\degr, $e^-$}}%
    }%
    \gplgaddtomacro\gplbacktext{%
      \csname LTb\endcsname
      \put(4320,852){\makebox(0,0)[r]{\strut{} }}%
      \put(4320,1528){\makebox(0,0)[r]{\strut{} }}%
      \put(4320,2204){\makebox(0,0)[r]{\strut{} }}%
      \put(4320,2880){\makebox(0,0)[r]{\strut{} }}%
      \put(4320,3555){\makebox(0,0)[r]{\strut{} }}%
      \put(4320,4231){\makebox(0,0)[r]{\strut{} }}%
      \put(4320,4907){\makebox(0,0)[r]{\strut{} }}%
      \put(4452,632){\makebox(0,0){\strut{}$0$}}%
      \put(5128,632){\makebox(0,0){\strut{}$60$}}%
      \put(5804,632){\makebox(0,0){\strut{}$120$}}%
      \put(6480,632){\makebox(0,0){\strut{}$180$}}%
      \put(7155,632){\makebox(0,0){\strut{}$240$}}%
      \put(7831,632){\makebox(0,0){\strut{}$300$}}%
      \put(8507,632){\makebox(0,0){\strut{}$360$}}%
    }%
    \gplgaddtomacro\gplfronttext{%
      \csname LTb\endcsname
      \put(3968,2879){\rotatebox{-270}{\makebox(0,0){\strut{}$\theta$}}}%
      \put(6479,302){\makebox(0,0){\strut{}$\phi$}}%
      \put(6479,5237){\makebox(0,0){\strut{}$\chi=120$\degr, $e^+$}}%
    }%
    \gplbacktext
    \put(0,0){\includegraphics{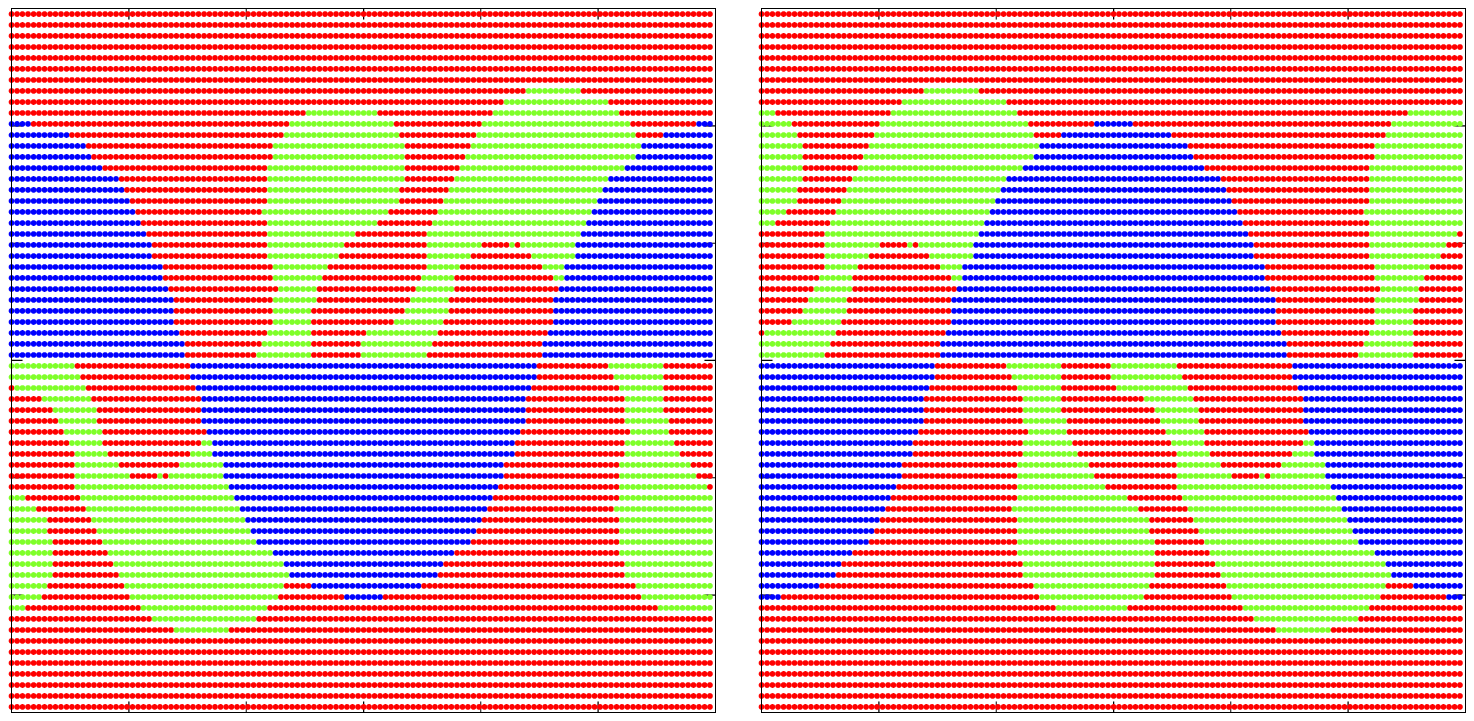}}%
    \gplfronttext
  \end{picture}%
\endgroup

%% file: trajectoire_fuite_cx.tex
\begingroup
  \makeatletter
  \providecommand\color[2][]{%
    \GenericError{(gnuplot) \space\space\space\@spaces}{%
      Package color not loaded in conjunction with
      terminal option `colourtext'%
    }{See the gnuplot documentation for explanation.%
    }{Either use 'blacktext' in gnuplot or load the package
      color.sty in LaTeX.}%
    \renewcommand\color[2][]{}%
  }%
  \providecommand\includegraphics[2][]{%
    \GenericError{(gnuplot) \space\space\space\@spaces}{%
      Package graphicx or graphics not loaded%
    }{See the gnuplot documentation for explanation.%
    }{The gnuplot epslatex terminal needs graphicx.sty or graphics.sty.}%
    \renewcommand\includegraphics[2][]{}%
  }%
  \providecommand\rotatebox[2]{#2}%
  \@ifundefined{ifGPcolor}{%
    \newif\ifGPcolor
    \GPcolortrue
  }{}%
  \@ifundefined{ifGPblacktext}{%
    \newif\ifGPblacktext
    \GPblacktextfalse
  }{}%
  \let\gplgaddtomacro\g@addto@macro
  \gdef\gplbacktext{}%
  \gdef\gplfronttext{}%
  \makeatother
  \ifGPblacktext
    \def\colorrgb#1{}%
    \def\colorgray#1{}%
  \else
    \ifGPcolor
      \def\colorrgb#1{\color[rgb]{#1}}%
      \def\colorgray#1{\color[gray]{#1}}%
      \expandafter\def\csname LTw\endcsname{\color{white}}%
      \expandafter\def\csname LTb\endcsname{\color{black}}%
      \expandafter\def\csname LTa\endcsname{\color{black}}%
      \expandafter\def\csname LT0\endcsname{\color[rgb]{1,0,0}}%
      \expandafter\def\csname LT1\endcsname{\color[rgb]{0,1,0}}%
      \expandafter\def\csname LT2\endcsname{\color[rgb]{0,0,1}}%
      \expandafter\def\csname LT3\endcsname{\color[rgb]{1,0,1}}%
      \expandafter\def\csname LT4\endcsname{\color[rgb]{0,1,1}}%
      \expandafter\def\csname LT5\endcsname{\color[rgb]{1,1,0}}%
      \expandafter\def\csname LT6\endcsname{\color[rgb]{0,0,0}}%
      \expandafter\def\csname LT7\endcsname{\color[rgb]{1,0.3,0}}%
      \expandafter\def\csname LT8\endcsname{\color[rgb]{0.5,0.5,0.5}}%
    \else
      \def\colorrgb#1{\color{black}}%
      \def\colorgray#1{\color[gray]{#1}}%
      \expandafter\def\csname LTw\endcsname{\color{white}}%
      \expandafter\def\csname LTb\endcsname{\color{black}}%
      \expandafter\def\csname LTa\endcsname{\color{black}}%
      \expandafter\def\csname LT0\endcsname{\color{black}}%
      \expandafter\def\csname LT1\endcsname{\color{black}}%
      \expandafter\def\csname LT2\endcsname{\color{black}}%
      \expandafter\def\csname LT3\endcsname{\color{black}}%
      \expandafter\def\csname LT4\endcsname{\color{black}}%
      \expandafter\def\csname LT5\endcsname{\color{black}}%
      \expandafter\def\csname LT6\endcsname{\color{black}}%
      \expandafter\def\csname LT7\endcsname{\color{black}}%
      \expandafter\def\csname LT8\endcsname{\color{black}}%
    \fi
  \fi
    \setlength{\unitlength}{0.0500bp}%
    \ifx\gptboxheight\undefined%
      \newlength{\gptboxheight}%
      \newlength{\gptboxwidth}%
      \newsavebox{\gptboxtext}%
    \fi%
    \setlength{\fboxrule}{0.5pt}%
    \setlength{\fboxsep}{1pt}%
\begin{picture}(8640.00,5760.00)%
    \gplgaddtomacro\gplbacktext{%
      \csname LTb\endcsname
      \put(132,874){\makebox(0,0)[r]{\strut{}$-2$}}%
      \put(132,1348){\makebox(0,0)[r]{\strut{}$-1.5$}}%
      \put(132,1822){\makebox(0,0)[r]{\strut{}$-1$}}%
      \put(132,2296){\makebox(0,0)[r]{\strut{}$-0.5$}}%
      \put(132,2770){\makebox(0,0)[r]{\strut{}$0$}}%
      \put(132,3243){\makebox(0,0)[r]{\strut{}$0.5$}}%
      \put(132,3717){\makebox(0,0)[r]{\strut{}$1$}}%
      \put(132,4191){\makebox(0,0)[r]{\strut{}$1.5$}}%
      \put(132,4665){\makebox(0,0)[r]{\strut{}$2$}}%
      \put(264,654){\makebox(0,0){\strut{}$-2$}}%
      \put(738,654){\makebox(0,0){\strut{}$-1.5$}}%
      \put(1212,654){\makebox(0,0){\strut{}$-1$}}%
      \put(1686,654){\makebox(0,0){\strut{}$-0.5$}}%
      \put(2160,654){\makebox(0,0){\strut{}$0$}}%
      \put(2633,654){\makebox(0,0){\strut{}$0.5$}}%
      \put(3107,654){\makebox(0,0){\strut{}$1$}}%
      \put(3581,654){\makebox(0,0){\strut{}$1.5$}}%
      \put(4055,654){\makebox(0,0){\strut{}$2$}}%
      \put(3676,4286){\makebox(0,0)[l]{\strut{}$e^-$}}%
    }%
    \gplgaddtomacro\gplfronttext{%
      \csname LTb\endcsname
      \put(2159,4995){\makebox(0,0){\strut{}Equatorial section $xy$}}%
    }%
    \gplgaddtomacro\gplbacktext{%
      \csname LTb\endcsname
      \put(4452,874){\makebox(0,0)[r]{\strut{}$-2$}}%
      \put(4452,1348){\makebox(0,0)[r]{\strut{}$-1.5$}}%
      \put(4452,1822){\makebox(0,0)[r]{\strut{}$-1$}}%
      \put(4452,2296){\makebox(0,0)[r]{\strut{}$-0.5$}}%
      \put(4452,2770){\makebox(0,0)[r]{\strut{}$0$}}%
      \put(4452,3243){\makebox(0,0)[r]{\strut{}$0.5$}}%
      \put(4452,3717){\makebox(0,0)[r]{\strut{}$1$}}%
      \put(4452,4191){\makebox(0,0)[r]{\strut{}$1.5$}}%
      \put(4452,4665){\makebox(0,0)[r]{\strut{}$2$}}%
      \put(4584,654){\makebox(0,0){\strut{}$-2$}}%
      \put(5058,654){\makebox(0,0){\strut{}$-1.5$}}%
      \put(5532,654){\makebox(0,0){\strut{}$-1$}}%
      \put(6006,654){\makebox(0,0){\strut{}$-0.5$}}%
      \put(6480,654){\makebox(0,0){\strut{}$0$}}%
      \put(6953,654){\makebox(0,0){\strut{}$0.5$}}%
      \put(7427,654){\makebox(0,0){\strut{}$1$}}%
      \put(7901,654){\makebox(0,0){\strut{}$1.5$}}%
      \put(8375,654){\makebox(0,0){\strut{}$2$}}%
      \put(7996,4286){\makebox(0,0)[l]{\strut{}$e^-$}}%
    }%
    \gplgaddtomacro\gplfronttext{%
      \csname LTb\endcsname
      \put(6479,4995){\makebox(0,0){\strut{}Meridional section $yz$}}%
    }%
    \gplbacktext
    \put(0,0){\includegraphics{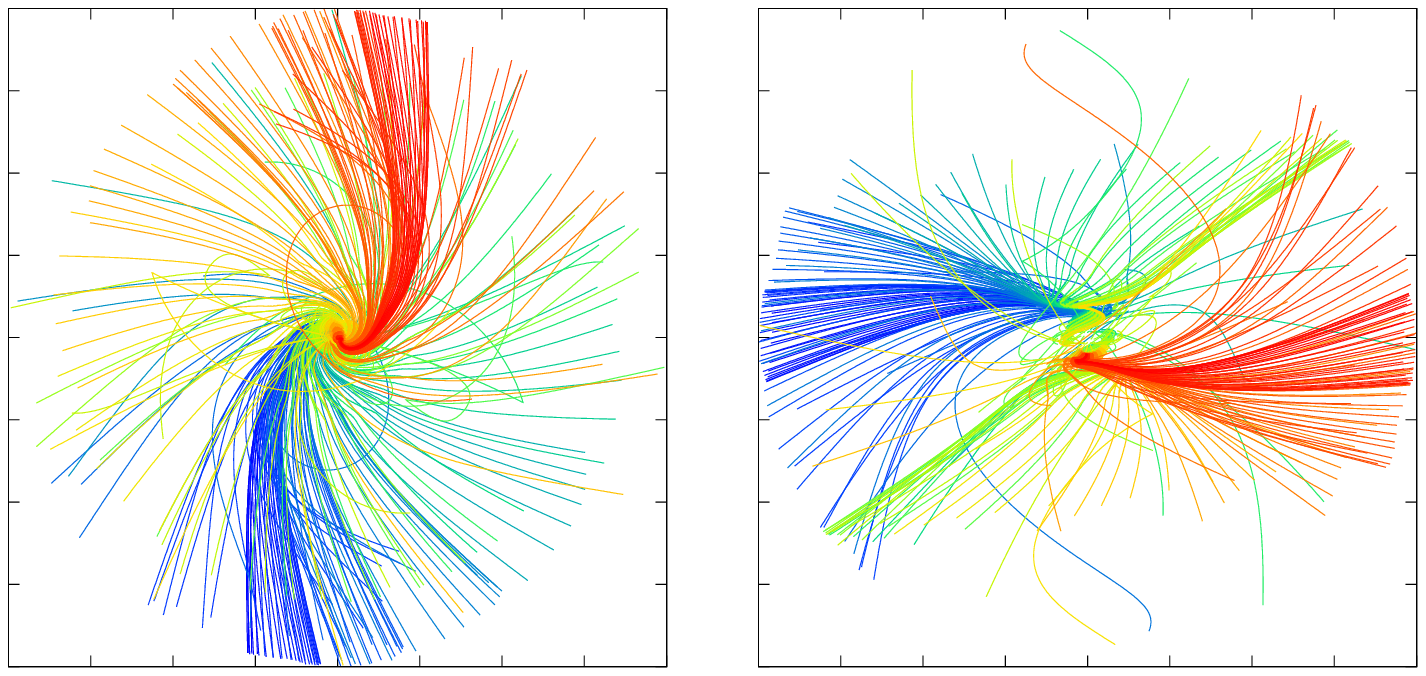}}%
    \gplfronttext
  \end{picture}%
\endgroup

%% file: carte_fuite_cx_onde.tex
\begingroup
  \makeatletter
  \providecommand\color[2][]{%
    \GenericError{(gnuplot) \space\space\space\@spaces}{%
      Package color not loaded in conjunction with
      terminal option `colourtext'%
    }{See the gnuplot documentation for explanation.%
    }{Either use 'blacktext' in gnuplot or load the package
      color.sty in LaTeX.}%
    \renewcommand\color[2][]{}%
  }%
  \providecommand\includegraphics[2][]{%
    \GenericError{(gnuplot) \space\space\space\@spaces}{%
      Package graphicx or graphics not loaded%
    }{See the gnuplot documentation for explanation.%
    }{The gnuplot epslatex terminal needs graphicx.sty or graphics.sty.}%
    \renewcommand\includegraphics[2][]{}%
  }%
  \providecommand\rotatebox[2]{#2}%
  \@ifundefined{ifGPcolor}{%
    \newif\ifGPcolor
    \GPcolortrue
  }{}%
  \@ifundefined{ifGPblacktext}{%
    \newif\ifGPblacktext
    \GPblacktextfalse
  }{}%
  \let\gplgaddtomacro\g@addto@macro
  \gdef\gplbacktext{}%
  \gdef\gplfronttext{}%
  \makeatother
  \ifGPblacktext
    \def\colorrgb#1{}%
    \def\colorgray#1{}%
  \else
    \ifGPcolor
      \def\colorrgb#1{\color[rgb]{#1}}%
      \def\colorgray#1{\color[gray]{#1}}%
      \expandafter\def\csname LTw\endcsname{\color{white}}%
      \expandafter\def\csname LTb\endcsname{\color{black}}%
      \expandafter\def\csname LTa\endcsname{\color{black}}%
      \expandafter\def\csname LT0\endcsname{\color[rgb]{1,0,0}}%
      \expandafter\def\csname LT1\endcsname{\color[rgb]{0,1,0}}%
      \expandafter\def\csname LT2\endcsname{\color[rgb]{0,0,1}}%
      \expandafter\def\csname LT3\endcsname{\color[rgb]{1,0,1}}%
      \expandafter\def\csname LT4\endcsname{\color[rgb]{0,1,1}}%
      \expandafter\def\csname LT5\endcsname{\color[rgb]{1,1,0}}%
      \expandafter\def\csname LT6\endcsname{\color[rgb]{0,0,0}}%
      \expandafter\def\csname LT7\endcsname{\color[rgb]{1,0.3,0}}%
      \expandafter\def\csname LT8\endcsname{\color[rgb]{0.5,0.5,0.5}}%
    \else
      \def\colorrgb#1{\color{black}}%
      \def\colorgray#1{\color[gray]{#1}}%
      \expandafter\def\csname LTw\endcsname{\color{white}}%
      \expandafter\def\csname LTb\endcsname{\color{black}}%
      \expandafter\def\csname LTa\endcsname{\color{black}}%
      \expandafter\def\csname LT0\endcsname{\color{black}}%
      \expandafter\def\csname LT1\endcsname{\color{black}}%
      \expandafter\def\csname LT2\endcsname{\color{black}}%
      \expandafter\def\csname LT3\endcsname{\color{black}}%
      \expandafter\def\csname LT4\endcsname{\color{black}}%
      \expandafter\def\csname LT5\endcsname{\color{black}}%
      \expandafter\def\csname LT6\endcsname{\color{black}}%
      \expandafter\def\csname LT7\endcsname{\color{black}}%
      \expandafter\def\csname LT8\endcsname{\color{black}}%
    \fi
  \fi
    \setlength{\unitlength}{0.0500bp}%
    \ifx\gptboxheight\undefined%
      \newlength{\gptboxheight}%
      \newlength{\gptboxwidth}%
      \newsavebox{\gptboxtext}%
    \fi%
    \setlength{\fboxrule}{0.5pt}%
    \setlength{\fboxsep}{1pt}%
\begin{picture}(8640.00,5760.00)%
    \gplgaddtomacro\gplbacktext{%
      \csname LTb\endcsname
      \put(0,852){\makebox(0,0)[r]{\strut{}$0$}}%
      \put(0,1528){\makebox(0,0)[r]{\strut{}$30$}}%
      \put(0,2204){\makebox(0,0)[r]{\strut{}$60$}}%
      \put(0,2880){\makebox(0,0)[r]{\strut{}$90$}}%
      \put(0,3555){\makebox(0,0)[r]{\strut{}$120$}}%
      \put(0,4231){\makebox(0,0)[r]{\strut{}$150$}}%
      \put(0,4907){\makebox(0,0)[r]{\strut{}$180$}}%
      \put(132,632){\makebox(0,0){\strut{}$0$}}%
      \put(808,632){\makebox(0,0){\strut{}$60$}}%
      \put(1484,632){\makebox(0,0){\strut{}$120$}}%
      \put(2160,632){\makebox(0,0){\strut{}$180$}}%
      \put(2835,632){\makebox(0,0){\strut{}$240$}}%
      \put(3511,632){\makebox(0,0){\strut{}$300$}}%
      \put(4187,632){\makebox(0,0){\strut{}$360$}}%
    }%
    \gplgaddtomacro\gplfronttext{%
      \csname LTb\endcsname
      \put(-616,2879){\rotatebox{-270}{\makebox(0,0){\strut{}$\theta$}}}%
      \put(2159,302){\makebox(0,0){\strut{}$\phi$}}%
      \put(2159,5237){\makebox(0,0){\strut{}$\chi=60$\degr, $e^-$}}%
    }%
    \gplgaddtomacro\gplbacktext{%
      \csname LTb\endcsname
      \put(4320,852){\makebox(0,0)[r]{\strut{} }}%
      \put(4320,1528){\makebox(0,0)[r]{\strut{} }}%
      \put(4320,2204){\makebox(0,0)[r]{\strut{} }}%
      \put(4320,2880){\makebox(0,0)[r]{\strut{} }}%
      \put(4320,3555){\makebox(0,0)[r]{\strut{} }}%
      \put(4320,4231){\makebox(0,0)[r]{\strut{} }}%
      \put(4320,4907){\makebox(0,0)[r]{\strut{} }}%
      \put(4452,632){\makebox(0,0){\strut{}$0$}}%
      \put(5128,632){\makebox(0,0){\strut{}$60$}}%
      \put(5804,632){\makebox(0,0){\strut{}$120$}}%
      \put(6480,632){\makebox(0,0){\strut{}$180$}}%
      \put(7155,632){\makebox(0,0){\strut{}$240$}}%
      \put(7831,632){\makebox(0,0){\strut{}$300$}}%
      \put(8507,632){\makebox(0,0){\strut{}$360$}}%
    }%
    \gplgaddtomacro\gplfronttext{%
      \csname LTb\endcsname
      \put(3968,2879){\rotatebox{-270}{\makebox(0,0){\strut{}$\theta$}}}%
      \put(6479,302){\makebox(0,0){\strut{}$\phi$}}%
      \put(6479,5237){\makebox(0,0){\strut{}$\chi=120$\degr, $e^+$}}%
    }%
    \gplbacktext
    \put(0,0){\includegraphics{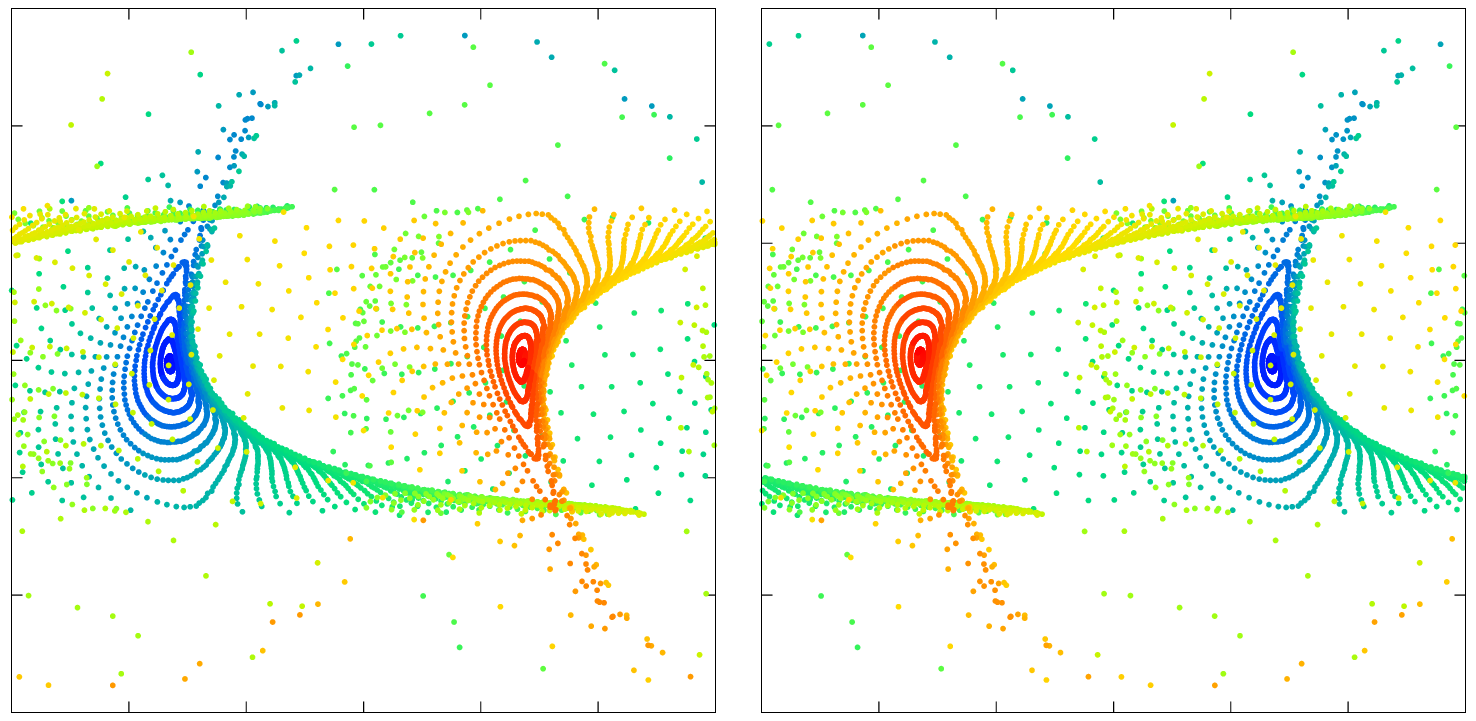}}%
    \gplfronttext
  \end{picture}%
\endgroup

%% file: invariant_c60.tex
\begingroup
  \makeatletter
  \providecommand\color[2][]{%
    \GenericError{(gnuplot) \space\space\space\@spaces}{%
      Package color not loaded in conjunction with
      terminal option `colourtext'%
    }{See the gnuplot documentation for explanation.%
    }{Either use 'blacktext' in gnuplot or load the package
      color.sty in LaTeX.}%
    \renewcommand\color[2][]{}%
  }%
  \providecommand\includegraphics[2][]{%
    \GenericError{(gnuplot) \space\space\space\@spaces}{%
      Package graphicx or graphics not loaded%
    }{See the gnuplot documentation for explanation.%
    }{The gnuplot epslatex terminal needs graphicx.sty or graphics.sty.}%
    \renewcommand\includegraphics[2][]{}%
  }%
  \providecommand\rotatebox[2]{#2}%
  \@ifundefined{ifGPcolor}{%
    \newif\ifGPcolor
    \GPcolortrue
  }{}%
  \@ifundefined{ifGPblacktext}{%
    \newif\ifGPblacktext
    \GPblacktextfalse
  }{}%
  \let\gplgaddtomacro\g@addto@macro
  \gdef\gplbacktext{}%
  \gdef\gplfronttext{}%
  \makeatother
  \ifGPblacktext
    \def\colorrgb#1{}%
    \def\colorgray#1{}%
  \else
    \ifGPcolor
      \def\colorrgb#1{\color[rgb]{#1}}%
      \def\colorgray#1{\color[gray]{#1}}%
      \expandafter\def\csname LTw\endcsname{\color{white}}%
      \expandafter\def\csname LTb\endcsname{\color{black}}%
      \expandafter\def\csname LTa\endcsname{\color{black}}%
      \expandafter\def\csname LT0\endcsname{\color[rgb]{1,0,0}}%
      \expandafter\def\csname LT1\endcsname{\color[rgb]{0,1,0}}%
      \expandafter\def\csname LT2\endcsname{\color[rgb]{0,0,1}}%
      \expandafter\def\csname LT3\endcsname{\color[rgb]{1,0,1}}%
      \expandafter\def\csname LT4\endcsname{\color[rgb]{0,1,1}}%
      \expandafter\def\csname LT5\endcsname{\color[rgb]{1,1,0}}%
      \expandafter\def\csname LT6\endcsname{\color[rgb]{0,0,0}}%
      \expandafter\def\csname LT7\endcsname{\color[rgb]{1,0.3,0}}%
      \expandafter\def\csname LT8\endcsname{\color[rgb]{0.5,0.5,0.5}}%
    \else
      \def\colorrgb#1{\color{black}}%
      \def\colorgray#1{\color[gray]{#1}}%
      \expandafter\def\csname LTw\endcsname{\color{white}}%
      \expandafter\def\csname LTb\endcsname{\color{black}}%
      \expandafter\def\csname LTa\endcsname{\color{black}}%
      \expandafter\def\csname LT0\endcsname{\color{black}}%
      \expandafter\def\csname LT1\endcsname{\color{black}}%
      \expandafter\def\csname LT2\endcsname{\color{black}}%
      \expandafter\def\csname LT3\endcsname{\color{black}}%
      \expandafter\def\csname LT4\endcsname{\color{black}}%
      \expandafter\def\csname LT5\endcsname{\color{black}}%
      \expandafter\def\csname LT6\endcsname{\color{black}}%
      \expandafter\def\csname LT7\endcsname{\color{black}}%
      \expandafter\def\csname LT8\endcsname{\color{black}}%
    \fi
  \fi
    \setlength{\unitlength}{0.0500bp}%
    \ifx\gptboxheight\undefined%
      \newlength{\gptboxheight}%
      \newlength{\gptboxwidth}%
      \newsavebox{\gptboxtext}%
    \fi%
    \setlength{\fboxrule}{0.5pt}%
    \setlength{\fboxsep}{1pt}%
\begin{picture}(5760.00,8640.00)%
    \gplgaddtomacro\gplbacktext{%
      \csname LTb\endcsname%
      \put(0,6221){\makebox(0,0)[r]{\strut{}-2}}%
      \put(0,6465){\makebox(0,0)[r]{\strut{}}}%
      \put(0,6710){\makebox(0,0)[r]{\strut{}-1}}%
      \put(0,6955){\makebox(0,0)[r]{\strut{}}}%
      \put(0,7200){\makebox(0,0)[r]{\strut{}0}}%
      \put(0,7444){\makebox(0,0)[r]{\strut{}}}%
      \put(0,7689){\makebox(0,0)[r]{\strut{}1}}%
      \put(0,7934){\makebox(0,0)[r]{\strut{}}}%
      \put(0,8178){\makebox(0,0)[r]{\strut{}2}}%
      \put(132,6001){\makebox(0,0){\strut{}-2}}%
      \put(376,6001){\makebox(0,0){\strut{}}}%
      \put(621,6001){\makebox(0,0){\strut{}-1}}%
      \put(866,6001){\makebox(0,0){\strut{}}}%
      \put(1111,6001){\makebox(0,0){\strut{}0}}%
      \put(1355,6001){\makebox(0,0){\strut{}}}%
      \put(1600,6001){\makebox(0,0){\strut{}1}}%
      \put(1845,6001){\makebox(0,0){\strut{}}}%
      \put(2089,6001){\makebox(0,0){\strut{}2}}%
    }%
    \gplgaddtomacro\gplfronttext{%
      \csname LTb\endcsname%
      \put(1110,8508){\makebox(0,0){\strut{}$\log(E_0)$}}%
      \csname LTb\endcsname%
      \put(2367,6221){\makebox(0,0)[l]{\strut{}-10}}%
      \put(2367,6500){\makebox(0,0)[l]{\strut{}-9}}%
      \put(2367,6780){\makebox(0,0)[l]{\strut{}-8}}%
      \put(2367,7059){\makebox(0,0)[l]{\strut{}-7}}%
      \put(2367,7339){\makebox(0,0)[l]{\strut{}-6}}%
      \put(2367,7618){\makebox(0,0)[l]{\strut{}-5}}%
      \put(2367,7898){\makebox(0,0)[l]{\strut{}-4}}%
      \put(2367,8178){\makebox(0,0)[l]{\strut{}-3}}%
    }%
    \gplgaddtomacro\gplbacktext{%
      \csname LTb\endcsname%
      \put(2880,6221){\makebox(0,0)[r]{\strut{}}}%
      \put(2880,6465){\makebox(0,0)[r]{\strut{}}}%
      \put(2880,6710){\makebox(0,0)[r]{\strut{}}}%
      \put(2880,6955){\makebox(0,0)[r]{\strut{}}}%
      \put(2880,7200){\makebox(0,0)[r]{\strut{}}}%
      \put(2880,7444){\makebox(0,0)[r]{\strut{}}}%
      \put(2880,7689){\makebox(0,0)[r]{\strut{}}}%
      \put(2880,7934){\makebox(0,0)[r]{\strut{}}}%
      \put(2880,8178){\makebox(0,0)[r]{\strut{}}}%
      \put(3012,6001){\makebox(0,0){\strut{}-2}}%
      \put(3256,6001){\makebox(0,0){\strut{}}}%
      \put(3501,6001){\makebox(0,0){\strut{}-1}}%
      \put(3746,6001){\makebox(0,0){\strut{}}}%
      \put(3991,6001){\makebox(0,0){\strut{}0}}%
      \put(4235,6001){\makebox(0,0){\strut{}}}%
      \put(4480,6001){\makebox(0,0){\strut{}1}}%
      \put(4725,6001){\makebox(0,0){\strut{}}}%
      \put(4969,6001){\makebox(0,0){\strut{}2}}%
    }%
    \gplgaddtomacro\gplfronttext{%
      \csname LTb\endcsname%
      \put(3990,8508){\makebox(0,0){\strut{}$\log(B_0), -\log(-B_0)$}}%
      \csname LTb\endcsname%
      \put(5247,6221){\makebox(0,0)[l]{\strut{}-8}}%
      \put(5247,6465){\makebox(0,0)[l]{\strut{}-6}}%
      \put(5247,6710){\makebox(0,0)[l]{\strut{}-4}}%
      \put(5247,6954){\makebox(0,0)[l]{\strut{}-2}}%
      \put(5247,7199){\makebox(0,0)[l]{\strut{}0}}%
      \put(5247,7444){\makebox(0,0)[l]{\strut{}2}}%
      \put(5247,7688){\makebox(0,0)[l]{\strut{}4}}%
      \put(5247,7933){\makebox(0,0)[l]{\strut{}6}}%
      \put(5247,8178){\makebox(0,0)[l]{\strut{}8}}%
    }%
    \gplgaddtomacro\gplbacktext{%
      \csname LTb\endcsname%
      \put(0,3341){\makebox(0,0)[r]{\strut{}-2}}%
      \put(0,3585){\makebox(0,0)[r]{\strut{}}}%
      \put(0,3830){\makebox(0,0)[r]{\strut{}-1}}%
      \put(0,4075){\makebox(0,0)[r]{\strut{}}}%
      \put(0,4320){\makebox(0,0)[r]{\strut{}0}}%
      \put(0,4564){\makebox(0,0)[r]{\strut{}}}%
      \put(0,4809){\makebox(0,0)[r]{\strut{}1}}%
      \put(0,5054){\makebox(0,0)[r]{\strut{}}}%
      \put(0,5298){\makebox(0,0)[r]{\strut{}2}}%
      \put(132,3121){\makebox(0,0){\strut{}-2}}%
      \put(376,3121){\makebox(0,0){\strut{}}}%
      \put(621,3121){\makebox(0,0){\strut{}-1}}%
      \put(866,3121){\makebox(0,0){\strut{}}}%
      \put(1111,3121){\makebox(0,0){\strut{}0}}%
      \put(1355,3121){\makebox(0,0){\strut{}}}%
      \put(1600,3121){\makebox(0,0){\strut{}1}}%
      \put(1845,3121){\makebox(0,0){\strut{}}}%
      \put(2089,3121){\makebox(0,0){\strut{}2}}%
    }%
    \gplgaddtomacro\gplfronttext{%
      \csname LTb\endcsname%
      \put(1110,5628){\makebox(0,0){\strut{}$\log \rho_c(e^-)$}}%
      \csname LTb\endcsname%
      \put(2367,3341){\makebox(0,0)[l]{\strut{}-2}}%
      \put(2367,3536){\makebox(0,0)[l]{\strut{}-1.5}}%
      \put(2367,3732){\makebox(0,0)[l]{\strut{}-1}}%
      \put(2367,3928){\makebox(0,0)[l]{\strut{}-0.5}}%
      \put(2367,4123){\makebox(0,0)[l]{\strut{}0}}%
      \put(2367,4319){\makebox(0,0)[l]{\strut{}0.5}}%
      \put(2367,4515){\makebox(0,0)[l]{\strut{}1}}%
      \put(2367,4710){\makebox(0,0)[l]{\strut{}1.5}}%
      \put(2367,4906){\makebox(0,0)[l]{\strut{}2}}%
      \put(2367,5102){\makebox(0,0)[l]{\strut{}2.5}}%
      \put(2367,5298){\makebox(0,0)[l]{\strut{}3}}%
    }%
    \gplgaddtomacro\gplbacktext{%
      \csname LTb\endcsname%
      \put(2880,3341){\makebox(0,0)[r]{\strut{}}}%
      \put(2880,3585){\makebox(0,0)[r]{\strut{}}}%
      \put(2880,3830){\makebox(0,0)[r]{\strut{}}}%
      \put(2880,4075){\makebox(0,0)[r]{\strut{}}}%
      \put(2880,4320){\makebox(0,0)[r]{\strut{}}}%
      \put(2880,4564){\makebox(0,0)[r]{\strut{}}}%
      \put(2880,4809){\makebox(0,0)[r]{\strut{}}}%
      \put(2880,5054){\makebox(0,0)[r]{\strut{}}}%
      \put(2880,5298){\makebox(0,0)[r]{\strut{}}}%
      \put(3012,3121){\makebox(0,0){\strut{}-2}}%
      \put(3256,3121){\makebox(0,0){\strut{}}}%
      \put(3501,3121){\makebox(0,0){\strut{}-1}}%
      \put(3746,3121){\makebox(0,0){\strut{}}}%
      \put(3991,3121){\makebox(0,0){\strut{}0}}%
      \put(4235,3121){\makebox(0,0){\strut{}}}%
      \put(4480,3121){\makebox(0,0){\strut{}1}}%
      \put(4725,3121){\makebox(0,0){\strut{}}}%
      \put(4969,3121){\makebox(0,0){\strut{}2}}%
    }%
    \gplgaddtomacro\gplfronttext{%
      \csname LTb\endcsname%
      \put(3990,5628){\makebox(0,0){\strut{}$\log \rho_c(e^+)$}}%
      \csname LTb\endcsname%
      \put(5247,3341){\makebox(0,0)[l]{\strut{}-1.5}}%
      \put(5247,3585){\makebox(0,0)[l]{\strut{}-1}}%
      \put(5247,3830){\makebox(0,0)[l]{\strut{}-0.5}}%
      \put(5247,4074){\makebox(0,0)[l]{\strut{}0}}%
      \put(5247,4319){\makebox(0,0)[l]{\strut{}0.5}}%
      \put(5247,4564){\makebox(0,0)[l]{\strut{}1}}%
      \put(5247,4808){\makebox(0,0)[l]{\strut{}1.5}}%
      \put(5247,5053){\makebox(0,0)[l]{\strut{}2}}%
      \put(5247,5298){\makebox(0,0)[l]{\strut{}2.5}}%
    }%
    \gplgaddtomacro\gplbacktext{%
      \csname LTb\endcsname%
      \put(0,461){\makebox(0,0)[r]{\strut{}-2}}%
      \put(0,705){\makebox(0,0)[r]{\strut{}}}%
      \put(0,950){\makebox(0,0)[r]{\strut{}-1}}%
      \put(0,1195){\makebox(0,0)[r]{\strut{}}}%
      \put(0,1440){\makebox(0,0)[r]{\strut{}0}}%
      \put(0,1685){\makebox(0,0)[r]{\strut{}}}%
      \put(0,1930){\makebox(0,0)[r]{\strut{}1}}%
      \put(0,2175){\makebox(0,0)[r]{\strut{}}}%
      \put(0,2419){\makebox(0,0)[r]{\strut{}2}}%
      \put(132,241){\makebox(0,0){\strut{}-2}}%
      \put(376,241){\makebox(0,0){\strut{}}}%
      \put(621,241){\makebox(0,0){\strut{}-1}}%
      \put(866,241){\makebox(0,0){\strut{}}}%
      \put(1111,241){\makebox(0,0){\strut{}0}}%
      \put(1355,241){\makebox(0,0){\strut{}}}%
      \put(1600,241){\makebox(0,0){\strut{}1}}%
      \put(1845,241){\makebox(0,0){\strut{}}}%
      \put(2089,241){\makebox(0,0){\strut{}2}}%
    }%
    \gplgaddtomacro\gplfronttext{%
      \csname LTb\endcsname%
      \put(1110,2749){\makebox(0,0){\strut{}$\log \gamma(e^-)$}}%
      \csname LTb\endcsname%
      \put(2367,461){\makebox(0,0)[l]{\strut{}6.8}}%
      \put(2367,656){\makebox(0,0)[l]{\strut{}7}}%
      \put(2367,852){\makebox(0,0)[l]{\strut{}7.2}}%
      \put(2367,1048){\makebox(0,0)[l]{\strut{}7.4}}%
      \put(2367,1244){\makebox(0,0)[l]{\strut{}7.6}}%
      \put(2367,1440){\makebox(0,0)[l]{\strut{}7.8}}%
      \put(2367,1635){\makebox(0,0)[l]{\strut{}8}}%
      \put(2367,1831){\makebox(0,0)[l]{\strut{}8.2}}%
      \put(2367,2027){\makebox(0,0)[l]{\strut{}8.4}}%
      \put(2367,2223){\makebox(0,0)[l]{\strut{}8.6}}%
      \put(2367,2418){\makebox(0,0)[l]{\strut{}8.8}}%
    }%
    \gplgaddtomacro\gplbacktext{%
      \csname LTb\endcsname%
      \put(2880,461){\makebox(0,0)[r]{\strut{}}}%
      \put(2880,705){\makebox(0,0)[r]{\strut{}}}%
      \put(2880,950){\makebox(0,0)[r]{\strut{}}}%
      \put(2880,1195){\makebox(0,0)[r]{\strut{}}}%
      \put(2880,1440){\makebox(0,0)[r]{\strut{}}}%
      \put(2880,1685){\makebox(0,0)[r]{\strut{}}}%
      \put(2880,1930){\makebox(0,0)[r]{\strut{}}}%
      \put(2880,2175){\makebox(0,0)[r]{\strut{}}}%
      \put(2880,2419){\makebox(0,0)[r]{\strut{}}}%
      \put(3012,241){\makebox(0,0){\strut{}-2}}%
      \put(3256,241){\makebox(0,0){\strut{}}}%
      \put(3501,241){\makebox(0,0){\strut{}-1}}%
      \put(3746,241){\makebox(0,0){\strut{}}}%
      \put(3991,241){\makebox(0,0){\strut{}0}}%
      \put(4235,241){\makebox(0,0){\strut{}}}%
      \put(4480,241){\makebox(0,0){\strut{}1}}%
      \put(4725,241){\makebox(0,0){\strut{}}}%
      \put(4969,241){\makebox(0,0){\strut{}2}}%
    }%
    \gplgaddtomacro\gplfronttext{%
      \csname LTb\endcsname%
      \put(3990,2749){\makebox(0,0){\strut{}$\log \gamma(e^+)$}}%
      \csname LTb\endcsname%
      \put(5247,461){\makebox(0,0)[l]{\strut{}6.5}}%
      \put(5247,787){\makebox(0,0)[l]{\strut{}7}}%
      \put(5247,1113){\makebox(0,0)[l]{\strut{}7.5}}%
      \put(5247,1440){\makebox(0,0)[l]{\strut{}8}}%
      \put(5247,1766){\makebox(0,0)[l]{\strut{}8.5}}%
      \put(5247,2092){\makebox(0,0)[l]{\strut{}9}}%
      \put(5247,2419){\makebox(0,0)[l]{\strut{}9.5}}%
    }%
    \gplbacktext
    \put(0,0){\includegraphics{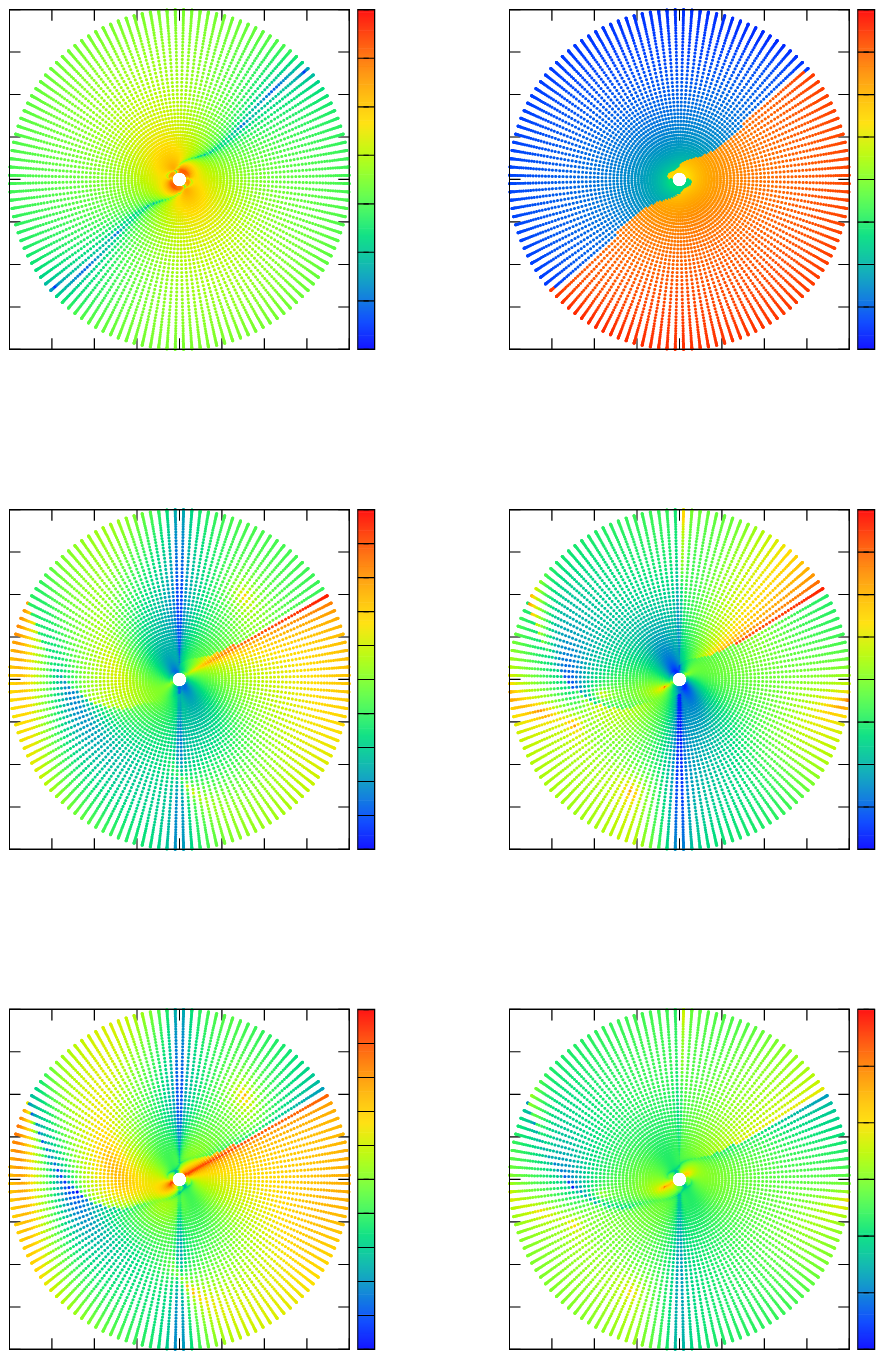}}%
    \gplfronttext
  \end{picture}%
\endgroup

%% file: spectre_moyen_r00025_b-3_q1_g10_dx.tex
\begingroup
  \makeatletter
  \providecommand\color[2][]{%
    \GenericError{(gnuplot) \space\space\space\@spaces}{%
      Package color not loaded in conjunction with
      terminal option `colourtext'%
    }{See the gnuplot documentation for explanation.%
    }{Either use 'blacktext' in gnuplot or load the package
      color.sty in LaTeX.}%
    \renewcommand\color[2][]{}%
  }%
  \providecommand\includegraphics[2][]{%
    \GenericError{(gnuplot) \space\space\space\@spaces}{%
      Package graphicx or graphics not loaded%
    }{See the gnuplot documentation for explanation.%
    }{The gnuplot epslatex terminal needs graphicx.sty or graphics.sty.}%
    \renewcommand\includegraphics[2][]{}%
  }%
  \providecommand\rotatebox[2]{#2}%
  \@ifundefined{ifGPcolor}{%
    \newif\ifGPcolor
    \GPcolortrue
  }{}%
  \@ifundefined{ifGPblacktext}{%
    \newif\ifGPblacktext
    \GPblacktextfalse
  }{}%
  \let\gplgaddtomacro\g@addto@macro
  \gdef\gplbacktext{}%
  \gdef\gplfronttext{}%
  \makeatother
  \ifGPblacktext
    \def\colorrgb#1{}%
    \def\colorgray#1{}%
  \else
    \ifGPcolor
      \def\colorrgb#1{\color[rgb]{#1}}%
      \def\colorgray#1{\color[gray]{#1}}%
      \expandafter\def\csname LTw\endcsname{\color{white}}%
      \expandafter\def\csname LTb\endcsname{\color{black}}%
      \expandafter\def\csname LTa\endcsname{\color{black}}%
      \expandafter\def\csname LT0\endcsname{\color[rgb]{1,0,0}}%
      \expandafter\def\csname LT1\endcsname{\color[rgb]{0,1,0}}%
      \expandafter\def\csname LT2\endcsname{\color[rgb]{0,0,1}}%
      \expandafter\def\csname LT3\endcsname{\color[rgb]{1,0,1}}%
      \expandafter\def\csname LT4\endcsname{\color[rgb]{0,1,1}}%
      \expandafter\def\csname LT5\endcsname{\color[rgb]{1,1,0}}%
      \expandafter\def\csname LT6\endcsname{\color[rgb]{0,0,0}}%
      \expandafter\def\csname LT7\endcsname{\color[rgb]{1,0.3,0}}%
      \expandafter\def\csname LT8\endcsname{\color[rgb]{0.5,0.5,0.5}}%
    \else
      \def\colorrgb#1{\color{black}}%
      \def\colorgray#1{\color[gray]{#1}}%
      \expandafter\def\csname LTw\endcsname{\color{white}}%
      \expandafter\def\csname LTb\endcsname{\color{black}}%
      \expandafter\def\csname LTa\endcsname{\color{black}}%
      \expandafter\def\csname LT0\endcsname{\color{black}}%
      \expandafter\def\csname LT1\endcsname{\color{black}}%
      \expandafter\def\csname LT2\endcsname{\color{black}}%
      \expandafter\def\csname LT3\endcsname{\color{black}}%
      \expandafter\def\csname LT4\endcsname{\color{black}}%
      \expandafter\def\csname LT5\endcsname{\color{black}}%
      \expandafter\def\csname LT6\endcsname{\color{black}}%
      \expandafter\def\csname LT7\endcsname{\color{black}}%
      \expandafter\def\csname LT8\endcsname{\color{black}}%
    \fi
  \fi
    \setlength{\unitlength}{0.0500bp}%
    \ifx\gptboxheight\undefined%
      \newlength{\gptboxheight}%
      \newlength{\gptboxwidth}%
      \newsavebox{\gptboxtext}%
    \fi%
    \setlength{\fboxrule}{0.5pt}%
    \setlength{\fboxsep}{1pt}%
\begin{picture}(5760.00,4320.00)%
    \gplgaddtomacro\gplbacktext{%
      \csname LTb\endcsname
      \put(814,704){\makebox(0,0)[r]{\strut{}$-20$}}%
      \put(814,1270){\makebox(0,0)[r]{\strut{}$-18$}}%
      \put(814,1836){\makebox(0,0)[r]{\strut{}$-16$}}%
      \put(814,2402){\makebox(0,0)[r]{\strut{}$-14$}}%
      \put(814,2967){\makebox(0,0)[r]{\strut{}$-12$}}%
      \put(814,3533){\makebox(0,0)[r]{\strut{}$-10$}}%
      \put(814,4099){\makebox(0,0)[r]{\strut{}$-8$}}%
      \put(946,484){\makebox(0,0){\strut{}$0$}}%
      \put(1682,484){\makebox(0,0){\strut{}$1$}}%
      \put(2418,484){\makebox(0,0){\strut{}$2$}}%
      \put(3155,484){\makebox(0,0){\strut{}$3$}}%
      \put(3891,484){\makebox(0,0){\strut{}$4$}}%
      \put(4627,484){\makebox(0,0){\strut{}$5$}}%
      \put(5363,484){\makebox(0,0){\strut{}$6$}}%
    }%
    \gplgaddtomacro\gplfronttext{%
      \csname LTb\endcsname
      \put(198,2401){\rotatebox{-270}{\makebox(0,0){\strut{}$\log(E^2 \, dN/dE\,dt) ~ (W/m^2)$}}}%
      \put(3154,154){\makebox(0,0){\strut{}$\log(E/\textrm{MeV})$}}%
      \csname LTb\endcsname
      \put(3321,1977){\makebox(0,0)[r]{\strut{}$r_{\rm in}=0.1$}}%
      \csname LTb\endcsname
      \put(3321,1757){\makebox(0,0)[r]{\strut{}0.2}}%
      \csname LTb\endcsname
      \put(3321,1537){\makebox(0,0)[r]{\strut{}0.5}}%
      \csname LTb\endcsname
      \put(3321,1317){\makebox(0,0)[r]{\strut{}q=1}}%
      \csname LTb\endcsname
      \put(3321,1097){\makebox(0,0)[r]{\strut{}2}}%
      \csname LTb\endcsname
      \put(3321,877){\makebox(0,0)[r]{\strut{}3}}%
    }%
    \gplbacktext
    \put(0,0){\includegraphics{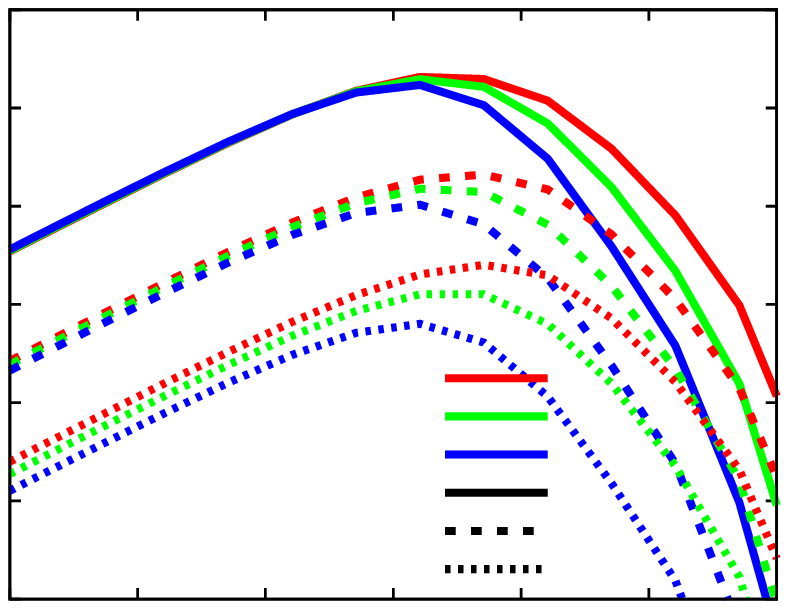}}%
    \gplfronttext
  \end{picture}%
\endgroup

%% file: skymaps_cl_r00025_b-3_c60_ri0.5_ro1_q1_g10_d3.tex
\begingroup
  \makeatletter
  \providecommand\color[2][]{%
    \GenericError{(gnuplot) \space\space\space\@spaces}{%
      Package color not loaded in conjunction with
      terminal option `colourtext'%
    }{See the gnuplot documentation for explanation.%
    }{Either use 'blacktext' in gnuplot or load the package
      color.sty in LaTeX.}%
    \renewcommand\color[2][]{}%
  }%
  \providecommand\includegraphics[2][]{%
    \GenericError{(gnuplot) \space\space\space\@spaces}{%
      Package graphicx or graphics not loaded%
    }{See the gnuplot documentation for explanation.%
    }{The gnuplot epslatex terminal needs graphicx.sty or graphics.sty.}%
    \renewcommand\includegraphics[2][]{}%
  }%
  \providecommand\rotatebox[2]{#2}%
  \@ifundefined{ifGPcolor}{%
    \newif\ifGPcolor
    \GPcolortrue
  }{}%
  \@ifundefined{ifGPblacktext}{%
    \newif\ifGPblacktext
    \GPblacktextfalse
  }{}%
  \let\gplgaddtomacro\g@addto@macro
  \gdef\gplbacktext{}%
  \gdef\gplfronttext{}%
  \makeatother
  \ifGPblacktext
    \def\colorrgb#1{}%
    \def\colorgray#1{}%
  \else
    \ifGPcolor
      \def\colorrgb#1{\color[rgb]{#1}}%
      \def\colorgray#1{\color[gray]{#1}}%
      \expandafter\def\csname LTw\endcsname{\color{white}}%
      \expandafter\def\csname LTb\endcsname{\color{black}}%
      \expandafter\def\csname LTa\endcsname{\color{black}}%
      \expandafter\def\csname LT0\endcsname{\color[rgb]{1,0,0}}%
      \expandafter\def\csname LT1\endcsname{\color[rgb]{0,1,0}}%
      \expandafter\def\csname LT2\endcsname{\color[rgb]{0,0,1}}%
      \expandafter\def\csname LT3\endcsname{\color[rgb]{1,0,1}}%
      \expandafter\def\csname LT4\endcsname{\color[rgb]{0,1,1}}%
      \expandafter\def\csname LT5\endcsname{\color[rgb]{1,1,0}}%
      \expandafter\def\csname LT6\endcsname{\color[rgb]{0,0,0}}%
      \expandafter\def\csname LT7\endcsname{\color[rgb]{1,0.3,0}}%
      \expandafter\def\csname LT8\endcsname{\color[rgb]{0.5,0.5,0.5}}%
    \else
      \def\colorrgb#1{\color{black}}%
      \def\colorgray#1{\color[gray]{#1}}%
      \expandafter\def\csname LTw\endcsname{\color{white}}%
      \expandafter\def\csname LTb\endcsname{\color{black}}%
      \expandafter\def\csname LTa\endcsname{\color{black}}%
      \expandafter\def\csname LT0\endcsname{\color{black}}%
      \expandafter\def\csname LT1\endcsname{\color{black}}%
      \expandafter\def\csname LT2\endcsname{\color{black}}%
      \expandafter\def\csname LT3\endcsname{\color{black}}%
      \expandafter\def\csname LT4\endcsname{\color{black}}%
      \expandafter\def\csname LT5\endcsname{\color{black}}%
      \expandafter\def\csname LT6\endcsname{\color{black}}%
      \expandafter\def\csname LT7\endcsname{\color{black}}%
      \expandafter\def\csname LT8\endcsname{\color{black}}%
    \fi
  \fi
    \setlength{\unitlength}{0.0500bp}%
    \ifx\gptboxheight\undefined%
      \newlength{\gptboxheight}%
      \newlength{\gptboxwidth}%
      \newsavebox{\gptboxtext}%
    \fi%
    \setlength{\fboxrule}{0.5pt}%
    \setlength{\fboxsep}{1pt}%
\begin{picture}(11520.00,5760.00)%
    \gplgaddtomacro\gplbacktext{%
    }%
    \gplgaddtomacro\gplfronttext{%
      \csname LTb\endcsname
      \put(577,2662){\makebox(0,0){\strut{}}}%
      \put(1297,2662){\makebox(0,0){\strut{}}}%
      \put(2016,2662){\makebox(0,0){\strut{}}}%
      \put(2735,2662){\makebox(0,0){\strut{}}}%
      \put(3455,2662){\makebox(0,0){\strut{}}}%
      \put(364,2939){\makebox(0,0)[r]{\strut{}}}%
      \put(364,3313){\makebox(0,0)[r]{\strut{}30}}%
      \put(364,3687){\makebox(0,0)[r]{\strut{}60}}%
      \put(364,4061){\makebox(0,0)[r]{\strut{}90}}%
      \put(364,4435){\makebox(0,0)[r]{\strut{}120}}%
      \put(364,4809){\makebox(0,0)[r]{\strut{}150}}%
      \put(-98,4061){\rotatebox{-270}{\makebox(0,0){\strut{}line of sight ($\zeta$)}}}%
      \put(3803,2939){\makebox(0,0)[l]{\strut{}0.5}}%
      \put(3803,3313){\makebox(0,0)[l]{\strut{} 1}}%
      \put(3803,3687){\makebox(0,0)[l]{\strut{} 2}}%
      \put(3803,4061){\makebox(0,0)[l]{\strut{} 2}}%
      \put(3803,4435){\makebox(0,0)[l]{\strut{} 2}}%
      \put(3803,4809){\makebox(0,0)[l]{\strut{} 3}}%
      \put(3803,5183){\makebox(0,0)[l]{\strut{} 4}}%
      \put(721,4959){\makebox(0,0)[l]{\strut{}161~MeV}}%
    }%
    \gplgaddtomacro\gplbacktext{%
    }%
    \gplgaddtomacro\gplfronttext{%
      \csname LTb\endcsname
      \put(4263,2662){\makebox(0,0){\strut{}}}%
      \put(4983,2662){\makebox(0,0){\strut{}}}%
      \put(5702,2662){\makebox(0,0){\strut{}}}%
      \put(6421,2662){\makebox(0,0){\strut{}}}%
      \put(7141,2662){\makebox(0,0){\strut{}}}%
      \put(4050,2939){\makebox(0,0)[r]{\strut{}}}%
      \put(4050,3313){\makebox(0,0)[r]{\strut{}}}%
      \put(4050,3687){\makebox(0,0)[r]{\strut{}}}%
      \put(4050,4061){\makebox(0,0)[r]{\strut{}}}%
      \put(4050,4435){\makebox(0,0)[r]{\strut{}}}%
      \put(4050,4809){\makebox(0,0)[r]{\strut{}}}%
      \put(7489,2939){\makebox(0,0)[l]{\strut{} 0}}%
      \put(7489,3259){\makebox(0,0)[l]{\strut{}0.5}}%
      \put(7489,3580){\makebox(0,0)[l]{\strut{} 1}}%
      \put(7489,3900){\makebox(0,0)[l]{\strut{} 2}}%
      \put(7489,4221){\makebox(0,0)[l]{\strut{} 2}}%
      \put(7489,4541){\makebox(0,0)[l]{\strut{} 2}}%
      \put(7489,4862){\makebox(0,0)[l]{\strut{} 3}}%
      \put(7489,5183){\makebox(0,0)[l]{\strut{} 4}}%
      \put(4407,4959){\makebox(0,0)[l]{\strut{}511~MeV}}%
    }%
    \gplgaddtomacro\gplbacktext{%
    }%
    \gplgaddtomacro\gplfronttext{%
      \csname LTb\endcsname
      \put(7950,2662){\makebox(0,0){\strut{}}}%
      \put(8670,2662){\makebox(0,0){\strut{}}}%
      \put(9389,2662){\makebox(0,0){\strut{}}}%
      \put(10108,2662){\makebox(0,0){\strut{}}}%
      \put(10828,2662){\makebox(0,0){\strut{}}}%
      \put(7737,2939){\makebox(0,0)[r]{\strut{}}}%
      \put(7737,3313){\makebox(0,0)[r]{\strut{}}}%
      \put(7737,3687){\makebox(0,0)[r]{\strut{}}}%
      \put(7737,4061){\makebox(0,0)[r]{\strut{}}}%
      \put(7737,4435){\makebox(0,0)[r]{\strut{}}}%
      \put(7737,4809){\makebox(0,0)[r]{\strut{}}}%
      \put(11176,2939){\makebox(0,0)[l]{\strut{} 0}}%
      \put(11176,3188){\makebox(0,0)[l]{\strut{}0.2}}%
      \put(11176,3437){\makebox(0,0)[l]{\strut{}0.4}}%
      \put(11176,3687){\makebox(0,0)[l]{\strut{}0.6}}%
      \put(11176,3936){\makebox(0,0)[l]{\strut{}0.8}}%
      \put(11176,4185){\makebox(0,0)[l]{\strut{} 1}}%
      \put(11176,4435){\makebox(0,0)[l]{\strut{} 1}}%
      \put(11176,4684){\makebox(0,0)[l]{\strut{} 1}}%
      \put(11176,4933){\makebox(0,0)[l]{\strut{} 2}}%
      \put(11176,5183){\makebox(0,0)[l]{\strut{} 2}}%
      \put(8094,4959){\makebox(0,0)[l]{\strut{}1.61~GeV}}%
    }%
    \gplgaddtomacro\gplbacktext{%
    }%
    \gplgaddtomacro\gplfronttext{%
      \csname LTb\endcsname
      \put(577,484){\makebox(0,0){\strut{}}}%
      \put(1297,484){\makebox(0,0){\strut{}}}%
      \put(2016,484){\makebox(0,0){\strut{}}}%
      \put(2735,484){\makebox(0,0){\strut{}}}%
      \put(3455,484){\makebox(0,0){\strut{}}}%
      \put(355,750){\makebox(0,0)[r]{\strut{}}}%
      \put(355,1095){\makebox(0,0)[r]{\strut{}30}}%
      \put(355,1440){\makebox(0,0)[r]{\strut{}60}}%
      \put(355,1785){\makebox(0,0)[r]{\strut{}90}}%
      \put(355,2130){\makebox(0,0)[r]{\strut{}120}}%
      \put(355,2475){\makebox(0,0)[r]{\strut{}150}}%
      \put(-107,1785){\rotatebox{-270}{\makebox(0,0){\strut{}line of sight ($\zeta$)}}}%
      \put(3803,750){\makebox(0,0)[l]{\strut{} 0}}%
      \put(3803,1164){\makebox(0,0)[l]{\strut{}0.05}}%
      \put(3803,1578){\makebox(0,0)[l]{\strut{}0.1}}%
      \put(3803,1992){\makebox(0,0)[l]{\strut{}0.2}}%
      \put(3803,2406){\makebox(0,0)[l]{\strut{}0.2}}%
      \put(3803,2820){\makebox(0,0)[l]{\strut{}0.2}}%
      \put(721,2613){\makebox(0,0)[l]{\strut{}5.11~GeV}}%
    }%
    \gplgaddtomacro\gplbacktext{%
    }%
    \gplgaddtomacro\gplfronttext{%
      \csname LTb\endcsname
      \put(4263,484){\makebox(0,0){\strut{}}}%
      \put(4983,484){\makebox(0,0){\strut{}}}%
      \put(5702,484){\makebox(0,0){\strut{}}}%
      \put(6421,484){\makebox(0,0){\strut{}}}%
      \put(7141,484){\makebox(0,0){\strut{}}}%
      \put(4041,750){\makebox(0,0)[r]{\strut{}}}%
      \put(4041,1095){\makebox(0,0)[r]{\strut{}}}%
      \put(4041,1440){\makebox(0,0)[r]{\strut{}}}%
      \put(4041,1785){\makebox(0,0)[r]{\strut{}}}%
      \put(4041,2130){\makebox(0,0)[r]{\strut{}}}%
      \put(4041,2475){\makebox(0,0)[r]{\strut{}}}%
      \put(7489,750){\makebox(0,0)[l]{\strut{} 0}}%
      \put(7489,1095){\makebox(0,0)[l]{\strut{}0.001}}%
      \put(7489,1440){\makebox(0,0)[l]{\strut{}0.002}}%
      \put(7489,1785){\makebox(0,0)[l]{\strut{}0.003}}%
      \put(7489,2130){\makebox(0,0)[l]{\strut{}0.004}}%
      \put(7489,2475){\makebox(0,0)[l]{\strut{}0.005}}%
      \put(7489,2820){\makebox(0,0)[l]{\strut{}0.006}}%
      \put(4407,2613){\makebox(0,0)[l]{\strut{}16.1~GeV}}%
    }%
    \gplgaddtomacro\gplbacktext{%
    }%
    \gplgaddtomacro\gplfronttext{%
      \csname LTb\endcsname
      \put(7950,484){\makebox(0,0){\strut{}}}%
      \put(8670,484){\makebox(0,0){\strut{}}}%
      \put(9389,484){\makebox(0,0){\strut{}}}%
      \put(10108,484){\makebox(0,0){\strut{}}}%
      \put(10828,484){\makebox(0,0){\strut{}}}%
      \put(7728,750){\makebox(0,0)[r]{\strut{}}}%
      \put(7728,1095){\makebox(0,0)[r]{\strut{}}}%
      \put(7728,1440){\makebox(0,0)[r]{\strut{}}}%
      \put(7728,1785){\makebox(0,0)[r]{\strut{}}}%
      \put(7728,2130){\makebox(0,0)[r]{\strut{}}}%
      \put(7728,2475){\makebox(0,0)[r]{\strut{}}}%
      \put(11176,750){\makebox(0,0)[l]{\strut{} 0}}%
      \put(11176,980){\makebox(0,0)[l]{\strut{}2e-05}}%
      \put(11176,1210){\makebox(0,0)[l]{\strut{}4e-05}}%
      \put(11176,1440){\makebox(0,0)[l]{\strut{}6e-05}}%
      \put(11176,1670){\makebox(0,0)[l]{\strut{}8e-05}}%
      \put(11176,1900){\makebox(0,0)[l]{\strut{}0.0001}}%
      \put(11176,2130){\makebox(0,0)[l]{\strut{}0.0001}}%
      \put(11176,2360){\makebox(0,0)[l]{\strut{}0.0001}}%
      \put(11176,2590){\makebox(0,0)[l]{\strut{}0.0002}}%
      \put(11176,2820){\makebox(0,0)[l]{\strut{}0.0002}}%
      \put(8094,2613){\makebox(0,0)[l]{\strut{}51.1~GeV}}%
    }%
    \gplbacktext
    \put(0,0){\includegraphics{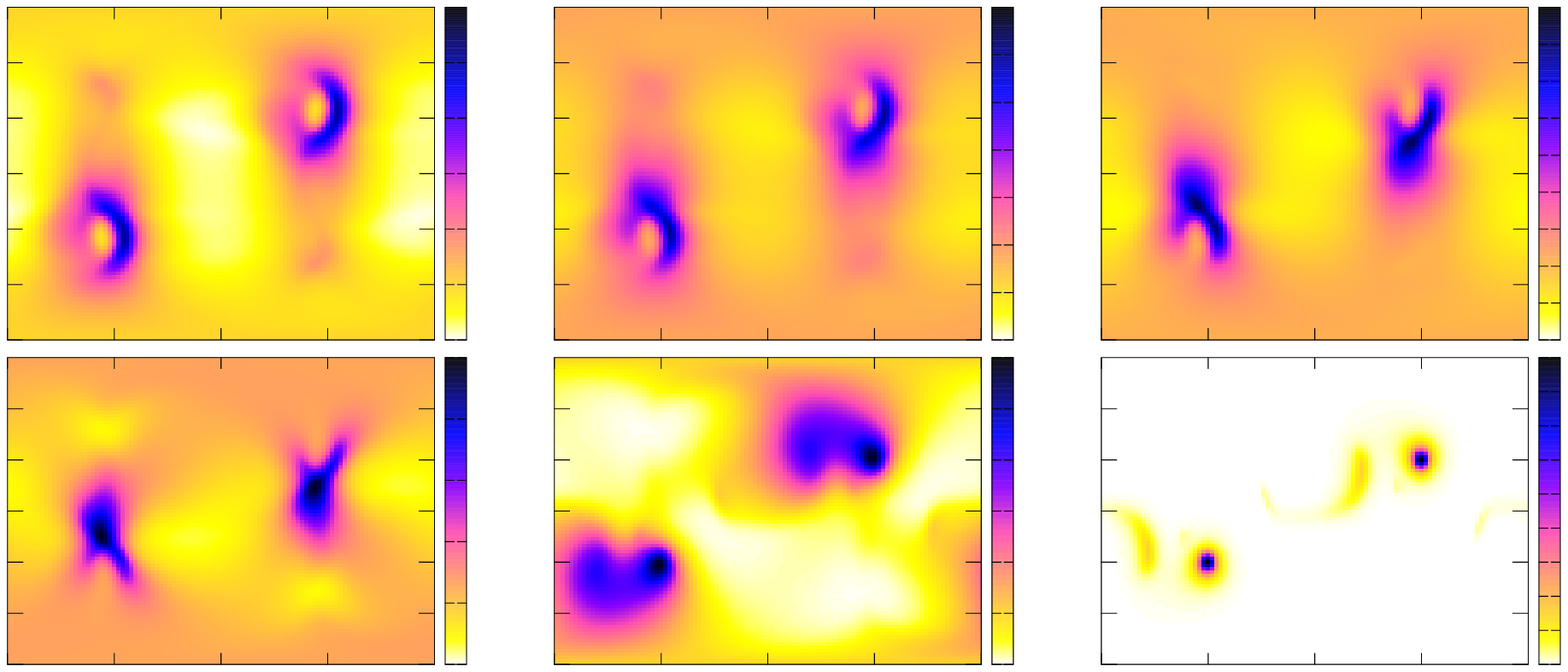}}%
    \gplfronttext
  \end{picture}%
\endgroup

%% file: spectre_moyen_r00025_b-3_q0_g10_d3_ro2.0.tex
\begingroup
  \makeatletter
  \providecommand\color[2][]{%
    \GenericError{(gnuplot) \space\space\space\@spaces}{%
      Package color not loaded in conjunction with
      terminal option `colourtext'%
    }{See the gnuplot documentation for explanation.%
    }{Either use 'blacktext' in gnuplot or load the package
      color.sty in LaTeX.}%
    \renewcommand\color[2][]{}%
  }%
  \providecommand\includegraphics[2][]{%
    \GenericError{(gnuplot) \space\space\space\@spaces}{%
      Package graphicx or graphics not loaded%
    }{See the gnuplot documentation for explanation.%
    }{The gnuplot epslatex terminal needs graphicx.sty or graphics.sty.}%
    \renewcommand\includegraphics[2][]{}%
  }%
  \providecommand\rotatebox[2]{#2}%
  \@ifundefined{ifGPcolor}{%
    \newif\ifGPcolor
    \GPcolortrue
  }{}%
  \@ifundefined{ifGPblacktext}{%
    \newif\ifGPblacktext
    \GPblacktextfalse
  }{}%
  \let\gplgaddtomacro\g@addto@macro
  \gdef\gplbacktext{}%
  \gdef\gplfronttext{}%
  \makeatother
  \ifGPblacktext
    \def\colorrgb#1{}%
    \def\colorgray#1{}%
  \else
    \ifGPcolor
      \def\colorrgb#1{\color[rgb]{#1}}%
      \def\colorgray#1{\color[gray]{#1}}%
      \expandafter\def\csname LTw\endcsname{\color{white}}%
      \expandafter\def\csname LTb\endcsname{\color{black}}%
      \expandafter\def\csname LTa\endcsname{\color{black}}%
      \expandafter\def\csname LT0\endcsname{\color[rgb]{1,0,0}}%
      \expandafter\def\csname LT1\endcsname{\color[rgb]{0,1,0}}%
      \expandafter\def\csname LT2\endcsname{\color[rgb]{0,0,1}}%
      \expandafter\def\csname LT3\endcsname{\color[rgb]{1,0,1}}%
      \expandafter\def\csname LT4\endcsname{\color[rgb]{0,1,1}}%
      \expandafter\def\csname LT5\endcsname{\color[rgb]{1,1,0}}%
      \expandafter\def\csname LT6\endcsname{\color[rgb]{0,0,0}}%
      \expandafter\def\csname LT7\endcsname{\color[rgb]{1,0.3,0}}%
      \expandafter\def\csname LT8\endcsname{\color[rgb]{0.5,0.5,0.5}}%
    \else
      \def\colorrgb#1{\color{black}}%
      \def\colorgray#1{\color[gray]{#1}}%
      \expandafter\def\csname LTw\endcsname{\color{white}}%
      \expandafter\def\csname LTb\endcsname{\color{black}}%
      \expandafter\def\csname LTa\endcsname{\color{black}}%
      \expandafter\def\csname LT0\endcsname{\color{black}}%
      \expandafter\def\csname LT1\endcsname{\color{black}}%
      \expandafter\def\csname LT2\endcsname{\color{black}}%
      \expandafter\def\csname LT3\endcsname{\color{black}}%
      \expandafter\def\csname LT4\endcsname{\color{black}}%
      \expandafter\def\csname LT5\endcsname{\color{black}}%
      \expandafter\def\csname LT6\endcsname{\color{black}}%
      \expandafter\def\csname LT7\endcsname{\color{black}}%
      \expandafter\def\csname LT8\endcsname{\color{black}}%
    \fi
  \fi
    \setlength{\unitlength}{0.0500bp}%
    \ifx\gptboxheight\undefined%
      \newlength{\gptboxheight}%
      \newlength{\gptboxwidth}%
      \newsavebox{\gptboxtext}%
    \fi%
    \setlength{\fboxrule}{0.5pt}%
    \setlength{\fboxsep}{1pt}%
\begin{picture}(5760.00,4320.00)%
    \gplgaddtomacro\gplbacktext{%
      \csname LTb\endcsname
      \put(814,704){\makebox(0,0)[r]{\strut{}$-21$}}%
      \put(814,1128){\makebox(0,0)[r]{\strut{}$-20$}}%
      \put(814,1553){\makebox(0,0)[r]{\strut{}$-19$}}%
      \put(814,1977){\makebox(0,0)[r]{\strut{}$-18$}}%
      \put(814,2402){\makebox(0,0)[r]{\strut{}$-17$}}%
      \put(814,2826){\makebox(0,0)[r]{\strut{}$-16$}}%
      \put(814,3250){\makebox(0,0)[r]{\strut{}$-15$}}%
      \put(814,3675){\makebox(0,0)[r]{\strut{}$-14$}}%
      \put(814,4099){\makebox(0,0)[r]{\strut{}$-13$}}%
      \put(946,484){\makebox(0,0){\strut{}$0$}}%
      \put(1682,484){\makebox(0,0){\strut{}$1$}}%
      \put(2418,484){\makebox(0,0){\strut{}$2$}}%
      \put(3155,484){\makebox(0,0){\strut{}$3$}}%
      \put(3891,484){\makebox(0,0){\strut{}$4$}}%
      \put(4627,484){\makebox(0,0){\strut{}$5$}}%
      \put(5363,484){\makebox(0,0){\strut{}$6$}}%
    }%
    \gplgaddtomacro\gplfronttext{%
      \csname LTb\endcsname
      \put(198,2401){\rotatebox{-270}{\makebox(0,0){\strut{}$\log(E^2 \, dN/dE\,dt) ~ (W/m^2)$}}}%
      \put(3154,154){\makebox(0,0){\strut{}$\log E  ~ (MeV)$}}%
      \put(1703,1757){\makebox(0,0){\strut{}$r_{\rm in}$}}%
      \csname LTb\endcsname
      \put(1474,1537){\makebox(0,0)[r]{\strut{}0.1}}%
      \csname LTb\endcsname
      \put(1474,1317){\makebox(0,0)[r]{\strut{}0.2}}%
      \csname LTb\endcsname
      \put(1474,1097){\makebox(0,0)[r]{\strut{}0.5}}%
      \csname LTb\endcsname
      \put(1474,877){\makebox(0,0)[r]{\strut{}1.0}}%
    }%
    \gplbacktext
    \put(0,0){\includegraphics{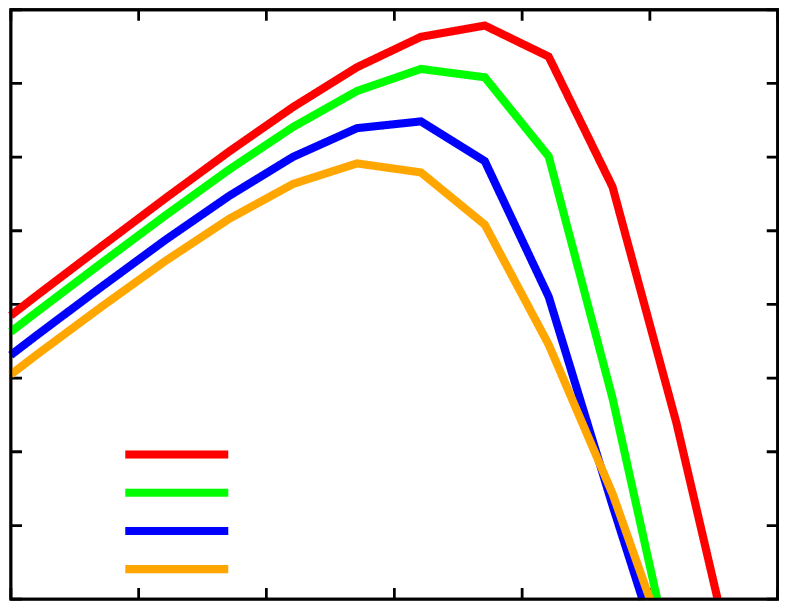}}%
    \gplfronttext
  \end{picture}%
\endgroup